\documentclass[prd,12pt,tightenlines,aps,letterpaper,amsmath,amssymb,preprintnumbers,showpacs,superscriptaddress,floatfix,longbibliography,nofootinbib]{revtex4-2}
\usepackage{graphicx}
\usepackage[export]{adjustbox}
\usepackage[caption=false]{subfig}
\usepackage{tikz}
\allowdisplaybreaks 
\usepackage[hypertexnames=true]{hyperref}
\hypersetup{
    colorlinks=true,       
    linkcolor=blue,          
    citecolor=blue,        
    filecolor=blue,      
    urlcolor=blue           
}

\usepackage{slashed}
\usepackage{mathtools}
\usepackage{listings}
\usepackage{pgf}
\usepackage{fancyvrb}
\usepackage{longtable}
\usepackage[normalem]{ulem}

\definecolor{listinggreen}{rgb}{0,0.6,0}
\definecolor{listinggray}{rgb}{0.5,0.5,0.5}
\definecolor{listingmauve}{rgb}{0.58,0,0.82}
\definecolor{listingkeywordcolor}{rgb}{1.0,0.4,0.0}
\definecolor{listinglightgray}{rgb}{0.8863,0.8863,0.8863}

\newcommand*\circled[1]{\tikz[baseline=(char.base)]{
            \node[shape=circle,draw,inner sep=1pt] (char) {\tiny #1};}}

\newcommand{\lsp}{\Lambda_{\cal S}}
\newcommand{\vsp}{V_{\cal S}}
\newcommand{\kappaOp}{Q}

\newcommand{\ket}[1]{\left| #1 \right>} 
\newcommand{\braket}[2]{\left< #1 \vphantom{#2} \right|
 \left. #2 \vphantom{#1} \right>} 
\newcommand{\mbraket}[3]{\left< #1 \vphantom{#2#3} \right|
 #2 \left| #3 \vphantom{#1#2} \right>} 

\newcommand{\Mod}[1]{\ (\mathrm{mod}\ #1)}

\begin{document}


\author{Saman~Amarasinghe} 
\affiliation{
	Computer Science and Artificial Intelligence Laboratory, 
	Massachusetts Institute of Technology, 
	Cambridge, MA 02139, USA}

\author{Riyadh~Baghdadi}
\affiliation{
	Computer Science and Artificial Intelligence Laboratory, 
	Massachusetts Institute of Technology, 
	Cambridge, MA 02139, USA}
\affiliation{New York University Abu Dhabi, PO Box 129188, Saadiyat Island, Abu Dhabi, United Arab Emirates}

\author{Zohreh~Davoudi}
\address{Department of Physics and Maryland Center for Fundamental Physics, University of Maryland, College Park, MD 20742, USA}

\author{William~Detmold} 
\affiliation{
	Center for Theoretical Physics, 
	Massachusetts Institute of Technology, 
	Cambridge, MA 02139, USA}
\affiliation{The NSF AI Institute for Artificial Intelligence and Fundamental Interactions}

\author{Marc~Illa}
\affiliation{Departament de F\'{\i}sica Qu\`{a}ntica i Astrof\'{\i}sica and Institut de Ci\`{e}ncies del Cosmos,	Universitat de Barcelona, Mart\'{\i} i Franqu\`es 1, E08028, Spain}

\author{Assumpta~Parre\~no}
\affiliation{Departament de F\'{\i}sica Qu\`{a}ntica i Astrof\'{\i}sica and Institut de Ci\`{e}ncies del Cosmos,	Universitat de Barcelona, Mart\'{\i} i Franqu\`es 1, E08028, Spain}

\author{Andrew~V.~Pochinsky}
\affiliation{
	Center for Theoretical Physics, 
	Massachusetts Institute of Technology, 
	Cambridge, MA 02139, USA}

\author{Phiala~E.~Shanahan } \affiliation{
	Center for Theoretical Physics, 
	Massachusetts Institute of Technology, 
	Cambridge, MA 02139, USA}
\affiliation{The NSF AI Institute for Artificial Intelligence and Fundamental Interactions}

\author{Michael~L.~Wagman}
\affiliation{Fermi National Accelerator Laboratory, Batavia, IL 60510, USA}

\begin{abstract}
  The low-energy spectrum and scattering of two-nucleon systems are studied with lattice quantum chromodynamics using a variational approach. A wide range of interpolating operators are used: dibaryon operators built from products of plane-wave nucleons, hexaquark operators built from six localized quarks, and quasi-local operators inspired by two-nucleon bound-state wavefunctions in low-energy effective theories. Sparsening techniques are used to compute the timeslice-to-all quark propagators required to form correlation-function matrices using products of these operators. Projection of these matrices onto irreducible representations of the cubic group, including spin-orbit coupling, is detailed. Variational methods are applied to constrain the low-energy spectra of two-nucleon systems in a single finite volume with quark masses corresponding to a pion mass of 806 MeV. Results for $S$- and $D$-wave phase shifts in the isospin singlet and triplet channels are obtained under the assumption that partial-wave mixing is negligible. Tests of interpolating-operator dependence are used to investigate the reliability of the energy spectra obtained and highlight both the strengths and weaknesses of variational methods. These studies and comparisons to previous studies using the same gauge-field ensemble demonstrate that interpolating-operator dependence can lead to significant effects on the two-nucleon energy spectra obtained using both variational and non-variational methods, including missing energy levels and other discrepancies. While this study is inconclusive regarding the presence of two-nucleon bound states at this quark mass, it provides robust upper bounds on two-nucleon energy levels that can be improved in future calculations using additional interpolating operators and is therefore a step toward reliable nuclear spectroscopy from the underlying Standard Model of particle physics.
\end{abstract}

\title{A variational study of two-nucleon systems with lattice QCD}

\preprint{FERMILAB-PUB-21-354-T, MIT-CTP/5320, UMD-PP-021-06}

\maketitle

\clearpage 

\tableofcontents

\section{Introduction}\label{sec:intro}

\noindent
Precise nuclear and hypernuclear forces are crucial inputs to state-of-the-art nuclear many-body studies of matter, from  the neutron-star equation of state, to stability of neutron-rich isotopes, to energetic reactions and exotic decays of various nuclei, to the scattering cross sections of nuclei with leptons and beyond-Standard-Model particles in experimental searches for new physics~\cite{Hebeler:2015hla, Engel:2016xgb, Navratil:2016ycn, Epelbaum:2019kcf, Drischler:2019xuo, Holt:2019gmc, Tews:2020hgp}. As \emph{ab initio} nuclear many-body investigations achieve reduced statistical and systematic uncertainties, uncertainties in the nuclear Hamiltonian that is input to these calculations become more essential to address~\cite{Tews:2020hgp}. One promising approach to constrain nuclear forces is to derive them from the underlying theory of the strong force, quantum chromodynamics (QCD), using lattice QCD (LQCD). Over the past two decades, initial steps toward this goal have been taken, leading to LQCD results~\cite{Beane:2006mx, Beane:2006gf, Beane:2009py, Beane:2010hg, Beane:2011zpa, Beane:2011iw, Beane:2012ey, Beane:2012vq, Beane:2013br, Orginos:2015aya, Wagman:2017tmp, Fukugita:1994ve, Yamazaki:2011nd, Yamazaki:2012hi, Yamazaki:2015asa, Nemura:2008sp, Inoue:2010es, Berkowitz:2015eaa, Francis:2018qch, Junnarkar:2019equ, Horz:2020zvv, Illa:2020nsi, Green:2021qol, Gongyo:2017fjb, Iritani:2018sra, Aoki:2020bew} that have constrained two-nucleon scattering amplitudes and effective-field-theory (EFT) representations of forces in the few-nucleon (and other few-baryon) sectors, although the use of unphysically large quark masses has prevented a complete quantification of uncertainties.  
The finite-volume (FV) spectrum of energies obtained in LQCD calculations provides input to mapping conditions that constrain scattering amplitudes~\cite{Luscher:1986pf, Luscher:1990ux, Rummukainen:1995vs, Beane:2003da, Kim:2005gf, He:2005ey,  Davoudi:2011md, Leskovec:2012gb, Hansen:2012tf, Briceno:2012yi, Gockeler:2012yj, Briceno:2013lba, Feng:2004ua, Lee:2017igf, Bedaque:2004kc, Luu:2011ep, Briceno:2013hya, Briceno:2013bda} and EFT descriptions of nuclear forces~\cite{Wagman:2017tmp, Illa:2020nsi, Barnea:2013uqa, Kirscher:2015yda, Contessi:2017rww, Bansal:2017pwn, Eliyahu:2019nkz, Detmold:2021oro}. The problem of accurate identification of ground- and excited-state energies of systems with multiple nucleons (baryons) using LQCD is, therefore, of great importance and constitutes a significant challenge in LQCD studies of nuclear systems.\footnote{Another approach to constraining two-body nuclear forces is to determine the Bethe-Salpeter wavefunction of multi-baryon systems from LQCD correlation functions, from which (hyper)nuclear potentials can be deduced~\cite{Ishii:2006ec, Aoki:2008hh, Aoki:2009ji, Aoki:2011ep, Aoki:2012tk, HALQCD:2012aa}. Later versions of this method do not rely on energy identification from Euclidean correlation functions and are suggested to be free from associated multi-baryon spectroscopy challenges (see Ref.~\cite{Aoki:2020bew} for a review). Such studies do, however, rely on the assumption that only elastic scattering states are present in correlation functions. This method is subject to various systematic uncertainties that are extensively discussed in the literature~\cite{Beane:2010em, Birse:2012ph, Walker-Loud:2014iea, Kawai:2017goq, Yamazaki:2017gjl, Davoudi:2017ddj, Iritani:2018zbt, Drischler:2019xuo}.}
The continued development of efforts to address this challenge is of particular importance for systems for which experimental data are scarce or non-existing, such as in multi-neutron systems and in systems with non-zero strangeness~\cite{Hebeler:2015hla, Tolos:2020aln}.

Constraining nuclear forces via LQCD is a computational challenge because LQCD path-integral representations of baryon correlation functions evaluated using Monte Carlo methods are plagued by a severe signal-to-noise problem~\cite{Parisi:1983ae, Lepage:1989hd, Beane:2009gs, Wagman:2016bam, Davoudi:2020ngi}. This has limited most LQCD studies of multi-baryon systems to date to larger-than-physical values of the quark masses, for which the growth of the statistical noise, as the Euclidean time separation in the correlation functions increases, is less severe than with the physical quark masses. Furthermore, the excitation spectrum of multi-nucleon (baryon) systems is rich, with energy gaps that are not set by the QCD scale and are often much smaller, which presents additional challenges. 
In principle, the ground-state energy, $E_0$, of a multi-nucleon system can be extracted from the large Euclidean-time behavior of any LQCD two-point correlation function with the quantum numbers of the state of interest. This is because the spectral decomposition of the correlation function is asymptotically dominated by the ground-state contribution proportional to $e^{-t E_0}$, where $t$ is the Euclidean-time separation. 
Excited-state effects are suppressed by $z_{\mathsf{n}} e^{-t\delta_{\mathsf{n}}}$, where $\delta_{\mathsf{n}} = E_{\mathsf{n}} - E_0$ is the energy gap between the ground state and the $\mathsf{n}$-th excited state, and $z_{\mathsf{n}}$ is the ratio of the interpolating-operator overlap factor of this excited state to that of the ground state.
Excited-state effects can therefore be neglected if either $t \gg \delta^{-1}_{\mathsf{n}}$ and/or $z_{\mathsf{n}}\ll 1$ for each excited state.
In practice, Euclidean times that satisfy $t \gg \delta_{\mathsf{n}}^{-1}$ cannot be achieved for multi-nucleon systems. In the case of two-nucleon systems, this is because excited states involving unbound nucleons correspond to small values of $\delta_{\mathsf{n}}$ for large spatial volumes and exponential signal-to-noise degradation limits $t$ to much smaller values than $\delta_{\mathsf{n}}^{-1}$ in calculations using current algorithms and computing resources (or those of the forseeable future).
For $t < \delta_{\mathsf{n}}^{-1}$, where signals can be resolved for multi-nucleon LQCD correlation functions, contributions from low-energy excited states can be significant, and calculations using a single correlation function (or a vector of correlation functions that have the same quantum numbers) face the challenging problem of performing multi-exponential fits in order to extract $E_0$.
This limitation could be ameliorated through the use of multi-nucleon interpolating operators that dominantly overlap with the ground state (or other states of interest) and are approximately orthogonal to other low-energy states ($z_{\mathsf{n}}\ll 1$ for all $\mathsf{n} > 0$).
However, the structure of the true QCD energy eigenstates is unknown \emph{a priori} and finding interpolating operators with maximal overlap onto states of interest is a challenging task.
Indirect tests, such as those enabled by L{\"u}scher's mapping from the FV spectrum to scattering amplitudes, can potentially signal issues with the extracted energies because scattering amplitudes need to satisfy certain constraints, see Refs.~\cite{Dudek:2012xn,Wilson:2015dqa,Baru:2016evv,Iritani:2018vfn,Wagman:2017tmp,Illa:2020nsi}; however, passing these tests is not sufficient to guarantee that the spectrum has been extracted reliably.

An approach for constructing interpolating operators with $z_{\mathsf{n}} \ll 1$ for a set of low-energy states $\{\mathsf{n}\}$ is provided by variational methods, in which a positive-definite Hermitian matrix of two-point correlation functions is formed using a set of interpolating operators and their conjugates, and a generalized eigenvalue problem (GEVP) is solved in order to obtain a mutually orthogonal set of approximate energy eigenstates~\cite{Fox:1981xz,Michael:1982gb,Luscher:1990ck}.
Symmetric correlation functions resulting from this diagonalization have spectral representations as sums of exponentials with positive-definite coefficients and are therefore convex functions of $t$.
It follows from this that logarithmic derivatives of symmetric correlation functions, called effective masses, provide variational upper bounds on ground-state energies that approach the true energy from above~\cite{Berg:1981zb,Ishikawa:1982tb}.
Further, if an interpolating-operator set that overlaps with all energy eigenstates below an energy threshold $\Delta \gg \delta$ can be identified, then variational methods can reduce excited-state contributions to ground-state energy determinations to $e^{-t \Delta}$~\cite{Luscher:1990ck}.
In this work, a large set of interpolating operators for two-nucleon systems is identified, and positive-definite Hermitian two-point correlation-function matrices are constructed using these interpolating operators.
The GEVP solutions for these correlation-function matrices are used to construct approximate energy eigenstates.
This procedure removes excited-state contamination from determinations of $E_0$ arising from the lowest-energy set of states that significantly overlap with the set of the interpolating operators that are considered.
Since variational methods result in convex correlation functions after diagonalization, cancellations between excited-state contributions that might conspire to be consistent with a single exponential (within uncertainties) over a significant range of $t$ for $t < \delta^{-1}$ cannot arise. 
Such a possibility, however, is present in calculations using a set of asymmetric LQCD correlation functions in which exchanging source and sink interpolating operators is not equivalent to Hermitian conjugation and is argued to be relevant for LQCD determinations of two-baryon energy spectra in  Refs.~\cite{Iritani:2016jie,Iritani:2018vfn}.
It is noteworthy, however, that the reliability of variational methods in extracting energies rather than only upper bounds on them requires the identification of an interpolating-operator set that overlaps strongly with the ground state and low-energy excited states.
Since the eigenstates of LQCD are not known \emph{a priori}, it is not possible to know beyond a doubt that this is the case for a given interpolating-operator set.

Variational methods have a long history of successful applications to LQCD studies of systems with baryon number $B=0$ and $B=1$.
Early studies used interpolating-operator sets consisting of color-singlet ``single-hadron'' operators describing glueballs, mesons, or baryons, with quark wavefunctions in the latter cases resembling Gaussians or similar functions centered about a common point  with one or more widths~\cite{Michael:1982gb,Allton:1993wc,Sasaki:2001nf,Melnitchouk:2002eg,Brommel:2003jm,Burch:2004he,Burch:2006dg,Burch:2006cc}.
These, and more recent studies, have generally found that Gaussian quark wavefunctions with widths $\lesssim 1$ fm have significant overlap with the nucleon ground state. Low-energy excited states can be described, for example, by using linear combinations of Gaussians containing nodes and qualitatively resembling quark-model radial excitations~\cite{Burch:2004he,Roberts:2013oea,Leinweber:2015kyz,Virgili:2019shg,Sun:2019aem}.
Further studies extended these interpolating-operator sets and included tens of single-hadron operators with (products of) gauge-covariant derivatives, as well as hybrid baryon operators with gluon fields carrying angular momentum~\cite{Dudek:2007wv,Basak:2007kj,Blossier:2009kd,Bulava:2009jb,Dudek:2009qf,Bulava:2010yg,Dudek:2010wm,Engel:2010my,Dudek:2012ag}.
The next extension was the inclusion of ``multi-hadron'' interpolating operators describing products of color-singlet hadron operators such as $\pi\pi$ or $N\pi$ each carrying definite relative momentum~\cite{Peardon:2009gh,Morningstar:2011ka,Dudek:2012gj,Dudek:2012xn,Mohler:2012na,Dudek:2013yja,Wilson:2014cna,Shultz:2015pfa,Wilson:2015dqa}.
It was noticed in studies of the $B=0$, isospin $I=1$ channel in the vicinity of the $\rho$ resonance that interpolating-operator sets including only single-hadron or multi-hadron operators lead to determinations of the energy spectrum with ``missing energy levels'' and other inconsistencies in energy-level results when compared to the energy spectrum determined using larger interpolating-operator sets including both types of operators~\cite{Dudek:2012xn,Wilson:2015dqa}.
For $t\rightarrow \infty$, any interpolating-operator set can be used to extract the energies of as many lowest-energy states as there are operators in the set. However, for the range $0.1 \lesssim t \lesssim $ 1 fm over which correlation functions are studied in these references, using an interpolating-operator set neglecting some single- or multi-hadron interpolating operators leads to a determination of the energy spectrum in which some energy levels identified using a larger interpolating set are missing but other higher-energy levels are present.
In the $B=1$ sector, multi-hadron interpolating operators such as $N\pi$ have also been included in interpolating-operator sets used for variational calculations~\cite{Lang:2012db,Lang:2016hnn,Andersen:2017una,Leskovec:2018lxb,Silvi:2021uya}. In this context, it has been noticed that omitting multi-hadron operators can lead to missing or displaced energy levels close to and above the pion production threshold~\cite{Lang:2012db} and that it is much more difficult (although possible) to resolve energy levels associated with $N\pi$ scattering states by only including local $qqq\overline{q}q$ operators with the same quark content as (plane-wave) $N\pi$ operators~\cite{Kiratidis:2015vpa}.
Analogous issues for $NN$ systems will be discussed at length below, see in particular Sec.~\ref{sec:discussion}.

Although the need for variational studies of multi-baryon systems has long been recognized, only recently with the advent of efficient algorithms for calculating approximate ``all-to-all'' quark propagators, such as the Laplacian Heaviside or ``distillation'' method~\cite{Peardon:2009gh}, stochastic Laplacian Heaviside~\cite{Morningstar:2011ka}, and sparsening methods~\cite{Detmold:2019fbk,Li:2020hbj}, has the application of variational methods to multi-baryon systems become computationally feasible, albeit still at unphysically large quark masses.
The first variational study of the two-nucleon isotriplet, ``dineutron'', channel and the $H$-dibaryon channel using multi-hadron interpolating operator was reported by Francis et al.\ in Ref.~\cite{Francis:2018qch}.
This reference presents studies of boosted two-baryon systems with several center-of-mass momenta using $2\times 2$ positive-definite Hermitian matrices of single-hadron interpolating operators, as well as a positive-definite multi-hadron correlation function and several other asymmetric correlation functions.
The two-nucleon isosinglet, ``deuteron'', channel as well as the dineutron channel have also been studied using calculations of $2\times 2$ multi-hadron correlation-function matrices for several values of the center-of-mass momentum by H{\"o}rz et al.\ in Ref.~\cite{Horz:2020zvv}.
Most recently, a variational study of the $H$-dibaryon channel was presented by Green et al.\ in Ref.~\cite{Green:2021qol} using correlation-function matrices with up to  3 multi-hadron interpolating operators in several boosted frames. This reference obtained consistent results with Ref.~\cite{Francis:2018qch} and quantified significant lattice artifacts in the finite-volume energy shifts of the $H$-dibaryon channel.
Interestingly, the results for ground- and excited-state energy levels for two-baryon systems calculated using variational methods in Refs.~\cite{Francis:2018qch,Horz:2020zvv,Green:2021qol} suggest tensions with earlier results obtained using sets of asymmetric correlation functions~\cite{Beane:2010hg,Beane:2011iw,Beane:2012vq,Yamazaki:2012hi,Beane:2013br,Berkowitz:2015eaa,Beane:2015yha,Chang:2015qxa,Detmold:2015daa,Yamazaki:2015asa,Parreno:2016fwu,Savage:2016kon,Shanahan:2017bgi,Tiburzi:2017iux,Wagman:2017tmp,Winter:2017bfs,Chang:2017eiq,Detmold:2020snb}, although these calculations use different discretizations and quark masses. Ref.~\cite{Green:2021qol} suggests that lattice-spacing artifacts may contribute to these discrepancies.
In this context, further variational studies of multi-baryon systems are clearly of great importance.

The goal of the present work is to perform a detailed study of two-nucleon systems using variational methods and using a significantly larger set of  single- and multi-hadron interpolating operators than the sets used in previous works.
To this end, the two-nucleon systems in both the isotriplet and isosinglet channels are studied at a single lattice spacing and lattice volume with larger-than-physical quark masses such that $m_\pi = 806$ MeV.
The largest set of two-baryon interpolating operators to date is constructed, including multiple types of  ``hexaquark'' interpolating operators built from a product of six quark fields with Gaussian wavefunctions centered around a common point and expected to strongly overlap with compact bound states, ``dibaryon'' interpolating operators built from products of momentum-projected baryons and expected to strongly overlap with unbound $NN$ scattering states, and ``quasi-local'' interpolating operators designed to somewhat resemble effective low-energy descriptions of the loosely-bound deuteron state present in nature. Through the use of recently-developed propagator sparsening techniques~\cite{Detmold:2019fbk} and highly optimized codes for constructing two-baryon correlation functions using the \verb!Tiramisu!~\cite{baghdadi2020tiramisu} compiler framework, positive-definite correlation-function matrices are constructed with dimensionalities as large as $16 \times 16$ for the dineutron channel and $42 \times 42$ for the deuteron channel.
The (upper bounds on the) ground-state energies obtained from the resulting GEVP solutions for these correlation-function matrices are significantly closer to threshold than the ground-state energies obtained using hexaquark sources and dibaryon sinks in this work and in previous studies using the same gauge-field ensemble~\cite{Beane:2012vq,Beane:2013br,Berkowitz:2015eaa,Wagman:2017tmp}. 

The results of this work do not provide a conclusive picture of nucleon-nucleon interactions with $m_\pi = 806$ MeV because the volume and lattice-spacing dependence of the two-nucleon energy spectra require further investigation. Perhaps more importantly, other states that have negligible overlap with the operator sets considered here may also be present in the spectrum, as demonstrated by the construction of a plausible model for overlap factors consistent with such behavior in Sec.~\ref{sec:I1}. 
Nonetheless, the variational method is an approach to the problem of excited-state contamination in two-baryon correlation functions that provides systematically improvable upper bounds on energy levels.
Future calculations exploring a range of lattice spacings and volumes with a wider variety of interpolating operators may lead to a conclusive understanding of nucleon-nucleon interactions at these unphysical values of the quark masses and provide the most robust available route to determinations of nucleon-nucleon interactions at the physical quark masses.
While the previous non-variational studies of multi-nucleon systems, including calculations of a range of important nuclear matrix elements~\cite{Beane:2015yha, Savage:2016kon, Shanahan:2017bgi, Tiburzi:2017iux, Winter:2017bfs, Chang:2017eiq, Detmold:2020snb, Parreno:2021ovq, Davoudi:2020ngi}, serve as milestones in accessing nuclear properties from QCD and have contributed to the development of the current suite of methods and algorithms, the era of precision LQCD calculations of multi-baryon systems is just beginning.

In order to introduce the LQCD technology for constructing two-nucleon interpolating operators and the associated correlation-function matrices for the variational approach, Sec.~\ref{sec:method} presents the relevant methods for evaluating correlation-function matrices and extracting the energy spectrum using variational methods for single- and two-nucleon systems.
In Sec.~\ref{sec:numerics}, this formalism is used to study two-nucleon correlation functions, the associated finite-volume spectra, and the $S$- and $D$-wave scattering phase shifts (assuming negligible partial-wave mixing) at quark masses corresponding to a pion mass of $806$ MeV. A total of 22 and 49 ground- and excited-state energy levels below the single-nucleon first-excited-state energy are identified for the two-nucleon systems with $I=1$ and $I=0$, respectively. The FV energy-spectrum results, as well as the corresponding scattering phase shifts, are compared with existing variational and non-variational LQCD results at similar quark masses.
Extensive studies of the interpolating-operator dependence of the results are performed, and the strengths and weaknesses of variational methods and implications of these results are summarized in Sec.~\ref{sec:concl}.
A number of appendices complement the formalism and numerical sections of the paper by providing further details. They are followed by a glossary of frequently used notation in Appendix~\ref{app:glossary}.

\section{Variational methods for two-nucleon systems}\label{sec:method}

\subsection{Interpolating operators} \label{sec:interp}

In the infinite-volume and continuum limits, interpolating operators for QCD energy eigenstates in Euclidean spacetime can be classified by their transformation properties under rotations, the $SO(3)$ subgroup of $SO(4)$ spacetime isometries valid at fixed Euclidean time $t$. Assuming a vanishing $\theta$-term, charge conjugation ($C$) and parity ($P$) are exact symmetries of the QCD action and interpolating operators are further classified by their $C$ and $P$ transformation properties.  For theories in a finite cubic spatial volume and/or cubically discretized lattice field theories, $SO(3)$ invariance is broken down to the cubic or octahedral group $O_h$ composed of the 48 symmetries of a cube.
Bosonic states in even baryon-number sectors, such as the vacuum and two-nucleon sectors, can be decomposed into direct sums of states that transform in the $O_h$ irreducible representations (irreps) $A_1^\pm,\ A_2^\pm,\ E^\pm,\ T_1^\pm$, and $T_2^\pm$, where $\pm$ denotes that states in the corresponding irrep are eigenstates of $P$ with eigenvalues $\pm 1$.\footnote{For a octahedral group representation $\Gamma_J$, $d_{\Gamma_J}$ denotes the dimension of the representation and the individual elements are referred to as ``rows'' of the irrep.}
Fermionic states in odd baryon-number sectors transform in direct sums of representations $G_1^\pm$, $G_2^\pm$, and $H^\pm$ of  the double-cover of the cubic group, $O_h^D$, which includes all elements of $O_h$ as well as the same elements composed with a $2\pi$ rotation about any axis.
LQCD actions also preserve $U(1)$ baryon-number symmetry exactly, and energy eigenstates can therefore be decomposed into irreps of $U(1)$ corresponding to baryon number $B \in \mathbb{Z}$.
Furthermore, the up, down, and strange quark masses are be chosen to be equal and electroweak interactions will be omitted throughout this work; therefore $SU(3)$ flavor symmetry and its $SU(2)$ isospin subgroup are exact symmetries in the study below.

Given these symmetries, FV LQCD energy eigenstates can be classified by their baryon number $B$, total isospin $I$, strangeness (only strangeness-0 systems are considered in this work)  and cubic irrep $\Gamma_J$, which plays the role of the continuum, infinite-volume total angular momentum, $J$.
Energy eigenstates with definite values of these quantum numbers will be denoted $\ket{\mathsf{n}^{(B,I,\Gamma_J)}}$, with $\mathsf{n}\in\{0,1,\ldots\}$ labeling discrete FV eigenstates within the spectrum, where  $\braket{\mathsf{n}^{(B,I,\Gamma_J)}}{\mathsf{n}^{(B,I,\Gamma_J)}} = 1$.
The corresponding energies are denoted $E_{\mathsf{n}}^{(B,I,\Gamma_J)}$, and states are ordered such that $E_{\mathsf{n}}^{(B,I,\Gamma_J)} \leq E_{\mathsf{m}}^{(B,I,\Gamma_J)}$ for $\mathsf{n} < \mathsf{m}$.
Energies of these states are exactly independent of the isospin component $I_z \in \{-I,\ldots,I\}$ and the row of the cubic irrep $\Gamma_J$ on ensemble average. To simplify notation, $\ket{\mathsf{n}^{(B,I,\Gamma_J)}}$ is defined to be averaged over $I_z$ and rows of $\Gamma_J$.
The nucleon is defined to be the ground state of the sector with baryon-number $B=1$, positive parity, and total isospin $I=1/2$, and transforms in the $G_1^+$ irrep associated with total angular momentum $J=1/2$ in the continuum, infinite-volume limit; see Appendices~\ref{app:weights}-\ref{app:contractions} and Refs.~\cite{Basak:2005aq,Basak:2005ir} for further discussion.
For boosted systems with non-zero total center-of-mass momentum $\vec{P}$, cubic symmetry is further broken down to the little group comprised of the elements of $O_h^D$ that leave $\vec{P}$ invariant~\cite{Thomas:2011rh}. 
Boosted systems are classified by their total center-of-mass momentum $\vec{P}$ and irrep $\Gamma_J^{\vec{P}}$ under the associated little group, and the energies and energy eigenstates of boosted systems will be denoted $E_{\mathsf{n}}^{(B,I,\vec{P},\Gamma_J^{\vec{P}})}$ and $\ket{\mathsf{n}^{(B,I,\vec{P},\Gamma_J^{\vec{P}})}}$  (the $\vec{P}$ label is dropped for $\vec{P}=0$), normalized to
$\braket{\mathsf{n}^{(B,I,\vec{P},\Gamma_J^{\vec{P}})}}{\mathsf{n}^{(B,I,\vec{P},\Gamma_J^{\vec{P}})}}=1$.

\subsubsection{Single-nucleon interpolating operators}\label{sec:onehadron}

This work uses standard nucleon interpolating operators whose properties are briefly reviewed. 
Point-like proton interpolating operators transforming in rows of the $G_1^+$ irrep indexed as $\sigma \in \{ 0,1\}$ can be constructed in the Dirac basis\footnote{For the definition of the Dirac basis and relations to other bases, see Appendix A of Ref.~\cite{Basak:2005ir}.} as
\begin{equation}
\begin{split}
  p_{\sigma}(x) &= \varepsilon^{abc} \frac{1}{\sqrt{2}}\left[ u_{\zeta}^{a}(x) (C \gamma_5 P_+)_{\zeta \xi} d_\xi^{b}(x) - d_{\zeta}^{a}(x) (C \gamma_5 P_+)_{\zeta \xi} u_{\xi}^{b}(x) \right] \\
   &\hspace{20pt} \times \left[ P_+ \left( 1 - (-1)^\sigma i \gamma_1 \gamma_2 \right) \right]_{\sigma \zeta} u_{\zeta}^{c}(x),
  \end{split}\label{eq:Ninterp}
\end{equation}
where $q_\zeta^a (x)$ denotes a quark field of flavor $q \in \{u,d\}$ with $a,b,c$ being $SU(3)$ color indices and $\zeta, \xi$ being Dirac spinor indices,\footnote{Note that repeated spinor and color indices are implicitly assumed to be summed throughout this work but that this summation convention will not be used for cubic irrep rows and other indices.} $\gamma_\mu$ are Euclidean gamma matrices satisfying $\gamma_\mu^\dagger =\gamma_\mu$ and $\{\gamma_\mu,\gamma_\nu\}=2\delta_{\mu\nu}$, $C=\gamma_2\gamma_4$, and $\gamma_5=\gamma_1 \gamma_2 \gamma_3 \gamma_4$ are used to build spin-singlet diquarks, and $P_+=\left( \frac{1+\gamma_4}{2} \right)$ is a positive-parity projector.
The application of $P_+$ projects each quark field from $G_1^+ \oplus G_1^-$ onto $G_1^+$, which allows for increases in computational efficiency, as discussed below.
The Dirac basis is convenient for expressing Eq.~\eqref{eq:Ninterp} because in this basis the 0 and 1 Dirac spinor components transform according to the $\sigma = 0$ and $\sigma = 1$ rows of $G_1^+$.
Point-like neutron interpolating operators $n_\sigma(x)$ are obtained by exchanging $u \leftrightarrow d$ in Eq.~\eqref{eq:Ninterp}.
These interpolating operators can be combined into a nucleon field $N_\sigma(x) \equiv (p_\sigma(x),\ n_\sigma(x))^T$ that transforms as a doublet under $SU(2)$ isospin symmetry, where $T$ denotes transpose.

Spin-color weights can be introduced in order to simplify expressions for the tensor contractions appearing in nucleon and multi-nucleon correlation functions as in Ref.~\cite{Detmold:2012eu}.  
Spin-color components of the quark field will be labelled with indices $i,j,k,\ldots$, where for example $i = (\zeta,a)$ denotes a compound spin-color index corresponding to spinor index $\zeta \in \{0,\ldots,3\}$ and color index $a \in \{0,1,2\}$.
In this way, the proton interpolating operator above can be expressed as a contraction of three quark fields,\footnote{A canonical ordering in which quark flavors occur lexicographically in all hadron interpolating operators can also be used.} $u^i(x) d^j(x) u^k(x)$, with a tensor of real-valued ``weights'' whose $ijk$ component is defined to be the coefficient of $u^i(x) d^j(x) u^k(x)$ in Eq.~\eqref{eq:Ninterp}.
The corresponding neutron interpolating operator can be expressed as a contraction of $d^i(x) u^j(x) d^k(x)$ with an identical tensor of weights.
The weights depend on the spin of the nucleon and will therefore be denoted $w^{[N]\sigma}$.
Most of the spin-color tensor components of $w^{[N]\sigma}$ are zero, and the numerical evaluation of spin-color contractions becomes significantly more efficient if the nucleon weights are represented as a sparse tensor $w_\alpha^{[N]\sigma}$ where $\alpha \in \{1,\ldots,\mathcal{N}_w^{[N]}\}$ runs over the $\mathcal{N}_w^{[N]}$ components of $u^{i(\alpha)}(x)d^{j(\alpha)}(x)u^{k(\alpha)}(x)$ with non-zero weight, where $i(\alpha)$, $j(\alpha)$, and $k(\alpha)$ are spin-color-index--valued maps from $\alpha$ to spin-color indices  such that
\begin{equation}
\begin{split}
  p_{\sigma}(x) &= \sum_{\alpha} w_\alpha^{[N]\sigma} u^{i(\alpha)}(x) d^{j(\alpha)}(x) u^{k(\alpha)}(x), \\
    n_{\sigma}(x) &= \sum_{\alpha}  w_\alpha^{[N]\sigma} d^{i(\alpha)}(x) u^{j(\alpha)}(x) d^{k(\alpha)}(x).
\end{split}\label{eq:Ndef}
\end{equation}
The nucleon weights $w_\alpha^{[N]\sigma}$ can be evaluated by choosing a particular basis for the spinor algebra in Eq.~\eqref{eq:Ninterp} and enforcing equivalence with the particular flavor ordering shown in Eq.~\eqref{eq:Ndef}.
The Dirac basis is particularly convenient because only two of the spinor components have non-zero weight (due to the application of $P_+$ to all quark fields), and the storage and computation time of quark propagator contractions can be significantly reduced by considering only these two spin components and treating spin-color indices as valued in $\{0,\ldots,5\}$.
In the Dirac basis, there are $\mathcal{N}_w^{[N]} = 12$ terms appearing in Eq.~\eqref{eq:Ndef} that are explicitly shown in Appendix~\ref{app:weights}.
Quark-exchange symmetry, arising from the presence of two identical quark fields in $N_\sigma(x)$, can be used to reduce the number of weights required to compute nucleon correlation functions to 9~\cite{Detmold:2012eu}.
However, this symmetry does not apply to Wick contractions involving non-trivial quark permutations in $B=2$ correlation functions with interpolating operators built from products of $N_{\sigma}(x_1)$ and $N_{\sigma'}(x_2)$ in which quark fields with distinct spatial labels appear.
For simplicity, quark-exchange symmetry is not used to reduce the number of nucleon weights throughout this work.

Non-point-like nucleon interpolating operators in the $G_1^+$ irrep can be constructed by including (gauge-covariant) derivatives or other functions of the gauge field into the interpolating operator of Eq.~\eqref{eq:Ninterp} as described in Refs.~\cite{Basak:2005aq,Basak:2005ir}.
Nucleon spectroscopy studies, for example in Refs.~\cite{Allton:1993wc,Burch:2004he,Burch:2006cc,Basak:2007kj,Bulava:2009jb,Bulava:2010yg,Dudek:2010wm,Engel:2010my,Dudek:2012ag,Roberts:2013oea,Leinweber:2015kyz,Virgili:2019shg,Sun:2019aem}, suggest that interpolating operators including derivatives or spatial wavefunctions with nodes dominantly overlap with nucleon excited states. 
Alternative spin structures are also found to overlap dominantly with nucleon excited states~\cite{Leinweber:1990dv,Melnitchouk:2002eg,Brommel:2003jm}.
Operators that have larger overlap with the nucleon ground state can be constructed by combining the spin-color tensor structure in Eq.~\eqref{eq:Ninterp} with spatial wavefunctions that ``smear'' the location of the quark field over a volume whose radius is of the order of the proton radius~\cite{Allton:1993wc,Burch:2004he}.
 Gauge-invariantly Gaussian-smeared quark fields are defined as~\cite{Gusken:1989ad,Gusken:1989qx}
\begin{equation}
  q^{i(\alpha)}_{g}(\vec{x},t) = \sum_{\vec{z} \in \Lambda}  \sum_{i'}  \Pi_g^{i(\alpha)i'}(\vec{x},\vec{z}) \ q^{i'}(\vec{z},t), 
   \label{eq:qsmear_def}
\end{equation}
where $i'$ is a spin-color index, $\Pi_g^{i(\alpha)i'}$ is a Gaussian smearing kernel with a width specified by the index $g$. The kernel is assumed here to be $O_h$ invariant and can be constructed, for example, by iteratively applying the gauge-covariant discrete Laplacian,\footnote{For systems with a large center-of-mass momentum, the kernel could be replaced with the momentum-smearing kernel introduced in Ref.~\cite{Bali:2016lva} to improve ground-state overlap.} and  $\Lambda = \{(x_1,x_2,x_3)\ |\ 0 \leq x_k < L\}$ is the set of spatial lattice sites where $k\in\{1,2,3\}$ labels the spatial dimensions of the lattice geometry.\footnote{The label $k$ for the spatial coordinate axes should not be confused with the spin-color index $k$.} Throughout this work units are used in which the lattice spacing is equal to unity, so $x_k, t \in \mathbb{Z}$.
Smeared proton and neutron interpolating operators are defined from the smeared quark fields as
\begin{equation}
\begin{split}
    p_{\sigma g}(x) &= \sum_{\alpha} w_\alpha^{[N]\sigma} u^{i(\alpha)}_{g}(x) d^{j(\alpha)}_{g}(x) u^{k(\alpha)}_{g}(x), \\
    n_{\sigma g}(x) &= \sum_{\alpha} w_\alpha^{[N]\sigma} d^{i(\alpha)}_{g}(x) u^{j(\alpha)}_{g}(x) d^{k(\alpha)}_{g}(x),
    \label{eq:Nsmear_def}
\end{split}
\end{equation}
and smeared isodoublet nucleon fields are defined as $N_{\sigma g}(x) \equiv (p_{\sigma g}(x),\ n_{\sigma g}(x))^T$.
Such smeared quark fields transform identically to unsmeared quark fields under $O_h^D$, and therefore smeared hadron fields transform identically to unsmeared hadron fields.

Projection to a definite center-of-mass momentum $\vec{P}_{\mathfrak{c}}$, where $\mathfrak{c}$ indexes the center-of-mass momenta included in an interpolating-operator set, is accomplished by multiplying $N_{\sigma g}(x)$ by $e^{i\vec{P}_{\mathfrak{c}}\cdot \vec{x}}$, where $\vec{x}$ denotes the spatial components of the coordinate $x = (\vec{x},t)$, and summing over the set of spatial lattice sites $\Lambda$.
In order to reduce the computational cost of performing this summation (and more costly volume sums for the two-hadron operators discussed below), the propagator sparsening algorithm introduced in Ref.~\cite{Detmold:2019fbk}  is applied.
In particular, ``sparsened'' plane-wave spatial wavefunctions  are used that only have support on a cubic sublattice $\lsp \subset \Lambda$ defined as 
\begin{equation}
    \lsp = \{(x_1,x_2,x_3)\ |\ 0 \leq x_k < L,\ x_k \equiv 0 \Mod{\mathcal{S}}\},
\label{eq:sparse_lattice}
\end{equation} 
where $L$ is the spatial extent of the cubic lattice geometry and $\mathcal{S} \in \mathbb{Z}$ is the ratio of the number of full and sparse lattice sites in each spatial dimension. Defining
\begin{equation}
  \psi_{\mathfrak{c}}^{[N]}(\vec{x}) = \left. e^{i\vec{P}_{\mathfrak{c}}\cdot\vec{x}}\right|_{\lsp}, \label{eq:psiNdef}
\end{equation}
where the bar denotes restriction of support to $\lsp$,
the nucleon interpolating operators including sparsened plane-wave spatial wavefunctions are defined as
\begin{equation}
  N_{\sigma\mathfrak{c}g}(t) = \sum_{\vec{x}\in \lsp} \psi_\mathfrak{c}^{[N]}(\vec{x}) N_{\sigma g}(\vec{x},t). \label{eq:Nprojdef}
\end{equation}
Momentum-projected two-point correlation functions for these nucleon interpolating operators are defined by
\begin{equation}
  C_{\sigma\mathfrak{c}g\sigma^\prime\mathfrak{c}^\prime g^\prime}^{[N,N]}(t) = \left< N_{\sigma\mathfrak{c}g} (t)\ \overline{N}_{\sigma^\prime \mathfrak{c}^\prime g'}^T(0) \right> ,
    \label{eq:CNsparse}
\end{equation}
where $\overline{N}_{\sigma^\prime \mathfrak{c}^\prime g^\prime}(t)$ is obtained from $N_{\sigma^\prime \mathfrak{c}^\prime g^\prime}(t)$ by replacing $q^i(x)$ with $\overline{q}^i(x)$ and reversing the order of the fields in Eq.~\eqref{eq:Ndef} (the weights should also be replaced by their complex conjugates but they satisfy $(w_\alpha^{(N)})^* = w_\alpha^{(N)}$ for the interpolating operators used here).
Two-point correlation functions only depend on the time difference between the two operators; for simplicity, the Euclidean time location of the ``source'' operator is denoted by $0$ and the Euclidean time location of the ``sink'' operator is denoted by $t$.
By isospin symmetry, $C_{\sigma\mathfrak{c}g\sigma^\prime\mathfrak{c}^\prime g^\prime}^{[N,N]}(t)$ is proportional to the identity matrix in flavor space.
Similarly, $C_{\sigma\mathfrak{c}g\sigma^\prime\mathfrak{c}^\prime g^\prime}^{[N,N]}(t)$ vanishes unless $\sigma = \sigma'$ and $\mathfrak{c} = \mathfrak{c}'$.
Sparsened two-point correlation functions of the form of Eq.~\eqref{eq:CNsparse} can be practically evaluated by computing point-to-all quark propagators with Gaussian-smeared quark sources at each of the $\vsp = (L/\mathcal{S})^3$ points on the sparse lattice, smearing the resulting quark propagators at the sink, and then for each $t$ restricting the propagators (obtained by solving the Dirac equation on the full lattice $\Lambda$) to have support on $\lsp$
and subsequently evaluating the $\vsp^2$ terms appearing in Eq.~\eqref{eq:CNsparse}.

Momentum projection with the sparsened plane-wave wavefunctions $\psi_{\mathfrak{c}}^{[N]}(\vec{x})$ leads to complete projection to states of momentum allowed by the lattice geometry for the case of trivial sparsening $\mathcal{S} = 1$, and it amounts to incomplete momentum projection for the case of interest where $1 < \mathcal{S} \ll L$.
The effects of this incomplete momentum projection on sparsened two-point correlation functions can be seen from the spectral representation 
\begin{equation}
  C_{\sigma\mathfrak{c}g\sigma^\prime\mathfrak{c}^\prime g^\prime}^{[N,N]}(t) = \sum_{\vec{P}, \Gamma_J^{\vec{P}}} \sum_\mathsf{n} Z^{(1,\frac{1}{2},\vec{P},\Gamma_J^{\vec{P}})}_{\mathsf{n} N_{\sigma\mathfrak{c}g}} \left( Z^{(1,\frac{1}{2},\vec{P},\Gamma_J^{\vec{P}})}_{\mathsf{n} N_{\sigma'\mathfrak{c}'g'}} \right)^* e^{-t E_\mathsf{n}^{(1,\frac{1}{2},\vec{P},\Gamma_J^{\vec{P}})} },
    \label{eq:nucSpec}
\end{equation}
where thermal effects arising from the finite Euclidean time extent of the lattice are neglected,  and $Z_{\mathsf{n} \chi}^{(1,\frac{1}{2},\vec{P}_{\mathfrak{c}},\Gamma_J^{\vec{P}_{\mathfrak{c}}})} \equiv  \mbraket{\mathsf{n}^{(1,\frac{1}{2},\vec{P}_{\mathfrak{c}},\Gamma_J^{\vec{P}_{\mathfrak{c}}})}}{\overline{\chi}(0)}{0}$ for a generic interpolating operator $\chi(x)$.
For $\vec{x} \in \Lambda_{\mathcal{S}}$, sparsened plane-wave wavefunctions satisfy $\psi_{\mathfrak{c}}^{[N]}(\vec{x}) = e^{i \vec{P}_{\mathfrak{c}} \cdot \vec{x}} = e^{i \left[ \vec{P}_{\mathfrak{c}} + \frac{2\pi}{\mathcal{S}} \hat{e}_k \right]\cdot \vec{x} } $, where $\hat{e}_k$ is a unit vector oriented along the $k$ axis, and therefore the sum over states in Eq.~\eqref{eq:nucSpec} includes not only states with momentum $\vec{P} = \vec{P}_{\mathfrak{c}}$, but also states with momenta that differ by multiples of $\frac{2\pi}{S} \hat{e}_k$.
In the numerical calculations presented below, only the center-of-mass rest frame $\vec{P}_{\mathfrak{c}} = \vec{0}$ will be considered. 
In this frame, sparsening results in contributions to Eq.~\eqref{eq:nucSpec} that are suppressed by $e^{-t \Delta E_{\mathcal{S}}^{(1,\frac{1}{2},G_1^+)} }$, where $\Delta E_{\mathcal{S}}^{(1,\frac{1}{2},G_1^+)} =\sqrt{ M_N^2 + \left( \frac{2\pi}{\mathcal{S}} \right)^2 } - M_N$, with $M_N \equiv E_0^{(1,\frac{1}{2},G_1^+)}$ being the mass of the nucleon.
Previous numerical investigations confirm that sparsening results in changes to the excited-state structure of correlation functions that can be treated analogously to other excited-state effects in nucleon and nuclear correlation functions~\cite{Detmold:2019fbk}.
Excited-state effects arising from sparsening can also be stochastically removed by using random source positions instead of a lattice of sources $\lsp$, but using random source positions has been found to decrease statistical precision~\cite{Li:2020hbj}.

The sum in Eq.~\eqref{eq:nucSpec} includes contributions from two- and more hadron states, e.g., ``elastic'' $N\pi$ states with the same quantum numbers as the nucleon that can be approximately described as sets of interacting color-singlet hadrons, as well as contributions from ``inelastic'' excited states that can be approximately described as a single localized color-singlet hadron with a different spatial and/or spin structure than the ground state.
Excited-state contamination from elastic $N\pi$ states is suppressed compared to the ground-state contribution by $e^{-t \delta}$, where $\delta \gtrsim m_\pi$.
These contributions can be neglected for Euclidean times $t \gg m_\pi^{-1}$, which can be easily achieved in LQCD calculations that use larger-than-physical quark masses such as those considered in Sec.~\ref{sec:numerics}.
A simple nucleon interpolating-operator set sufficient for describing the nucleon ground state and one or more inelastic excited states with energies below $m_\pi$ can be obtained by using a set of two or more smeared nucleon interpolating operators.
It is noteworthy, however, that the highest-energy state obtained using variational methods must describe a linear combination of a whole tower of higher-energy excited states and is unlikely to be reliable.

\subsubsection{Two-nucleon interpolating operators}\label{sec:twohadron}

In the two-nucleon sector, the presence of a closely-spaced tower of elastic two-nucleon excitations in addition to inelastic excitations of the two nucleons leads to the expectation that larger sets of operators are necessary to find combinations that strongly overlap onto the low-energy eigenstates than in the single-nucleon sector. One possibility for two-nucleon interpolating operators is to construct single-hadron operators that describe spatially local color-singlet products of quark fields analogous to Eq.~\eqref{eq:Ninterp}. Due to the quark spin, ``hexaquark'' operators of this form transform under the cubic group according to a sixfold tensor product of $G_1^+$ irreps (only positive-parity quark components are used for simplicity and computational expediency).
The six quark spin representations can be grouped into the product $(G_1^+ \otimes G_1^+ \otimes G_1^+) \otimes (G_1^+ \otimes G_1^+ \otimes G_1^+)$ using an arbitrary quark-field ordering and, after evaluating the products in parentheses, can be represented as a product of two spins that can be associated with $B=1$ operators.
Hexaquark operators can be constructed that transform according to the $G_1^+ \otimes G_1^+$ irrep associated with $NN$ operator products, as well as the $H^+ \otimes H^+$ irrep associated with $\Delta \Delta$ operator products and other combinations.\footnote{Additional operators that do not factorize into products of color-singlet baryons can be constructed~\cite{Harvey:1981udr} but are not used in this work.}
Since $M_N < M_\Delta \equiv E_0^{(1,\frac{3}{2},H^+)}$, both in nature and in LQCD calculations at unphysically large quark masses, it is expected that $\Delta \Delta$ operators and other combinations will dominantly overlap with higher-energy states than $NN$ operators.
For simplicity, only $NN$ hexaquark operators constructed from products of color-singlet nucleons that transform under the cubic group as $G_1^+ \otimes G_1^+$ are considered in this work.

Hexaquark operators will be denoted as $H_{\rho \mathfrak{c}g}(t)$ below, where $\rho \in \{0,\ldots,3\}$ labels the row of $G_1^+ \otimes G_1^+ = A_1^+ \oplus T_1^+$, $\mathfrak{c}$ labels the center-of-mass momentum $\vec{P}_{\mathfrak{c}}$, and $g$ labels the quark smearing, as in the single-nucleon case.
Quark antisymmetry and the $A_1^+$ spatial symmetry of hexaquark operators force $I=1$ operators to transform in the one-dimensional $A_1^+$ irrep associated with spin-singlet states, and they similarly force $I=0$ operators to transform in the $T_1^+$ irrep associated with spin-triplet states.
The $I=1$ hexaquark operators have $\rho=0$, and for the $I_z = 0$ case  they are defined by
\begin{equation}
\begin{split}
  H_{0\mathfrak{c}g}(t) &= \sum_{\vec{x} \in \lsp} \psi_{\mathfrak{c}}^{[H]}(\vec{x}) \frac{1}{2} \left[ p_{0g}(\vec{x},t) n_{1g}(\vec{x},t) - p_{1g}(\vec{x},t) n_{0g}(\vec{x},t) \right. \\
    &\hspace{120pt} \left. + n_{0g}(\vec{x},t) p_{1g}(\vec{x},t) - n_{1g}(\vec{x},t) p_{0g}(\vec{x},t)\right] ,
\end{split}\label{eq:H0def}
\end{equation}
where $g$ specifies the quark-field smearing (chosen to be the same for all quarks),
and the same sparsened plane-wave wavefunctions $\psi_{\mathfrak{c}}^{[H]}(\vec{x}) \equiv \psi_{\mathfrak{c}}^{[N]}(\vec{x}) = \left. e^{i\vec{P}_{\mathfrak{c}}\cdot\vec{x}}\right|_{\lsp}$ are used as for the nucleon above. 
The spectra of $pp$ and $nn$ states with $I=1$ and $I_z = \pm 1$ are identical to those of $pn$ states with $I=1$ and $I_z=0$ by isospin symmetry, and it is therefore sufficient to only consider $pn$ operators in isospin-symmetric calculations of the two-nucleon spectrum.
Hexaquark operators for $pn$ systems with $I=0$ are defined as
\begin{equation}
\begin{split}
    H_{1\mathfrak{c}g}(t) &= \sum_{\vec{x} \in \lsp} \psi_{\mathfrak{c}}^{[H]}(\vec{x}) \frac{1}{\sqrt{2}} \left[ p_{0g}(\vec{x},t) n_{0g}(\vec{x},t) - n_{0g}(\vec{x},t) p_{0g}(\vec{x},t) \right], \\
    H_{2\mathfrak{c}g}(t) &= \sum_{\vec{x} \in \lsp} \psi_{\mathfrak{c}}^{[H]}(\vec{x}) \frac{1}{2} \left[ p_{0g}(\vec{x},t) n_{1g}(\vec{x},t)  + p_{1g}(\vec{x},t) n_{0g}(\vec{x},t) \right. \\
    &\hspace{120pt} \left. - n_{0g}(\vec{x},t) p_{1g}(\vec{x},t) - n_{1g}(\vec{x},t) p_{0g}(\vec{x},t) \right], \\
    H_{3\mathfrak{c}g}(t) &= \sum_{\vec{x} \in \lsp} \psi_{\mathfrak{c}}^{[H]}(\vec{x}) \frac{1}{\sqrt{2}} \left[ p_{1g}(\vec{x},t) n_{1g}(\vec{x},t) - n_{1g}(\vec{x},t) p_{1g}(\vec{x},t) \right].
    \end{split}\label{eq:H3def}
\end{equation}
Quark-level representations of hexaquark operators can be derived from Eqs.~\eqref{eq:H0def}-\eqref{eq:H3def} by inserting the representations of $p_{\sigma g}$ and $n_{\sigma g}$ in terms of the quark fields $q^{i(\alpha)}_g$.
These quark-level representations can be used to define spin-color weights and associated spin-color-index-valued maps\footnote{The same notation is used for index maps $i(\alpha),j(\alpha),\ldots,$ for different interpolating operators since the labels carried by the corresponding weights are sufficient to specify the interpolating operators in all contexts.} analogous to the weights and index maps defined for the nucleon in Eq.~\eqref{eq:Ndef} as
\begin{equation}
\begin{split}
  H_{\rho\mathfrak{c}g}(t) &= \sum_{\vec{x}\in \lsp} \psi_\mathfrak{c}^{[H]}(\vec{x}) \sum_{\alpha} w_\alpha^{[H]\rho} u^{i(\alpha)}_{g}(\vec{x},t) d^{j(\alpha)}_{g}(\vec{x},t) u^{k(\alpha)}_{g}(\vec{x},t) \\
    &\hspace{110pt}\times d^{l(\alpha)}_{g}(\vec{x},t) u^{m(\alpha)}_{g}(\vec{x},t) d^{n(\alpha)}_{g}(\vec{x},t).
\end{split}\label{eq:Hweightdef}
\end{equation}
Quark-exchange symmetries can be used to greatly reduce the number of independent spin-color weights required to construct local multi-baryon operators as described in Ref.~\cite{Detmold:2012eu}. For $pn$ systems, the number of non-zero elements of $w_\alpha^{[H]\rho}$ can be reduced to $\mathcal{N}_w^{[H]\rho}$ with $\mathcal{N}_w^{[H]0} = \mathcal{N}_w^{[H]2} = 32$ and $\mathcal{N}_w^{[H]1} = \mathcal{N}_w^{[H]3} = 21$. An explicit representation of these reduced weights is presented in Appendix~\ref{app:weights}. 
Hexaquark operators with one or more values of the quark-field smearing radius can be included in an interpolating-operator set that describes two-nucleon systems (here all quarks are smeared in the same way, although more general constructions are possible).

In addition to hexaquark operators that are expected to strongly overlap with compact bound states, operators constructed from products of pairs of nucleon operators at non-zero spatial separation may be expected to have larger overlap with unbound states of two-nucleon systems.
``Dibaryon'' interpolating operators, constructed from products of nucleon interpolating operators with factorizable plane-wave wavefunctions that are symmetric under exchange of the nucleon positions, are defined as
\begin{equation}
\begin{split}
  D_{\rho\mathfrak{m}g}(t) &= \sum_{\vec{x}_1,\vec{x}_2 \in \lsp} \psi_\mathfrak{m}^{[D]}(\vec{x}_1,\vec{x}_2) \sum_{\sigma,\sigma'} v^\rho_{\sigma \sigma^\prime}\, \frac{1}{\sqrt{2}}\left[ p_{\sigma g}(\vec{x}_1, t) n_{\sigma' g}(\vec{x}_2, t) \right. \\
    &\hspace{180pt} \left. + (-1)^{1-\delta_{\rho 0}} n_{\sigma g}(\vec{x}_1, t) p_{\sigma' g}(\vec{x}_2, t)\right], 
\end{split}\label{eq:Dinterp}
\end{equation}
where $v^\rho_{\sigma\sigma^\prime}$ is a weight tensor that projects the two-nucleon system into a row $\rho \in \{0,\ldots, 3\}$ of the $A_1^+ \oplus T_1^+$ two-nucleon spin representation, analogous to the hexaquark operators in Eqs.~\eqref{eq:H0def}-\eqref{eq:H3def}. Explicitly,
\begin{equation}
\begin{split}
    v^{0}_{\sigma\sigma^\prime} &= \frac{1}{\sqrt{2}}(\delta_{\sigma 0} \delta_{\sigma^\prime 1} - \delta_{\sigma 1} \delta_{\sigma^\prime 0}), \\
    v^1_{\sigma\sigma^\prime} &= \delta_{\sigma 0} \delta_{\sigma^\prime 0}, \\
    v^{2}_{\sigma\sigma^\prime} &= \frac{1}{\sqrt{2}}(\delta_{\sigma 0} \delta_{\sigma^\prime 1} + \delta_{\sigma 1} \delta_{\sigma^\prime 0}), \\
    v^3_{\sigma\sigma^\prime} &= \delta_{\sigma 1} \delta_{\sigma^\prime 1}.
\end{split}\label{eq:veights}
\end{equation}
The spatial wavefunctions appearing in Eq.~\eqref{eq:Dinterp} are chosen to be symmetric in $\vec{x}_1 \leftrightarrow \vec{x}_2$ and are labeled by $\mathfrak{m}$, where $\vec{P}_{\mathfrak{m}}$ is the center-of-mass momentum and $\pm\vec{k}_{\mathfrak{m}}$ is the momentum carried by each nucleon in the center-of-mass frame, and are given by
\begin{equation}
    \psi_\mathfrak{m}^{[D]}(\vec{x}_1,\vec{x}_2) = \frac{1}{\sqrt{2}} \left[ e^{i\left( \frac{\vec{P}_{\mathfrak{m}}}{2} + \vec{k}_{\mathfrak{m}}\right) \cdot \vec{x}_1 } e^{i\left( \frac{\vec{P}_{\mathfrak{m}}}{2} - \vec{k}_{\mathfrak{m}}\right) \cdot \vec{x}_2 } + e^{i\left( \frac{\vec{P}_{\mathfrak{m}}}{2} + \vec{k}_{\mathfrak{m}}\right) \cdot \vec{x}_2 } e^{i\left( \frac{\vec{P}_{\mathfrak{m}}}{2} - \vec{k}_{\mathfrak{m}}\right) \cdot \vec{x}_1 } \right].
\label{eq:psiDdef}
\end{equation}
Quark-level representations for dibaryon interpolating operators can be derived analogously to the hexaquark case and are given by
\begin{equation}
\begin{split}
  D_{\rho\mathfrak{m}g}(t) &= \sum_{\vec{x}_1,\vec{x}_2 \in \lsp} \psi_\mathfrak{m}^{[D]}(\vec{x}_1,\vec{x}_2) \sum_{\alpha} w_\alpha^{[D]\rho} u^{i(\alpha)}_{g}(\vec{x}_1,t) d^{j(\alpha)}_{g}(\vec{x}_1,t) u^{k(\alpha)}_{g}(\vec{x}_1,t) \\
    &\hspace{110pt}\times d^{l(\alpha)}_{g}(\vec{x}_2,t) u^{m(\alpha)}_{g}(\vec{x}_2,t) d^{n(\alpha)}_{g}(\vec{x}_2,t),
\end{split}\label{eq:Dweights}
\end{equation}
where the weights $w_\alpha^{[D]\rho}$, with $\alpha \in \{1,\ldots,\mathcal{N}_w^{[D]\rho}\}$, with $\mathcal{N}_w^{[D]0} = \mathcal{N}_w^{[D]2} = 288$ and $\mathcal{N}_w^{[D]1} = \mathcal{N}_w^{[D]3} = 144$, are obtained from products of $w_\alpha^{[N]\sigma}$, $w_\alpha^{[N]\sigma^\prime}$, and $v^\rho_{\sigma\sigma^\prime}$ and are explicitly shown in Appendix~\ref{app:weights}.
The dibaryon weights $w_\alpha^{[D]\rho}$ differ from the hexaquark weights $w_\alpha^{[H]\rho}$, since quark exchange symmetries can be used to reduce the number of independent weights in the latter case.
For boosted systems with $\vec{P}_{\mathfrak{m}} \neq \vec{0}$, cubic symmetry is broken and the two-nucleon operators transform under the appropriate little groups as described in Ref.~\cite{Briceno:2013bda}.
This work specializes to two-nucleon systems with $\vec{P}_{\mathfrak{m}} =\vec{0}$, where cubic symmetry and parity can be used to simplify interpolating-operator construction. Further, only positive-parity systems are considered in this work since the ground states of $I=1$ and $I=0$ two-nucleon systems in nature are known to be of positive parity.
Although products of two negative-parity nucleon excitations have the same quantum numbers, they are expected to correspond to higher-energy states than those studied here and are therefore not considered.

The plane-wave wavefunctions $\psi_{\mathfrak{m}}^{[D]}(\vec{x}_1,\vec{x}_2)$ with $\vec{P}_{\mathfrak{m}}=\vec{0}$ for all $\vec{k}_{\mathfrak{m}}$ would be energy eigenfunctions in the absence of strong interactions between the nucleons, with energies given by $2\sqrt{M_N^2 + | \vec{k}_{\mathfrak{m}} |^2}$.
For FV systems with spatial periodic boundary conditions (PBCs), which will be assumed below, the set of dibaryon operators  $D_{\rho\mathfrak{m}g}(t)$ with $\vec{k}_{\mathfrak{m}} = \left( \frac{2\pi}{L} \right)\vec{n}_{\mathfrak{m}}$, where $\vec{n}_{\mathfrak{m}} \in \mathbb{Z}^3$, and relative momentum ``shell'' $\mathfrak{s}(\mathfrak{m}) \equiv |\vec{n}_{\mathfrak{m}}|^2 \leq K$ provides a complete basis for relative wavefunctions of non-interacting two-nucleon scattering states (neglecting internal structure of the nucleon) with relative  momentum less than a cutoff set by $K$.
For notational simplicity, the functional dependence of $\mathfrak{s}$ on $\mathfrak{m}$ will be dropped below.
For non-interacting nucleons, the energy spectrum is therefore given by $2 \sqrt{M_N^2 + \mathfrak{s} \left( \frac{2\pi}{L} \right)^2 }$ for all integer $\mathfrak{s}$ that can be written as a sum of three integer squares (multiplicities can be enumerated for a given $\mathfrak{s}$~\cite{Grosswald1985}).
Including strong interactions, one expects the spectrum to resemble a tower of states whose energies approach these non-interacting energy levels as the infinite-volume limit is approached, as well as additional energy levels associated with bound states or resonances.
Although in an interacting theory such as QCD, $\vec{k}_{\mathfrak{m}}$ is not a quantum number of energy eigenstates, the non-interacting two-nucleon wavefunction basis still provides a useful way to enumerate linearly independent two-nucleon interpolating operators.
Further, the transformations of plane-wave wavefunctions under the cubic group are straightforward, see Ref.~\cite{Luu:2011ep} and Appendix~\ref{app:FV}.
The total cubic transformation representation, $\Gamma_J$, of a dibaryon interpolating operator with $\vec{P}_{\mathfrak{m}}=\vec{0}$ depends on the product of the cubic representation of the spin, $\Gamma_S$, with the cubic representation of the relative spatial wavefunction, $\Gamma_\ell$, associated with the continuum, infinite-volume orbital angular momentum, $\ell$.
As discussed in Sec.~\ref{sec:FV}, linear combinations of $D_{\rho\mathfrak{m}g}(t)$ can be constructed that transform with definite $\Gamma_J$.
It is therefore well-motivated to include the dibaryon operators $D_{\rho\mathfrak{m}g}(t)$ with $\mathfrak{s} \leq K$, and with the same two choices of quark smearings $g$ used for single-hadron interpolating operators, in an interpolating-operator set that can be diagonalized to obtain the two-nucleon low-lying energies using variational methods as described below.

\begin{figure}[t]
   \includegraphics[width=\columnwidth]{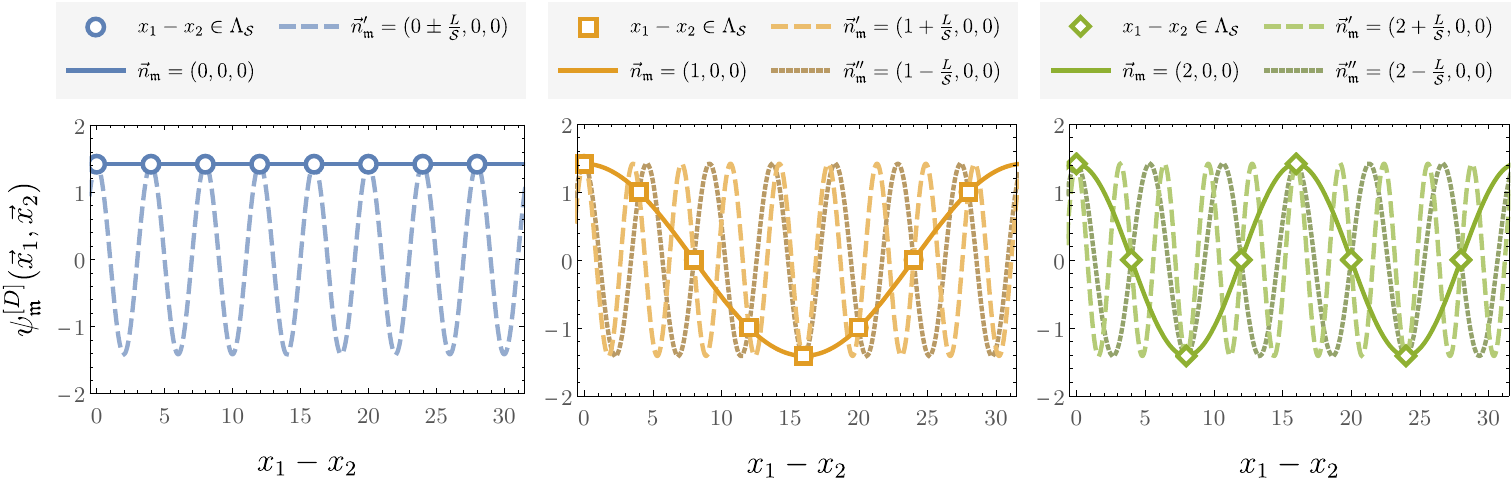}
   \caption{\label{fig:planewave_wvfs} Examples of wavefunctions demonstrating how sparsening leads to the coincidence on $\lsp$ of plane-wave dibaryon wavefunctions $\psi^{[D]}_\mathfrak{m}(\vec{x}_1,\vec{x}_2)$ with zero center-of-mass momenta and relative momenta proportional to $\vec{n}_{\mathfrak{m}}$ and $\vec{n}_{\mathfrak{m}} \pm \frac{L}{\mathcal{S}}\hat{e}_k$. Each momentum component can be analyzed independently, and for simplicity the relative position of the nucleons is taken to be $\vec{x}_1-\vec{x}_2 = (x_1-x_2,0,0)$ with momenta corresponding to $\vec{n}_{\mathfrak{m}}=(0,0,0)$, $\vec{n}_{\mathfrak{m}}=(1,0,0)$, and $\vec{n}_{\mathfrak{m}}=(2,0,0)$ shown as solid lines in the left, center, and right panels, respectively. Positions satisfying $\vec{x}_1-\vec{x}_2 \in \lsp$ are shown as open shapes. Analogous spatial wavefunctions for momenta $\vec{n}_{\mathfrak{m}} \pm \frac{L}{\mathcal{S}}\hat{e}_1$ that are identical to  $\psi^{[D]}_\mathfrak{m}(\vec{x}_1,\vec{x}_2)$ for $\vec{x}_1-\vec{x}_2 \in \lsp$ are shown as dashed and dotted lines (degenerate for the case $\vec{n}_{\mathfrak{m}}=(0,0,0)$). For concreteness, the values $L=32$ and $\mathcal{S}=4$ used in the numerical calculations in Sec.~\ref{sec:numerics} are chosen.}
\end{figure}

Sparsening leads to incomplete momentum projection in the 
sum over $\lsp$ in Eq.~\eqref{eq:Dweights}, as described for the single-nucleon sector in Sec.~\ref{sec:onehadron}. Specializing to $\vec{P}_{\mathfrak{m}}=\vec{0}$, dibaryon interpolating operators with $\vec{n}_{\mathfrak{m}}$ related by shifts of $\pm L/\mathcal{S}$ (and integer multiples of these) along any lattice axis are identical since $e^{i \left( \frac{2\pi}{L}\right) \left( \vec{n}_{\mathfrak{m}} \pm \frac{L}{\mathcal{S}}\hat{e}_k \right) \cdot (\vec{x}_1 - \vec{x}_2)} = e^{i \left( \frac{2\pi}{L}\right) \cdot \vec{n}_{\mathfrak{m}} (\vec{x}_1 - \vec{x}_2)}$ for $\vec{x}_1-\vec{x}_2 \in \lsp$, where $\hat{e}_k$ is a spatial unit vector. 
This equivalence is illustrated in Fig.~\ref{fig:planewave_wvfs} for particular examples of $\vec{n}_{\mathfrak{m}}$ with $L=32$ and $\mathcal{S}=4$, as relevant to the numerical calculations discussed in Sec.~\ref{sec:numerics}. 
For an interpolating-operator set including dibaryon operators with $\mathfrak{s} \leq K$, sparsening effects therefore lead to additional contamination from operators with $\mathfrak{s} > K$.
Assuming that $\mathcal{S} \ll L$ and $\mathfrak{s} \ll K$, the excited-state contamination from higher $\mathfrak{s}$-shell interpolating operators introduced by sparsening is suppressed in comparison to excited-state contamination from states strongly overlapping with interpolating operators with $\mathfrak{s}$ just above $K$. 
For the same example parameters, sparsening leads to the identification of the single-nucleon momentum vector $\vec{k}_{\mathfrak{m}} = (2,0,0)$ with $\vec{k}_{\mathfrak{m}} - \frac{L}{\mathcal{S}}\hat{e}_1 = (-6,0,0)$ as seen in Fig.~\ref{fig:planewave_wvfs}.
For non-interacting nucleons, this leads to an excited-state energy gap $\Delta E_{\mathcal{S}}^{(2,I,\Gamma_J)} = \sqrt{M_N^2 + 4 \left(\frac{2\pi}{L}\right)^2 } +  \sqrt{M_N^2 + 36 \left(\frac{2\pi}{L}\right)^2 } - 2M_N $ for the $I,\ \Gamma_J$ pairs that have overlap with dibaryon operators with $\mathfrak{s}=0$.
This is coincident with the non-interacting energy of states with $\mathfrak{s} = 20$ in the $M_N \rightarrow \infty$ limit (for the quark masses in Sec.~\ref{sec:numerics}, it is closest to the non-interacting energy with $\mathfrak{s} = 19$), and $t  \gg 1/\Delta E_{\mathcal{S}}^{(2,I,\Gamma_J)}$ is achievable in practical calculations as seen below.
For the choice of $K=6$ used in Sec.~\ref{sec:numerics},  non-interacting energy levels with $\mathfrak{s} = 8$  and many other relative-momentum shells with $20 > \mathfrak{s} > K$ will lead to excited-state contamination from states that at a given level of statistical precision are outside the subspace spanned by the interpolating-operator set and with smaller excitation energies than $\Delta E_{\mathcal{S}}^{(2,I,\Gamma_J)}$.
Excited-state effects arising from sparsening are therefore expected to be suppressed compared to other excited-state effects present in two-nucleon correlation functions and are not given any special significance below.

\begin{figure}[t]
    \includegraphics[width=\columnwidth]{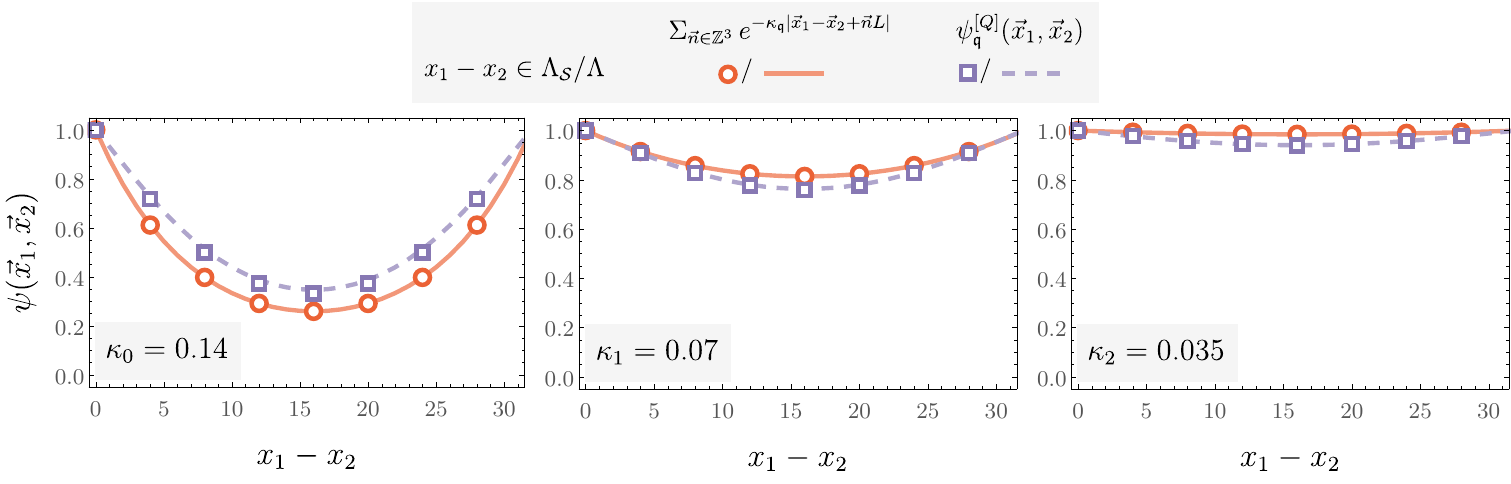}
   \caption{\label{fig:exponential_wvfs}Spatial wavefunctions $\psi^{[\kappaOp]}_\mathfrak{q}(\vec{x}_1,\vec{x}_2)$ associated with quasi-local interpolating operators with relative nucleon positions parallel to a lattice axis $\vec{x}_1-\vec{x}_2 = (x_1-x_2,0,0)$ and localization scales $\kappa_{\mathfrak{q}} \in \{ 0.14, 0.07, 0.035\}$ are shown in the left, center, and right panels, respectively. The open squares demonstrate the sparsened wavefunction with positions satisfying $\vec{x}_1-\vec{x}_2 \in \lsp$, while the dashed line is obtained by setting $\mathcal{S}=1$ to show sparsening effects. Spatial wavefunctions of the form $\sum_{\vec{n} \in \mathbb{Z}^3} e^{-\kappa_{\mathfrak{q}}|\vec{x}_1-\vec{x}_2 + \vec{n}L|}$ are also shown with solid lines and open circles for comparison. For concreteness, the figure corresponds to $L=32$, $\mathcal{S}=4$, and the values of $\kappa_{\mathfrak{q}}$ used in the numerical calculations in Sec.~\ref{sec:numerics}.}
\end{figure}

For large volumes, both plane-wave dibaryon operators and compact hexaquark operators may have small overlap with the loosely-bound deuteron state found in nature.
Within low-energy EFTs and phenomenological nuclear models  with nucleon degrees of freedom, the deuteron is described by a wavefunction that for large $|\vec{x}_1 - \vec{x}_2|$ and a cubic volume with PBCs is proportional to $\sum_{\vec{n} \in \mathbb{Z}^3} e^{-\kappa_d |\vec{x}_1 - \vec{x}_2 + \vec{n} L|}$ times a polynomial in $1/|\vec{x}_1-\vec{x}_2+\vec{n} L|$~\cite{Luscher:1986pf,Luscher:1990ux,Konig:2011ti,Briceno:2013bda}, where the deuteron binding momentum is $\kappa_d = \sqrt{M_N B_d}$ in terms of the nucleon mass, $M_N$, and the deuteron binding energy, $B_d$.
However, interpolating operators proportional to $ \sum_{\vec{n} \in \mathbb{Z}^3} e^{-\kappa_d |\vec{x}_1 - \vec{x}_2 + \vec{n}L|}$ do not factorize into products of functions of $\vec{x}_1$ and $\vec{x}_2$.\footnote{Using fast-Fourier transform (FFT) techniques, correlation functions built from such interpolating operators could be computed using $(\vsp\ln \vsp)^2$ operations. An FFT approach may be useful for calculations of two-nucleon correlation functions, although it has less favorable scaling for $B>2$ systems than the baryon-block methods discussed in Sec.~\ref{sec:contractions} and Ref.~\cite{Detmold:2012eu}.}
In Sec.~\ref{sec:contractions}, the factorization of $\psi_\mathfrak{m}^{[D]}(\vec{x}_1,\vec{x}_2)$ into a sum of two products of functions of $\vec{x}_1$ and $\vec{x}_2$ shown in Eq.~\eqref{eq:psiDdef} is exploited using baryon-block algorithms that efficiently compute $C^{[D,D]}_{\rho\mathfrak{m}g\rho^\prime\mathfrak{m}^\prime g'}(t)$ using $\mathcal{O}(\vsp^3)$ operations.
``Quasi-local'' interpolating operators that can be efficiently computed using baryon-block algorithms are defined by
\begin{equation}
\begin{split}
    \kappaOp_{\rho\mathfrak{q}g}(t) &= \sum_{\vec{x}_1,\vec{x}_2 \in \lsp} \psi_{\mathfrak{q}}^{[\kappaOp]}(\vec{x}_1,\vec{x}_2,\vec{R}) \sum_{\sigma,\sigma'} v^\rho_{\sigma \sigma^\prime}\, \frac{1}{\sqrt{2}}\left[ p_{\sigma g}(\vec{x}_1, t) n_{\sigma' g}(\vec{x}_2, t) \right. \\
    &\hspace{180pt} \left. + (-1)^{1-\delta_{\rho 0}} n_{\sigma g}(\vec{x}_1, t) p_{\sigma' g}(\vec{x}_2, t)\right], 
\end{split}\label{eq:Einterp}
\end{equation}
with wavefunctions
\begin{equation}
  \psi_{\mathfrak{q}}^{[\kappaOp]}(\vec{x}_1,\vec{x}_2,\vec{R})  = \frac{1}{\vsp} \sum_{\tau \in \mathbb{T}_{\mathcal{S}}} e^{ - \kappa_{\mathfrak{q}} |\tau(\vec{x}_1) - \vec{R}|}e^{ - \kappa_{\mathfrak{q}} |\tau(\vec{x}_2) - \vec{R}|},
  \label{eq:Edef}
\end{equation}
where $\mathfrak{q}$ labels the various localization scales $\kappa_\mathfrak{q}$ included in an interpolating-operator set, $\vec{R}$ is an arbitrary parameter\footnote{Quasi-local wavefunctions are invariant under shifts of $\vec{R}$ by $\mathcal{S}\hat{e}_k$ but depend on $\vec{R} \Mod{\mathcal{S}}$. } specifying the location of the center of the two-nucleon system in the lattice volume (before translation averaging over $\lsp$), and $\mathbb{T}_{\mathcal{S}}$ is the set of translations by multiples of the sparse lattice spacing, which are defined to act on coordinate vectors by $\tau(\vec{x}) = \vec{x} + \mathcal{S}(\tau_1 \hat{e}_1 + \tau_2 \hat{e}_2 + \tau_3 \hat{e_3})$ with $\tau_1,\tau_2,\tau_3 \in \{0,\ldots, L/\mathcal{S} - 1\}$, with each component of $\tau(\vec{x})$ defined modulo $L$ to respect PBCs.
Quark-level spin-color weights for quasi-local interpolating operators are identical to $w_\alpha^{[D]\rho}$, defined in Appendix~\ref{app:weights}.
The sum over translations in Eq.~\eqref{eq:Edef} introduces correlations between the positions of the nucleons and leads to an entangled two-nucleon wavefunction describing a pair of nucleons exponentially localized around a common point.
Applying the same translation to $\vec{x}_1$ and $\vec{x}_2$ in Eq.~\eqref{eq:Edef} ensures that the wavefunction is independent of $\vec{x}_1 + \vec{x}_2$ and therefore has definite center-of-mass momentum $\vec{P}_{\mathfrak{q}} = \vec{0} \pmod{\frac{2\pi}{\mathcal{S}}}$.
These quasi-local wavefunctions are qualitatively similar although quantitatively different\footnote{Since $\sum_{\vec{n} \in \mathbb{Z}^3} e^{-\kappa_{\mathfrak{q}}|\vec{x}_1-\vec{x}_2 + \vec{n} L|}$ is not an exact description of the FV QCD two-nucleon wavefunction,  maximizing the quantitative similarity of an interpolating operator wavefunction with this expression does not guarantee maximal overlap with loosely bound two-nucleon systems in QCD.} from the periodic EFT expectation $\sum_{\vec{n} \in \mathbb{Z}^3} e^{-\kappa_{\mathfrak{q}}|\vec{x}_1-\vec{x}_2 + \vec{n}L|}$, as illustrated in Fig.~\ref{fig:exponential_wvfs}. 
Linear combinations of $\psi_{\mathfrak{q}}^{[\kappaOp]}(\vec{x}_1,\vec{x}_2,\vec{R})$ with different $\kappa_{\mathfrak{q}}$ can be used to construct more general wavefunctions for quasi-local two-nucleon systems.
Quasi-local wavefunctions $\psi_{\mathfrak{q}}^{[\kappaOp]}(\vec{x}_1,\vec{x}_2,\vec{R})$  are linearly independent from a truncated set of dibaryon wavefunctions $\psi_\mathfrak{m}^{[D]}(\vec{x}_1,\vec{x}_2)$ with $\mathfrak{s} = |\vec{n}_{\mathfrak{m}}|^2 < 3L^2$. They can be included in a variational interpolating-operator set in an attempt to describe loosely bound states, with spatially correlated pairs of nucleons, more efficiently than a set including only dibaryon operators. 
Quasi-local interpolating operators therefore provide a well-motivated extension to a set of dibaryon operators (that approximately describe unbound two-nucleon systems) and hexaquark interpolating operators (that approximately describe tightly bound two-nucleon systems).

Two-nucleon correlation functions using this interpolating-operator set are defined by
\begin{equation}
\begin{split}
  C^{[\mathcal{T}, \mathcal{T}^\prime]}_{\rho \mathfrak{t} s \rho^\prime \mathfrak{t}^\prime g' }(t) &= \left< \mathcal{T}_{\rho \mathfrak{t} s}(t) \left( \mathcal{T}^\prime_{\rho^\prime \mathfrak{t}^\prime  g'}\right)^\dagger \!\! (0) \right>,
  \end{split}  \label{eq:C2Ndef}
\end{equation}
for all $\mathcal{T}\in\lbrace H,D,\kappaOp \rbrace$ with corresponding wavefunction indices $\mathfrak{t}, \mathfrak{t}' \in \lbrace \mathfrak{c},\mathfrak{m},\mathfrak{q} \rbrace$. 
Calculations of the correlation-function matrix with elements given by Eq.~\eqref{eq:C2Ndef} generalize previous LQCD calculations in the two-nucleon sector including positive-definite $[D,D]$ correlation functions of the form $\left< D_{\rho \mathfrak{m}g}(t) D_{\rho^\prime \mathfrak{m}^\prime  g'}^\dagger (0) \right>$, which have been recently studied in Refs.~\cite{Francis:2018qch,Horz:2020zvv,Green:2021qol}, as well as asymmetric $[D,H]$ correlation functions of the form $\left< D_{\rho \mathfrak{m}g}(t) H_{\rho^\prime \mathfrak{c}^\prime  g'}^\dagger (0) \right>$, which have identical structure (up to differences in quark-field smearing) to the correlation functions studied in Refs.~\cite{Beane:2010hg,Beane:2011iw,Beane:2012vq,Yamazaki:2012hi,Beane:2013br,Berkowitz:2015eaa,Beane:2015yha,Chang:2015qxa,Detmold:2015daa,Yamazaki:2015asa,Parreno:2016fwu,Savage:2016kon,Shanahan:2017bgi,Tiburzi:2017iux,Wagman:2017tmp,Winter:2017bfs,Chang:2017eiq,Detmold:2020snb}.

A modified form of Eq.~\eqref{eq:C2Ndef} is required for calculating correlation functions involving quasi-local operators using generalized baryon-block algorithms that assume factorizability of two-nucleon spatial wavefunctions.
The sum in Eq.~\eqref{eq:Edef} projects $\kappaOp_{\rho \mathfrak{q}g}$ to total momentum $0 \Mod{2\pi/\mathcal{S}}$ while introducing correlations in the two-nucleon wavefunctions between the positions of the two quasi-local nucleon interpolating operators.
The same sum prevents correlation functions involving $\kappaOp_{\rho\mathfrak{q}g}$ from factorizing into products of single-nucleon wavefunctions; however, it is possible to approximate such correlation functions using factorized wavefunctions by relying on ensemble averaging to impose translational invariance.
A ``factorized'' quasi-local interpolating operator can be defined as
\begin{equation}
\begin{split}
 F_{\rho\mathfrak{q}g}(t)  &= \sum_{\vec{x}_1,\vec{x}_2 \in \lsp} \psi_{\mathfrak{q}}^{[F]}(\vec{x}_1,\vec{x}_2,\vec{R})   \sum_{\sigma,\sigma'} v^\rho_{\sigma \sigma^\prime}\, \frac{1}{\sqrt{2}}\left[ p_{\sigma g}(\vec{x}_1, t) n_{\sigma' g}(\vec{x}_2, t) \right. \\
  &\hspace{180pt} \left. + (-1)^{1-\delta_{\rho 0}}  n_{\sigma g}(\vec{x}_1, t) p_{\sigma' g}(\vec{x}_2, t)\right], 
  \end{split}\label{eq:Finterp}
  \end{equation}
  where
  \begin{equation}
  \psi_{\mathfrak{q}}^{[F]}(\vec{x}_1,\vec{x}_2,\vec{R})  =   e^{ - \kappa_{\mathfrak{q}} |\vec{x}_1 - \vec{R}|}e^{ - \kappa_{\mathfrak{q}} |\vec{x}_2 - \vec{R} |}.
  \label{eq:Fdef}
    \end{equation}
The quasi-local wavefunction $\psi_{\mathfrak{q}}^{[\kappaOp]}(\vec{x}_1,\vec{x}_2,\vec{R})$ can be obtained by averaging the factorized wavefunction  $\psi_{\mathfrak{q}}^{[F]}(\vec{x}_1,\vec{x}_2,\vec{R})$ over sparse-lattice translations,
\begin{equation}
  \psi_{\mathfrak{q}}^{[\kappaOp]}(\vec{x}_1,\vec{x}_2,\vec{R})  = \frac{1}{\vsp} \sum_{\tau \in \mathbb{T}_{\mathcal{S}}} \psi_\mathfrak{q}^{[F]}(\tau(\vec{x}_1),\tau(\vec{x}_2)) .
\end{equation}
Translation invariance of (gauge-field-averaged) correlation functions therefore implies that correlation functions involving quasi-local interpolating operators can be computed using factorized quasi-local sources,
\begin{equation}
\begin{split}
  C^{[\mathcal{T}, \kappaOp]}_{\rho \mathfrak{t} g \rho^\prime \mathfrak{q}^\prime g' }(t) &= \left< \mathcal{T}_{\rho \mathfrak{t} g}(t) \kappaOp_{\rho^\prime \mathfrak{q}^\prime  g'}^\dagger (0) \right> = \left< \mathcal{T}_{\rho \mathfrak{t} g}(t) F_{\rho^\prime \mathfrak{q}^\prime  g'}^\dagger (0) \right>,
  \end{split}  \label{eq:CEEdef}
\end{equation}
for all $\mathcal{T}\in\lbrace H,D,\kappaOp \rbrace$.
Use of factorized interpolating operators and Eq.~\eqref{eq:CEEdef} allows the full correlation-function matrix given in Eq.~\eqref{eq:C2Ndef} to be efficiently computed using local and bilocal baryon blocks as described in the next section.

\subsection{Contraction algorithm}\label{sec:contractions}
Quark propagators, defined as
\begin{equation}
  S_{gg'}^{ij}(\vec{x},t;\vec{y},0) = \left< q_g^i(\vec{x},t) \overline{q}_{g'}^j(\vec{y},0) \right>, \label{eq:propdef}
\end{equation}
are computed by solving the Dirac equation associated with the LQCD quark action (using the Dirac operator with support on the entire lattice geometry $\Lambda \times \{0,\ldots,T -1\}$, where $T$ is the length of the Euclidean time direction) for a set of Gaussian-smeared quark sources located at each sparse lattice site $\vec{y} \in \lsp$ for some chosen source timeslice(s).
In order to construct positive-definite Hermitian correlation-function matrices, the same set of smearings used at the source is applied to the solution at each sink point $\vec{x} \in \Lambda$ and the resulting sink-smeared propagators are subsequently restricted to $\vec{x} \in \lsp$.
Proton correlation functions can be expressed in terms of a sum over permutations of products of quark propagators using Wick's theorem as
\begin{equation}
\begin{split}
  C^{[p,p]}_{\sigma 0g\sigma' 0 g'}(t) &=  - \sum_{\vec{x},\vec{y} \in \lsp} \psi^{[N]}_{0}(\vec{x})  \left\lbrace \psi^{[N]}_{0}(\vec{y}) \right\rbrace^* \sum_{\alpha,\alpha'} w_\alpha^{[N]\sigma} w_{\alpha'}^{[N]\sigma'} \\
    &\hspace{20pt} \times \left< u_{g}^{i(\alpha)}(\vec{x},t) d_{g}^{j(\alpha)}(\vec{x},t) u_{g}^{k(\alpha)}(\vec{x},t) \overline{u}_{g'}^{i'(\alpha')}(\vec{y},0) \overline{d}_{g'}^{j'(\alpha')}(\vec{y},0) \overline{u}_{g'}^{k'(\alpha')}(\vec{y},0) \right>   \\
    &= \sum_{\vec{x},\vec{y} \in \lsp} \psi^{[N]}_{0}(\vec{x})  \left\lbrace \psi^{[N]}_{0}(\vec{y}) \right\rbrace^* \sum_{\alpha,\alpha'} w_\alpha^{[N]\sigma} w_{\alpha'}^{[N]\sigma'} \sum_{\mathcal{P} \in P^{[N]}} \text{sign}(\mathcal{P}) \\
    &\hspace{20pt} \times S_{gg'}^{\mathcal{P}[i(\alpha)] i'(\alpha')}\left( \vec{x}, t; \vec{y}, 0\right) S_{gg'}^{\mathcal{P}[j(\alpha)] j'(\alpha')}\left(\vec{x}, t;\vec{y}, 0\right) S_{gg'}^{\mathcal{P}[k(\alpha)] k'(\alpha')}\left(\vec{x}, t;\vec{y}, 0\right),
\end{split}\label{eq:N-Ncontract}
\end{equation}
where $\mathfrak{c}=\mathfrak{c}'=0$ corresponds to the center-of-mass frame $\vec{P}_{\mathfrak{c}} = \vec{P}_{\mathfrak{c}'} = \vec{0}$ used throughout this section for simplicity.\footnote{Generalizations of the results in this section to non-zero center-of-mass momentum are straightforward. Calculations of dibaryon interpolating operators with $\vec{P}_{\mathfrak{m}} \neq 0$ can reduce computational cost by reusing baryon blocks between calculations with different $\vec{P}_{\mathfrak{m}}$.} The nucleon quark permutations $\mathcal{P} \in P^{[N]}$ act on spin-color index functions and are given in Cauchy's two-line notation by
\begin{equation}
    P^{[N]} = \left\lbrace \begin{pmatrix} i & j & k \\ i & j & k \end{pmatrix}, \begin{pmatrix} i & j & k \\ k & j & i \end{pmatrix}  \right\rbrace.
\end{equation}
Neutron correlation functions are obtained by exchanging $u \leftrightarrow d$ in Eq.~\eqref{eq:N-Ncontract} and are identical in the isospin-symmetric limit.
Hexaquark correlation functions are similarly given by Wick's theorem as
\begin{equation}
\begin{split}
    C^{[H,H]}_{\rho 0g \rho' 0 g'}(t) &= \sum_{\vec{x}, \vec{y} \in \lsp} \psi^{[H]}_{0}(\vec{x})  \left\lbrace \psi_{0}^{[H]}(\vec{y}) \right\rbrace^* \sum_{\alpha,\alpha'} w_\alpha^{[H]\rho} w_{\alpha'}^{[H]\rho'} \sum_{\mathcal{P} \in P^{[pn]}} \text{sign}(\mathcal{P})   \\
    &\hspace{5pt} \times S_{gg'}^{\mathcal{P}[i(\alpha)] i'(\alpha')}\left(\vec{x}, t;\vec{y}, 0\right) S_{gg'}^{\mathcal{P}[j(\alpha)] j'(\alpha')}\left(\vec{x}, t;\vec{y}, 0\right) S_{gg'}^{\mathcal{P}[k(\alpha)] k'(\alpha')}\left(\vec{x}, t;\vec{y}, 0\right)  \\
    &\hspace{5pt} \times S_{gg'}^{\mathcal{P}[l(\alpha)] l'(\alpha')}\left(\vec{x}, t;\vec{y}, 0\right) S_{gg'}^{\mathcal{P}[m(\alpha)] m'(\alpha')}\left(\vec{x}, t;\vec{y}, 0\right) S_{gg'}^{\mathcal{P}[n(\alpha)] n'(\alpha')}\left(\vec{x}, t;\vec{y}, 0\right),
\end{split}\label{eq:H-Hcontract}
\end{equation}
where the 36 $pn$ quark permutations are given by
\begin{equation}
    P^{[pn]} = \left\lbrace \begin{pmatrix} i & j & k & l & m & n \\ \mathcal{P}_u(i) & \mathcal{P}_d(j) & \mathcal{P}_u(k) & \mathcal{P}_d(l) & \mathcal{P}_u(m) & \mathcal{P}_d(n)  \end{pmatrix} \ \bigg| \ \mathcal{P}_u, \mathcal{P}_d \in S_3 \right\rbrace,
\end{equation}
where $S_3$ denotes the symmetric group acting over a set of three elements.
Both Eq.~\eqref{eq:N-Ncontract} and Eq.~\eqref{eq:H-Hcontract} can be evaluated by computing sums with $\mathcal{O}(\vsp^2)$ terms for each $t$ for which correlation functions are evaluated.
Linear dependence of the cost of evaluating correlation functions on the number of $t$  for which correlation functions are evaluated applies to all correlation functions considered in this work and will not be repeated below.

\begin{figure}[t]
  \includegraphics[width=0.49\textwidth]{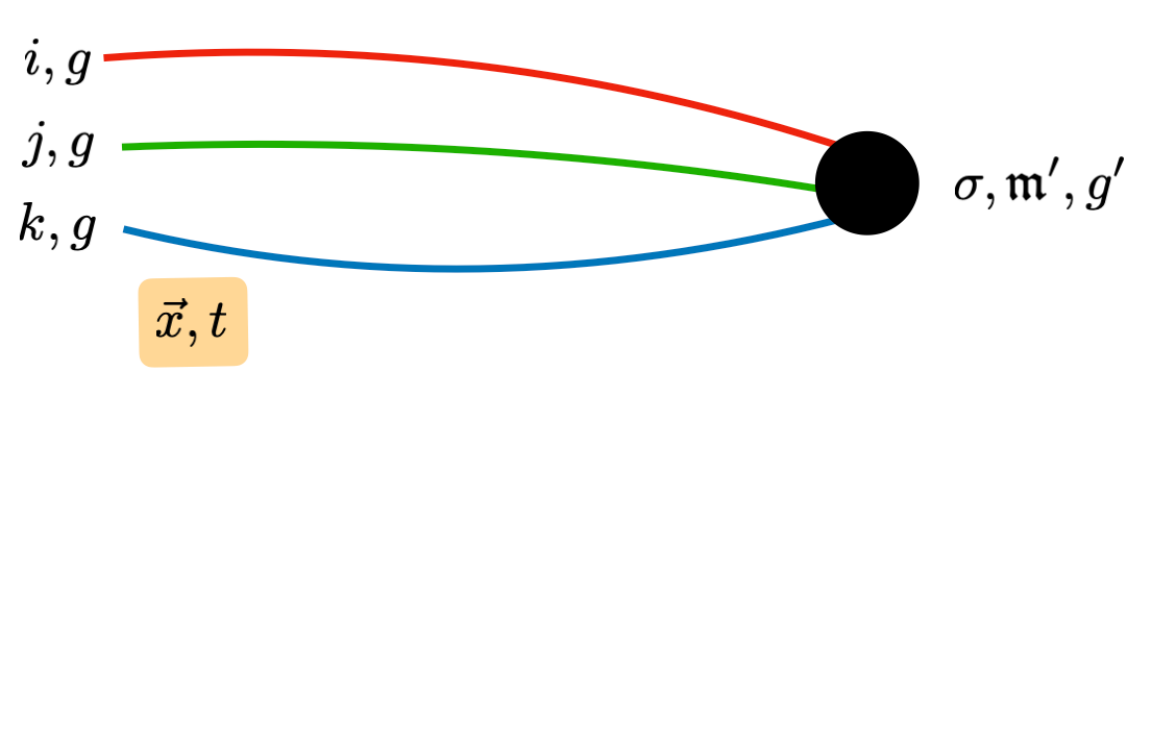}
   \includegraphics[width=0.49\textwidth]{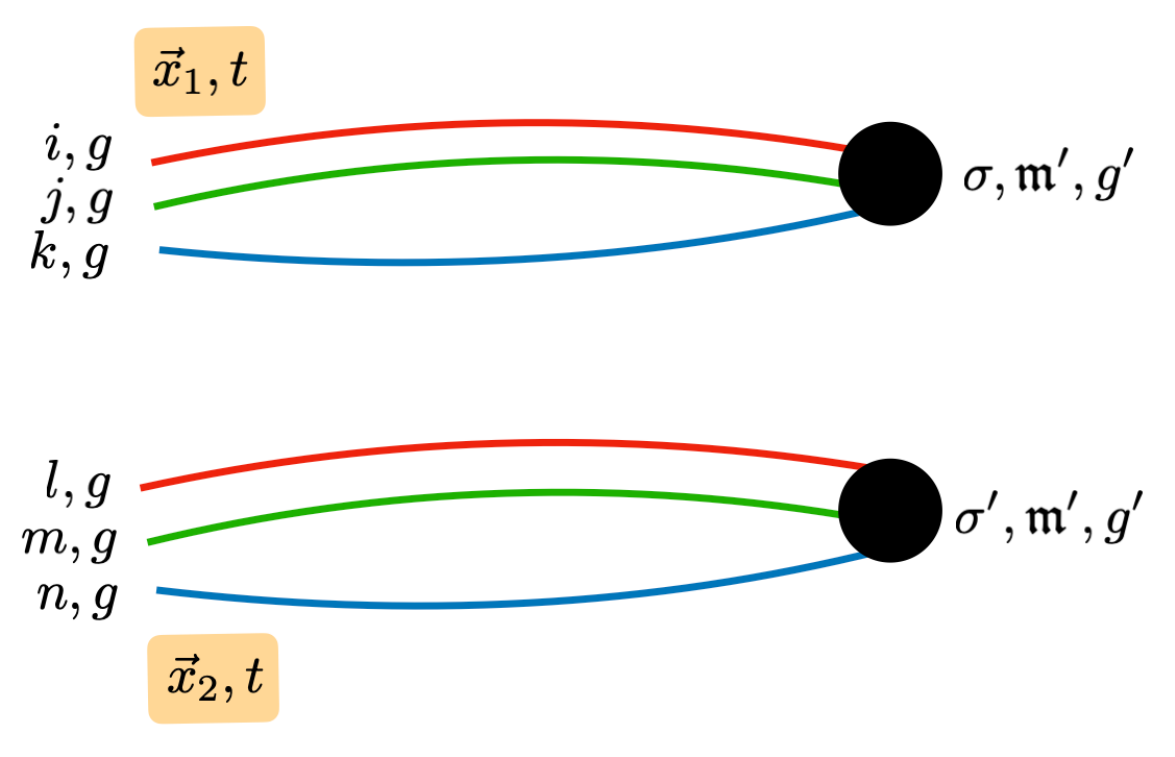}\\
   (a) \hspace*{7cm} (b)
   \caption{\label{fig:contraction_diagrams} Diagrams illustrating (a) the tensor structure of local baryon blocks and (b) contributions to Eq.~\eqref{eq:H-NNcontract2} involving such blocks. Diagram (a) describes $\mathcal{B}^{(\pm)ijk}_{g\sigma \mathfrak{m}'g'}(\vec{x},t)$, which includes a wavefunction $e^{\pm i \vec{k}_{\mathfrak{m}'} \cdot \vec{y}}$, weights $w_{\alpha'}^{[N]\sigma}$, and sums over spatial points $\vec{y} \in \lsp$ and weight index $\alpha'$, all collectively denoted by the black circle. Free indices $\sigma$, $\mathfrak{m}'$, and $g'$ label the source baryon spin, wavefunction momentum, and quark-field smearing, respectively. Free spin-color indices $i,j,k$ are associated with the sink position $\vec{x}$, and $g$ denotes the sink quark-field smearing. Diagram (b) similarly illustrates the product $\mathcal{B}^{(\pm)ijk}_{g\sigma \mathfrak{m}'g'}(\vec{x}_1,t) \mathcal{B}^{(\mp)lmn}_{g\sigma' \mathfrak{m}'g'}(\vec{x}_2,t)$ appearing in  Eq.~\eqref{eq:NN-NNblocks}.}
\end{figure}

Correlation functions involving a hexaquark sink and a dibaryon source are analogously given by
\begin{equation}
\begin{split}
  C^{[H,D]}_{\rho 0g \rho' \mathfrak{m}'g'}(t) &= \sum_{\vec{x}, \vec{y}_1, \vec{y}_2 \in \lsp} \psi^{[H]}_{0}(\vec{x})  \left\lbrace \psi_{\mathfrak{m}'}^{[D]}(\vec{y}_1,\vec{y}_2) \right\rbrace^* \sum_{\alpha,\alpha'} w_\alpha^{[H]\rho} w_{\alpha'}^{[D]\rho'} \sum_{\mathcal{P} \in P^{[pn]}} \text{sign}(\mathcal{P})   \\
    &\hspace{5pt} \times S_{gg'}^{\mathcal{P}[i(\alpha)] i'(\alpha')}\left(\vec{x}, t;\vec{y}_1, 0\right) S_{gg'}^{\mathcal{P}[j(\alpha)] j'(\alpha')}\left(\vec{x}, t;\vec{y}_1, 0\right) S_{gg'}^{\mathcal{P}[k(\alpha)] k'(\alpha')}\left(\vec{x}, t;\vec{y}_1, 0\right)  \\
    &\hspace{5pt} \times S_{gg'}^{\mathcal{P}[l(\alpha)] l'(\alpha')}\left(\vec{x}, t;\vec{y}_2, 0\right) S_{gg'}^{\mathcal{P}[m(\alpha)] m'(\alpha')}\left(\vec{x}, t;\vec{y}_2, 0\right) S_{gg'}^{\mathcal{P}[n(\alpha)] n'(\alpha')}\left(\vec{x}, t;\vec{y}_2, 0\right) .
\end{split}\label{eq:H-NNcontract}
\end{equation}
Direct evaluation of Eq.~\eqref{eq:H-NNcontract} requires evaluating $\mathcal{O}(\vsp^3 )$ propagator products; however, for factorizable two-nucleon wavefunctions, it is possible to reduce this  by introducing local baryon blocks, defined by
\begin{equation}
\begin{split}
    \mathcal{B}^{ (\pm)ijk}_{g\sigma \mathfrak{m}'g'}(\vec{x},t) &= \sum_{\alpha'} w^{[N]\sigma}_{\alpha'} \sum_{\vec{y} \in \lsp} e^{\pm i \vec{k}_{\mathfrak{m}'} \cdot \vec{y}}  \\
    &\hspace{20pt} \times S_{gg'}^{i i'(\alpha')}\left(\vec{x},t;\vec{y},0\right) S_{gg'}^{jj'(\alpha')}\left(\vec{x},t;\vec{y},0\right) S_{gg'}^{kk'(\alpha')}\left(\vec{x},t;\vec{y},0\right).
\end{split}\label{eq:blockdef}
\end{equation}
The tensor structure of these local baryon blocks is illustrated in Fig.~\ref{fig:contraction_diagrams}.
Evaluating these local baryon blocks for all $\vec{x} \in \lsp$ requires $\mathcal{O}(\vsp^2 )$ operations.
The $[H,D]$ correlation functions in Eq.~\eqref{eq:H-NNcontract} can be expressed in terms of local baryon blocks using the relation between dibaryon weights $w_\alpha^{[D]\rho}$ and products of nucleon weights $w_\alpha^{[N]\sigma}$ detailed in Appendix~\ref{app:weights}, and the tensor $v^\rho_{\sigma\sigma'}$ defined in Eq.~\eqref{eq:veights}.
The baryon-block components appearing in each term in the sum depend on the permutation index $\mathcal{P}$.
The required components can be denoted by
\begin{equation}
    \mathcal{B}^{(\pm,\mathcal{P},\alpha)ijk}_{g\sigma  \mathfrak{m}'g'}(\vec{x},t) \equiv \mathcal{B}^{(\pm)\mathcal{P}[i(\alpha)]\mathcal{P}[j(\alpha)]\mathcal{P}[k(\alpha)]}_{g\sigma  \mathfrak{m}'g'}(\vec{x},t),
\label{eq:localshort}
\end{equation}
in terms of which Eq.~\eqref{eq:H-NNcontract} can be re-expressed as
\begin{equation}
\begin{split}
  C^{[H,D]}_{\rho 0g \rho' \mathfrak{m}'g'}(t) &= \sum_{\vec{x} \in \lsp} \psi^{[H]}_{0}(\vec{x})   \sum_{\alpha} w_\alpha^{[H]\rho}  \sum_{\mathcal{P} \in P^{[pn]}} \text{sign}(\mathcal{P})  \sum_{\sigma,\sigma'} \frac{1}{\sqrt{2}}  v_{\sigma\sigma'}^{\rho'} \\
    &\hspace{20pt} \times  \left\lbrace  \mathcal{B}^{(+,\mathcal{P},\alpha)ijk}_{g\sigma  \mathfrak{m}'g'}(\vec{x},t) \mathcal{B}^{(-,\mathcal{P},\alpha)lmn}_{g\sigma' \mathfrak{m}'g'}(\vec{x},t) + \mathcal{B}^{(-,\mathcal{P},\alpha)ijk}_{g\sigma  \mathfrak{m}'g'}(\vec{x},t) \mathcal{B}^{(+,\mathcal{P},\alpha)lmn}_{g\sigma' \mathfrak{m}'g'}(\vec{x},t) \right\rbrace.
\end{split}\label{eq:H-NNcontract2}
\end{equation}
After determining the local baryon blocks, Eq.~\eqref{eq:H-NNcontract2} can be evaluated in $\mathcal{O}(\vsp )$ operations.
Analogous results can be derived for $[D,H]$ correlation functions by defining baryon blocks with the spatial sum performed at the sink instead of the source in Eq.~\eqref{eq:blockdef}.
Similarly, results can be derived for $[H,F]$ and $[F,H]$ correlation functions (which are equivalent to $[H,\kappaOp]$ and $[\kappaOp, H]$ correlation functions on ensemble average) by defining baryon blocks with $e^{\pm i k_{\mathfrak{m}} \cdot \vec{y}}$ in Eq.~\eqref{eq:blockdef} replaced by $e^{-\kappa_{\mathfrak{q}}|\vec{y} - \vec{R}|}$.

\begin{figure}[t]
   \includegraphics[width=0.49\textwidth]{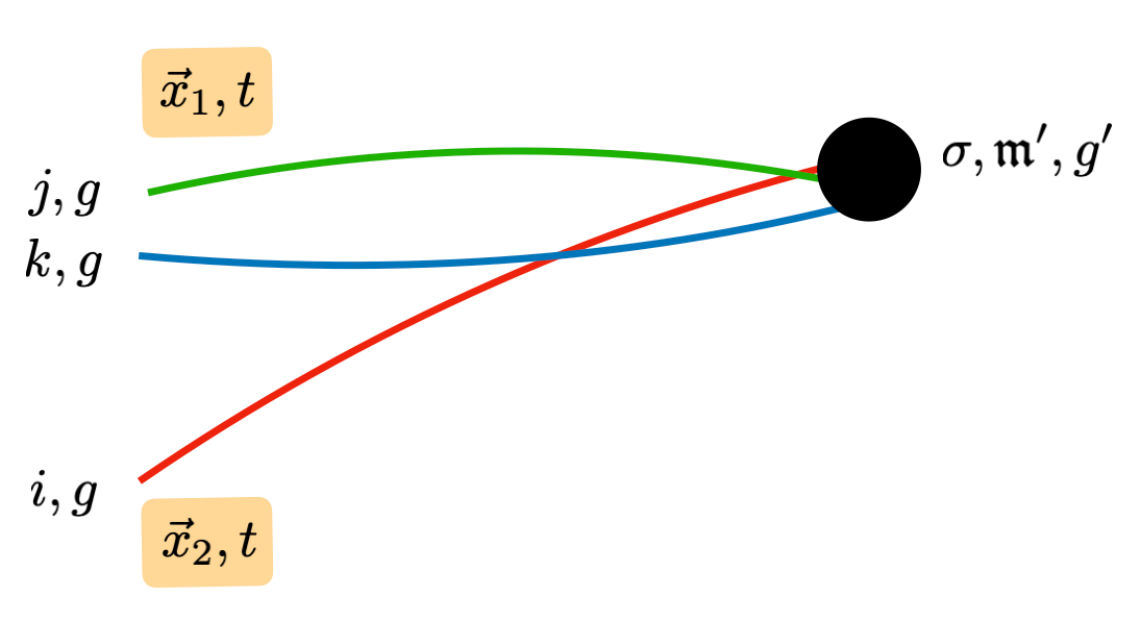}
   \includegraphics[width=0.49\textwidth]{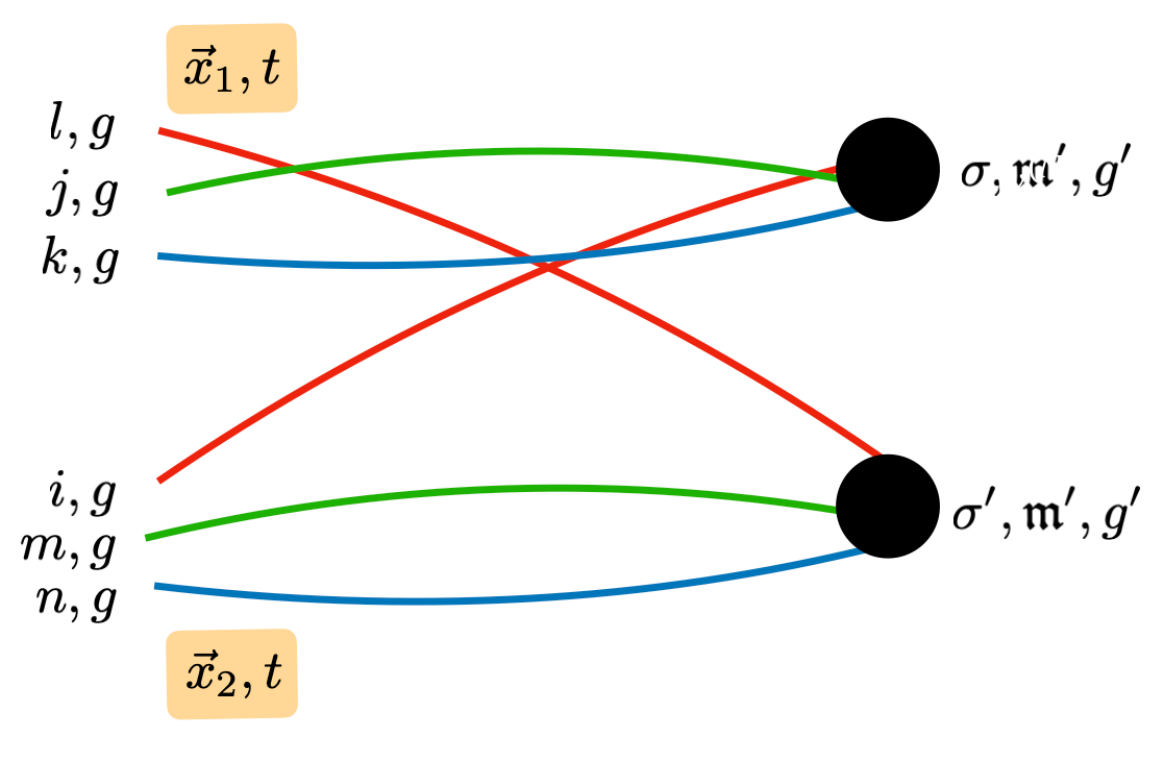}\\
   (a) \hspace*{7cm} (b)
   \caption{\label{fig:contraction_diagrams2} Diagrams illustrating (a) the tensor structure of bilocal baryon blocks and (b) contributions to Eq.~\eqref{eq:NN-NNblocks} involving them. Diagram (a) describes $\mathcal{E}^{(q\pm)ijk}_{g\sigma \mathfrak{m}'g'}(\vec{x}_1,\vec{x}_2,t)$ with $q=1$, which includes a wavefunction $e^{\pm i \vec{k}_{\mathfrak{m}'} \cdot \vec{y}}$, weights $w_{\alpha'}^{[N]\sigma}$, and sums over spatial points $\vec{y} \in \lsp$ and weight index $\alpha'$ collectively denoted by the black circle. Free indices $\sigma$, $\mathfrak{m}'$, and $g'$ label the source baryon spin, wavefunction momentum, and quark-field smearing, respectively. Free spin-color indices $i$ and $j,k$ are associated with the sink positions $\vec{x}_2$ and $\vec{x}_1$ respectively (other choices of $q$ have different spin-color indices associated with $\vec{x}_2$), and $g$ denotes the sink quark-field smearing. Diagram (b) similarly illustrates the product $\mathcal{E}^{(1 \pm)ijk}_{g\sigma \mathfrak{m}'g'}(\vec{x}_1,\vec{x}_2,t) \mathcal{E}^{(1 \mp)lmn}_{g\sigma' \mathfrak{m}'g'}(\vec{x}_2,\vec{x}_1,t)$ appearing in  Eq.~\eqref{eq:NN-NNblocks}.}
\end{figure}

Correlation functions with dibaryon sources and sinks involve the same quark permutations and an additional sum over sparse lattice sites,
\begin{equation}
  \begin{split}
    C^{[D,D]}_{\rho \mathfrak{m}g \rho' \mathfrak{m}'g'}(t) &= \sum_{\vec{x}_1, \vec{x}_2, \vec{y}_1, \vec{y}_2 \in \lsp} \psi^{[D]}_{\mathfrak{m}}(\vec{x}_1,\vec{x}_2)  \left\lbrace \psi_{\mathfrak{m}'}^{[D]}(\vec{y}_1,\vec{y}_2) \right\rbrace^* \sum_{\alpha,\alpha'} w_\alpha^{[D]\rho} w_{\alpha'}^{[D]\rho'} \sum_{\mathcal{P} \in P^{[pn]}} \text{sign}(\mathcal{P})   \\
    &\hspace{10pt} \times S_{gg'}^{\mathcal{P}[i(\alpha)] i'(\alpha')}\left(\vec{x}_{b_1(\mathcal{P})}, t;\vec{y}_1, 0\right) S_{gg'}^{\mathcal{P}[j(\alpha)] j'(\alpha')}\left(\vec{x}_{b_2(\mathcal{P})}, t;\vec{y}_1, 0\right)  \\
    &\hspace{10pt} \times S_{gg'}^{\mathcal{P}[k(\alpha)] k'(\alpha')}\left(\vec{x}_{b_3(\mathcal{P})}, t;\vec{y}_1, 0\right) S_{gg'}^{\mathcal{P}[l(\alpha)] l'(\alpha')}\left(\vec{x}_{b_4(\mathcal{P})}, t;\vec{y}_2, 0\right)  \\
    &\hspace{10pt} \times  S_{gg'}^{\mathcal{P}[m(\alpha)] m'(\alpha')}\left(\vec{x}_{b_5(\mathcal{P})}, t;\vec{y}_2, 0\right) S_{gg'}^{\mathcal{P}[n(\alpha)] n'(\alpha')}\left(\vec{x}_{b_6(\mathcal{P})}, t;\vec{y}_2, 0\right),
\end{split}\label{eq:NN-NNcontract}
\end{equation}
where $b_q(\mathcal{P}) \in \{1,2\}$ indexes the spatial position where the sink connected to the $q$-th source quark is located in permutation $\mathcal{P}$.
Direct evaluation of Eq.~\eqref{eq:NN-NNcontract} requires $\mathcal{O}(\vsp^4 )$ products, which can be prohibitive in practice. In particular, it can dominate the $\mathcal{O}(\vsp V )$ cost of calculating timeslice-to-all quark propagators on $\lsp$ by orders of magnitude.
The computational cost can be reduced by using a combination of the local baryon blocks discussed above and bilocal baryon blocks required for permutations in which quarks are exchanged between the baryons, defined by
\begin{equation}
\begin{split}
    \mathcal{E}^{(q\pm)ijk}_{g\sigma \mathfrak{m}'g'}(\vec{x}_1,\vec{x}_2,t) &= \sum_{\alpha'} w_{\alpha'}^{[N]\sigma} \sum_{\vec{y}\in \lsp} e^{\pm i \vec{k}_{\mathfrak{m}'} \cdot \vec{y}}\  S_{gg'}^{ii'(\alpha')}\left(\vec{x}_{1+\delta_{q,1}},t;\vec{y},0\right)  \\
    &\hspace{20pt} \times   S_{gg'}^{jj'(\alpha')}\left(\vec{x}_{1+\delta_{q,2}},t;\vec{y},0\right) S_{gg'}^{kk'(\alpha')}\left(\vec{x}_{1+\delta_{q,3}},t;\vec{y},0\right).
\end{split} \label{eq:exblock1}
\end{equation}
The notation $\vec{x}_{1+\delta_{q,r}}$ denotes $\vec{x}_2$ if $q=r$ and $\vec{x}_1$ if $q \neq r$, and therefore $\mathcal{E}^{(q\pm)ijk}_{g\sigma \mathfrak{m}'g'}(\vec{x}_1,\vec{x}_2,t)$ includes two quark propagators connected to the sink at $\vec{x}_1$ and one unpaired quark propagator connected to the sink at $\vec{x}_2$.
The position of the source quark field connected to the unpaired sink quark field at $\vec{x}_2$ is denoted by $q \in \{1,2,3\}$. Figure \ref{fig:contraction_diagrams2} shows the structure of these bilocal blocks.
The components of bilocal baryon blocks appearing for each permutation can be denoted analogously to Eq.~\eqref{eq:localshort} as
\begin{equation}
    \mathcal{E}^{(q\pm,\mathcal{P},\alpha)ijk}_{g\sigma  \mathfrak{m}'g'}(\vec{x}_1,\vec{x}_2,t) \equiv \mathcal{E}^{(q\pm)\mathcal{P}[i(\alpha)]\mathcal{P}[j(\alpha)]\mathcal{P}[k(\alpha)]}_{g\sigma  \mathfrak{m}'g'}(\vec{x}_1,\vec{x}_2,t).
\label{eq:bilocalshort}
\end{equation}
The dibaryon-dibaryon correlation functions given in Eq.~\eqref{eq:NN-NNcontract} can then be represented in terms of local and bilocal baryon blocks as
\begin{equation}
  \begin{split}
  &C^{[D,D]}_{\rho \mathfrak{m}g \rho' \mathfrak{m}'g'}(t) = \sum_{\vec{x}_1, \vec{x}_2 \in \lsp} \psi_{\mathfrak{m}}^{[D]}(\vec{x}_1,\vec{x}_2) \sum_{\alpha} w_\alpha^{[D]\rho} \sum_{\sigma,\sigma'}  \frac{1}{\sqrt{2}}  v^{\rho'}_{\sigma \sigma'}  \\
&\hspace{15pt}  \times  \left\lbrace \sum_{\mathcal{P} \in P^{[pn]}_{\rm ex0}} \text{sign}(\mathcal{P}) \left[ \mathcal{B}^{(+,\mathcal{P},\alpha)i j k}_{g\sigma \mathfrak{m}'g'}(\vec{x}_1,t) \mathcal{B}^{(-,\mathcal{P},\alpha)lmn }_{g\sigma' \mathfrak{m}'  g'}(\vec{x}_2,t) \right. \right. \\
    &\hspace{140pt} \left. + \mathcal{B}^{(-,\mathcal{P},\alpha)i j k}_{g\sigma \mathfrak{m}'g'}(\vec{x}_1,t) \mathcal{B}^{(+,\mathcal{P},\alpha)lmn }_{g\sigma' \mathfrak{m}'  g'}(\vec{x}_2,t) \right]  \\
    &\hspace{40pt} + \sum_{\mathcal{P} \in P^{[pn]}_{\rm ex1}} \text{sign}(\mathcal{P}) \left[ \mathcal{E}^{(q_{\mathcal{P}}^1 +,\mathcal{P},\alpha)ijk}_{g\sigma \mathfrak{m}'g'}(\vec{x}_1,\vec{x}_2,t) \mathcal{E}^{(q_{\mathcal{P}}^2-,\mathcal{P},\alpha)lmn }_{g\sigma' \mathfrak{m}'g'}(\vec{x}_2,\vec{x}_1,t) \right. \\
    &\hspace{140pt}  \left. +  \mathcal{E}^{(q_{\mathcal{P}}^2-,\mathcal{P},\alpha)ijk}_{g\sigma \mathfrak{m}'g'}(\vec{x}_1,\vec{x}_2,t) \mathcal{E}^{(q_{\mathcal{P}}^1 +,\mathcal{P},\alpha)lmn }_{g\sigma' \mathfrak{m}'g'}(\vec{x}_2,\vec{x}_1,t) \right] \\
    &\hspace{40pt} + \sum_{\mathcal{P} \in P^{[pn]}_{\rm ex2}} \text{sign}(\mathcal{P}) \left[ \mathcal{E}^{(q_{\mathcal{P}}^1 +,\mathcal{P},\alpha)ijk}_{g\sigma \mathfrak{m}'g'}(\vec{x}_2,\vec{x}_1,t) \mathcal{E}^{(q_{\mathcal{P}}^2-,\mathcal{P},\alpha)lmn }_{g\sigma' \mathfrak{m}'g'}(\vec{x}_1,\vec{x}_2,t) \right. \\
    &\hspace{140pt} \left. \left. + \mathcal{E}^{(q_{\mathcal{P}}^2-,\mathcal{P},\alpha)ijk}_{g\sigma \mathfrak{m}'g'}(\vec{x}_2,\vec{x}_1,t) \mathcal{E}^{(q_{\mathcal{P}}^1 +,\mathcal{P},\alpha)lmn }_{g\sigma' \mathfrak{m}'g'}(\vec{x}_1,\vec{x}_2,t) \right] \vphantom{\sum_{\mathcal{P} \in P^{[pn]}_{\rm ex0}}} \right\rbrace,
\end{split}\label{eq:NN-NNblocks}
\end{equation}
where $P^{[pn]}_{\rm ex0}$, $P^{[pn]}_{\rm ex1}$, $P^{[pn]}_{\rm ex2} \subset P^{[pn]}$ are the sets of permutations in which zero, one, and two quarks respectively are exchanged between the source and sink baryons, and $q_{\mathcal{P}}^1, q_{\mathcal{P}}^2 \in \{1,2,3\}$ denote the positions of the quark fields within each source baryon (labeled by superscripts $1,2$) that  is connected to the unpaired quark field at the sink.
Note that the second-to-last line of Eq.~\eqref{eq:NN-NNblocks} includes $\mathcal{E}^{(q_{\mathcal{P}}^1 +,\mathcal{P},\alpha)ijk}_{g\sigma \mathfrak{m}'g'}(\vec{x}_1,\vec{x}_2,t)$, which has paired quark fields connected to $\vec{x}_1$ and an unpaired quark field $q_{\mathcal{P}}^1$ ``exchanged'' to $\vec{x}_2$, while the last line includes $\mathcal{E}^{(q_{\mathcal{P}}^1 +,\mathcal{P},\alpha)ijk}_{g\sigma \mathfrak{m}'g'}(\vec{x}_2,\vec{x}_1,t)$ which has two paired quark fields exchanged to $\vec{x}_2$ and an unpaired quark field $q_{\mathcal{P}}^1$ connected to $\vec{x}_1$.
Results for $[\kappaOp, D]$ correlation functions are obtained by replacing $\psi_{\mathfrak{m}}^{[D]}(\vec{x}_1,\vec{x}_2)$ with $\psi_{\mathfrak{q}}^{[\kappaOp]}(\vec{x}_1,\vec{x}_2,\vec{R})$ in Eq.~\eqref{eq:NN-NNblocks}.
Results for $[\kappaOp, F]$ and $[D,F]$ correlation functions (equivalent to $[\kappaOp, \kappaOp]$ and $[D, \kappaOp]$ correlation functions on ensemble average) are obtained by defining bilocal baryon blocks with $e^{\pm i k_{\mathfrak{m}} \cdot \vec{y}}$ in Eq.~\eqref{eq:exblock1} replaced by $e^{-\kappa_{\mathfrak{q}}|\vec{y} - \vec{R}|}$.

Once the local and bilocal baryon blocks are constructed, Eq.~\eqref{eq:NN-NNblocks} can be computed by evaluating $\mathcal{O}(\vsp^2 )$ terms. 
Bilocal block construction, which requires evaluating $\mathcal{O}(\vsp^3 )$ terms, provides the dominant cost for large $\vsp$; however, block construction is linear in the number of interpolating operators used while evaluation of Eq.~\eqref{eq:NN-NNblocks} is quadratic.
When both the number of interpolating operators and $\vsp$ are large, this allows contractions to be performed significantly more efficiently than direct evaluation of Eq.~\eqref{eq:NN-NNcontract}, which is both quadratic in the number of interpolating operators and requires evaluating sums with $\mathcal{O}(\vsp^4 )$ terms.

\subsection{Projection to cubic irreps}\label{sec:FV}

Cubic symmetry implies that correlation-function matrices have a block diagonal decomposition, with blocks corresponding to each irrep of the cubic group that do not mix under (Euclidean) time evolution in LQCD.
Nucleon interpolating operators, $N_{\sigma \mathfrak{c}g}$, with $\vec{P}_{\mathfrak{c}}=\vec{0}$ transform in the $G_1^+$ irrep of the cubic group with rows indexed by $\sigma \in \{0,1\}$ and $C^{[N,N]}_{\sigma\mathfrak{c}g\sigma^\prime\mathfrak{c}^\prime g'}(t)$ is therefore already in this block-diagonal form.
Spin-averaged nucleon correlation-function matrices with definite $(B,I,\Gamma_J)$ quantum numbers are therefore given by
\begin{equation}
\begin{split}
    C^{\left(1,\frac{1}{2},G_1^+\right)}_{N_g N_{g'} }(t) &= \frac{1}{2} \sum_{\sigma} \left< N_{\sigma 0g}(t)  \overline{N}_{\sigma 0 g'} (0) \right>,
\end{split}\label{eq:CNNJz}
\end{equation}
where $N_g N_{g'}$ labels the sink and source interpolating-operator structures. On the right-hand-side  $\mathfrak{c}=\mathfrak{c}'=0$ corresponds to $\vec{P}_{\mathfrak{c}} = \vec{P}_{\mathfrak{c}'} = \vec{0}$.

Hexaquark interpolating operators $H_{\rho 0g}$ with zero center-of-mass momentum have wavefunctions that transform in the $A_1^+$ irrep and spin that transforms in the $A_1^+$ and $T_1^+$ irreps for $\rho=0$ and $\rho \in \{1,2,3\}$, respectively, as described in Sec.~\ref{sec:twohadron}.
Hexaquark interpolating operators with definite quantum numbers $H^{(B,I,\Gamma_J,J_z)}_g$ can therefore be identified as $H^{(2,1,A_1^+,0)}_g \equiv H_{0 0g}$ and $H^{(2,0,T_1^+,S_z(\rho))}_g \equiv H_{\rho 0g}$, where
\begin{equation}
  S_z(\rho) = \begin{cases} 1, & \rho=1 \\ 0, & \rho=2 \\ 3, & \rho=3 \end{cases} \label{eq:Szrho}
\end{equation}
is related to the eigenvalue $e^{i J_z \pi/2}$ of the cubic transformation\footnote{Note that for the cubic group, the conserved charge $J_z$ is only defined modulo 4 but is otherwise analogous to the continuum, infinite-volume quantum number $J_z$. In particular $S_z(3) = -1 = 3 \pmod 4$.} corresponding to a rotation by $\pi/2$ about the $z$-axis acting on states created by $\overline{H}^{(2,0,T_1^+,S_z(\rho))}_g$ with $J_z = S_z(\rho)$, since hexaquark operators have spatial wavefunctions with $A_1^+$ symmetry.
Hexaquark correlation-function matrices can be averaged over $J_z$ as
\begin{equation}
  C^{(2,I,\Gamma_J)}_{H_g H_{g'}} = \frac{1}{d_{\Gamma_J}} \sum_{J_z} \left< H_g^{(2,I,\Gamma_J,J_z)}(t) \left( H_{g'}^{(2,I,\Gamma_J,J_z)} \right)^\dagger\!\! (0) \right>,
\end{equation}
where $d_{\Gamma_J}$ is the dimension of the $\Gamma_J$ irrep and $H_g H_{g'}$ labels the sink and source hexaquark operators.

The quasi-local two-nucleon interpolating operators defined above have spatial wavefunctions transforming in the $A_1^+$ irrep after ensemble averaging, and quasi-local interpolating operators with definite quantum numbers $\kappaOp^{(2,I,\Gamma_J,J_z)}_{\mathfrak{q}g}$ can be similarly identified as $\kappaOp^{(2,1,A_1^+,0)}_{\mathfrak{q}g} \equiv \kappaOp_{0 \mathfrak{q}g}$ and $\kappaOp^{(2,0,T_1^+,S_z(\rho))}_{\mathfrak{q}g} \equiv \kappaOp_{\rho \mathfrak{q}g}$.
Analogous $J_z$-averaged correlation-function matrices can be defined for quasi-local two-nucleon interpolating operators as
\begin{equation}
    C^{(2,I,\Gamma_J)}_{\kappaOp_{\mathfrak{q}g} \kappaOp_{\mathfrak{q}'g'}} = \frac{1}{d_{\Gamma_J}} \sum_{J_z} \left< \kappaOp_{\mathfrak{q}g}^{(2,I,\Gamma_J,J_z)}(t) \left( \kappaOp_{\mathfrak{q}'g'}^{(2,I,\Gamma_J,J_z)} \right)^\dagger\!\!(0) \right>,
\end{equation}
where $\kappaOp_{\mathfrak{q}g} \kappaOp_{\mathfrak{q}'g'}$ labels the sink and source quasi-local interpolating operators.
Off-diagonal correlation-function matrix elements $C^{(2,I,\Gamma_J)}_{H_g \kappaOp_{\mathfrak{q}'g'}}$ and $C^{(2,I,\Gamma_J)}_{ \kappaOp_{\mathfrak{q}g} H_{g'}}$ are defined analogously.

Dibaryon interpolating-operator wavefunctions $\psi^{[D]}_{\mathfrak{m}}(\vec{x}_1,\vec{x}_2)$ defined in Eq.~\eqref{eq:psiDdef} do not transform irreducibly under the cubic group when either the relative or center-of-mass momentum of the two-nucleon system is non-zero.
The rest of this section summarizes the change-of-basis relation between the dibaryon interpolating operators of Eq.~\eqref{eq:Dinterp}, which are convenient for efficient computation of correlation-function matrices as described above, and an interpolating-operator set in which each operator transforms in a definite cubic irrep.
It is convenient to express the total cubic irrep $\Gamma_J$ as
\begin{equation}
  \Gamma_J = \Gamma_\ell \otimes \Gamma_S, \label{eq:JLS}
\end{equation}
where $\Gamma_S \in \{A_1^+,T_1^+\}$ labels the cubic irrep of the two-nucleon spin determined by $\rho$ and $\Gamma_\ell$ labels the cubic irrep of the two-nucleon spatial wavefunction.
One could instead project nucleon operators onto irreps of the little groups that leave $\pm \vec{k}_{\mathfrak{m}}$ invariant and then form two-nucleon operators from products of these operators as done for two-meson systems in Ref.~\cite{Thomas:2011rh}.
Since only the center-of-mass rest frame is considered here, it is simple to construct two-nucleon interpolating operators with definite $\Gamma_J$ from the product of two cubic irreps in Eq.~\eqref{eq:JLS} without introducing the little groups of $\pm \vec{k}_{\mathfrak{m}}$ as an intermediate step.

\begin{table}[!t]
\begin{ruledtabular}
\begin{tabular}{cccc}
            $\Gamma_J$     & $\Gamma_\ell$ & $\Gamma_S$      & $\ell$  \\\hline
            $A_1^+$   & $A_1^+ $ & $ A_1^+$  & $0,4,6,\ldots$   \\
            $E^+$     & $E^+    $ & $  A_1^+$  & $2,4,6,\ldots$   \\
            $T_2^+$   & $T_2^+  $ & $  A_1^+$  & $2,4,6,\ldots$   \\
            $T_1^+$   & $T_1^+  $ & $  A_1^+$  & $4,6,\ldots$   \\
            $A_2^+$   & $A_2^+  $ & $  A_1^+$  & $6,\ldots$   \\
\end{tabular}
\caption{ The cubic irreps for $I=1$ $NN$ systems with positive-parity spatial wavefunctions and their dominant partial-wave contributions in the infinite-volume limit. The left column shows the cubic irrep possible for the total-angular-momentum representation obtained from the spin-orbit product representation $\Gamma_J \subseteq \Gamma_\ell \otimes \Gamma_S$, the middle columns show the orbital-angular-momentum and spin irreps contributing to this irrep, and the right column shows the lowest orbital-angular-momentum representations associated with the infinite-volume behavior of states in each irrep.
\label{tab:I1reps}}
   \end{ruledtabular}
\end{table}

For $I=1$ (flavor-symmetric) dibaryon operators, positive-parity spatial wavefunctions can only be combined with antisymmetric spin wavefunctions while satisfying fermion antisymmetry, and they therefore have $\Gamma_S = A_1^+$. The dominant large-volume contribution associated with orbital angular momentum $\ell = 0$ states arises with the $\Gamma_\ell = A_1^+$ irrep and therefore has $\Gamma_J = A_1^+$. Other irreps include contributions from partial waves with $\ell \geq 2$ as summarized in Table~\ref{tab:I1reps}. For $I=0$ (flavor-antisymmetric) dibaryon operators, conversely, positive-parity spatial wavefunctions can only be combined with symmetric spin wavefunctions and therefore have $\Gamma_S = T_1^+$. In this case, spatial wavefunctions in the $A_1^+$ irrep (relevant for orbital angular momentum $\ell = 0$ states) lead to total-angular-momentum cubic irrep $\Gamma_J = A_1^+ \otimes T_1^+ = T_1^+$. This total-angular-momentum irrep also includes contributions from spatial wavefunctions in the $E^+$, $T_1^+$, and $T_2^+$ irreps, since each of these projects onto $T_1^+$ in the product representation $\Gamma_J = \Gamma_\ell \otimes T_1^+$. Other total-angular-momentum irreps similarly include contributions from several orbital angular momentum irreps as shown in Table~\ref{tab:I0reps}.

\begin{table}[!t]
\begin{ruledtabular}
\begin{tabular}{cccc} 
            $\Gamma_J$& $\Gamma_\ell$ & $\Gamma_S$      & $\ell$  \\\hline
            $T_1^+$   & $A_1^+ \oplus E^+ \oplus T_1^+ \oplus T_2^+ $ & $T_1^+$  & $0,2,4,6,\ldots$   \\
            $E^+$     & $T_1^+ \oplus T_2^+$ & $T_1^+$  & $2,4,6,\ldots$   \\
            $T_2^+$   & $A_2^+ \oplus E^+ \oplus T_1^+ \oplus T_2^+$ & $T_1^+$  & $2,4,6,\ldots$   \\
            $A_2^+$   & $T_2^+$ & $  T_1^+$  & $2,4,6,\ldots$   \\
            $A_1^+$   & $T_1^+$ & $  T_1^+$  & $4,6,\ldots$   \\
\end{tabular}
\caption{ The cubic irreps for $I=0$ $NN$ systems with positive-parity spatial wavefunctions and their dominant partial-wave contributions in the infinite-volume limit. As in Table~\ref{tab:I1reps}, the left column shows the cubic irrep possible for the total-angular-momentum representation obtained from the spin-orbit product representation $\Gamma_J \subseteq \Gamma_\ell \otimes \Gamma_S$, the middle columns show the orbital-angular-momentum and spin irreps contributing to this irrep, and the right column shows the lowest orbital-angular-momentum representations associated with the infinite-volume behavior of states in each irrep.
\label{tab:I0reps}}
   \end{ruledtabular}
\end{table}

The spatial wavefunctions transforming in each positive-parity cubic irrep can be denoted by $\psi^{[D](\Gamma_\ell,\ell_z)}_{\mathfrak{s} \mathfrak{k}}$, where $\Gamma_\ell$ labels the cubic irrep of the spatial wavefunction, $\ell_z$ labels the eigenvalue $e^{i \ell_z \pi/2}$ of a rotation by $\pi/2$ about the $z$-axis applied to $\left( \psi^{[D](\Gamma_\ell,\ell_z)}_{\mathfrak{s} \mathfrak{k}} \right)^*$ (or equivalently to states created by Hermitian conjugates of operators involving the spatial wavefunctions), $\mathfrak{s}$ labels the squared magnitude of the relative momentum, and $\mathfrak{k}$ runs from 1 to  $\mathcal{N}_{\mathfrak{s}}^{(1,\Gamma_\ell)}$ (the multiplicity of wavefunctions with this set of labels), as detailed below.
These wavefunctions are linear combinations of the plane-wave wavefunctions introduced above,
\begin{equation}
\begin{split}
    \psi^{[D](\Gamma_\ell,\ell_z)}_{\mathfrak{s} \mathfrak{k} }(\vec{x}_1,\vec{x}_2)  &= \sum_{\mathfrak{m}} G^{(\Gamma_\ell,\ell_z)}_{\mathfrak{s} \mathfrak{k}\mathfrak{m}}  \psi^{[D]}_{\mathfrak{m}}(\vec{x}_1,\vec{x}_2)\\
    &= \sum_{\mathfrak{m}} G^{(\Gamma_\ell,\ell_z)}_{\mathfrak{s} \mathfrak{k}\mathfrak{m}} \frac{1}{\sqrt{2}}\left[ e^{ i \left( \frac{2\pi}{L} \right)  \vec{n}_{\mathfrak{m}} \cdot (\vec{x}_1 - \vec{x}_2)} + e^{- i \left( \frac{2\pi}{L} \right)  \vec{n}_{\mathfrak{m}} \cdot (\vec{x}_1 - \vec{x}_2)} \right], \label{eq:cubicwvfdef}
\end{split}
\end{equation}
where the sums include non-zero contributions from $\vec{n}_{\mathfrak{m}}$ satisfying $\vec{n}_{\mathfrak{m}}\cdot\vec{n}_{\mathfrak{m}} = \mathfrak{s}$ and $G^{(\Gamma_\ell,\ell_z)}_{\mathfrak{s} \mathfrak{k}\mathfrak{m}}$ are coefficients projecting the wavefunction to definite cubic irreps. These coefficients for plane waves with relative momentum corresponding to $\mathfrak{s} \leq 6$ are presented and compared with results from Ref.~\cite{Luu:2011ep} in Appendix~\ref{app:FV}.

The coefficients $G^{(\Gamma_\ell,\ell_z)}_{\mathfrak{s} \mathfrak{k}\mathfrak{m}}$ can be used to define operators $D^{(2,I,\Gamma_J,J_z)}_{\mathfrak{s}\mathfrak{k}g}(t)$ with definite isospin ($I$), total angular momentum ($\Gamma_J$), and $J_z$ quantum numbers, relative-momentum shell indexed by $\mathfrak{s}$, multiplicity indexed by $\mathfrak{k}$, and smearing label $g$.
For $I=1$, these dibaryon operators are simply given by
\begin{equation}
    D^{(2,1,\Gamma_J,J_z)}_{\mathfrak{s}\mathfrak{k}g}(t) =   \sum_{\mathfrak{m}} G^{(\Gamma_J,J_z)}_{\mathfrak{s}  \mathfrak{k}\mathfrak{m}} D_{0\mathfrak{m}g}(t).
\label{eq:D1def}
\end{equation}
For the $I=0$ channel, interpolating operators with definite total angular momentum are obtained by taking products of the orbital wavefunctions in all the $\Gamma_\ell$ irreps shown in Table~\ref{tab:I0reps} for a given $\Gamma_J$ as
\begin{equation}
    D^{(2,0,\Gamma_J,J_z)}_{\mathfrak{s}\mathfrak{k}g}(t) = \sum_{\rho,\Gamma_\ell,\ell_z} \mathcal{C}^{(\Gamma_J,J_z,\Gamma_\ell,\ell_z)}_\rho \sum_{\mathfrak{k}^\prime} M^{(\Gamma_J,\Gamma_\ell)}_{\mathfrak{s}\mathfrak{k}\mathfrak{k}^\prime} \sum_{\mathfrak{m}} G^{(\Gamma_\ell,\ell_z)}_{\mathfrak{s} \mathfrak{k}^\prime \mathfrak{m}} D_{\rho\mathfrak{m}g}(t),
\label{eq:D0def}
\end{equation}
where the $\mathcal{C}^{(\Gamma_J,J_z,\Gamma_\ell,\ell_z)}_\rho$ are Clebsch-Gordan coefficients for $\Gamma_J = \Gamma_\ell \otimes \Gamma_S$ presented for example in Ref.~\cite{Basak:2005ir} and summarized in Appendix~\ref{app:FV}, and $\mathfrak{k}\in\{1,\ldots,\mathcal{N}_{\mathfrak{s}}^{(0,\Gamma_J)}\}$, where $\mathcal{N}_{\mathfrak{s}}^{(0,\Gamma_J)}$ is the total multiplicity of a given total-angular-momentum irrep in the spin-orbit product irrep for each $\mathfrak{s}$, which are shown for $\mathfrak{s} \leq 6$ in Table~\ref{tab:Imult}.
The multiplicity-label tensor $M^{(\Gamma_J,\Gamma_\ell)}_{\mathfrak{s}\mathfrak{k}\mathfrak{k}^\prime}$ converts from the individual multiplicity labels for each irrep to the total multiplicity label needed for the product.
As detailed in Appendix~\ref{app:FV}, $M^{(\Gamma_J,\Gamma_\ell)}_{\mathfrak{s}\mathfrak{k}\mathfrak{k}^\prime}$ with fixed $\mathfrak{s}$ and $\mathfrak{k}$ is equal to one for a single value of $\mathfrak{k}^\prime$ and equal to zero otherwise.

\begin{table}[!t]
\begin{ruledtabular}
\begin{tabular}{c|ccccc|ccccc} 
  $\mathfrak{s}$ & $\mathcal{N}_{\mathfrak{s}}^{(1,A_1^+)}$ & $\mathcal{N}_{\mathfrak{s}}^{(1,A_2^+)}$ & $\mathcal{N}_{\mathfrak{s}}^{(1,E^+)}$ & $\mathcal{N}_{\mathfrak{s}}^{(1,T_1^+)}$ & $\mathcal{N}_{\mathfrak{s}}^{(1,T_2^+)}$ & $\mathcal{N}_{\mathfrak{s}}^{(0,A_1^+)}$ & $\mathcal{N}_{\mathfrak{s}}^{(0,A_2^+)}$&  $\mathcal{N}_{\mathfrak{s}}^{(0,E^+)}$ & $\mathcal{N}_{\mathfrak{s}}^{(0,T_1^+)}$ & $\mathcal{N}_{\mathfrak{s}}^{(0,T_2^+)}$  \\\hline\hline
            $0$   & 1 & 0 & 0 & 0 & 0 &  0 & 0 & 0 & 1 & 0   \\\hline
            $1$   & 1 & 0 & 1 & 0 & 0 &  0 & 0 & 0 & 2 & 1   \\\hline
            $2$   & 1 & 0 & 1 & 0 & 1 &  0 & 1 & 1 & 3 & 2   \\\hline
            $3$   & 1 & 0 & 0 & 0 & 1 &  0 & 1 & 1 & 2 & 1   \\\hline
            $4$   & 1 & 0 & 1 & 0 & 0 &  0 & 0 & 0 & 2 & 1   \\\hline
            $5$   & 1 & 1 & 2 & 1 & 1 &  1 & 1 & 2 & 5 & 5  \\\hline
            $6$   & 1 & 0 & 1 & 1 & 2 &  1 & 2 & 3 & 5 & 4   \\\hline\hline
            Total   & 7 & 1 & 6 & 2 & 5 &  2 & 5 & 7 & 20 & 14   \\
\end{tabular}
  \caption{ Multiplicities $\mathcal{N}_{\mathfrak{s}}^{(I,\Gamma_J)}$ of linearly independent dibaryon interpolating operators with fixed $J_z$ (each independent spin-orbit product gives rise to operators in each row of $\Gamma_J$) and the total-isospin and total-angular-momentum irrep $(I,\Gamma_J)$ shown in each column and relative-momentum shell $\mathfrak{s}$ shown in each row.
\label{tab:Imult}}
   \end{ruledtabular}
\end{table}

Since correlation functions are independent of the $J_z$ quantum number upon ensemble averaging, $J_z$-averaged correlation functions can be defined as
\begin{equation}
\begin{split}
    C^{(2,I,\Gamma_J)}_{D_{\mathfrak{s}\mathfrak{k}g} D_{\mathfrak{s}'\mathfrak{k}'g'} }(t) &= \frac{1}{d_{\Gamma_J}} \sum_{J_z} \left< D_{\mathfrak{s}\mathfrak{k}g}^{(2,I,\Gamma_J,J_z)}(t) \left( D_{\mathfrak{s}'\mathfrak{k}'g'}^{(2,I,\Gamma_J,J_z)} \right)^\dagger\!\! (0) \right>,
\end{split}\label{eq:CDDJz}
\end{equation}
where $d_{\Gamma_J}$ is the dimension of $\Gamma_J$.
Off-diagonal correlation-function matrix elements involving dibaryon operators and either hexaquark or quasi-local operators can be defined analogously; for example the two-nucleon correlation functions with hexaquark sources and dibaryon sinks used in Refs.~\cite{Beane:2010hg,Beane:2011iw,Beane:2012vq,Yamazaki:2012hi,Beane:2013br,Berkowitz:2015eaa,Beane:2015yha,Chang:2015qxa,Detmold:2015daa,Yamazaki:2015asa,Parreno:2016fwu,Savage:2016kon,Shanahan:2017bgi,Tiburzi:2017iux,Wagman:2017tmp,Winter:2017bfs,Chang:2017eiq,Detmold:2020snb} are given by
\begin{equation}
    C^{(2,I,\Gamma_J)}_{D_{\mathfrak{s}\mathfrak{k}g} H_{g'} }(t) = \frac{1}{d_{\Gamma_J}} \sum_{J_z} \left< D_{\mathfrak{s}\mathfrak{k}g}^{(2,I,\Gamma_J,J_z)}(t) \left( H_{g'}^{(2,I,\Gamma_J,J_z)} \right)^\dagger\!\! (0) \right>.
\end{equation}

\subsection{Variational analysis of correlation functions}\label{sec:GEVP}

Correlation functions with baryon number $B$ and isospin $I$, projected to the cubic irrep $\Gamma_J$, and averaged over rows as above can generically be denoted $C_{\chi\chi'}^{(B,I,\Gamma_J)}(t)$, where $\chi$ and $\chi'$ denote the sink and source interpolating operators respectively.
For $B=1$, interpolating operators are of the form
$\{\chi,\chi'\} \in \{ N_g \}$, while for $B=2$, $\{\chi,\chi'\} \in \{  D_{\mathfrak{s}\mathfrak{k}g}^{(2,I,\Gamma_J)}, \  H_g^{(2,I,\Gamma_J)}, \ \kappaOp_{\mathfrak{q}g}^{(2,I,\Gamma_J)} \}$, where $J_z$ is averaged as described above and omitted from operator labels here and below.
Correlation function matrices have spectral representations
\begin{equation}
  C_{\chi \chi^\prime}^{(B,I,\Gamma_J)}(t) = \sum_{\mathsf{n}=0}^{\infty} Z_{\mathsf{n} \chi}^{(B,I,\Gamma_J)} \left(Z_{\mathsf{n} \chi'}^{(B,I,\Gamma_J)} \right)^* e^{- t E_{\mathsf{n}}^{(B,I,\Gamma_J)}},
\label{eq:spectral}
\end{equation}
where $E_{\mathsf{n}}^{(B,I,\Gamma_J)}$ is the energy of the ${\mathsf{n}}$-th QCD energy eigenstate with the quantum numbers indicated. Thermal effects are neglected,\footnote{Correlation-function fits are restricted to $t \leq \frac{3 T}{8}$ in order to avoid non-negligible thermal effects as discussed in Appendix~\ref{app:fits}.} and $Z_{\mathsf{n}\chi}^{(B,I,\Gamma_J)}$ describes the overlap of interpolating operator $\chi$ with this state, as in Eq.~\eqref{eq:nucSpec}.

Given a set $\mathbb{S}$ of $\mathcal{I}$ interpolating operators that have maximum overlaps with an equal number of energy eigenstates, it is possible to construct a set of approximately orthogonal interpolating operators that each dominantly overlap with a single energy eigenstate by solving the GEVP~\cite{Michael:1982gb,Luscher:1990ck},
\begin{equation}
    \sum_{\chi'} C_{\chi\chi'}^{(B,I,\Gamma_J)}(t) v^{(B,I,\Gamma_J,\mathbb{S})}_{{\mathsf{n}}\chi'}(t,t_0)  = \lambda_{\mathsf{n}}^{(B,I,\Gamma_J,\mathbb{S})}(t,t_0) \sum_{\chi'} C_{\chi\chi'}^{(B,I,\Gamma_J)}(t_0) v^{(B,I,\Gamma_J,\mathbb{S})}_{{\mathsf{n}}\chi'}(t,t_0),
\label{eq:GEVP}
\end{equation}
where $\mathsf{n} \in \{0,\ldots,\mathcal{I}-1\}$, $\lambda_{\mathsf{n}}^{(B,I,\Gamma_J,\mathbb{S})}(t,t_0)$ are the eigenvalues, $v^{(B,I,\Gamma_J,\mathbb{S})}_{{\mathsf{n}}\chi'}(t,t_0)$ are the eigenvectors, and $t_0$ is a reference time that can be for example a fixed $t$-independent value or a fixed fraction of $t$.
If the infinite sum in Eq.~\eqref{eq:spectral} can be approximately truncated to include $\mathcal{I}$ states, then the eigenvalues satisfy
\begin{equation}
  \lambda_{\mathsf{n}}^{(B,I,\Gamma_J,\mathbb{S})}(t,t_0) \approx e^{-(t - t_0) E_{\mathsf{n}}^{(B,I,\Gamma_J)}  }, \hspace{20pt} \mathsf{n} = 0,1,\ldots, \mathcal{I}-1 ,
\label{eq:GEVPeigenvalues}
\end{equation}
and the energy levels $E_{\mathsf{n}}^{(B,I,\Gamma_J)}$ can be obtained from fits to the $t$-dependence of the GEVP eigenvalues.
In general, truncating spectral representations to include $\mathcal{I}$ states may not be a good approximation, and it is important to understand how GEVP results are related to the energy spectrum without this approximation.

Correlation functions associated with a set of approximately orthogonal interpolating operators can be explicitly constructed using the GEVP eigenvectors as 
\begin{equation}
    \widehat{C}_{\mathsf{n}}^{(B,I,\Gamma_J,\mathbb{S})}(t) = \sum_{\chi\chi'} v_{{\mathsf{n}}\chi}^{(B,I,\Gamma_J,\mathbb{S})}(t_{\rm ref},t_0)^*  C_{\chi\chi'}^{(B,I,\Gamma_J)}(t) v^{(B,I,\Gamma_J,\mathbb{S})}_{{\mathsf{n}}\chi'}(t_{\rm ref},t_0),
\label{eq:GEVPcorrelators}
\end{equation}
where the eigenvectors are obtained from the solution to the GEVP Eq.~\eqref{eq:GEVP} with $t_0$ as shown and $t$ set equal to $t_{\rm ref}$ (the dependence of the left-hand side on these parameters is suppressed).
When analyzing the GEVP correlation functions   defined by Eq.~\eqref{eq:GEVPcorrelators}, the ideal scenario is that both $t_0$ and $t_{\rm ref}$ can be chosen large enough that contributions from states outside the subspace spanned by the interpolating-operator set can be neglected from  $v_{{\mathsf{n}}\chi}^{(B,I,\Gamma_J,\mathbb{S})}(t_{\rm ref},t_0)$.
If this can be achieved, then contributions from such states can be neglected from the energy spectrum obtained from fits to $\widehat{C}_{\mathsf{n}}^{(B,I,\Gamma_J,\mathbb{S})}(t)$.
However, in systems with small excitation energies, $\delta^{(B,I,\Gamma_J)} \equiv E_1^{(B,I,\Gamma_J)} - E_0^{(B,I,\Gamma_J)}$, this condition is difficult to achieve and requires the introduction of a large interpolating-operator set that has significant overlap with all states in a low-energy subspace of Hilbert space. 
The dependence of $\widehat{C}_{\mathsf{n}}^{(B,I,\Gamma_J,\mathbb{S})}(t)$ and associated fit results on $t_0$ and $t_{\rm ref}$ should be studied in numerical calculations in order to verify the stability of $v_{{\mathsf{n}}\chi}^{(B,I,\Gamma_J,\mathbb{S})}(t_{\rm ref},t_0)$ under these choices and assign systematic uncertainties if the dependence is not negligible.
The $t_0$ dependence of results can also be used to study excited-state effects as discussed in Ref.~\cite{Blossier:2009kd}.

Treating $t_0$ and $t_{\rm ref}$ as fixed parameters independent of $t$ guarantees that $\widehat{C}_{\mathsf{n}}^{(B,I,\Gamma_J,\mathbb{S})}(t)$ is a linear combination of LQCD correlation functions and therefore has a simple spectral representation (irrespective of the (in)completeness of the interpolating-operator set as a basis for energy eigenstates) that is given by
\begin{equation}
\begin{split}
    \widehat{C}_{\mathsf{n}}^{(B,I,\Gamma_J,\mathbb{S})}(t) &= \sum_{\mathsf{m}}  \left|\sum_{\chi} v_{{\mathsf{n}}\chi}^{(B,I,\Gamma_J,\mathbb{S})}(t_{\rm ref},t_0)^* Z_{\mathsf{m}\chi}^{(B,I,\Gamma_J)}\right|^2 e^{-tE_{\mathsf{m}}^{(B,I,\Gamma_J)} } \\ 
    &\equiv \sum_{{\mathsf{m}}}  \left|\widehat{Z}_{{\mathsf{n}}\mathsf{m}}^{(B,I,\Gamma_J,\mathbb{S})}(t_{\rm ref},t_0) \right|^2 e^{-t E_{\mathsf{m}}^{(B,I,\Gamma_J)} }. \\ 
\end{split}\label{eq:GEVPspectral}
\end{equation}
Fits of $\widehat{C}_{{\mathsf{n}}}^{(B,I,\Gamma_J,\mathbb{S})}(t)$ to positive-definite sums of exponentials can therefore be used to constrain the energy spectrum.
As will be discussed in detail below, the energy spectrum determined from fits to correlation-function results with finite $t$ depends on the choice of interpolating-operator set $\mathbb{S}$.
The extracted spectrum will therefore be denoted $E_{\mathsf{n}}^{(B,I,\Gamma_J,\mathbb{S})}$, and the $\mathbb{S}$-dependence of $E_{\mathsf{n}}^{(B,I,\Gamma_J,\mathbb{S})}$ and the associated systematic uncertainties involved in determining $E_{\mathsf{n}}^{(B,I,\Gamma_J)}$ using $E_{\mathsf{n}}^{(B,I,\Gamma_J,\mathbb{S})}$ are discussed at length in Sec.~\ref{sec:numerics}. 
The excitation energy gap determined using a particular interpolating-operator set will similarly be denoted $\delta^{(B,I,\Gamma_J,\mathbb{S})} \equiv E_1^{(B,I,\Gamma_J,\mathbb{S})} - E_0^{(B,I,\Gamma_J,\mathbb{S})}$.

If the interpolating-operator set was a basis for the full Hilbert space, then orthogonality of the GEVP eigenvectors would imply that $\widehat{Z}_{\mathsf{n}\mathsf{m}}^{(B,I,\Gamma_J,\mathbb{S})}(t_{\rm ref},t_0)$ are the components of a diagonal matrix.
In calculations with a finite interpolating-operator set of size $\mathcal{I}$ overlapping with the $\mathcal{I}$ lowest energy eigenstates, this orthogonality should approximately hold within the subspace spanned by the interpolating-operator set. 
In the limit $t\rightarrow \infty$, it can be shown that excited-state effects are exponentially suppressed by $e^{-t \left[ E_{\mathcal{I}}^{(B,I,\Gamma_J)} - E_{{\mathsf{n}}}^{(B,I,\Gamma_J)} \right] }$ provided that the interpolating operators considered are not effectively orthogonal (at a given statistical precision) to any of the lowest $\mathcal{I}$ energy eigenstates and $t_0$ is chosen to be sufficiently large~\cite{Luscher:1990ck,Blossier:2009kd}. 
This allows variational methods to achieve exponential suppression of excited-state effects on ground-state energy determinations with a suppression scale $E_{\mathcal{I}}^{(B,I,\Gamma_J)} - E_{0}^{(B,I,\Gamma_J)}$ that can be made much larger than the excitation energy $\delta^{(B,I,\Gamma_J)} \equiv E_1^{(B,I,\Gamma_J)} - E_0^{(B,I,\Gamma_J)}$  that controls the size of excited-state effects for individual correlation functions with asymptotically large $t$.
However, it is noteworthy that achieving excited-state suppression of the form $e^{-t \left[ E_{\mathcal{I}}^{(B,I,\Gamma_J)} - E_{{\mathsf{n}}}^{(B,I,\Gamma_J)} \right] }$ at finite $t$ requires that the overlap of the interpolating operators onto the lowest $\mathcal{I}$ levels is not too small\footnote{The precise condition depends on the structure of the energy spectrum. A simple model is discussed in Sec.~\ref{sec:I1} below in which the ground state has overlap $\epsilon \ll 1$ with interpolating operators dominantly overlapping with excited states that are separated from the ground state by a gap $\Delta$. In this model, $\epsilon^2 > e^{-t \Delta}$ is required for exponential excited-state suppression.   }  compared to the overlaps with higher-energy states. 
In practical application of the variation method, contamination can be from states much lower in the spectrum that the interpolating-operator set is only weakly coupled to (including the ground state), see Sec.~\ref{sec:numerics}.

In order to study the finite-$t$ behavior of correlation functions, effective energies can be constructed as
\begin{equation}
  E_{\chi\chi'}^{(B,I,\Gamma_J)}(t) =  \ln \left( \frac{ C_{\chi\chi'}^{(B,I,\Gamma_J)}(t) }{ C_{\chi \chi'}^{(B,I,\Gamma_J)}(t+1) } \right), \label{eq:Echichip}
\end{equation}
which approach $E_0^{(B,I,\Gamma_J)}$ for $t\rightarrow \infty$ and include additional contributions from excited states at finite $t$.
Effective energies can also be constructed from the GEVP correlation functions
\begin{equation}
    E_{\mathsf{n}}^{(B,I,\Gamma_J,\mathbb{S})}(t) =  \ln \left( \frac{ \widehat{C}_{\mathsf{n}}^{(B,I,\Gamma_J,\mathbb{S})}(t) }{ \widehat{C}_{\mathsf{n}}^{(B,I,\Gamma_J,\mathbb{S})}(t+1) } \right),
\label{eq:GEVPEM}
\end{equation}
which for large but finite $t$ are equal to $E_{\mathsf{n}}^{(B,I,\Gamma_J)}$ up to corrections from states outside the subspace spanned by the interpolating-operator set considered.
It follows from the positivity of the spectral representation in Eq.~\eqref{eq:GEVPspectral} that $E_{0}^{(B,I,\Gamma_J,\mathbb{S})}(t)~\geq~E_{0}^{(B,I,\Gamma_J)}$, and in this sense GEVP solutions provide a variational method for bounding the ground-state energy~\cite{Berg:1981zb,Ishikawa:1982tb}.
Applying analogous arguments to the subspaces orthogonal to states $\mathsf{m}$ with $\mathsf{m} < \mathsf{n}$ shows that $E_{\mathsf{n}}^{(B,I,\Gamma_J,\mathbb{S})}(t)~\geq~E_{\mathsf{n}}^{(B,I,\Gamma_J)}$.
For $B=2$ systems, it is also useful to form correlated differences of effective energies with twice the nucleon ground-state effective energy:
\begin{equation}
    \Delta E_{\mathsf{n}}^{(2,I,\Gamma_J,\mathbb{S})}(t) = E_{\mathsf{n}}^{(2,I,\Gamma_J,\mathbb{S})}(t) - 2 E_0^{(1,\frac{1}{2},G_1^+,\mathbb{S}')}(t),
\label{eq:GEVPDeltaEM}
\end{equation}
where $\mathbb{S}'$ labels the interpolating-operator set used in the single-nucleon sector (note that $\Delta E_{\mathsf{n}}^{(2,I,\Gamma_J,\mathbb{S})}(t)$ implicitly depends on $\mathbb{S}'$ as well as $\mathbb{S}$).
These correlated differences involve ratios of correlation functions; these ratios do not share the convexity of individual correlation functions and do not provide variational bounds.
Correlated differences between fit results for $E_{\mathsf{n}}^{(B,I,\Gamma_J,\mathbb{S})}$ can be defined analogously,
\begin{equation}
    \Delta E_{\mathsf{n}}^{(2,I,\Gamma_J,\mathbb{S})} = E_{\mathsf{n}}^{(2,I,\Gamma_J,\mathbb{S})} - 2 E_0^{(1,\frac{1}{2},G_1^+,\mathbb{S}')}.
\label{eq:GEVPDeltaM}
\end{equation}
Below, fits are performed to individual one- and two-nucleon correlation functions rather than to the ratios entering $\Delta E_{\mathsf{n}}^{(2,I,\Gamma_J,\mathbb{S})}(t)$, and the effective energies of each correlation function provide variational bounds that are consistent with the (not strictly variational) results of multi-state fits.
Results for $\Delta E_{\mathsf{n}}^{(2,I,\Gamma_J,\mathbb{S})}$ are not strictly variational because the single-nucleon ground-state energy could be over-estimated; however, the statistical and systematic uncertainties on the fitted single-nucleon mass are much smaller than the corresponding uncertainties in two-nucleon energies below and for simplicity results for $\Delta E_{\mathsf{n}}^{(2,I,\Gamma_J,\mathbb{S})}$ are interpreted as variational bounds below.
These FV energy shifts can be used to constrain infinite-volume scattering amplitudes through quantization conditions~\cite{Luscher:1986pf, Luscher:1990ux, Rummukainen:1995vs, Beane:2003da, Kim:2005gf, He:2005ey,  Davoudi:2011md, Leskovec:2012gb, Hansen:2012tf, Briceno:2012yi, Gockeler:2012yj, Briceno:2013lba, Feng:2004ua, Lee:2017igf, Bedaque:2004kc, Luu:2011ep, Briceno:2013hya, Briceno:2013bda}, as discussed in Sec.~\ref{sec:summary}.
For large but finite $t$, effective FV energy shifts defined by Eq.~\eqref{eq:GEVPDeltaEM} are equal to the FV energy shifts up to corrections from states outsides the subspace spanned by the interpolating-operator set.

It is also possible to define effective energies from the GEVP eigenvalues analogously as $\ln \left( \frac{ \lambda_{\mathsf{n}}^{(B,I,\Gamma_J,\mathbb{S})}(t,t_0) }{ \lambda_{\mathsf{n}}^{(B,I,\Gamma_J,\mathbb{S})}(t+1,t_0) } \right)$.
The effective energy based on Eq.~\eqref{eq:GEVPcorrelators} has the advantage that smooth $t$ dependence of the form of Eq.~\eqref{eq:GEVPspectral} is guaranteed even when $\mathcal{I}$ is much smaller than the dimension of the Hilbert space as discussed in Refs.~\cite{Bulava:2010yg,Bulava:2016mks}, and this definition is used for all results in the main text.
Both effective-energy definitions are compared for $B=1$ and $B=2$ systems in Appendix~\ref{app:plots} and found to give consistent results, although more precise results are obtained using Eq.~\eqref{eq:GEVPcorrelators} and Eq.~\eqref{eq:GEVPEM} in cases where there are closely spaced energy levels.

Given a determination of the GEVP energy levels from the GEVP correlation functions or eigenvalues, the eigenvectors can be used to determine the corresponding overlap factors from Eq.~\eqref{eq:spectral}  as~\cite{Bulava:2016mks}
\begin{equation}
    Z_{\mathsf{n}\chi}^{(B,I,\Gamma_J,\mathbb{S})} = \frac{ \sum_{\chi'} v^{(B,I,\Gamma_J,\mathbb{S})}_{{\mathsf{n}}\chi'}(t_{\rm ref},t_0) C^{(B,I,\Gamma_J,\mathbb{S})}_{\chi \chi'}(t_{\rm ref}) }{e^{-(t_0/2)E^{(B,I,\Gamma_J,\mathbb{S})}_{\mathsf{n}} }  \sqrt{ \widehat{C}_{{\mathsf{n}}}^{(B,I,\Gamma_J,\mathbb{S})}(t_{\rm ref}) }}.
\label{eq:GEVPeigenvectors}
\end{equation}
The relative contributions of each interpolating operator $\chi$ in the original set to a particular GEVP eigenstate $\mathsf{n}$ can be obtained as 
\begin{equation}
    \mathcal{Z}_{\mathsf{n}\chi}^{(B,I,\Gamma_J,\mathbb{S})} =  \frac{ \left| Z_{\mathsf{n}\chi}^{(B,I,\Gamma_J,\mathbb{S})} \right|} { \sum_{\chi'}  \left| Z_{\mathsf{n}\chi'}^{(B,I,\Gamma_J,\mathbb{S})} \right| } .
\label{eq:GEVPZrelative}
\end{equation}
GEVP results for the energies and overlap factors can also be used to reconstruct an estimate of the original set of correlation functions through Eq.~\eqref{eq:spectral}.
The relative contributions of each GEVP eigenstate $\mathsf{n}$ to the real part of the correlation function with interpolating operators $\chi$ and $\chi'$ is given by 
\begin{equation}
  \widetilde{\mathcal{Z}}_{\mathsf{n}\chi\chi'}^{(B,I,\Gamma_J,\mathbb{S})} = \frac{ \text{Re}\left[ Z_{\mathsf{n}\chi}^{(B,I,\Gamma_J,\mathbb{S})} \left(Z_{\mathsf{n}\chi'}^{(B,I,\Gamma_J,\mathbb{S})}\right)^* \right]}{ \sum_{\mathsf{m}} \left| Z_{\mathsf{m}\chi}^{(B,I,\Gamma_J,\mathbb{S})} \left(Z_{\mathsf{m}\chi'}^{(B,I,\Gamma_J,\mathbb{S})}\right)^* \right| }.
\label{eq:GEVPZtilde}
\end{equation}

\section{Numerical study with \texorpdfstring{$m_\pi = 806$}{mpi = 800}  MeV}\label{sec:numerics}

This section presents a variational study of one- and two-nucleon systems with $m_\pi = 806$ MeV using a gauge-field ensemble with $L=32$, which was previously used in Refs.~\cite{Beane:2012vq,Beane:2013br,Berkowitz:2015eaa,Wagman:2017tmp} to study multi-baryon systems with $[D,H]$ correlation functions built from hexaquark source and dibaryon sink interpolating operators.
The LQCD action used is the tadpole-improved~\cite{Lepage:1992xa} L{\"u}scher-Weisz gauge-field action~\cite{Luscher:1984xn} and the Wilson quark action including the Sheikholeslami-Wohlert (clover) improvement term~\cite{Sheikholeslami:1985ij}, with parameters shown in Table~\ref{tab:gauge_param}, and one step of four-dimensional stout smearing~\cite{Morningstar:2003gk} with $\rho=0.125$ applied to the gauge field.
An ensemble of $N_{\rm cfg} = 727$ gauge fields corresponding to a subset of the gauge-field ensemble studied in Ref.~\cite{Wagman:2017tmp} is used for this variational study.

Sparsened timeslice-to-all quark propagators with $\mathcal{S} = (V / V_{\mathcal{S}})^{1/3}=4$ are computed on each gauge-field configuration from  point-to-all quark propagators computed at each site of a $V_{\mathcal{S}} = 8^3 = 512$  sparse lattice, as described in Sec.~\ref{sec:contractions}. 
For more details on the effects of this sparsening on correlation functions, see Refs~\cite{Detmold:2019fbk,Li:2020hbj}.
Each source position is separated by $\mathcal{S} = 4$ lattice sites from its nearest neighbors, which corresponds to $\approx 0.6$ fm in physical units and $m_\pi \mathcal{S} \approx 2.4$ lattice units.
Correlations between sources, which are suppressed by at least $e^{- m_\pi \mathcal{S}}$, are therefore expected to be small.
We calculate sparsened timeslice-to-all smeared-smeared quark propagators using the \verb!Chroma!~\cite{Edwards:2004sx} LQCD software framework with Dirac-operator inversions performed using the conjugate gradient inverter with a residual tolerance of $10^{-12}$.
Two different Gaussian-smearing radii are used for both the propagator source and sink: the first smearing is denoted $T$ for ``thin'' and is defined by 20 steps of gauge-invariant Gaussian smearing~\cite{Gusken:1989ad,Gusken:1989qx,Alexandrou:1990dq} with \verb!Chroma! smearing parameter 2.1, which corresponds to a Gaussian smearing width of $\approx 1.24$ lattice units ($\approx 0.18$ fm).
The second smearing is denoted $W$ for ``wide'' and is defined by 100 steps of gauge-invariant Gaussian smearing with \verb!Chroma! smearing width 4.7, corresponding to a Gaussian smearing width of $\approx 2.77$ lattice units ($\approx 0.40$ fm).\footnote{The $T$ and $W$ smearings are constructed using identical Gaussian smearing kernels with width $\varepsilon \approx 0.055$ in the notation of Ref.~\cite{Bali:2016lva} and differ in the number of iterative applications of this kernel.}
Link smearing corresponding to 15 steps of stout smearing~\cite{Morningstar:2003gk} with $\rho=0.1$ is also applied to the spatial links of the gauge field used for constructing smeared sources and sinks.
We subsequently calculate one- and two-nucleon correlation-function matrices using the \verb!Qlua! LQCD software framework~\cite{qlua} and a \verb!C++! implementation of the contraction algorithm described in Sec.~\ref{sec:contractions} that includes significant scheduling and memory optimizations facilitated by the polyhedral compiler \verb!Tiramisu!~\cite{baghdadi2020tiramisu}, as described in Appendix~\ref{app:contractions}.

In a preliminary study~\cite{Wagman:2021spu}, we calculated timeslice-to-all smeared-smeared quark propagators using \verb!Qlua! on a subset of 167 gauge-field configurations and further calculated one- and two-nucleon correlation-function matrices.
We analyzed these results independently because slightly different quark-field smearings and inverter tolerances were used, and we found results that are consistent at (1-2)$\sigma$ for all energy levels with the significantly more precise results from the full set of gauge-field configurations  presented below.

\begin{table}[t!]
  \caption{Parameters of the gauge-field ensembles used in this work. $L$ and $T$ are the spatial and temporal dimensions of the hypercubic lattice, respectively, $\beta$ is related to the strong coupling, $a$ is the lattice spacing as determined from $\Upsilon$ spectroscopy~\cite{Beane:2012vq}, $m_{q}$ is the bare quark mass, $N_{\rm{cfg}}$ is the number of configurations used and $N_{\rm src} = V_{\mathcal{S}} N_{\rm{cfg}}$ is the total number of source points used for propagator calculations. The pion mass $m_\pi = 0.59445(15)(17)$ is taken from Ref.~\cite{Beane:2012vq}. }
\label{tab:gauge_param}
\begin{ruledtabular}
\renewcommand{\arraystretch}{1.2}
\begin{tabular}{cccccccccc}
$L^3\times T$ & $\beta$ & $m_q$ & $a$ [fm] & $L$ [fm] & $T$ [fm] & $m_{\pi}L$ & $m_{\pi}T$ & $N_{\text{cfg}}$ & $N_{\text{src}}$ \\ \hline 
  $32^3\times 48$ & $6.1$ & $-0.2450$ & $0.1453(16)$ & $4.5$ & $6.7$ & $19.0$ & $28.5$ & $727$ & $3.72\times 10^5$  \\
\end{tabular} 
\renewcommand{\arraystretch}{1}
\end{ruledtabular}
\end{table}

\subsection{The nucleon channel}\label{sec:nucresults}

\begin{figure}[!t]
	\includegraphics[width=0.47\columnwidth]{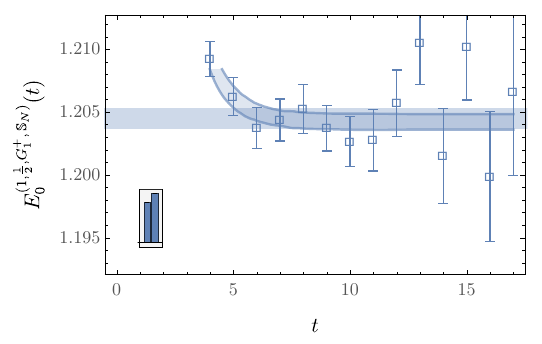}
	\includegraphics[width=0.47\columnwidth]{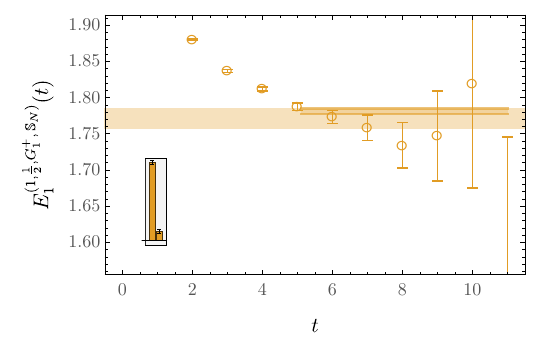}
   \caption{ Single-nucleon GEVP effective energies defined by Eq.~\eqref{eq:GEVPEM} and Eq.~\eqref{eq:GEVPcorrelators} with $t_0 =5$ and $t_{\rm ref} = 10$ for the ground state and first excited state. Points with error bars show the central values and $68\%$ bootstrap confidence intervals for GEVP correlation-function results. Shaded colored bands show the total statistical plus fitting systematic uncertainties added in quadrature (the outlined regions show the statistical uncertainty of the highest-weight fit entering the weighted average of acceptable fit results along with its corresponding fit time interval) as described in Appendix~\ref{app:fits}.
   The maximum $t$ shown corresponds to the largest $t$ less than $tol_{\rm therm} = 3T/8$ for which the signal-to-noise ratio of $E_{\mathsf{n}}^{(1,\frac{1}{2},G_1^+,\mathbb{S}_N)}(t)$ is greater than $tol_{\rm noise} = 0.1$; see Fig.~\ref{fig:B1G1_stability} for larger $t$ results for $E_{1}^{(1,\frac{1}{2},G_1^+,\mathbb{S}_N)}(t)$.
   The inset histograms show the relative overlap factors $\mathcal{Z}_{\mathsf{n}\chi}^{(1,\frac{1}{2},G_1^+,\mathbb{S}_N)}$ for $\chi \in \mathbb{S}_N$ corresponding to thin (left bar of each histogram) and wide (right bar of each histogram) Gaussian smearings. The error bars on the histograms denote the bootstrap $68\%$ confidence intervals.
   \label{fig:B1G1_fits}}
\end{figure}

For the nucleon, the quark-wavefunction smearing radius is the only parameter varied in the single-hadron interpolating-operator construction described in Sec.~\ref{sec:onehadron}. This results in an interpolating-operator set $\mathbb{S}_N = \{N_T,N_W\}$ where $N_T$ ($N_W$) denotes the  ``thin'' (``wide'') nucleon interpolating operator with the smaller- (larger-)radius quark-field smearing. 
The $2\times 2$ correlation-function matrix $C^{(1,\frac{1}{2},G_1^+)}_{\chi\chi'}(t)$ for this interpolating-operator set is diagonalized by solving the GEVP in Eq.~\eqref{eq:GEVP} to obtain  $\widehat{C}_{\mathsf{n}}^{(1,\frac{1}{2},G_1^+,\mathbb{S}_N)}(t)$.
The effective energies $E_{\mathsf{n}}^{(1,\frac{1}{2},G_1^+,\mathbb{S}_N)}(t)$ defined by Eq.~\eqref{eq:GEVPEM} are shown in Fig.~\ref{fig:B1G1_fits}. 
The GEVP correlation functions are fit to truncated spectral representations using a variety of minimum source/sink separations and numbers of excited states, and a weighted average of these fit results is used to obtain $E_{\mathsf{n}}^{(1,\frac{1}{2},G_1^+,\mathbb{S}_N)}$  as described in Appendix~\ref{app:fits}.
The central values and uncertainties (which here and below correspond to $68\%$ confidence intervals determined using bootstrap methods) on the GEVP energies determined using this fitting procedure, with fixed values of $t_0 = 5$ and $t_{\rm ref} = 10$ in the notation of Eq.~\eqref{eq:GEVPcorrelators}, are
\begin{align}
  E_0^{(1,\frac{1}{2},G_1^+,\mathbb{S}_N)} &= 1.20446(83) = 1.636(1)(18) \text{ GeV} , \\
  E_1^{(1,\frac{1}{2},G_1^+,\mathbb{S}_N)} &= 1.770(14) = 2.404(19)(26) \text{ GeV},
\end{align}
where the first uncertainties show statistical and fitting systematic uncertainties added in quadrature and the second uncertainties for the energies in physical units are associated with the uncertainties in the determination of $a = 0.1453(16)$ fm~\cite{Beane:2012vq} (ambiguities in defining the lattice spacing away from the physical values of the quark masses are not quantified).
The ground-state energy is consistent with previous calculations using a larger ensemble of gauge-field configurations with identical parameters, which obtained $M_N = 1.20467(57) =  1.636(18)$ GeV~\cite{Wagman:2017tmp}.
The gap between the ground state and first excited-state energies is given by $\delta^{(1,\frac{1}{2},G_1^+,\mathbb{S}_N)} = 0.566(14)$, which is smaller than the gap to the non-interacting $P$-wave pion production threshold for this volume $\sqrt{M_N^2 + (2\pi/L)^2} + \sqrt{m_\pi^2 + (2\pi/L)^2} - M_N = 0.6410(12)$.

Negligible sensitivity is seen to either the choices of $t_0 \in [3, 7]$ and $t_{\rm ref} \in [8, 12]$ or the choice of GEVP effective-energy definition.
The variation in $t_0$ and $t_{\rm ref}$ is discussed in Appendix~\ref{app:plots}.
Alternative effective-energy definitions based on GEVP eigenvalues rather than GEVP correlation functions computed using the eigenvectors are also shown in Appendix~\ref{app:plots} for comparison.
Given this insensitivity, the value $t_0 = 5$ is chosen for the final results.
This choice is motivated by the fact that excited-state contamination is clearly visible in GEVP correlation function results for $t < 5$. Therefore, the approximation that spectral representations can be truncated to only include the states overlapping with the chosen interpolating-operator set may be suspect for $t< 5$, even though fitted GEVP energy levels are insensitive.
The value $t_{\rm ref} = 2 t_0 = 10$ is then motivated by the arguments of Ref.~\cite{Blossier:2009kd}, where under the assumption that an interpolating-operator set dominantly overlaps with the lowest $\mathcal{I}$ states, it is shown in perturbation theory that $t_0 \geq t/2$ is sufficient to demonstrate that excited-state effects from states outside the subspace spanned by the GEVP eigenvectors are suppressed by $e^{-t\left[ E_{\mathcal{I}}^{(B,I,\Gamma_J)} - E_{\mathsf{n}}^{(B,I,\Gamma_J)} \right]}$, where $\mathcal{I}$ is the size of the interpolating-operator set.

\begin{figure}[!t]
	\includegraphics[width=0.47\columnwidth]{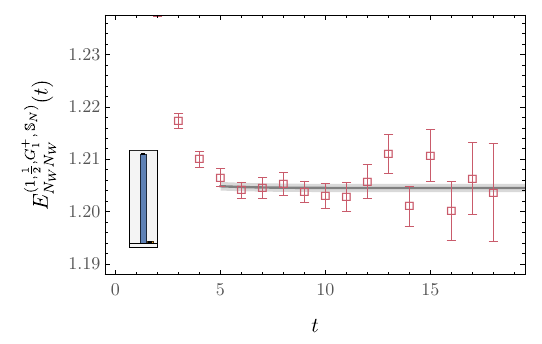}
	\includegraphics[width=0.47\columnwidth]{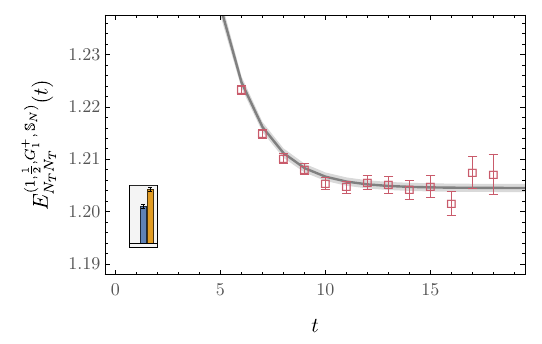}
	\includegraphics[width=0.47\columnwidth]{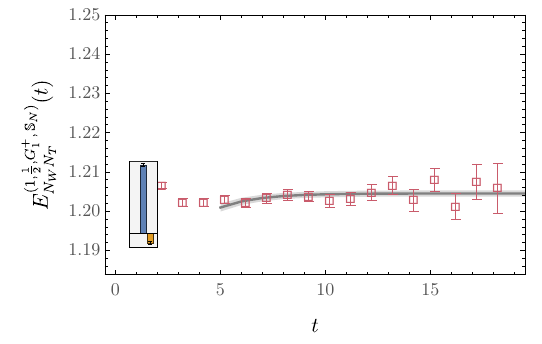}
   \caption{\label{fig:B1G1_reconstruction} Points with error bars show single-nucleon effective masses for several correlation functions built from the interpolating-operator set $\mathbb{S}_N$. Gray bands are GEVP reconstructions determined by inserting GEVP energies and overlap factors into truncated spectral representations for the correlation functions shown. In particular, the gray bands are not obtained using fits to the individual correlation functions shown. The GEVP reconstructions are displayed for $t \geq t_0$. Inset histograms show the central value and $68\%$ confidence intervals (the error bars) for the relative contribution of each GEVP eigenstate to the corresponding $\chi,\,\chi'$ element of the correlation-function matrix, that is $\widetilde{\mathcal{Z}}_{\mathsf{n}\chi\chi'}^{(1,\frac{1}{2},G_1^+,\mathbb{S}_N)}$ as defined in Eq.~\eqref{eq:GEVPZtilde}. They are colored as in Fig.~\eqref{eq:GEVPcorrelators}, with blue bars ($\mathsf{n}=0$) in the left and orange bars ($\mathsf{n}=1$) in the right.}
\end{figure}

The overlap factors $Z_{\mathsf{n}\chi}^{(1,\frac{1}{2},G_1^+,\mathbb{S}_N)}$ are obtained from Eq.~\eqref{eq:GEVPeigenvectors} using fit results for the GEVP energy levels and eigenvectors with the same fixed $t_0$ and $t_{\rm ref}$.
Normalized relative overlap factors $\mathcal{Z}_{\mathsf{n}\chi}^{(1,\frac{1}{2},G_1^+,\mathbb{S}_N)}$ are then obtained from Eq.~\eqref{eq:GEVPZrelative} and are given by
\begin{equation}
  \begin{split}
    \mathcal{Z}_{0 N_T}^{(1,\frac{1}{2},G_1^+,\mathbb{S}_N)}= 0.4488(3), \hspace{50pt} \mathcal{Z}_{0N_W}^{(1,\frac{1}{2},G_1^+,\mathbb{S}_N)} &= 0.5512(3), \\
    \mathcal{Z}_{1 N_T}^{(1,\frac{1}{2},G_1^+,\mathbb{S}_N)} = 0.89(3),  \hspace{55pt} \mathcal{Z}_{1 N_W}^{(1,\frac{1}{2},G_1^+,\mathbb{S}_N)} &= 0.11(3).
  \end{split}
\end{equation}
These GEVP results for $E_{\mathsf{n}}^{(1,\frac{1}{2},G_1^+,\mathbb{S}_N)}(t)$ and $\mathcal{Z}_{\mathsf{n}\chi}^{(1,\frac{1}{2},G_1^+,\mathbb{S}_N)}$ can be used to reconstruct the correlation-function matrix in the $\{N_T, N_W \}$ basis using Eq.~\eqref{eq:spectral}.
These GEVP reconstructions can be compared to numerical results for $C_{\chi\chi'}^{(1,\frac{1}{2},G_1^+,\mathbb{S}_N)}(t)$ to check how well a spectral representation including only the GEVP energy levels can describe the full correlation function.
The effective masses $E_{\chi\chi'}^{(1,\frac{1}{2},G_1^+,\mathbb{S}_N)}(t)$ obtained from the diagonal and off-diagonal elements of the correlation-function matrix are shown in Fig.~\ref{fig:B1G1_reconstruction}, along with the effective masses obtained from the corresponding GEVP reconstructions using Eq.~\eqref{eq:spectral}.
The GEVP reconstructions reproduce the effective masses obtained from the $\{N_T, N_W\}$ interpolating operators within $2\sigma$ uncertainties for $t \gtrsim t_0 = 5$.
The central value of the GEVP reconstruction of the $[N_W, N_T]$  effective mass approaches the GEVP ground-state energy from below because $\widetilde{\mathcal{Z}}_{1 N_W N_T}^{(1,\frac{1}{2},G_1^+,\mathbb{S}_N)}=-0.12(2)$ is negative.
These excited-state effects in $[N_W, N_T]$ are suppressed by both the small excited-state overlap of $N_W$ and by $e^{-t \delta^{(1,\frac{1}{2},G_1^+,\mathbb{S}_N)}}$ with  $t \gg 1/\delta^{(1,\frac{1}{2},G_1^+,\mathbb{S}_N)} = 1.77(4)$, and it is clear from Fig.~\ref{fig:B1G1_reconstruction} that excited-state effects in $[N_W, N_T]$ correlation functions are negligible for $t \gtrsim 5$.

\subsection{The dineutron channel}\label{sec:I1}

The LQCD energy spectrum for two-nucleon systems with $I=1$ is independent of $I_z$ in this isospin-symmetric calculation and results for $nn$, $pp$, and spin-singlet $pn$ systems are equivalent; this channel will be referred to as the dineutron channel below in order to distinguish it from the $I=0$, spin-triplet deuteron channel.
$S$-wave two-nucleon wavefunctions in the $I=1$ channel are associated with the $\Gamma_J = A_1^+$ representation in the infinite-volume limit, as discussed in Sec.~\ref{sec:FV}.
The FV analogs of $S$-wave dineutron operators are dibaryon operators $D_{\mathfrak{s}\mathfrak{k}g}^{(2,1,A_1^+)}$, where $\mathfrak{s}$ labels the relative-momentum shell with all positive-parity plane-wave wavefunctions with $\mathfrak{s} \in \{0,\ldots,6\}$ included in our calculations, $\mathfrak{k}$ labels the multiplicity index of the relative wavefunction (which for the $\Gamma_J = A_1^+$ dineutron channel is simply $1$), and $g\in \{T, W\}$ describes whether thin or wide Gaussian smearing is used on all six quark fields.
In addition, we include hexaquark operators $H_g^{(2,1,A_1^+)}$ and quasi-local operators $\kappaOp_{\mathfrak{q}g}^{(2,1,A_1^+)}$ with exponential localization scales $\kappa_{\mathfrak{q}} \in \{\kappa_1,\kappa_2,\kappa_3\} = \{0.035,0.070,0.14 \}$ in lattice units, corresponding to $\{ 48,95,190\}$~MeV.
These quasi-local wavefunctions resemble FV two-nucleon bound-state wavefunctions with binding momenta corresponding to $\kappa_{\mathfrak{q}}$ that are associated with binding energies of $\{0.0010,0.0041,0.016\}$ in lattice units, corresponding to $\{ 1.4,\ 5.5,\ 22\}$ MeV.
This ranges from less than the binding energy of the deuteron in nature to the average of the more deeply-bound results for negative FV energy shifts for the  dineutron ($0.0111(21)$, $0.0127(21)$, and $0.0137(17)$, corresponding to $15.1(2.8)$, $17.3(2.9)$,  and $18.6(2.3)$ MeV) and deuteron ($0.0165(26)$, $0.0206(22)$, and  $0.0194(23)$, corresponding to $22.5(3.5)$, $28.0(3.0)$, and $26.3(3.1)$ MeV) channels obtained in  Refs.~\cite{Beane:2012vq,Berkowitz:2015eaa,Wagman:2017tmp} for the same gauge field ensemble.

The set of two-nucleon interpolating operators computed in this work with $I=1$ and $\Gamma_J=A_1^+$ includes 22 dibaryon, hexaquark, and quasi-local operators.
However, a correlation-function matrix including all 22 interpolating operators leads to large statistical noise and $E_{0}^{(2,1,A_1^+,\mathbb{S})}(t)$ that are consistent with zero at (1-3)$\sigma$ for all $t$.
This suggests that at our current statistical precision, this is effectively a degenerate set of operators that has statistically significant overlap with fewer than 22 LQCD energy eigenstates.
In fact, all $I=1$, $\Gamma_J=A_1^+$ interpolating-operator sets with more than 16 interpolating operators are similarly found to be degenerate. 
Sets with 16 or fewer interpolating operators can be constructed that satisfy $E_{0}^{(2,1,A_1^+,\mathbb{S})}(t) \neq 0$ and are therefore non-degenerate at high statistical significance.

\begin{figure}[!ht]
	\includegraphics[width=0.47\columnwidth]{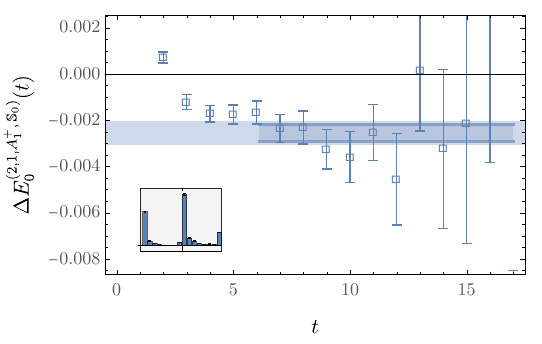}
	\includegraphics[width=0.47\columnwidth]{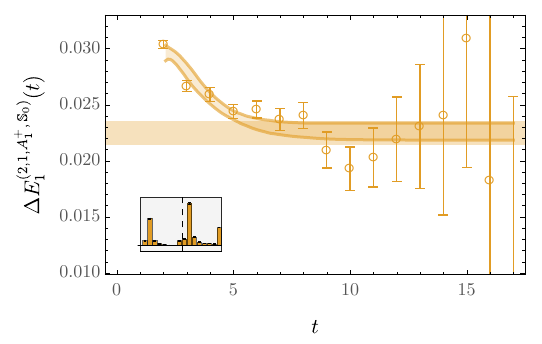}
	\includegraphics[width=0.47\columnwidth]{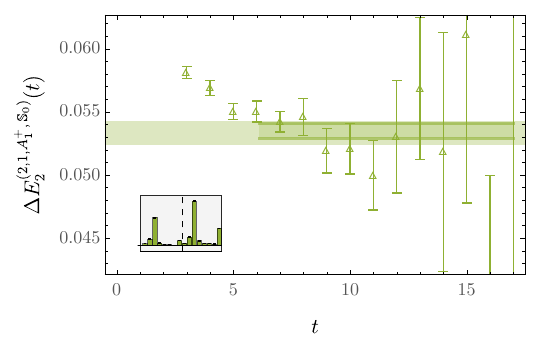}
	\includegraphics[width=0.47\columnwidth]{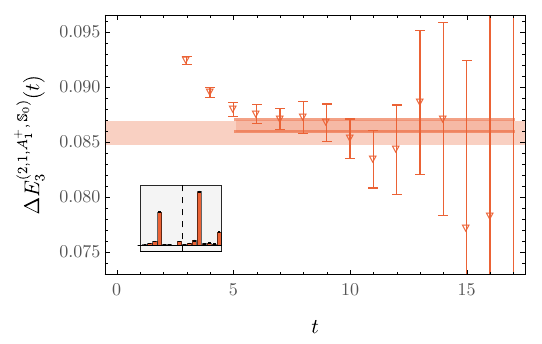}
	\includegraphics[width=0.47\columnwidth]{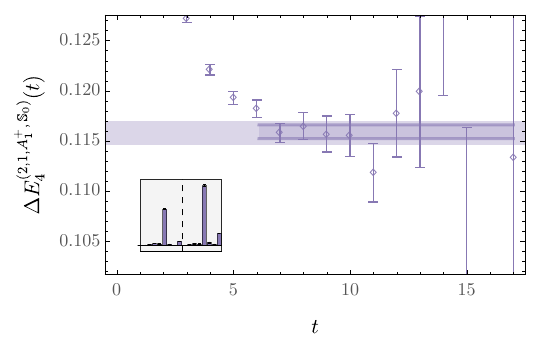}
	\includegraphics[width=0.47\columnwidth]{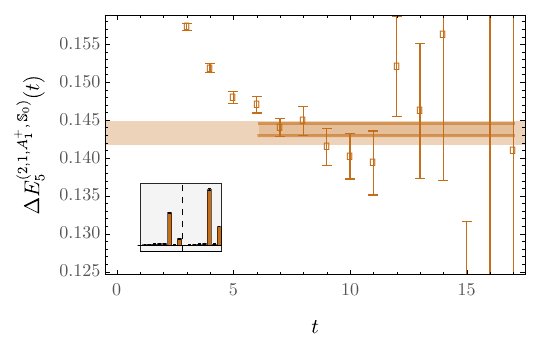}
	\includegraphics[width=0.47\columnwidth]{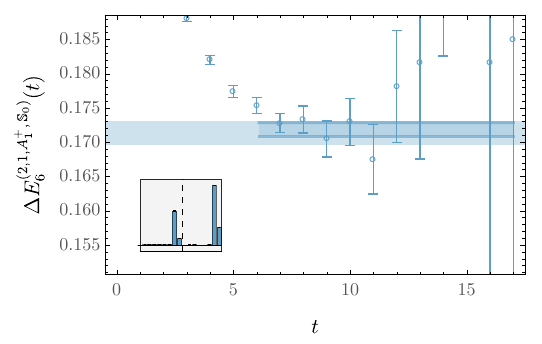}
	\includegraphics[width=0.47\columnwidth]{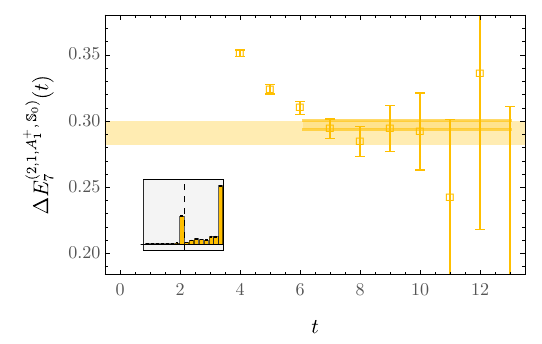}
   \caption{\label{fig:B2I1A1_FV_fits} Results for two-nucleon GEVP effective FV energy shifts for the eight lowest-energy states with $I=1$ and $\Gamma_J = A_1^+$ obtained using the interpolating-operator set $\mathbb{S}_{\protect \circled{0}}^{(2,1,A_1^+)}$ defined in Eq.~\eqref{eq:B0defI1A1}. Histograms show $\mathcal{Z}_\mathsf{n}^{(2,1,A_1^+,\mathbb{S}_0)}$ and are analogous to those in Fig.~\ref{fig:B1G1_fits} with the addition of a dashed line separating interpolating operators with thin (left) and wide (right) Gaussian quark-field smearing; the operators with each smearing are ordered with $D_{\mathfrak{s}1g}^{(2,1,A_1^+)}$ with $\mathfrak{s}$ increasing from left to right, and with $H_g^{(2,1,A_1^+)}$ operators rightmost.}
\end{figure}

\begin{figure}[!t]
	\includegraphics[width=0.6\columnwidth]{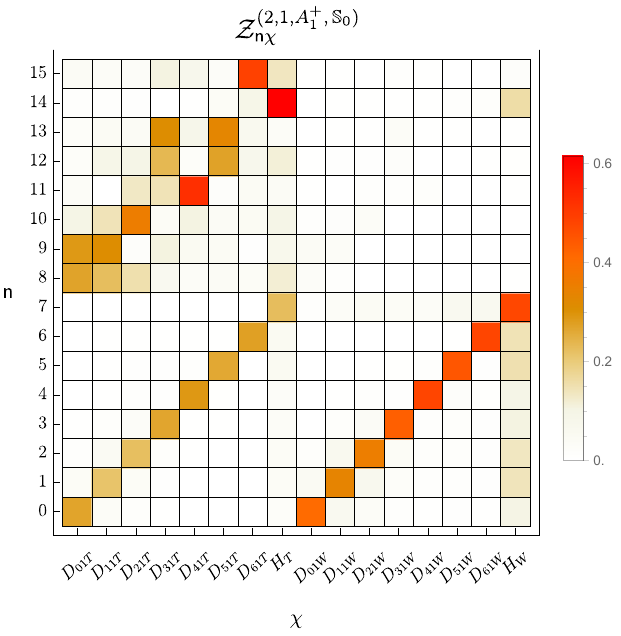}
   \caption{\label{fig:B2I1A1_Zplot} Results for overlap factors $\mathcal{Z}_{\mathsf{n} \chi}^{(2,1,A_1^+,\mathbb{S}_0)}$ defined in Eq.~\eqref{eq:GEVPZrelative}. Uncertainties, shown by error bars in histograms for the eight lowest-energy states in Fig.~\ref{fig:B2I1A1_FV_fits}, are not shown here.}
\end{figure}

\begin{figure}[!t]
	\includegraphics[width=0.49\columnwidth]{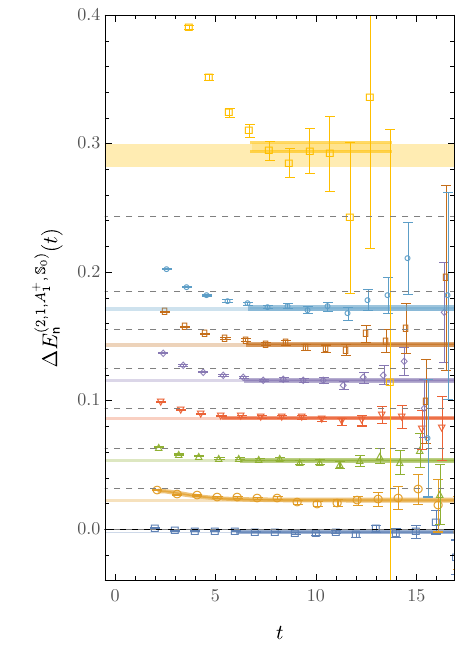}
	\includegraphics[width=0.49\columnwidth]{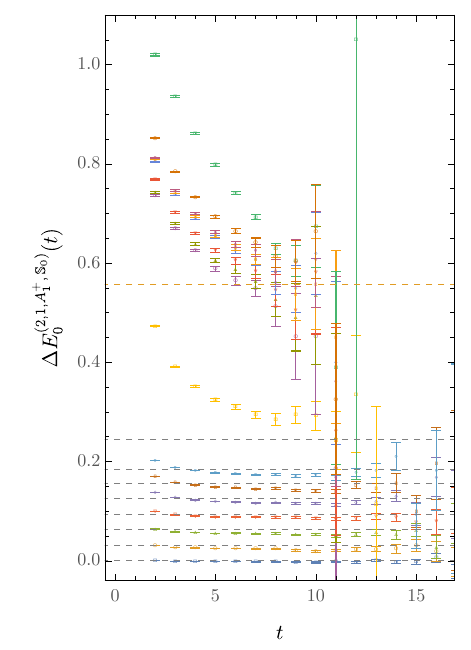}
   \caption{\label{fig:B2I1A1_rainbow} 
  Compilations of the $\mathsf{n}\in\{0,\ldots,7\}$ GEVP effective FV energy shifts shown in Fig.~\ref{fig:B2I1A1_FV_fits} (left) and of the full set of $\mathsf{n}\in\{0,\ldots,15\}$ results (right) obtained with the interpolating-operator set $\mathbb{S}_{\protect \circled{0}}^{(2,1,A_1^+)}$. Non-interacting two-nucleon FV energy shifts $2\sqrt{M_N^2 + \mathfrak{s} (2\pi/L)^2  } - 2M_N$ with $\mathfrak{s}\in\{0,1,2,3,4,5,6,8\}$ are shown as dashed gray lines, and the central values of the single-nucleon excited-state energy gap $\delta^{(1,\frac{1}{2},G_1^+,\mathbb{S}_N)}$ is shown as a dashed orange line. Horizontal offsets are applied to points and fit bands in the left panel for clarity.}
\end{figure}

One such non-degenerate interpolating-operator set is given by the 16 operators 
\begin{equation}
  \mathbb{S}_{\circled{0}}^{(2,1,A_1^+)} = \bigg \{ D_{\mathfrak{s}\mathfrak{k}g}^{(2,1,A_1^+)},\ H_g^{(2,1,A_1^+)} \ | \ \mathfrak{s}\in \{0,\ldots,6\},\ \mathfrak{k}\in \{1\}, \  g\in\{ T,W\} \bigg \},
\label{eq:B0defI1A1}
\end{equation}
where the multiplicity index $\mathfrak{k}$ is necessarily unity because $\mathcal{N}_{\mathfrak{s}}^{(1,A_1^+)} = 1$.
This set includes two copies (one for each quark-field smearing) of the seven dibaryon operators required to describe non-interacting nucleons with relative-momentum magnitude less than $\sqrt{8} \left( \frac{2\pi}{L} \right)$ in the $A_1^+$ irrep as well as two hexaquark operators (for the two quark-field smearings).
Alternative interpolating-operator sets including different sets of 16 or fewer interpolating operators are investigated below.
The same fitting procedure described in Sec.~\ref{sec:nucresults} is used to determine $E_{\mathsf{n}}^{(2,1,A_1^+,\mathbb{S}_0)}$.
The FV energy shifts $\Delta E_{\mathsf{n}}^{(2,1,A_1^+,\mathbb{S}_0)}$ are determined from analogous weighted averages of correlated differences of fit results for $E_{\mathsf{n}}^{(2,1,A_1^+,\mathbb{S}_0)}$ and $2 E_{0}^{(1,\frac{1}{2},G_1^+,\mathbb{S}_N)}$ with independent fits performed for one- and two-baryon correlation functions as detailed in Appendix~\ref{app:fits}.
Results for the GEVP effective FV energy shifts as well as fit results for $\Delta E_{\mathsf{n}}^{(2,1,A_1^+,\mathbb{S}_0)}$ and $Z_{\mathsf{n}}^{(2,1,A_1^+,\mathbb{S}_0)}$ for the eight lowest-energy states obtained with $\mathbb{S}_{\circled{0}}^{(2,1,A_1^+)}$ are shown in Figs.~\ref{fig:B2I1A1_FV_fits}-\ref{fig:B2I1A1_Zplot}.
The ground- and first-excited-state FV energy shifts obtained using $\mathbb{S}_{\circled{0}}^{(2,1,A_1^+)}$ with this fitting procedure are
\begin{equation}
  \begin{split}
    \Delta E_{0}^{(2,1,A_1^+,\mathbb{S}_0)} &= -0.00255(51) = -3.46(69)(4) \text{ MeV}, \\
    \Delta E_1^{(2,1,A_1^+,\mathbb{S}_0)} &= 0.0225(11) = 30.5(1.4)(0.3) \text{ MeV}.
  \end{split}
\end{equation}
Results for the eight lowest-energy states obtained with this interpolating-operator set are tabulated in Appendix~\ref{app:tabs}.
There is a clear gap between the lowest- and highest-energy halves of the GEVP energy levels obtained with $\mathbb{S}_{\circled{0}}^{(2,1,A_1^+)}$, as seen in Fig.~\ref{fig:B2I1A1_rainbow}.
The higher-energy levels ($\mathsf{n}\geq 8$) satisfy $\Delta E_{\mathsf{n}}^{(2,1,A_1^+,\mathbb{S}_0)}(t) \gtrsim \delta^{(1,\frac{1}{2},G_1^+,\mathbb{S}_N)}$ for all $t$ where signals can be resolved despite the presence of many non-interacting two-nucleon energy shifts between $\Delta E_7^{(2,1,A_1^+,\mathbb{S}_0)}$ and $\Delta E_8^{(2,1,A_1^+,\mathbb{S}_0)}$.
One scenario consistent with these results, which is investigated further below, is that several QCD energy levels exist between $E_7^{(2,1,A_1^+,\mathbb{S}_0)}$ and $E_8^{(2,1,A_1^+,\mathbb{S}_0)}$, but $\mathbb{S}_{\circled{0}}^{(2,1,A_1^+)}$ does not include interpolating operators with significant overlap onto these states.
The dineutron-channel excitation energy gap corresponds to $1/\delta^{(2,1,A_1^+,\mathbb{S}_0)} = 40(2)$ lattice units or $5.8(3)$ fm.
Degrading statistical precision limits the results to the regime $t  \ll 1/\delta^{(2,1,A_1^+,\mathbb{S}_0)}$ where it should not be surprising that a scenario could arise in which the energy states making the largest contributions to the correlation-function matrix used are not simply the lowest-energy states.

Overlap-factor results demonstrate that states $\mathsf{n} \in \{0,\ldots, 7\}$ each dominantly overlap with either dibaryon operators with a single $\mathfrak{s}$ or with hexaquark operators as shown in Figs.~\ref{fig:B2I1A1_FV_fits} and~\ref{fig:B2I1A1_Zplot}. 
States corresponding to $\mathsf{n} \in \{0,\ldots,6\}$ have maximum overlap with $D_{\mathfrak{s}1g}$ operators with $\mathfrak{s} = \mathsf{n}$, while the $\mathsf{n} = 7$ state has maximum overlap with $H_g$ operators.
In particular, the ground state has maximum overlap with $\mathfrak{s}=0$ dibaryon operators, with $ \sum_{g} \mathcal{Z}_{0 D_{01g}}^{(2,1,A_1^+,\mathbb{S}_0)}~=~0.682(8)$, and it has next largest overlaps with hexaquark operators, with $ \sum_{g} \mathcal{Z}_{0 H_{g}}^{(2,1,A_1^+,\mathbb{S}_0)}~=~0.127(1)$, and $\mathfrak{s}=1$ dibaryon operators, with $ \sum_{g} \mathcal{Z}_{0 D_{11g}}^{(2,1,A_1^+,\mathbb{S}_0)}~=~0.093(4)$.

The eight lowest energies obtained using $\mathbb{S}_{\circled{0}}^{(2,1,A_1^+)}$ are shown compared to the locations of the non-interacting two-nucleon energy levels $2\sqrt{M_N^2 + \mathfrak{s} (2\pi/L)^2 }$ in Fig.~\ref{fig:B2I1A1_rainbow}.
The GEVP energy levels $\mathsf{n}\in\{0,\ldots,6\}$ each appear slightly below one of the $\mathfrak{s}\in\{0,\ldots,6\}$ non-interacting levels.
The $\mathsf{n}=7$ state is determined much less precisely than the other states and dominantly overlaps with  $H_g^{(2,1,A_1^+)}$ operators.
This state appears close to the $\mathfrak{s}=8$ non-interacting threshold, which is the first non-interacting threshold that is not associated with a dibaryon operator explicitly included in this interpolating-operator set.

\begin{figure}[!t]
	\includegraphics[width=0.47\columnwidth]{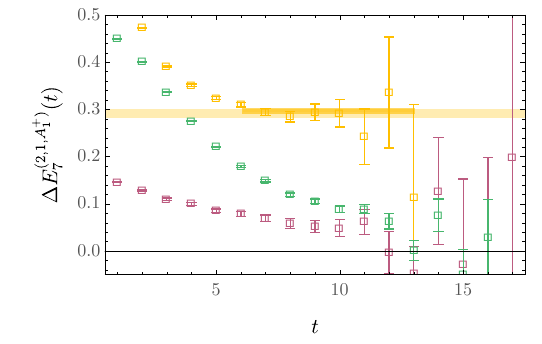}
   \caption{\label{fig:B2I1A1_hex_comp}  The GEVP correlation functions for the $\mathsf{n}=7$ energy level dominantly overlapping with hexaquark operators, also shown in Fig.~\ref{fig:B2I1A1_FV_fits}, is displayed in  yellow. It is compared with diagonal correlation functions for $H^{(2,1,A_1^+)}_W$, in purple, and $H^{(2,1,A_1^+)}_T$, in green, to demonstrate the differences between hexaquark correlation functions and the GEVP principle correlation function with the largest hexaquark component.
   }
\end{figure}

The $\mathsf{n} \in \{0,\ldots,6\}$ energy levels are all statistically consistent with the apparent effective mass plateaus of the diagonal correlation functions for interpolating operator that has maximum overlap with a given level.
The $\mathsf{n}=7$ state is different in this respect.
Although it has maximum overlap with hexaquark operators, the diagonal hexaquark correlation functions display apparent plateaus at lower energies roughly commensurate with the $\mathsf{n} \in \{1,2\}$ levels as shown in Fig.~\ref{fig:B2I1A1_hex_comp}.
This suggests that hexaquark operators have relatively large ovelap with the $\mathsf{n} \in \{1,2\}$ levels, which is consistent with the overlap factor results in Fig.~\ref{fig:B2I1A1_Zplot}; however the $\mathfrak{s}=\{1,2\}$ dibaryon operators have still larger overlap with the same energy levels.
When both hexaquark and dibaryon operators are included in a GEVP analysis, an additional energy level close to the $\mathfrak{s}=8$ non-interacting level is resolved beyond the $\mathsf{n} \in \{0,\ldots,6\}$ levels overlapping predominantly with dibaryon operators.

\begin{figure}[!tp]
	\includegraphics[width=0.47\columnwidth]{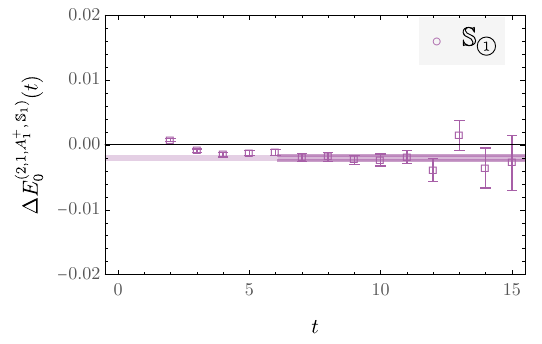}
	\includegraphics[width=0.47\columnwidth]{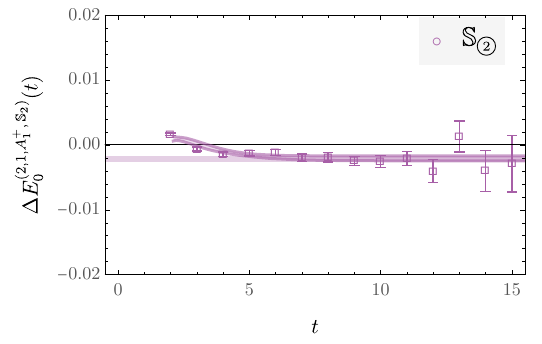}
	\includegraphics[width=0.47\columnwidth]{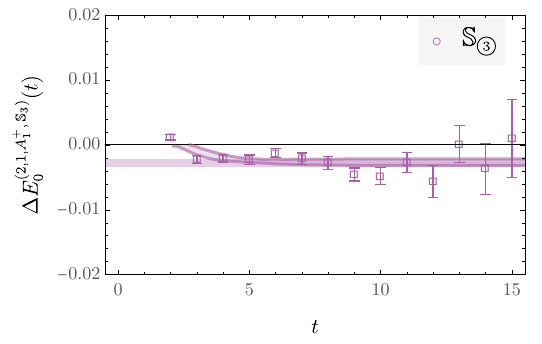}
	\includegraphics[width=0.47\columnwidth]{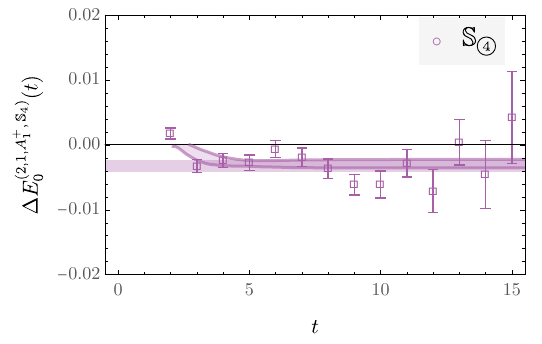}
	\includegraphics[width=0.47\columnwidth]{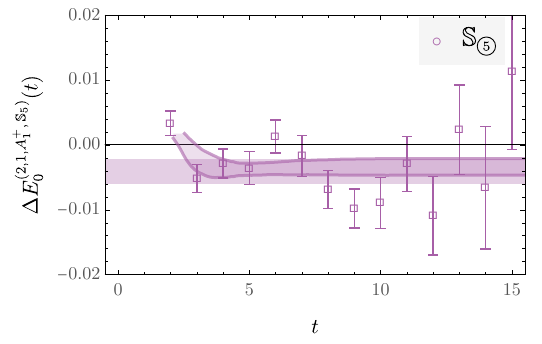}
	\includegraphics[width=0.47\columnwidth]{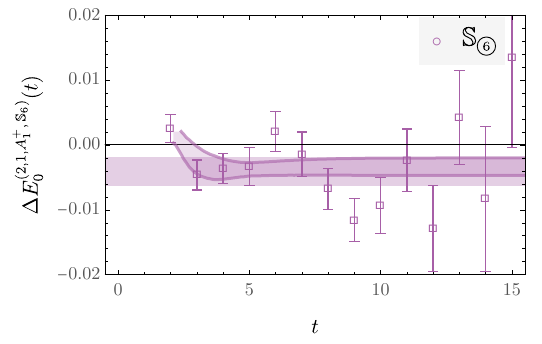}
	\includegraphics[width=0.47\columnwidth]{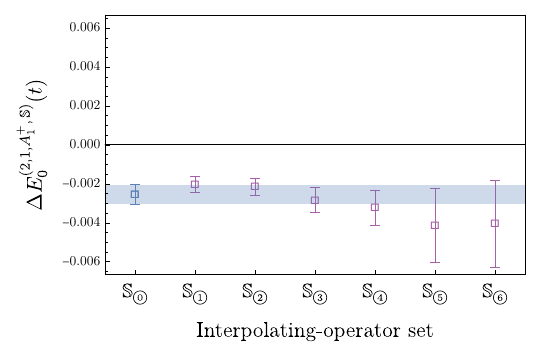}
   \caption{ \label{fig:B2I1A1_goodbasis_fits} The bottom panel shows the ground-state FV energy shifts obtained using GEVP correlation functions with interpolating-operator set $\mathbb{S}_{\protect \circled{0}}^{(2,1,A_1^+)}$ shown in Fig.~\ref{fig:B2I1A1_FV_fits} (blue) compared to results obtained using seven other interpolating-operator sets defined in Eq.~\eqref{eq:BA1def2} (purple). Other panels show the ground-state effective FV energy shift for each interpolating-operator set as indicated in the corresponding legend.}
\end{figure}

The higher-energy GEVP energy levels $\mathsf{n}\in\{8,\ldots,16\}$ obtained using $\mathbb{S}_{\circled{0}}^{(2,1,A_1^+)}$ are all consistent with $\Delta E_\mathsf{n}^{(2,1,A_1^+,\mathbb{S}_0)}~\gtrsim~\delta^{(1,\frac{1}{2},G_1^+,\mathbb{S}_N)}$.
Given the presence of GEVP energy levels close to the non-interacting $\mathfrak{s} \leq 8$ levels, one expects there to be QCD energy levels close to the $\mathfrak{s} \geq 9$  non-interacting states, and therefore at lower energies than these higher-energy excited states.
It is plausible that such energy eigenstates exist, but $\mathbb{S}_{\circled{0}}^{(2,1,A_1^+)}$ does not have any interpolating operators with significant overlap onto them (in comparison to the overlaps with other states).
In this case, the higher-energy excited states are not free from contamination from lower-energy states.
The GEVP energy levels associated with these higher-energy states are therefore not considered to be reliable and are not reported.

Other non-degenerate sets with 16 or fewer operators can also be constructed.
Results for the ground-state FV energy shifts for a variety of interpolating-operator sets,
\begin{equation}
\begin{split}
    \mathbb{S}_{\circled{1}}^{(2,1,A_1^+)} &= \bigg \{ D_{\mathfrak{s}\mathfrak{k}g}^{(2,1,A_1^+)} \ | \ \mathfrak{s}\in \{0\} \bigg \}, \\
    \mathbb{S}_{\circled{2}}^{(2,1,A_1^+)} &= \bigg \{ D_{\mathfrak{s}\mathfrak{k}g}^{(2,1,A_1^+)},\ H_g^{(2,1,A_1^+)} \ | \ \mathfrak{s}\in \{0\} \bigg \}, \\
    \mathbb{S}_{\circled{3}}^{(2,1,A_1^+)} &= \bigg \{ \kappaOp_{1g}^{(2,1,A_1^+)},\ D_{\mathfrak{s}\mathfrak{k}g}^{(2,1,A_1^+)},\ H_g^{(2,1,A_1^+)} \ | \ \mathfrak{s}\in \{1,\ldots,6\} \bigg \}, \\
    \mathbb{S}_{\circled{4}}^{(2,1,A_1^+)} &= \bigg \{ \kappaOp_{2g}^{(2,1,A_1^+)},\ D_{\mathfrak{s}\mathfrak{k}g}^{(2,1,A_1^+)},\ H_g^{(2,1,A_1^+)} \ | \ \mathfrak{s}\in \{1,\ldots,6\} \bigg \}, \\
    \mathbb{S}_{\circled{5}}^{(2,1,A_1^+)} &= \bigg \{ \kappaOp_{3g}^{(2,1,A_1^+)},\ D_{\mathfrak{s}\mathfrak{k}g}^{(2,1,A_1^+)},\ H_g^{(2,1,A_1^+)} \ | \ \mathfrak{s}\in \{1,\ldots,6\} \bigg \}, \\
    \mathbb{S}_{\circled{6}}^{(2,1,A_1^+)} &= \bigg \{ \kappaOp_{3g}^{(2,1,A_1^+)},\ D_{\mathfrak{s}\mathfrak{k}g}^{(2,1,A_1^+)},\ H_g^{(2,1,A_1^+)} \ | \ \mathfrak{s}\in \{0,2,\ldots,6\} \bigg \}, \\
\end{split}\label{eq:BA1def2}
\end{equation}
leading to relatively precise GEVP correlation functions are shown in Fig.~\ref{fig:B2I1A1_goodbasis_fits}.\footnote{ Relatively precise GEVP results can also be extracted from the correlation-function matrices studied in Ref.~\cite{Wagman:2021spu} using an additional operator set related to $\mathbb{S}_{\circled{6}}^{(2,1,A_1^+)}$ by replacing  $\kappaOp_{3g}^{(2,1,A_1^+)}$ with $\kappaOp_{2g}^{(2,1,A_1^+)}$.  However, using the larger set of correlation-function matrices included in the analysis presented here, the corresponding correlation-function matrices are found to be degenerate within statistical uncertainties.}
For all of these sets, $\mathfrak{k}\in \{1\}$ and $g\in\{ T,W\}$.
The ground-state FV energy shifts for these sets are consistent with the results for $\mathbb{S}_{\circled{0}}^{(2,1,A_1^+)}$ at $1\sigma$.
It is noteworthy that ground-state energy results from $\mathbb{S}_{\circled{1}}^{(2,1,A_1^+)}$, which only includes $D_{01g}$ operators, are consistent at this level of precision with ground-state energy results using $\mathbb{S}_{\circled{0}}^{(2,1,A_1^+)}$.
It is also noteworthy that $\kappaOp_{\mathfrak{q}g}^{(2,1,A_1^+)}$ operators with all three values of $\kappa_{\mathfrak{q}}$ can be included in place of $D_{01g}^{(2,1,A_1^+)}$ operators without significantly affecting the ground-state FV energy-shift results, although the ground-state energy precision decreases as $\kappa_{\mathfrak{q}}$ is increased.
Further, the $\kappaOp_{3g}^{(2,1,A_1^+)}$ operators can be included alongside $D_{01g}^{(2,1,A_1^+)}$ operators in place of $D_{11g}^{(2,1,A_1^+)}$ operators; this significantly decreases the precision of the ground-state FV energy shift and leads to some distortions of the higher-energy spectrum  but leads to consistent results for the ground-state energy.
Replacing $D_{\mathfrak{s}1g}^{(2,1,A_1^+)}$ operators with larger $\mathfrak{s}$ with $\kappaOp_{\mathfrak{q}g}^{(2,1,A_1^+)}$ operators leads to even less precise results, suggesting that the $\kappaOp_{\mathfrak{q}g}^{(2,1,A_1^+)}$ operators dominantly overlap with the same ground and first excited states identified by GEVP calculations using $\mathbb{S}_{\circled{0}}^{(2,1,A_1^+)}$.
This is consistent with a scenario in which there are 16 LQCD energy eigenstates that have significant overlap with the 22 operators of the forms $D_{\mathfrak{s}\mathfrak{k}g}^{(2,1,A_1^+)}$, $H_g^{(2,1,A_1^+)}$, and $\kappaOp_{\mathfrak{q}g}^{(2,1,A_1^+)}$ used here.

\begin{figure}[!t]
	\includegraphics[width=0.47\columnwidth]{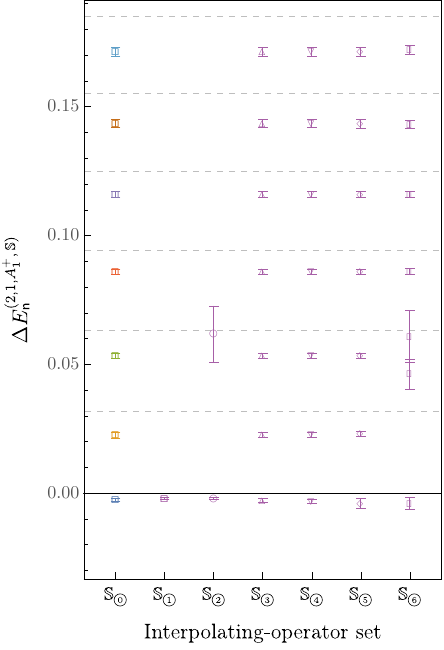} \hspace{4mm}
	\includegraphics[width=0.47\columnwidth]{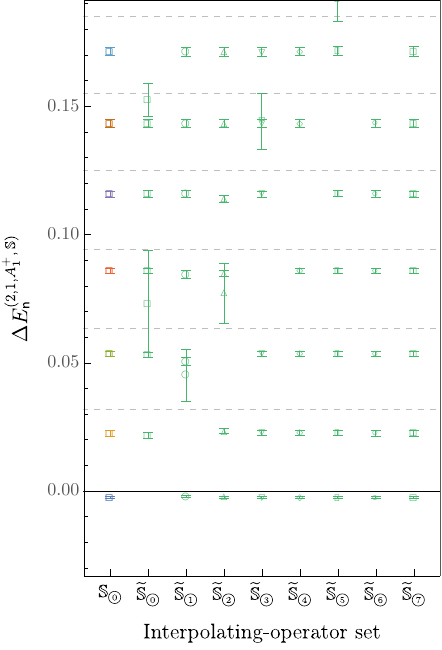}
   \caption{\label{fig:B2I1A1_badbasis_summary} FV energy shifts obtained using GEVP correlation functions with interpolating-operator set $\mathbb{S}_{\protect \circled{0}}^{(2,1,A_1^+)}$ (left column in each panel and corresponding to Fig.~\ref{fig:B2I1A1_FV_fits}) are compared to results obtained using
   the interpolating-operator sets defined in Eq.~\eqref{eq:BA1def2} with ground-state energy shifts featured in Fig.~\ref{fig:B2I1A1_goodbasis_fits} (left), as well as
   the eight other interpolating-operator sets defined in Eq.~\eqref{eq:BA1def3} (right). The color coding for $\mathbb{S}_{\protect \circled{0}}^{(2,1,A_1^+)}$ energies follows that in Fig.~\ref{fig:B2I1A1_FV_fits}, and the dashed lines show non-interacting two-nucleon energies as in Fig.~\ref{fig:B2I1A1_rainbow}.}
\end{figure}

In general, operator sets with fewer than 16 operators do not give results that are qualitatively or quantitatively consistent with results obtained using $ \mathbb{S}_{\circled{0}}^{(2,1,A_1^+)}$.
Results for sets of 14 operators,
\begin{equation}
\begin{split}
  \widetilde{\mathbb{S}}_{\circled{m}}^{(2,1,A_1^+)} &= \left\{  D_{\mathfrak{s}\mathfrak{k}g}^{(2,1,A_1^+)},\ H_g^{(2,1,A_1^+)} \ | \ \mathfrak{s}\in \{0,\ldots,6\} \setminus \text{m} \right\},\  \text{m} \in \{0,\ldots,6\}, \\
    \widetilde{\mathbb{S}}_{\circled{7}}^{(2,1,A_1^+)} &= \left\{  D_{\mathfrak{s}\mathfrak{k}g}^{(2,1,A_1^+)} \ | \ \mathfrak{s}\in \{0,\ldots,6\} \right\} ,
\end{split}\label{eq:BA1def3}
\end{equation}
obtained by removing one of the two-nucleon operators with both smearings or the hexaquark operators with both smearings, are shown in Fig.~\ref{fig:B2I1A1_badbasis_summary}.
It is noteworthy that the lowest-energy state obtained using $\widetilde{\mathbb{S}}_{\circled{0}}^{(2,1,A_1^+)}$ is consistent with the first excited state obtained using $\mathbb{S}_{\circled{0}}^{(2,1,A_1^+)}$ rather than the corresponding ground state.
Although $\widetilde{\mathbb{S}}_{\circled{0}}^{(2,1,A_1^+)}$ includes interpolating operators that are not orthogonal to the ground state identified by $\mathbb{S}_{\circled{0}}^{(2,1,A_1^+)}$ (the overlap factors $\mathcal{Z}_{0\chi}^{(2,1,A_1^+,\mathbb{S}_0)}$ computed using $\mathbb{S}_{\circled{0}}^{(2,1,A_1^+)}$ are non-zero at high statistical significance for several of the interpolating operators included in $\widetilde{\mathbb{S}}_{\circled{0}}^{(2,1,A_1^+)} \subset \mathbb{S}_{\circled{0}}^{(2,1,A_1^+)}$), each operator in $\widetilde{\mathbb{S}}_{\circled{0}}^{(2,1,A_1^+)}$ overlaps more strongly with some excited state than with the ground state. Therefore, over the range of $t$ considered here, the GEVP results resemble a spectrum that is missing the $\mathbb{S}_{\circled{0}}^{(2,1,A_1^+)}$ ground state, while results for the other levels are largely unaffected (note that the lowest energy obtained using $\widetilde{\mathbb{S}}_{\circled{0}}^{(2,1,A_1^+)}$ is still a valid bound on the ground-state energy). 
Results for sets $\widetilde{\mathbb{S}}_{\circled{1}}^{(2,1,A_1^+)},\ldots,\widetilde{\mathbb{S}}_{\circled{6}}^{(2,1,A_1^+)}$ similarly resemble results for $\mathbb{S}_{\circled{0}}^{(2,1,A_1^+)}$ with one low-lying energy level ``missing'' while the remaining energy levels are largely unaffected; although in several cases the $\mathsf{n}=7$ level dominantly overlapping with the $H_W^{(2,1,A_1^+)}$ operators is replaced by a lower-energy level somewhat above the ``missing'' level.
Results for $\widetilde{\mathbb{S}}_{\circled{7}}^{(2,1,A_1^+)}$ similarly resemble results for $\mathbb{S}_{\circled{0}}^{(2,1,A_1^+)}$ without the $\mathsf{n=7}$ level.
These results demonstrate the importance of using an interpolating-operator set with significant overlap onto all energy levels of interest and further demonstrate that having a large interpolating-operator set is not sufficient to guarantee that a set will have good overlap onto the ground state or a particular excited state.
It is also not guaranteed that adding linearly independent interpolating operators will allow a larger non-degenerate correlation-function matrix to be resolved at a given statistical ensemble size (as demonstrated here by the onset of degeneracy if quasi-local interpolating operators are added to $\mathbb{S}_{\circled{0}}^{(2,1,A_1^+)}$).  
These results suggest that it is difficult to diagnose the presence of missing energy levels, since GEVP results for interpolating-operator sets that give rise to missing energy levels do not show obvious signs of inconsistency without comparing them to results from a more complete set.
Because the positions of the energy levels that are identified by these smaller interpolating-operator sets are consistent with the energy levels appearing in more complete sets, the scattering phase shifts associated with these energy levels will also be consistent and the appearance of a missing level, for example, in results for $\widetilde{\mathbb{S}}_{\circled{7}}^{(2,1,A_1^+)}$, is no more obvious from a scattering amplitude analysis than from inspection of the FV energy shifts, see Sec.~\ref{sec:summary}.

\begin{figure}[!t]
	\includegraphics[width=0.47\columnwidth]{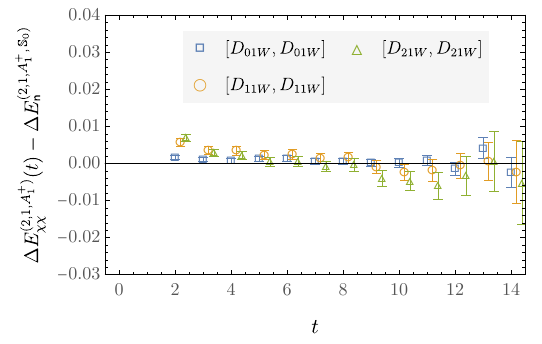}
	\includegraphics[width=0.47\columnwidth]{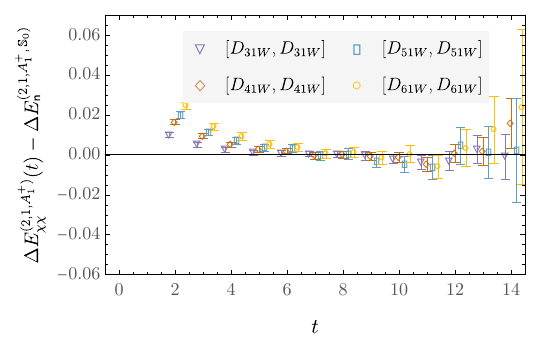}
	\includegraphics[width=0.47\columnwidth]{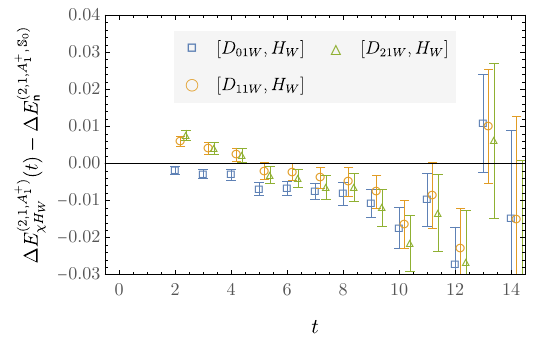}
	\includegraphics[width=0.47\columnwidth]{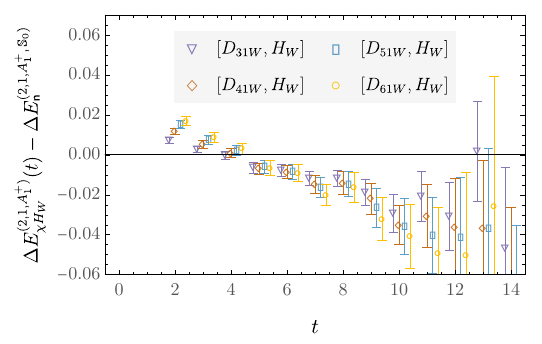}
   \caption{\label{fig:B2I1A1_diffs} The upper (lower) panels show differences between the effective FV energy shifts computed using correlation functions with  dibaryon sources and sinks $[D_{\mathfrak{s}1W}, D_{\mathfrak{s}1W}]$ (hexaquark sources and dibaryon sinks $[D_{\mathfrak{s}1W}, H_W]$) and the fitted FV energy shifts for the $\mathbb{S}_{\protect \circled{0}}^{(2,1,A_1^+)}$ GEVP energy levels that dominantly overlap with $D_{\mathfrak{s}1W}^{(2,1,A_1^+)}$, which are $\mathsf{n} \in \{0,1,2\}$ for the left panels and $\mathsf{n} \in  \{3,4,5,6\}$ for the right panels. }
\end{figure}

\begin{figure}[!t]
	\includegraphics[width=0.47\columnwidth]{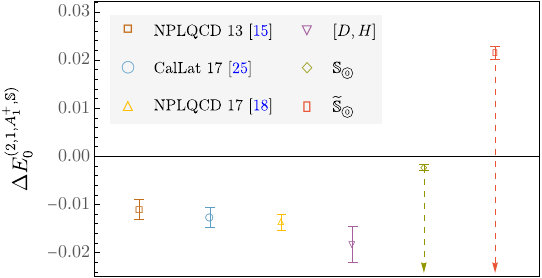}
	\includegraphics[width=0.47\columnwidth]{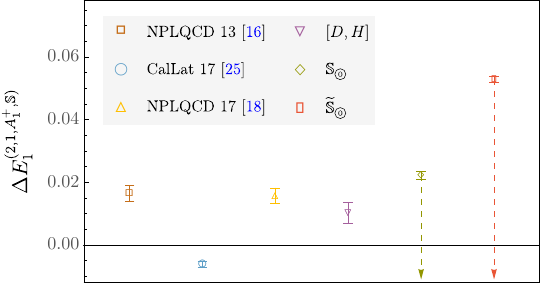}
   \caption{\label{fig:B2I1A1_spectrum_comp} Comparisons of FV energy-shift results for the $\mathsf{n}=0$ and $\mathsf{n}=1$ states between previous results obtained in Refs.~\cite{Beane:2012vq,Beane:2013br,Berkowitz:2015eaa,Wagman:2017tmp} using the same gauge field ensemble with $[D,H]$ correlation function (as well as displaced sources in Ref.~\cite{Berkowitz:2015eaa}) and results of this work obtained using variational methods with interpolating-operator set $\mathbb{S}_{\protect \circled{0}}^{(2,1,A_1^+)}$,  interpolating-operator set  $\widetilde{\mathbb{S}}_{\protect \circled{0}}^{(2,1,A_1^+)}$ (in which the operators dominantly overlapping with the ground state are omitted), and non-variational results obtained using $[D, H]$ correlation functions. Arrows emphasize the fact that the variational method provides (stochastic) upper bounds on energy levels and therefore FV energy shifts under the assumption that the nucleon mass is accurately identified, while $[D, H]$ correlation functions provide an estimate of the energy but have systematic uncertainties in both directions that are difficult to estimate.
   }
\end{figure}

The individual correlation functions associated with the interpolating operator most strongly overlapping with each GEVP energy level provide good approximations to the GEVP correlation functions for $t \gtrsim t_0 = 5$, as shown in Fig.~\ref{fig:B2I1A1_diffs}. In all cases the FV energy shifts associated with $[D_{\mathfrak{s}1W}, D_{\mathfrak{s}1W}]$ correlation functions approach the GEVP results from above. 
The FV energy shifts associated with $[D_{\mathfrak{s}1W}, H_{W}]$ correlation functions are also shown in Fig.~\ref{fig:B2I1A1_diffs}. Both the effective energies and the effective FV energy shifts for all $[D_{\mathfrak{s}1W}, H_{W}]$ correlation functions are consistently (1-4)$\sigma$ below the corresponding GEVP results for $t \gtrsim 5$ until signal-to-noise degradation makes the difference consistent with zero.
Applying the same fit range selection and weighted averaging procedure described in Appendix~\ref{app:fits} to a combined fit of $[D_{01g}, H_{g}]$ correlation functions with $s \in \{T, W\}$ gives a ground-state energy result of $-0.0186(28)$. 
The same fit procedure applied to $[D_{11g}, H_{g}]$ gives a result for the first excited state of $0.0097(33)$.
Higher-statistics analyses of $[D_{01g}, H_{g'}]$ and $[D_{11g}, H_{g'}]$ correlation functions with smeared sources and both smeared and point-like sinks in Ref.~\cite{Beane:2012vq}, Ref.~\cite{Berkowitz:2015eaa}, and Ref.~\cite{Wagman:2017tmp} give more precise results of $-0.0111(21)$, $-0.0127(21)$, and $-0.0137(17)$ for the ground-state FV energy shift, respectively, and Ref.~\cite{Beane:2013br} and Ref.~\cite{Wagman:2017tmp} give results of $0.0165(26)$ and $0.0157(25)$ for the first-excited-state FV energy shifts, respectively. These are consistent with the $[D_{\mathfrak{s}1g}, H_{g}]$ results of this work at $2\sigma$.
The first excited-state energy determined in Ref.~\cite{Berkowitz:2015eaa}, $-0.0061(9)$, is obtained using asymmetric correlation-functions with spatially displaced two-nucleon sources and $D_{01g}$ sinks. 
There is therefore consistency between results from different studies using similar interpolating operators, as shown in Fig.~\ref{fig:B2I1A1_spectrum_comp} for the ground state, but interpolating-operator dependence is significantly larger than the statistical and fitting systematic errors obtained using single- or multi-exponential fits with a given interpolating-operator set. 
Interpolating-operator dependence also leads to differences between $\Delta E_{\mathsf{n}}^{(2,1,A_1^+,\mathbb{S})}$ obtained using variational methods with different $\mathbb{S}$ that are larger than their uncertainties, in particular $\Delta E_0^{(2,1,A_1^+,\widetilde{\mathbb{S}}_0)} = 0.0216(12)$ differs from $\Delta E_0^{(2,1,A_1^+,\mathbb{S}_0)}$ by 18$\sigma$, and it is important to recall that the correct interpretation of results obtained using variational methods is as upper bounds on energies.
In other words, the ground-state energy bound obtained using $\widetilde{\mathbb{S}}_{\circled{0}}^{(2,1,A_1^+)}$ is valid, but it is less constraining than the bound obtained using $\mathbb{S}_{\circled{0}}^{(2,1,A_1^+)}$.

\begin{figure}[!t]
	\includegraphics[width=0.47\columnwidth]{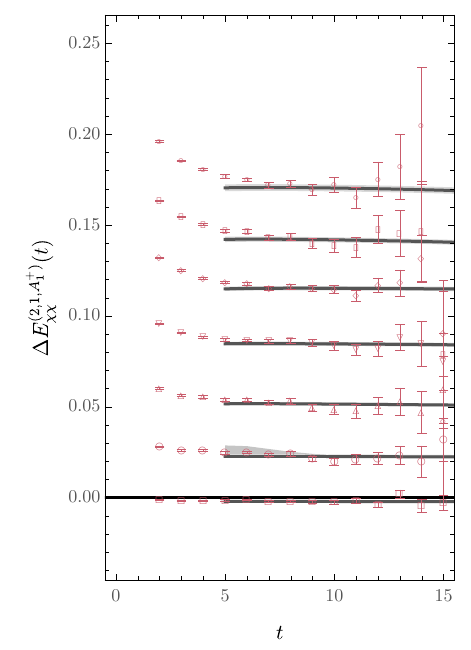}
	\includegraphics[width=0.47\columnwidth]{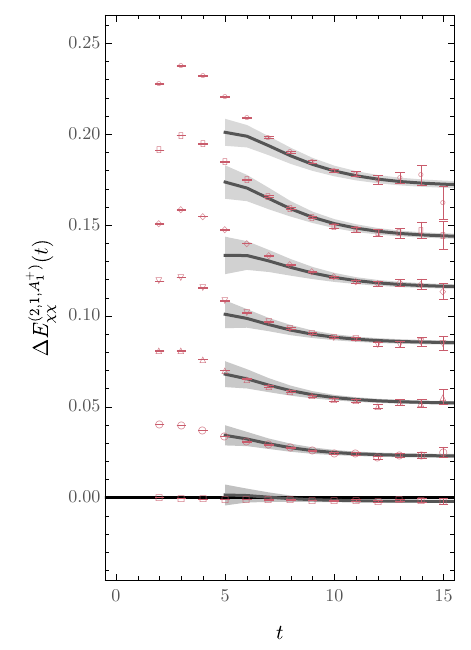}
   \caption{\label{fig:B2I1A1_rainbow_reconstruction} Points with error bars show effective FV energy shifts for positive-definite dibaryon correlation functions $[D_{\mathfrak{s}1g}, D_{\mathfrak{s}1g}]$ with $\mathfrak{s} \in \{0,\ldots,6\}$ from bottom to top with wide (thin) quark-field smearing in the left (right) panel. Gray curves and bands show the central values and $68\%$ confidence intervals for GEVP reconstructions that are obtained by inserting GEVP energy-level and overlap-factor results obtained using the interpolating-operator set $\mathbb{S}_{\protect \circled{0}}^{(2,1,A_1^+)}$ into the spectral representations for these correlation functions for all $t \geq t_0$ in analogy to Fig.~\ref{fig:B1G1_reconstruction}. In particular, gray curves and bands are not obtained directly from fits to the correlation functions shown. Relatively large uncertainties on the higher-energy half of the spectrum lead to larger uncertainties in GEVP reconstructions at small $t$, particularly for the narrow smearing case in the right panel.}
\end{figure}

\begin{figure}[!t]
	\includegraphics[width=0.47\columnwidth]{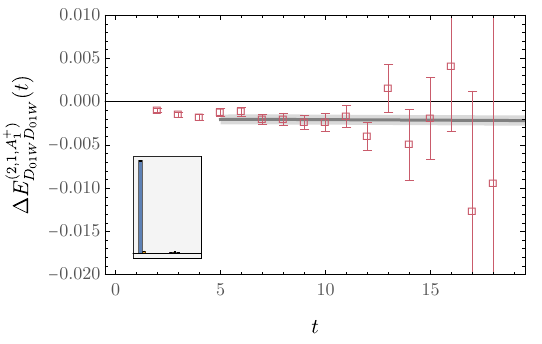}
	\includegraphics[width=0.47\columnwidth]{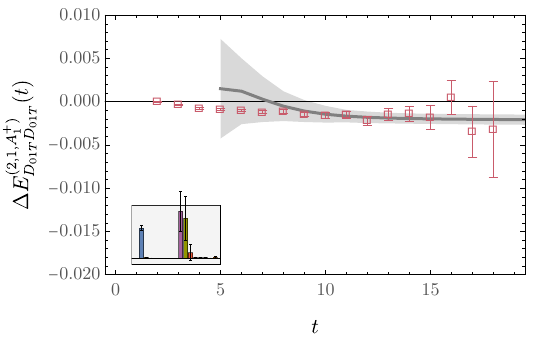}
	\includegraphics[width=0.47\columnwidth]{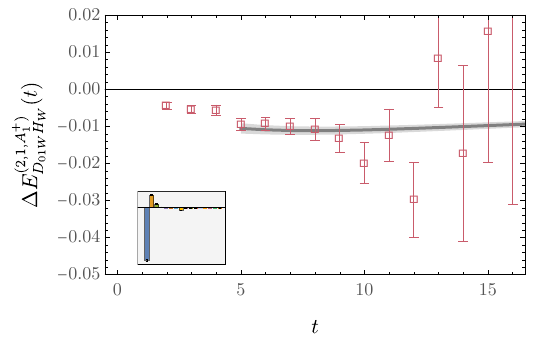}
	\includegraphics[width=0.47\columnwidth]{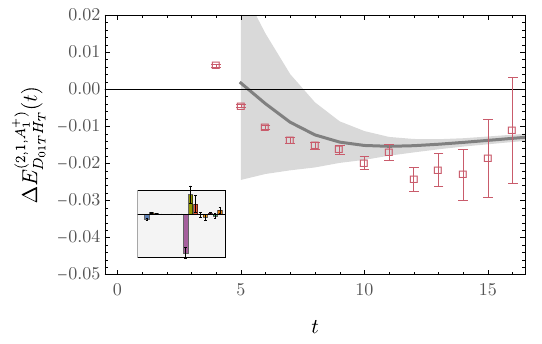}
	\includegraphics[width=0.47\columnwidth]{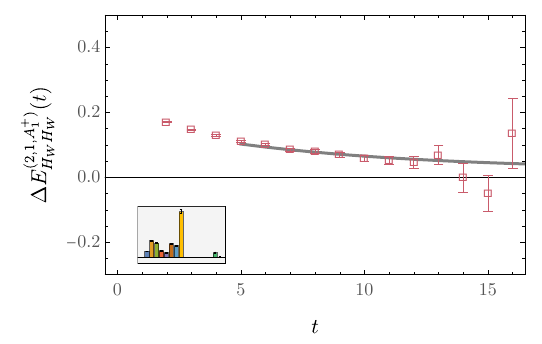}
	\includegraphics[width=0.47\columnwidth]{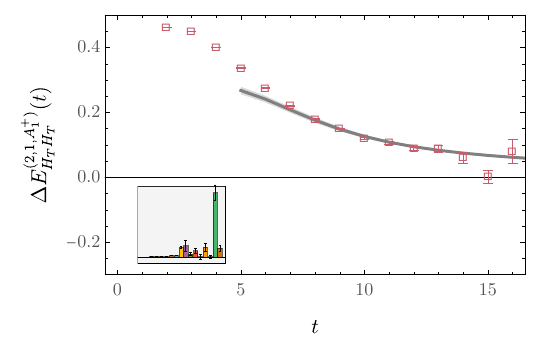}
   \caption{\label{fig:B2I1A1_reconstruction_1} Effective FV energy shifts and GEVP reconstructions obtained using $\mathbb{S}_{\protect \circled{0}}^{(2,1,A_1^+)}$ for $[D_{01g}, D_{01g}]$, $[D_{01g}, H_g]$, and $[H_g, H_g]$ correlation functions with wide (thin) quark-field smearing are shown in the left (right) panels. Histograms show the central values and $68\%$ confidence intervals for the relative contributions $\widetilde{\mathcal{Z}}_{\mathsf{n}\chi\chi'}^{(2,1,A_1^+,\mathbb{S}_0)}$ of each GEVP energy level to the correlation function. Bars correspond to $\mathsf{n}$ values increasing from left to right, with coloring consistent with Fig.~\ref{fig:B2I1A1_rainbow}. Overlap factors for thin smearing interpolating operators are more sensitive to less precisely determined high-energy excited-state energies, which leads to the relatively large uncertainties in $\widetilde{\mathcal{Z}}_{\mathsf{n}D_{01T} H_T}^{(2,1,A_1^+,\mathbb{S}_0)}$.}
\end{figure}

Results for $E_\mathsf{n}^{(2,1,A_1^+,\mathbb{S}_0)}$ and $\mathcal{Z}_\mathsf{n}^{(2,1,A_1^+,\mathbb{S}_0)}$ can be used to reconstruct correlation functions for particular interpolating operators as discussed for the nucleon in Sec.~\ref{sec:nucresults}.
The interpolating-operator set $\mathbb{S}_{\circled{0}}^{(2,1,A_1^+)}$ includes $D_{01g}^{(2,1,A_1^+)}$ and $H_g^{(2,1,A_1^+)}$ operators and the correlation-function matrix for $\mathbb{S}_{\circled{0}}^{(2,1,A_1^+)}$ therefore includes $[D_{01g}, D_{01g}]$, $[D_{01g}, H_g]$, and $[H_g, H_g]$ correlation functions.
Reconstructions of the FV energy shifts associated with positive-definite $[D_{\mathfrak{s}1g}, D_{\mathfrak{s}1g}]$ correlation functions with $\mathfrak{s}\in\{0,\ldots,6\}$ and $g\in \{T,W\}$ are shown in Fig.~\ref{fig:B2I1A1_rainbow_reconstruction}.
Consistency between correlation function and GEVP reconstruction results is found at the (1-2)$\sigma$ level for $t \gtrsim 6$. Significant deviations for $[D_{\mathfrak{s}1T}, D_{\mathfrak{s}1T}]$ can be seen at $t=5$, which suggests that contributions to correlation functions with thin smearing from excited states outside the GEVP energy levels can only be neglected for $t \gtrsim 6$.

Analogous GEVP reconstructions for $[D_{01g}, D_{01g}]$, $[D_{01g}, H_g]$, and $[H_g, H_g]$ correlation functions are shown in Fig.~\ref{fig:B2I1A1_reconstruction_1}.
There is agreement at the (1-2)$\sigma$ level between the GEVP reconstructions and correlation function results for $t \gtrsim 6$.
The $[D_{01W}, D_{01W}]$ correlation function is dominated by the GEVP ground-state contribution with overlap $\widetilde{\mathcal{Z}}_{0 D_{01W} D_{01W}}^{(2,1,A_1^+,\mathbb{S}_0)}=0.956(7)$  and includes a few-percent-level contribution from the GEVP first excited state. 
The apparent plateau below threshold in the $[D_{01W}, H_W]$ correlation function is reproduced by the GEVP reconstruction with $\widetilde{\mathcal{Z}}_{0  D_{01W} H_W}^{(2,1,A_1^+,\mathbb{S}_0)}=-0.72(1)$, a significant opposite-sign contribution from the first GEVP excited state $\widetilde{\mathcal{Z}}_{1  D_{01W} H_W}^{(2,1,A_1^+,\mathbb{S}_0)}=0.17(1)$, a smaller contribution from the second GEVP excited state $\widetilde{\mathcal{Z}}_{2  D_{01W} H_W}^{(2,1,A_1^+,\mathbb{S}_0)}=0.05(1)$, and even smaller contributions from higher-energy GEVP excited states. 
The GEVP reconstruction predicts that the $[D_{01W}, H_W]$ effective FV energy shifts approach the GEVP ground-state FV energy shift for large $t$, and in particular that the GEVP reconstruction of the $[H_W, D_{01W}]$ effective FV energy shift reaches $1\sigma$ consistency with $\Delta E_{0}^{(2,1,A_1^+,\mathbb{S}_0)}$ only for unachievable $t\gtrsim 80$.
It has been previously argued in Refs.~\cite{Iritani:2016jie,Iritani:2018vfn}, based on calculations using the HAL QCD potential method, that $[D,H]$ correlation functions exhibit large FV energy shifts at $\sim 1$ fm that will approach values closer to threshold for similar $t$ to those predicted by the GEVP analysis here.
Although the results of Ref.~\cite{Iritani:2018vfn} and the GEVP analysis here both indicate significant opposite-sign contributions to $[D,H]$ correlation functions from the ground state and low-energy excited states, the pattern of $\widetilde{\mathcal{Z}}_{0 H_W D_{\mathfrak{s}1W}}^{(2,1,A_1^+,\mathbb{S}_0)}$  predicted in Ref.~\cite{Iritani:2018vfn} is relatively constant in $\mathfrak{s}$ while the overlap factors calculated here decrease rapidly with increasing $\mathfrak{s}$.
Further, there is no evidence for the extra level associated with hexaquark operators found here in the potential-method-based results of Ref.~\cite{Iritani:2018vfn}.

\begin{figure}[!t]
	\includegraphics[width=0.47\columnwidth]{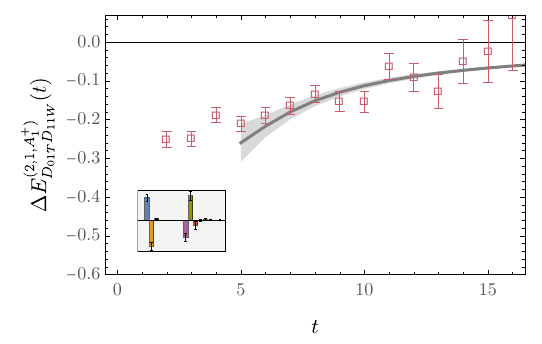}
	\includegraphics[width=0.47\columnwidth]{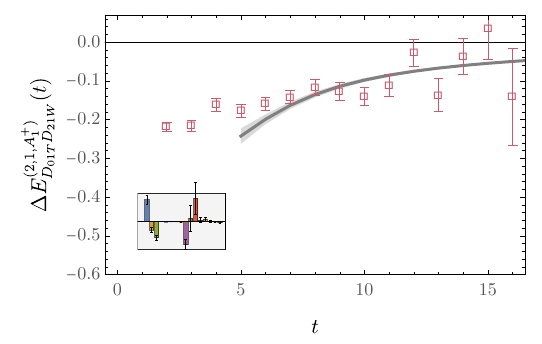}
	\includegraphics[width=0.47\columnwidth]{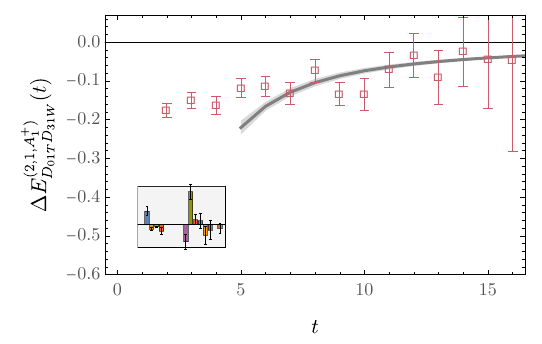}
	\includegraphics[width=0.47\columnwidth]{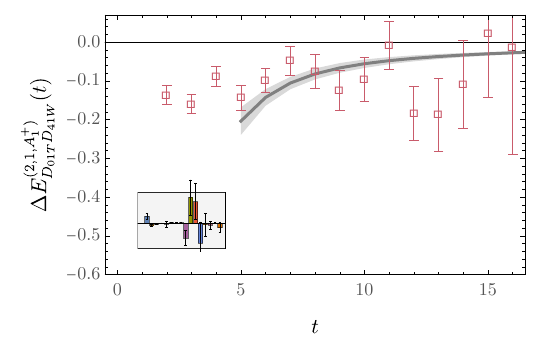}
	\includegraphics[width=0.47\columnwidth]{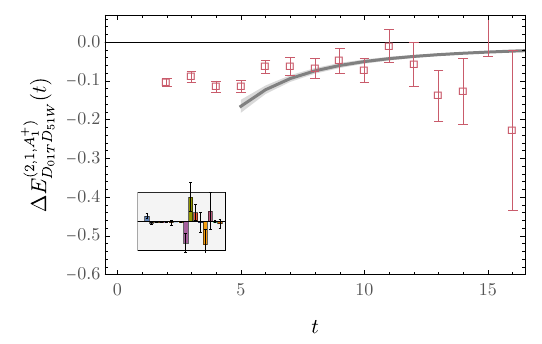}
	\includegraphics[width=0.47\columnwidth]{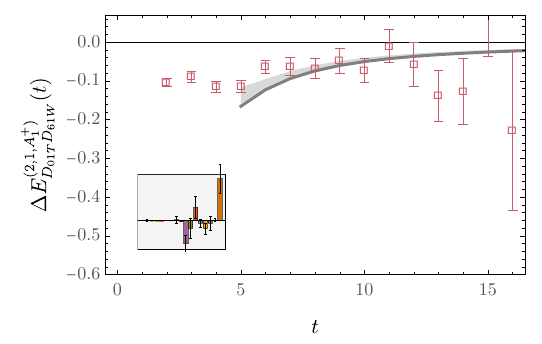}
   \caption{\label{fig:B2I1A1_reconstruction_3} Effective FV energy shifts and GEVP reconstructions obtained using $\mathbb{S}_{\protect \circled{0}}^{(2,1,A_1^+)}$ $[D_{01T}, D_{\mathfrak{s}1W}]$ correlation functions with $\mathfrak{s} \in \{1,\ldots, 6\}$. Histograms of $\widetilde{\mathcal{Z}}_{\mathsf{n}\chi \chi'}^{(2,1,A_1^+,\mathbb{S}_0)}$ are as in Fig.~\ref{fig:B2I1A1_reconstruction_1}.}
\end{figure}

Other asymmetric correlation functions exhibit effective FV energy shifts significantly below threshold with mild $t$ dependence over the range $ t \gtrsim 5$ where signals can be resolved.
Results for $[D_{01T}, D_{\mathfrak{s}1W}]$ correlation functions with $\mathfrak{s} \neq 0$ are shown in Fig.~\ref{fig:B2I1A1_reconstruction_3}.  
These correlation functions show FV energy shifts $\sim 0.1$ below threshold with mild, although in some cases statistically significant,  $t$ dependence.
Applying the same fitting procedure described above to these correlation functions leads to statistically significant FV energy shifts in several cases: $-0.065(14)$, $-0.084(21)$, $-0.088(18)$, $-0.049(27)$, and $-0.053(10)$ for $\mathfrak{s} = 2$ to $\mathfrak{s} = 6$, respectively.
Conversely, a large fitting systematic uncertainty is obtained for $\mathfrak{s}=1$ with a result $-0.1(1.1)$.
The GEVP reconstructions using $\mathbb{S}_{\circled{0}}^{(2,1,A_1^+)}$ reproduce the sub-threshold behavior of all these correlation functions over the range of $t$ where signals can be resolved through opposite-sign linear combinations of ground-state contributions and contributions from excited states above and below the single-nucleon first excited state.
One possible interpretation of the fact that GEVP results using $\mathbb{S}_{\circled{0}}^{(2,1,A_1^+)}$ with a ground-state energy of $-0.00255(51)$ can reproduce the sub-threshold behavior of these correlation functions is that these sub-threshold effective FV energy shifts are not associated with physical states at these energies but are $t$-dependent combinations of energy levels.
Another possible interpretation is that some or all of these asymmetric correlation functions are overlapping with states below threshold that are not visible in the GEVP correlation functions obtained using $\mathbb{S}_{\circled{0}}^{(2,1,A_1^+)}$.
In particular, it is straightforward to construct toy examples of overlap factors for which asymmetric correlation functions constructed from nearly orthogonal source and sink interpolating operators overlap with deeply bound states that do not contribute to the corresponding positive-definite diagonal correlation functions. Consider for example a three-state system and a pair of interpolating operators $A$ and $B$, with the true energy spectrum for this toy model given by
\begin{equation}
  E_0^{(AB)} = \eta - \Delta, \hspace{20pt} E_1^{(AB)} = \eta, \hspace{20pt} E_2^{(AB)} = \eta + \delta,
\end{equation}
and normalized overlap factors for operators $A$ and $B$  onto these three states (in increasing energy order) given by
\begin{equation}
  \mathcal{Z}_A = (\epsilon,\sqrt{1 - \epsilon^2},0), \hspace{20pt} \mathcal{Z}_B = (\epsilon,0,\sqrt{1 - \epsilon^2}),
  \label{eq:badZ}
\end{equation}
with $\epsilon\ll 1$ and real. Positive-definite $[A,A]$ and $[B,B]$ correlation functions will overlap dominantly with the first and second excited states, respectively, while asymmetric $[A,B]$ correlation functions will overlap perfectly with the ground state and may or may not be positive definite depending on $\epsilon$. 
In the $\epsilon \rightarrow 0$ limit (or in practical situations where $\epsilon \approx 0$ within statistical uncertainty), the GEVP solution to the $2\times 2$ correlation-function matrix using the interpolating-operator set $\{A,B\}$ will simply be the $[A,A]$ and $[B,B]$ correlation functions.
Corrections arising from non-zero $\epsilon$ can be calculated exactly and admit a simple perturbative expansion in $\epsilon$. To $\mathcal{O}(\epsilon^3)$ accuracy, the GEVP eigenvalues are given by
\begin{equation}
  \begin{split}
  \lambda_0^{(AB)} &= e^{-(t-t_0)\eta}\left[ 1 + \epsilon^2  \left(  e^{t \Delta} - e^{t_0 \Delta} \right) + \mathcal{O}(\epsilon^4) \right], \\
    \lambda_1^{(AB)} &= e^{-(t-t_0)(\eta + \delta) }\left[ 1 + \epsilon^2 \left(  e^{t (\Delta + \delta)} - e^{t_0 (\Delta + \delta)} \right) + \mathcal{O}(\epsilon^4) \right],
  \end{split}
\end{equation}
demonstrating that variational methods will not recover the true ground-state energy that dominates the $[A,B]$ correlation function unless $t$ is large enough that $e^{t \Delta}$ compensates for the $\mathcal{O}(\epsilon^2)$ overlap-factor suppression of the ground-state contribution to the GEVP eigenvalues. 
This example can be trivially generalized to include more states.
If states with energies much further below threshold than $E_0^{(2,1,A_1^+,\mathbb{S}_0)}$ are present in QCD, additional interpolating operators that are nearly orthogonal to the dibaryon, hexaquark, and quasi-local interpolating operators included in this calculation will be required to construct an interpolating-operator set that strongly overlaps with them. 

\begin{figure}[!t]
	\includegraphics[width=0.47\columnwidth]{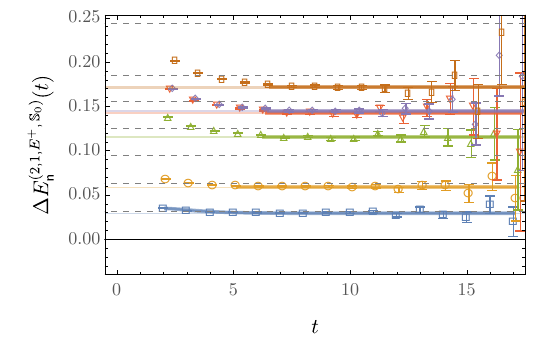}
	\includegraphics[width=0.47\columnwidth]{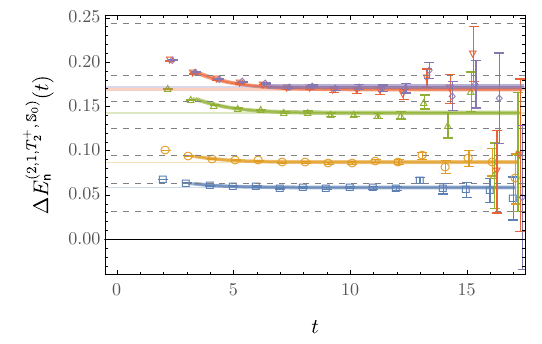}
	\includegraphics[width=0.47\columnwidth]{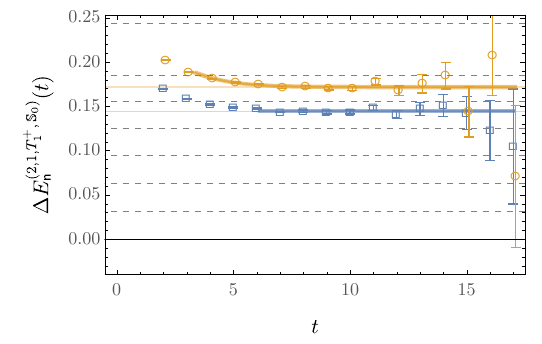}
	\includegraphics[width=0.47\columnwidth]{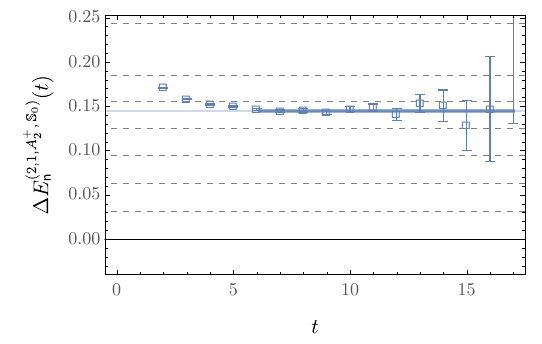}
   \caption{\label{fig:B2I1other_rainbow} GEVP effective FV energy shifts for two-nucleon systems computed using the complete sets of interpolating operators with $I=1$ and $\Gamma_J \in \{ E^+,T_2^+,T_1^+,A_2^+ \}$. As in the left side of Fig.~\ref{fig:B2I1A1_rainbow}, only GEVP energy levels below the single-nucleon first excited state are shown and the non-interacting two-nucleon energy shifts are shown as dashed gray lines. Horizontal offsets are applied to points and fit bands for clarity.}
\end{figure}

Correlation functions for two-nucleon systems with $I=1$ and other cubic transformation properties associated with non-zero angular momentum can also be constructed using linear combinations of plane-wave dibaryon operators as described in Sec.~\ref{sec:FV}. 
Interpolating operators with $I=1$ and $\Gamma_J \in \{E^+,\ T_2^+,\ T_1^+,\ A_2^+\}$ require non-translationally-invariant wavefunctions and therefore only the $D_{\mathfrak{s}\mathfrak{k}g}$ operators with $\mathfrak{s} \neq 0$ contribute to these irreps.
The set of interpolating operators computed for these irreps does not include additional operators besides two copies (for the two quark-field smearings) of each plane-wave dibaryon operator required to describe non-interacting nucleons, and for these irreps the full sets of interpolating operators are non-degenerate.
Results for GEVP FV energy shifts obtained using the same fitting methods applied to the $A_1^+$ irrep are tabulated in Appendix~\ref{app:tabs}.
The lower-energy half of the GEVP energy levels are in one-to-one correspondence with the non-interacting two-nucleon energy levels with the same $\mathfrak{s}$, as shown in Fig.~\ref{fig:B2I1other_rainbow}.\footnote{Note that the non-interacting two-nucleon energy levels have non-trivial multiplicities $\mathcal{N}_{\mathfrak{s}}^{(1,\Gamma_J)}$, as shown in Table~\ref{tab:Imult}, and in particular vanish for some choices of $\mathfrak{s}$ and $\Gamma_J$. This leads to the absence of energy levels for some combinations of $\mathfrak{s}$ and $\Gamma_J$ in Fig.~\ref{fig:B2I1other_rainbow} as well as the presence of multiple closely spaced energy levels near some other combinations of $\mathfrak{s}$ and $\Gamma_J$.}
Also, as in the $A_1^+$ irrep, the higher-energy half of the GEVP energy levels satisfy $\Delta E_{\mathsf{n}}^{(2,1,\Gamma_J,\mathbb{S}_0)}~\gtrsim~\delta^{(1,\frac{1}{2},G_1^+,\mathbb{S}_N)}$ and appear above non-interacting energy levels not associated with interpolating operators present in these sets.
These higher-energy GEVP correlation functions, therefore, can be expected to contain contamination from lower-energy states that weakly overlap with the interpolating operators present in this calculation, and the associated GEVP energy levels are considered unreliable and not presented.
Since only dibaryon operators are included in these irreps, states that are analogous to the relatively noisy state in the $A_1^+$ irrep for which $\mathcal{Z}_{\mathsf{n} H_g}^{(2,I,\Gamma_J)} \gg \mathcal{Z}_{\mathsf{n} D_{\mathfrak{s}\mathfrak{k}g}}^{(2,I,\Gamma_J)}$ may well exist in these irreps but their extraction would require additional localized operators to be included in the interpolating-operator set.

\subsection{The deuteron channel}\label{sec:I0}

\begin{figure}[!t]
	\includegraphics[width=0.47\columnwidth]{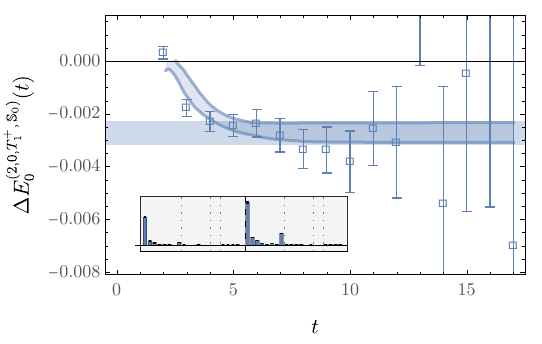}
	\includegraphics[width=0.47\columnwidth]{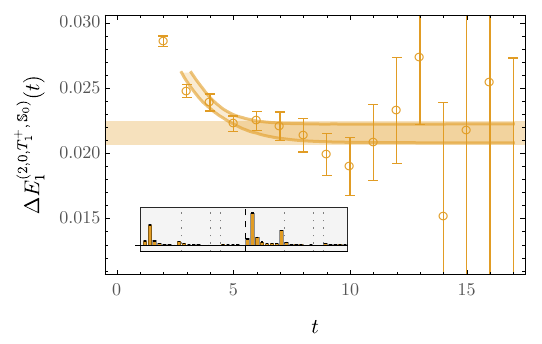}
	\includegraphics[width=0.47\columnwidth]{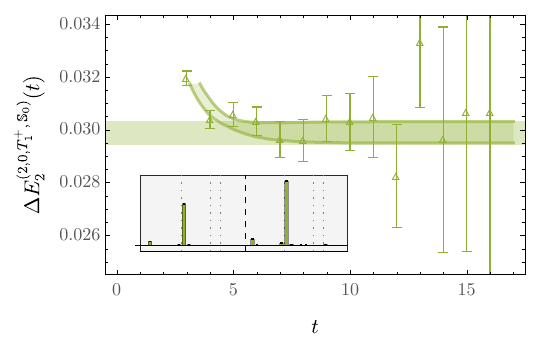}
	\includegraphics[width=0.47\columnwidth]{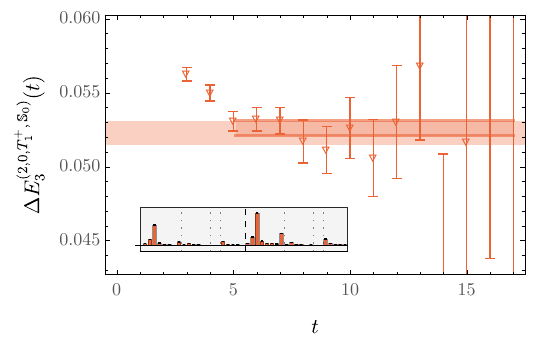}
	\includegraphics[width=0.47\columnwidth]{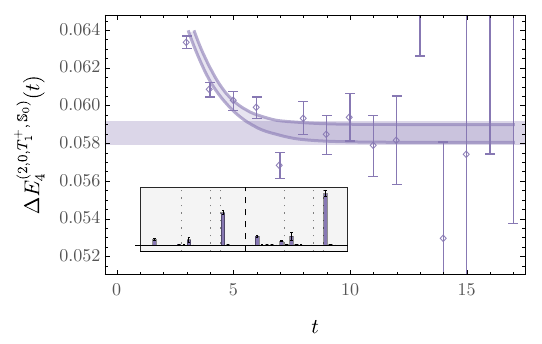}
	\includegraphics[width=0.47\columnwidth]{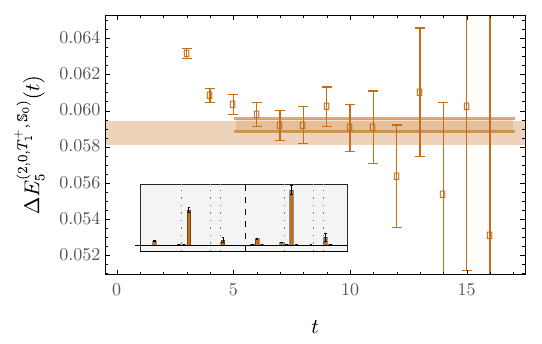}
	\includegraphics[width=0.47\columnwidth]{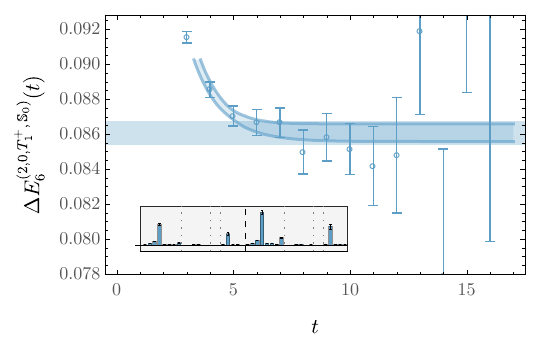}
	\includegraphics[width=0.47\columnwidth]{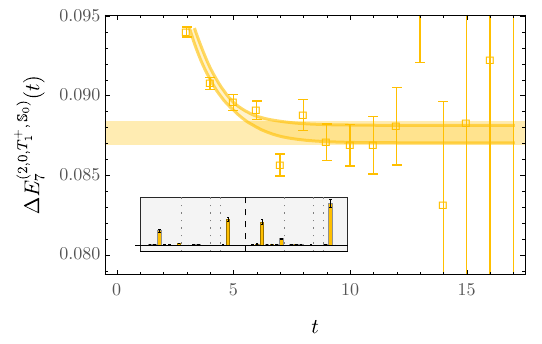}
   \caption{\label{fig:B2I0T1_FV_fits} Results for two-nucleon GEVP effective FV energy shifts for the lowest-energy GEVP correlation functions, $\mathsf{n}\in\{0,\ldots,7\}$, with $I=0, \Gamma_J = T_1^+$ obtained using the set $\mathbb{S}_{\protect \circled{0}}^{(2,0,T_1^+)}$. Statistical and systematic fit uncertainties are shown as in Fig.~\ref{fig:B2I1A1_FV_fits}. Histograms of $\mathcal{Z}_{\mathsf{n}\chi}^{(2,0,T_1^+,\mathbb{S}_0)}$ are analogous to the histograms in that figure and include a dashed line separating thin (left) and wide (right) quark-field smearing as well as dotted lines that further divide the operators into subsets corresponding to $\Gamma_\ell = A_1^+,\ E^+,\ T_1^+,\ T_2^+$ from left to right. Hexaquark operators appear rightmost among operators with $\Gamma_\ell = A_1^+$.}
\end{figure}

\begin{figure}[!tp]
	\includegraphics[width=0.47\columnwidth]{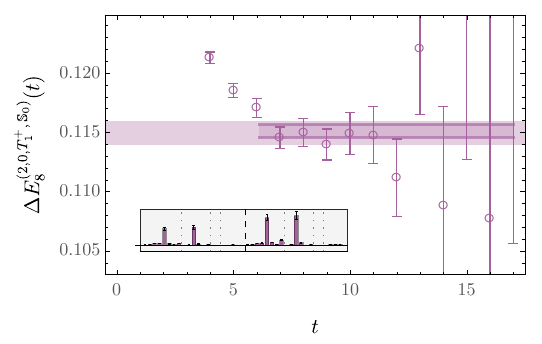}
	\includegraphics[width=0.47\columnwidth]{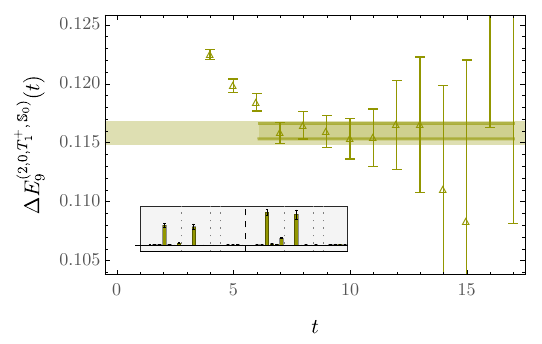}
	\includegraphics[width=0.47\columnwidth]{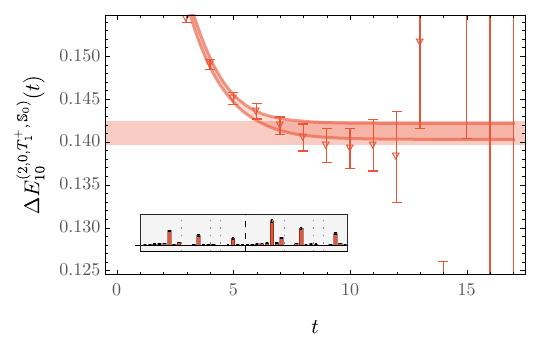}
	\includegraphics[width=0.47\columnwidth]{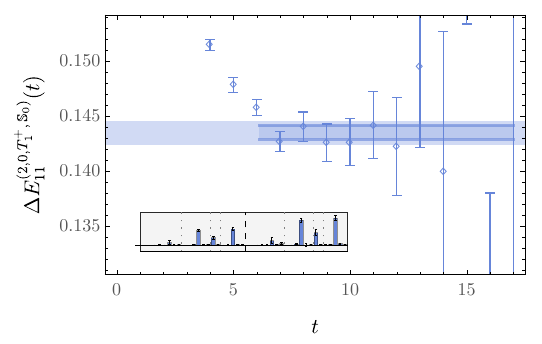}
	\includegraphics[width=0.47\columnwidth]{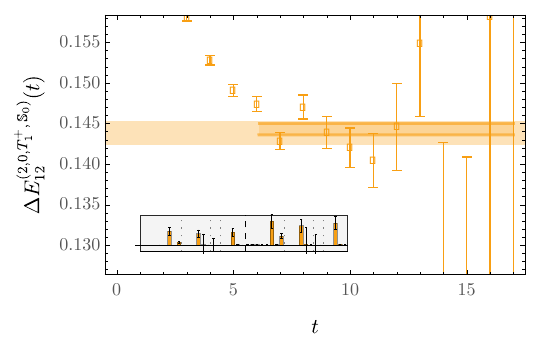}
	\includegraphics[width=0.47\columnwidth]{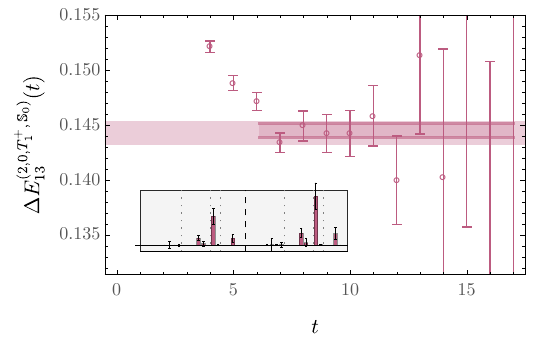}
	\includegraphics[width=0.47\columnwidth]{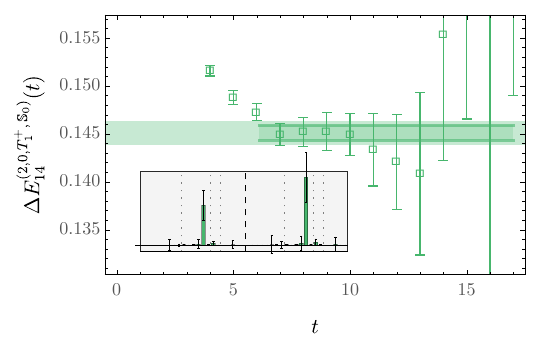}
	\includegraphics[width=0.47\columnwidth]{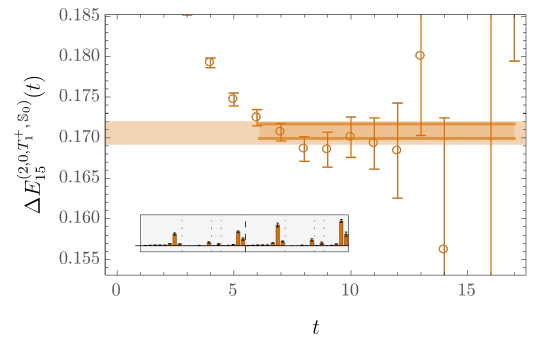}
  \caption{\label{fig:B2I0T1_FV_fits_2} Results for two-nucleon GEVP effective FV energy shifts for states $\mathsf{n}\in\{8,\ldots,15\}$ obtained using the set $\mathbb{S}_{\protect \circled{0}}^{(2,0,T_1^+)}$. Symbols and bands are analogous to those in Fig.~\ref{fig:B2I0T1_FV_fits}.}
\end{figure}

\begin{figure}[!t]
	\includegraphics[width=0.47\columnwidth]{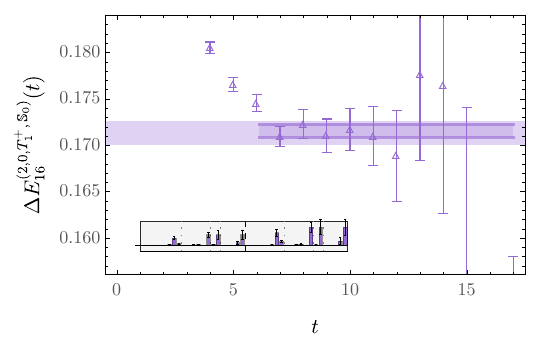}
	\includegraphics[width=0.47\columnwidth]{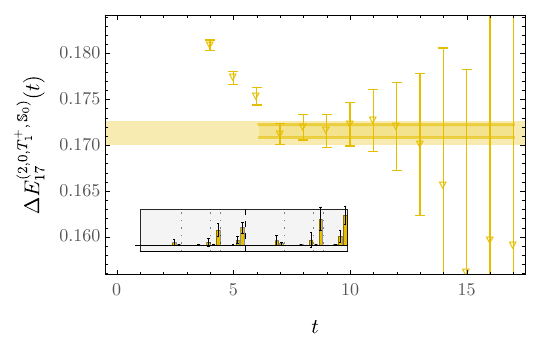}
	\includegraphics[width=0.47\columnwidth]{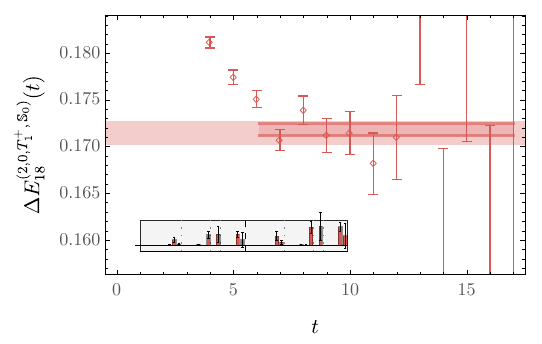}
	\includegraphics[width=0.47\columnwidth]{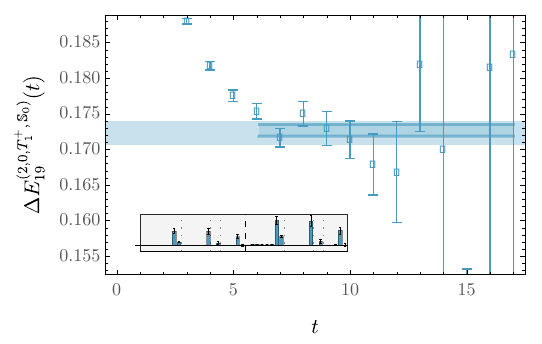}
	\includegraphics[width=0.47\columnwidth]{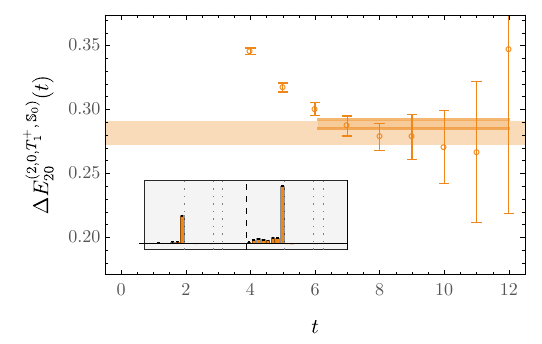}
  \caption{\label{fig:B2I0T1_FV_fits_3} Results for two-nucleon GEVP effective FV energy shifts for states $\mathsf{n}\in\{16,\ldots,20\}$ obtained using the interpolating-operator set $\mathbb{S}_{\protect \circled{0}}^{(2,0,T_1^+)}$. Symbols and bands are  analogous to those in Fig.~\ref{fig:B2I0T1_FV_fits}.}
\end{figure}

The $I=0$ deuteron channel includes flavor-antisymmetric $pn$ systems that for positive-parity spatial wavefunctions must be antisymmetric in spin and therefore in a cubic FV have $\Gamma_S = T_1^+$.\footnote{This restriction does not apply to $N \Delta$, $\Delta \Delta$, and other operators with the same quantum numbers that are not constructed from products of two nucleon operators.}
Hexaquark, quasi-local, and $\mathfrak{s}=0$ dibaryon operators in this channel have $\Gamma_\ell=A_1^+$ and therefore $\Gamma_J = T_1^+$.
Due to spin-orbit coupling, the $\Gamma_J = T_1^+$ irrep includes not only $\mathfrak{s} \geq 1$ dibaryon operators with $\Gamma_\ell = A_1^+$ but also dibaryon operators with $\Gamma_\ell \in \{E^+, T_1^+, T_2^+\}$.
Since the deuteron bound state in nature can be described as an admixture of $S$- and $D$-wave states, operators with $\Gamma_\ell \in \{E^+, T_2^+\}$ associated with $D$-wave wavefunctions in the infinite-volume limit are of particular interest.
Results for $\Gamma_J = T_1^+$, including operators with all relevant $\Gamma_\ell$, will be discussed first.
Total-angular-momentum cubic irreps $\Gamma_J \in \{T_2^+, E^+, A_2^+, A_1^+\}$ are associated with $\Gamma_\ell \neq A_1^+$ and therefore $\ell > 0$ partial waves in the infinite-volume limit. These will be discussed at the end of this section.

A total of 48 linearly independent interpolating operators with $I=0$ and $\Gamma_J = T_1^+$ are included in our calculations: two copies for the two quark-field smearings of each of $\sum_{\mathfrak{s}=0}^6 \mathcal{N}_{\mathfrak{s}}^{(0,T_1^+)}~=~20$ dibaryon operators, three quasi-local operators, and one hexaquark operator.
Including all interpolating operators leads to a correlation-function matrix whose determinant is consistent with zero at $1\sigma$ for all $t$ and whose eigenvalues cannot be reliably determined at the precision of our calculation.
As in the dineutron case, operators can be removed until a non-degenerate set  is obtained (defined as a set where the principle correlation-function effective masses are resolved from zero).
The largest non-degenerate interpolating-operator sets constructed in this way include 42 interpolating operators.

\begin{figure}[!t]
	\includegraphics[width=0.47\columnwidth]{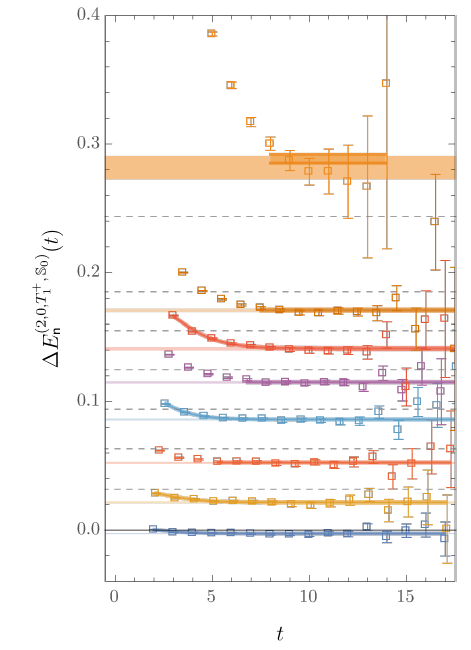}
	\includegraphics[width=0.47\columnwidth]{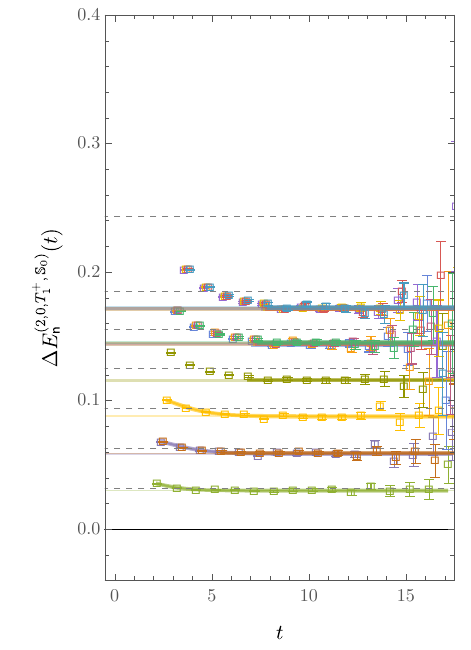}
   \caption{\label{fig:B2I0T1_rainbow} Compilations of the GEVP effective FV energy shifts shown in Figs.~\ref{fig:B2I0T1_FV_fits}-\ref{fig:B2I0T1_FV_fits_3} corresponding to energy levels below the single-nucleon first excited state and comparisons to non-interacting two-nucleon FV energy shifts (dashed lines) as in Fig.~\ref{fig:B2I1A1_rainbow}. The left (right) figures show the GEVP energy levels dominantly overlapping with interpolating operators with $\Gamma_\ell = A_1^+$ associated with $S$-wave states in the infinite-volume limit ($\Gamma_\ell \in \{E^+, T_1^+, T_2^+\}$ associated with $D$-wave and higher-partial wave states in the infinite-volume limit). The multiplicities of approximately degenerate energy levels are equal to the multiplicities $\mathcal{N}_{\mathfrak{s}}^{(0,T_1^+)}$ tabulated in Table~\ref{tab:Imult} for levels near the non-interacting levels associated with the $\mathfrak{s} \in \{0,\ldots,6\}$ shells. }
\end{figure}

\begin{figure}[!t]
	\includegraphics[width=0.6\columnwidth]{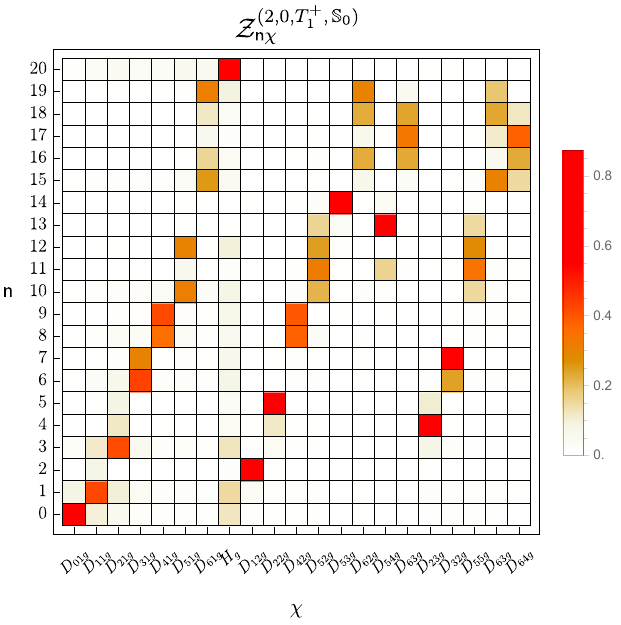}
   \caption{\label{fig:B2I0T1_Zplot} Results for overlap factors $\mathcal{Z}_{\mathsf{n} \chi}^{(2,0,T_1^+,\mathbb{S}_0)}$ for the lower half of the GEVP energy eigenstates, which are shown as histograms in Figs.~\ref{fig:B2I0T1_FV_fits}-\ref{fig:B2I0T1_FV_fits_3}. Each column represents the sum of the overlap factors corresponding to the two quark-field smearings, $\sum_g \mathcal{Z}_{\mathsf{n} D_{\mathfrak{s}\mathfrak{k}g}}^{(2,0,T_1^+,\mathbb{S}_0)}$ or $\sum_g \mathcal{Z}_{\mathsf{n} H_{g}}^{(2,0,T_1^+,\mathbb{S}_0)}$, for each dibaryon or hexaquark operator. Uncertainties, shown by error bars in histograms in Figs.~\ref{fig:B2I0T1_FV_fits}-\ref{fig:B2I0T1_FV_fits_3}, are not shown here.}
\end{figure}

One choice of a non-degenerate set of 42 interpolating operators includes all 20 $D_{\mathfrak{s}\mathfrak{k}g}^{(2,0,T_1^+)}$ interpolating operators with each smearing required to describe non-interacting nucleons with relative momentum less than $\sqrt{8}\left(\frac{2\pi}{L}\right)$ as well as the two $H_g^{(2,0,T_1^+)}$ interpolating operators corresponding to thin and wide smearing widths,
\begin{equation}
    \mathbb{S}_{\circled{0}}^{(2,0,T_1^+)} = \bigg \{ D_{\mathfrak{s}\mathfrak{k}g}^{(2,0,T_1^+)},\ H_g^{(2,0,T_1^+)} \ | \ \mathfrak{s}\in\{0,\ldots,6\},\ \mathfrak{k}\in \{1,\ldots,\mathcal{N}_{\mathfrak{s}}^{(0,T_1^+)}\},\ g\in\{T,W\} \bigg  \},
\label{eq:B0defI0T1}
\end{equation}
where the multiplicities of dibaryon operators with $\mathfrak{s}$-shell relative momenta $\mathcal{N}_{\mathfrak{s}}^{(0,T_1^+)}$ are shown in Table~\ref{tab:Imult}.
Results for GEVP effective FV energy shifts using $\mathbb{S}_{\circled{0}}^{(2,0,T_1^+)}$ and the corresponding fit results for $\Delta E_{\mathsf{n}}^{(2,0,T_1^+,\mathbb{S}_0)}$ obtained using the same fitting methods as before are shown for the levels $\mathsf{n}\in\{0,\ldots,20\}$ in Figs.~\ref{fig:B2I0T1_FV_fits}-\ref{fig:B2I0T1_FV_fits_3}.
The low-energy GEVP energy levels include one state that dominantly overlaps with the $H_W^{(2,0,T_1^+)}$ operator as well as energy levels that dominantly overlap with each $D_{\mathfrak{s}\mathfrak{k}W}^{(2,0,T_1^+)}$ operator for each $\mathfrak{s}$ and $\mathfrak{k}$.
The ground-state and first two excited-state FV energy shifts obtained with this interpolating-operator set are given by 
\begin{equation}
  \begin{split}
    \Delta E_{0}^{(2,0,T_1^+,\mathbb{S}_0)} &= -0.00273(45) = -3.70(61)(4) \text{ MeV}, \\
    \Delta E_{1}^{(2,0,T_1^+,\mathbb{S}_0)} &= 0.02156(90) = 29.3(1.2)(0.3) \text{ MeV}.
  \end{split}
\end{equation}
The inverse excitation-energy gap for the deuteron channel corresponds to Euclidean time  $1/\delta^{(2,0,T_1^+,\mathbb{S}_0)} = 41(2)$ in lattice units or $6.0(3)$ fm, and degrading statistical precision limits our results to the regime $t  \ll 1/\delta^{(2,0,T_1^+,\mathbb{S}_0)}$.
Further results for $\Delta E_\mathsf{n}^{(2,0,T_1^+,\mathbb{S}_0)}$ are presented for $\mathsf{n}\in\{0,\ldots,20\}$ in Appendix~\ref{app:tabs} and a compilation of the GEVP effective FV energy shifts for these states is shown in Fig.~\ref{fig:B2I0T1_rainbow}.
The lowest-energy half of the GEVP energy levels are below the $\mathfrak{s} = 6$ non-interacting energy level, except for one level close to the $\mathfrak{s}=8$ non-interacting energy level, while the highest-energy half of the GEVP energy levels are at significantly higher energies with $\Delta E_{\mathsf{n}}^{(2,0,T_1^+,\mathbb{S}_0)}~\gtrsim~\delta^{(1,\frac{1}{2},G_1^+,\mathbb{S}_N)}$.
Several non-interacting energy levels lie in the gap between these two halves of the GEVP energy levels, and as in the dineutron channel it is very likely that there are missing energy levels associated with states in this energy range that do not have significant overlap with the interpolating operators in $\mathbb{S}_{\circled{0}}^{(2,0,T_1^+)}$. 
The higher-energy half of the GEVP energy levels are therefore considered unreliable and are not reported.

Overlap-factor results demonstrate that states $\mathsf{n} \in \{0,\ldots, 20\}$ each dominantly overlap with either dibaryon operators with a single $\mathfrak{s}$ or with hexaquark operators as shown in Figs.~\ref{fig:B2I0T1_FV_fits}-\ref{fig:B2I0T1_FV_fits_3} and \ref{fig:B2I0T1_Zplot}.
As in the dineutron channel, the ground state has maximum overlap with $\mathfrak{s}=0$ dibaryon operators with $ \sum_{g} \mathcal{Z}_{0 D_{01g}}^{(2,0,T_1^+,\mathbb{S}_0)}~=~ 0.57(1)$, and has next largest overlaps with $\mathfrak{s}=1$ dibaryon operators with  $ \sum_{g} \mathcal{Z}_{0 D_{11g}}^{(2,0,T_1^+,\mathbb{S}_0)}~=~0.101(4)$,  and hexaquark operators with  $ \sum_{g} \mathcal{Z}_{0 H_{g}}^{(2,0,T_1^+,\mathbb{S}_0)}~=~ 0.123(1)$. 
The excited-state spectrum includes sets of approximately degenerate energy levels with multiplicity $\mathcal{N}_{\mathfrak{s}}^{(0,T_1^+)}$ that are close to the non-interacting $\mathfrak{s}$-shell energies and separated by comparatively large energy gaps.
States within each closely spaced set can be clearly distinguished by their overlap factors and dominantly overlap with either $\Gamma_\ell = A_1^+$ or $\Gamma_\ell \neq A_1^+$ operators, although the size of subdominant contributions increases with increasing $\mathfrak{s}$.

\begin{figure}[!t]
	\includegraphics[width=0.47\columnwidth]{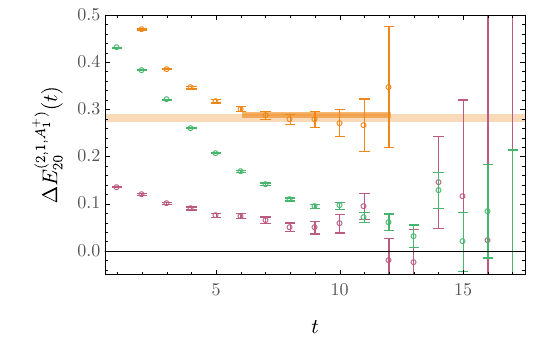}
   \caption{\label{fig:B2I0T1_hex_comp} The GEVP correlation functions for the  $\mathfrak{n}=20$ energy level dominantly overlapping with hexaquark operators, also shown in Fig.~\ref{fig:B2I0T1_FV_fits}, is displayed in
    orange with a horizontal offset for clarity. It is compared with diagonal correlation functions for $H^{(2,0,T_1^+)}_W$, in purple, and $H^{(2,0,T_1^+)}_T$, in green, to demonstrate the differences between hexaquark correlation functions and the GEVP principle correlation function with the largest hexaquark component.
   }
\end{figure}

The $\mathsf{n}=20$ energy level has maximum overlap with $H_g^{(2,0,T_1^+)}$ operators and, as in the dineutron channel, appears around the $\mathfrak{s} = 8$ non-interacting energy level (the smallest $\mathfrak{s}$ for which $D_{\mathfrak{sk}g}^{(2,0,T_1^+)}$ operators are not included).
As shown in Fig.~\ref{fig:B2I0T1_hex_comp}, the symmetric correlation functions associated with $H_g^{(2,0,T_1^+)}$ operators have effective masses that reach much lower energies roughly corresponding to the $\mathfrak{s} \in \{1,2\}$ shells.
Orthogonalization provided by solving the GEVP is needed to see that $H_g^{(2,0,T_1^+)}$ symmetric correlation functions can be described using the same states overlapping with dibaryon operators plus additional higher-energy states.

\begin{figure}[!tp]
	\includegraphics[width=0.47\columnwidth]{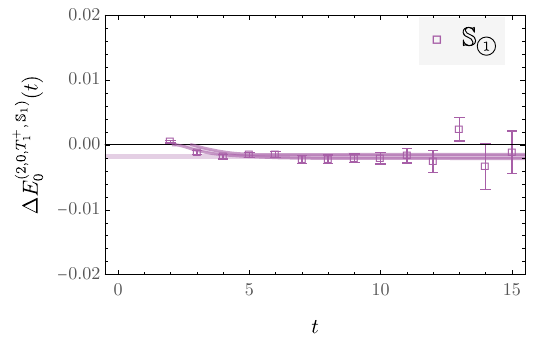}
	\includegraphics[width=0.47\columnwidth]{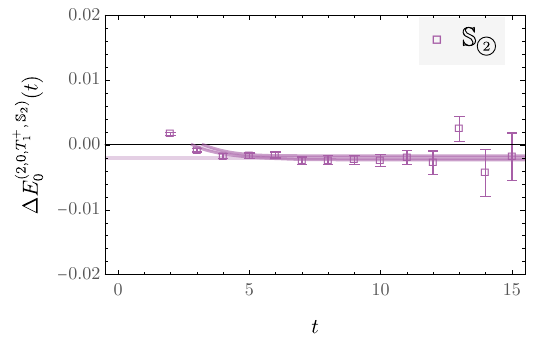}
	\includegraphics[width=0.47\columnwidth]{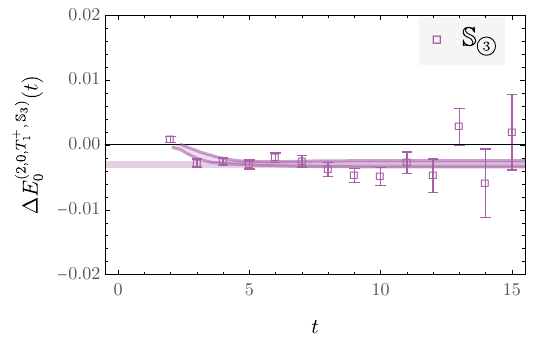}
	\includegraphics[width=0.47\columnwidth]{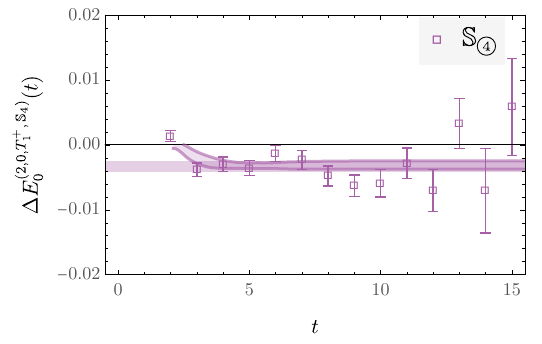}
	\includegraphics[width=0.47\columnwidth]{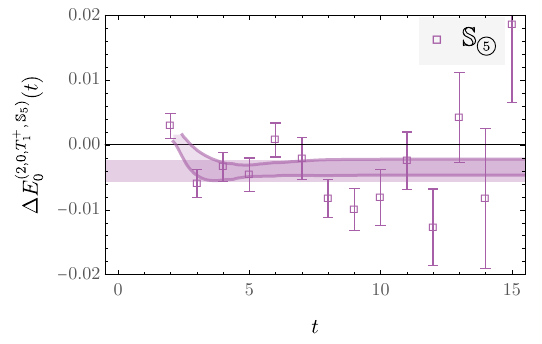}
	\includegraphics[width=0.47\columnwidth]{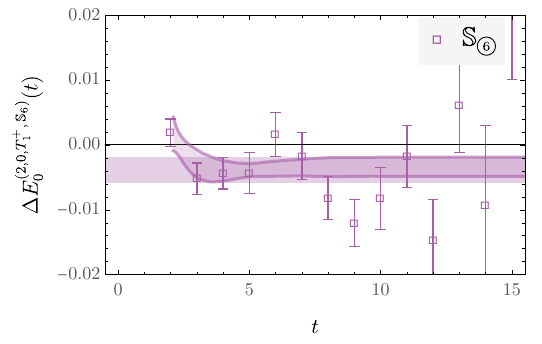}
	\includegraphics[width=0.47\columnwidth]{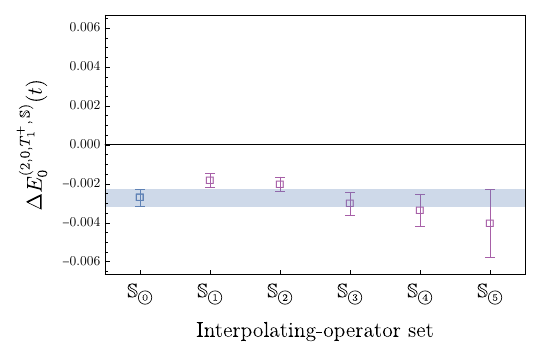}
   \caption{\label{fig:B2I0T1_goodbasis_fits} The bottom panel shows the ground-state FV energy shifts obtained using GEVP correlation functions with interpolating-operator set $\mathbb{S}_{\protect \circled{0}}^{(2,0,T_1^+)}$ shown in Fig.~\ref{fig:B2I0T1_FV_fits} (blue) compared to results obtained using seven other interpolating-operator sets defined in Eq.~\eqref{eq:BT1def2} (purple). Other panels show the ground-state effective FV energy shift for each interpolating-operator set as indicated in the corresponding legend.}
\end{figure}

\begin{figure}[!t]
	\includegraphics[width=0.47\columnwidth]{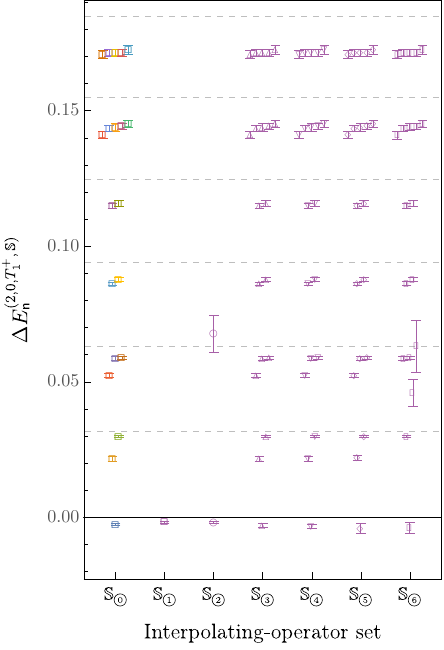}\hspace{3mm}
	\includegraphics[width=0.47\columnwidth]{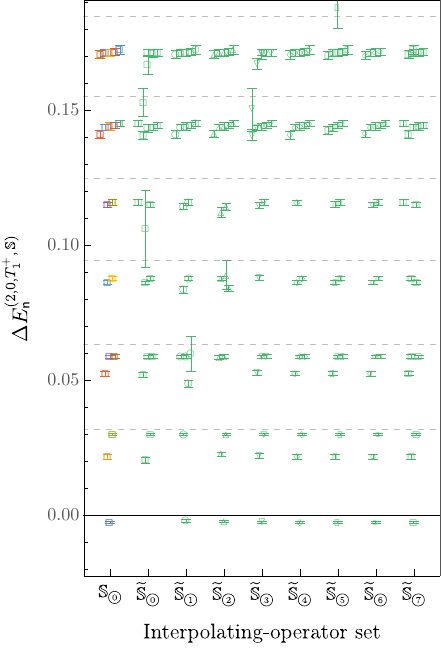}
   \caption{\label{fig:B2I0T1_badbasis_summary} FV energy shifts obtained using GEVP correlation functions with interpolating-operator set $\mathbb{S}_{\protect \circled{0}}^{(2,0,T_1^+)}$ (left column in each panel and corresponding to Figs.~\ref{fig:B2I0T1_FV_fits}-\ref{fig:B2I0T1_FV_fits_3}) are compared to results obtained using
   the interpolating-operator sets defined in Eq.~\eqref{eq:BT1def2} and featured in Fig.~\ref{fig:B2I0T1_goodbasis_fits} (left) as well as
   the eight other interpolating-operator sets defined in Eq.~\eqref{eq:BT1def3} (right).
   The color coding for $\mathbb{S}_{\protect \circled{0}}^{(2,0,T_1^+)}$ energies follows that in Figs.~\ref{fig:B2I0T1_FV_fits}-\ref{fig:B2I0T1_FV_fits_3}, and the dashed lines show non-interacting two-nucleon energies as in Fig.~\ref{fig:B2I0T1_rainbow}.}
\end{figure}

The interpolating-operator dependence of the other GEVP energy-level results in this channel can also be studied by removing or replacing interpolating operators in $\mathbb{S}_{\circled{0}}^{(2,0,T_1^+)}$ as in the dineutron channel.
Relatively precise GEVP correlation functions are obtained using the interpolating-operator sets
\begin{equation}
\begin{split}
  \mathbb{S}_{\circled{1}}^{(2,0,T_1^+)} &= \bigg \{ D_{\mathfrak{s}\mathfrak{k}g}^{(2,0,T_1^+)} \ | \ \mathfrak{s}\in \{0\} \bigg \}, \\
    \mathbb{S}_{\circled{2}}^{(2,0,T_1^+)} &= \bigg \{ D_{\mathfrak{s}\mathfrak{k}g}^{(2,0,T_1^+)},\ H_g^{(2,0,T_1^+)} \ | \ \mathfrak{s}\in \{0\} \bigg \}, \\
    \mathbb{S}_{\circled{3}}^{(2,0,T_1^+)} &= \bigg \{ \kappaOp_{1g}^{(2,0,T_1^+)},\ D_{\mathfrak{s}\mathfrak{k}g}^{(2,0,T_1^+)},\ H_g^{(2,0,T_1^+)} \ | \ \mathfrak{s}\in \{1,\ldots,6\} \bigg \}, \\
    \mathbb{S}_{\circled{4}}^{(2,0,T_1^+)} &= \bigg \{ \kappaOp_{2g}^{(2,0,T_1^+)},\ D_{\mathfrak{s}\mathfrak{k}g}^{(2,0,T_1^+)},\ H_g^{(2,0,T_1^+)} \ | \ \mathfrak{s}\in \{1,\ldots,6\} \bigg \}, \\
    \mathbb{S}_{\circled{5}}^{(2,0,T_1^+)} &= \bigg \{ \kappaOp_{3g}^{(2,0,T_1^+)},\ D_{\mathfrak{s}\mathfrak{k}g}^{(2,0,T_1^+)},\ H_g^{(2,0,T_1^+)} \ | \ \mathfrak{s}\in \{1,\ldots,6\} \bigg \}, \\
    \mathbb{S}_{\circled{6}}^{(2,0,T_1^+)} &= \bigg \{ \kappaOp_{3g}^{(2,0,T_1^+)},\ D_{1\mathfrak{k}'g}^{(2,0,T_1^+)},\ D_{\mathfrak{s}\mathfrak{k}g}^{(2,0,T_1^+)} ,\ H_g^{(2,0,T_1^+)} \ | \ \mathfrak{s}\in \{0,2,\ldots,6\} \bigg \}, \\
\end{split}\label{eq:BT1def2}
\end{equation}
where in all cases $g~\in~\{T,W\}$ and $\mathfrak{k}~\in~\{1,\ldots, \mathcal{N}_{\mathfrak{s}}^{(0,T_1^+)}\}$.\footnote{ As in the dineutron channel, relatively precise GEVP results can be extracted for an additional operator set related to $\mathbb{S}_{\circled{6}}^{(2,0,T_1^+)}$ by replacing  $\kappaOp_{3g}^{(2,0,T_1^+)}$ with $\kappaOp_{2g}^{(2,0,T_1^+)}$ using the ensemble of correlation-function matrices studied in Ref.~\cite{Wagman:2021spu} but not using the larger ensemble studied here. }
For the last set $\mathfrak{k}'~\in~\{2,\ldots,\allowbreak \mathcal{N}_{1}^{(0,T_1^+)}\}$. Results for the GEVP effective ground-state FV energy shifts obtained using these interpolating-operator sets are shown in Fig.~\ref{fig:B2I0T1_goodbasis_fits} and compared to results obtained using $\mathbb{S}_{\circled{0}}^{(2,0,T_1^+)}$.
As in the dineutron channel, interpolating-operator sets obtained by replacing $D_{01g}^{(2,0,T_1^+)}$ operators with $\kappaOp_{\mathfrak{q}g}^{(2,0,T_1^+)}$ operators lead to consistent results for the GEVP ground-state FV energy shift with uncertainties that increase with increasing $\kappa_{\mathfrak{q}}$.
Sets obtained by replacing the $D_{11g}^{(2,0,T_1^+)}$ operators with $\kappaOp_{\mathfrak{q}g}^{(2,0,T_1^+)}$ operators also give consistent results for the ground-state and other low-lying excited-state FV energy shifts although the uncertainties on the ground-state FV energy shift are much larger than those obtained using $\mathbb{S}_{\circled{0}}^{(2,0,T_1^+)}$.
This is consistent with a scenario in which the lower-energy half of the GEVP energy levels identified by $\mathbb{S}_{\circled{0}}^{(2,0,T_1^+)}$ are in one-to-one correspondence with LQCD energy eigenstates, and the $\kappaOp_{\mathfrak{q}g}^{(2,0,T_1^+)}$ operators only have significant overlap with the same set of energy eigenstates at the statistical precision of this calculation.

\begin{figure}[!t]
	\includegraphics[width=0.47\columnwidth]{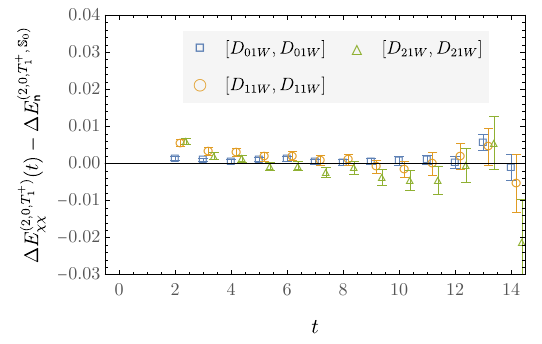}
	\includegraphics[width=0.47\columnwidth]{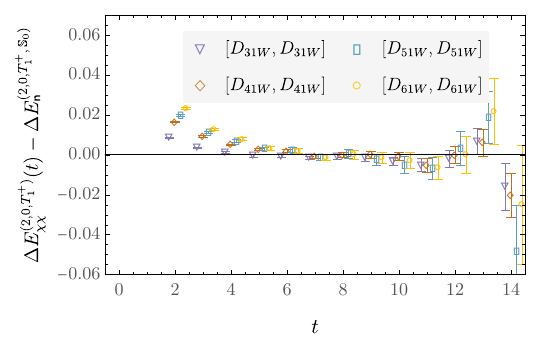}
	\includegraphics[width=0.47\columnwidth]{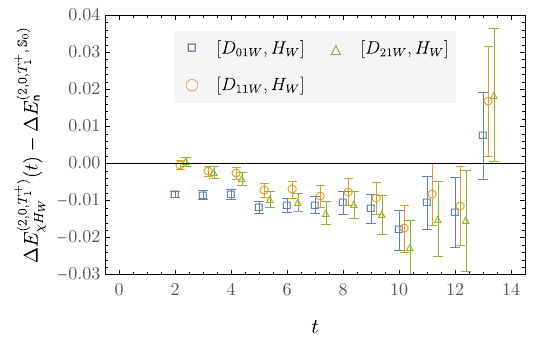}
   \includegraphics[width=0.47\columnwidth]{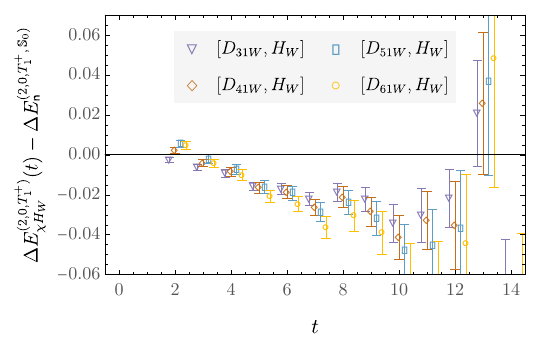}
   \caption{\label{fig:B2I0T1_diffs} The upper (lower) panels show differences between the effective FV energy shifts computed using correlation functions with  dibaryon sources and sinks $[D_{\mathfrak{s}1W}, D_{\mathfrak{s}1W}]$ (hexaquark sources and dibaryon sinks $[D_{\mathfrak{s}1W}, H_W]$) and the fitted FV energy shifts for the $\mathbb{S}_{\protect \circled{0}}^{(2,0,T_1^+)}$ GEVP energy levels that dominantly overlap with $D_{\mathfrak{s}1W}^{(2,0,T_1^+)}$, which are $\mathsf{n} \in \{0,1,3\}$ for the left panels and $\mathsf{n} \in  \{6,8,10,15\}$ for the right panels. }
\end{figure}

\begin{figure}[!t]
	\includegraphics[width=0.47\columnwidth]{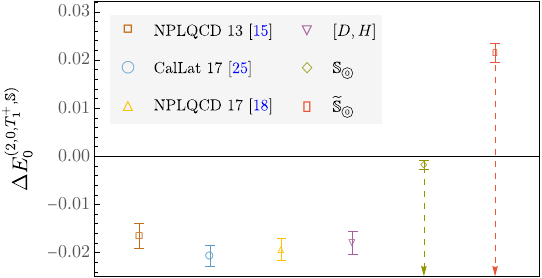}
	\includegraphics[width=0.47\columnwidth]{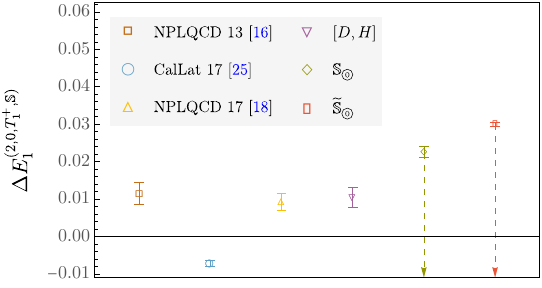}
   \caption{\label{fig:B2I0T1_spectrum_comp} Comparisons of FV energy-shift results for the $\mathsf{n}=0$ and $\mathsf{n}=1$ states between previous results obtained in Refs.~\cite{Beane:2012vq,Beane:2013br,Berkowitz:2015eaa,Wagman:2017tmp} using the same gauge field ensemble with $[D,H]$ correlation function (as well as displaced sources in Ref.~\cite{Berkowitz:2015eaa}) and results of this work obtained using variational methods with interpolating-operator set $\mathbb{S}_{\protect \circled{0}}^{(2,0,T_1^+)}$,  interpolating-operator set  $\widetilde{\mathbb{S}}_{\protect \circled{0}}^{(2,0,T_1^+)}$, and non-variational results obtained using $[D, H]$ correlation functions. Arrows emphasize the fact that the variational method provides (stochastic) upper bounds on energy levels and therefore FV energy shifts under the assumption that the nucleon mass is accurately identified, while $[D, H]$ correlation functions provide an estimate of the energy but have systematic uncertainties in both directions that are difficult to estimate. }
\end{figure}

\begin{figure}[!tp]
	\includegraphics[width=0.47\columnwidth]{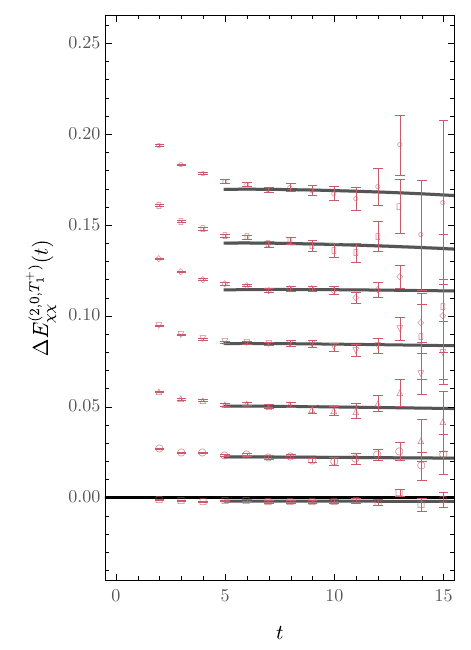}
	\includegraphics[width=0.47\columnwidth]{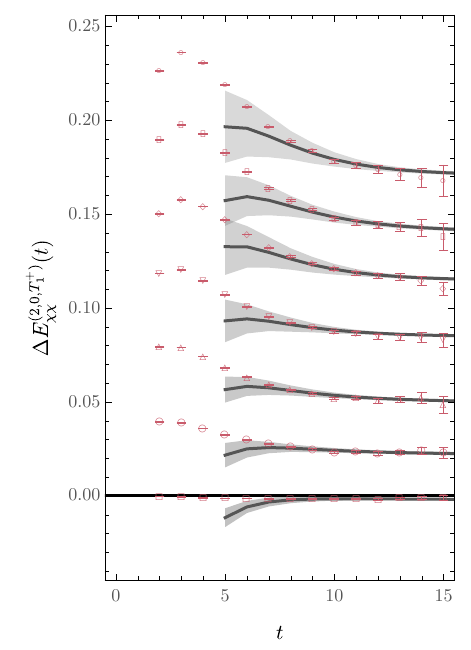}
\caption{\label{fig:B2I0T1_rainbow_reconstruction} GEVP reconstructions of $[D_{\mathfrak{s}1g}, D_{\mathfrak{s}1g}]$ with $\mathfrak{s} \in \{0,\ldots,6\}$ from bottom to top for wide (thin) quark-field smearing are shown in the left (right) panel. As in Fig.~\ref{fig:B2I1A1_rainbow_reconstruction}, gray curves and bands show the central values and uncertainties for GEVP reconstructions for $t \geq t_0$. These are obtained by inserting GEVP energy level and overlap-factor results into spectral representations for the correlation functions and not from directly fitting the correlation functions shown.}
\end{figure}

Additional interpolating-operator sets can be obtained by removing a (thin- and wide-smeared) pair of interpolating operators from $\mathbb{S}_{\circled{0}}^{(2,0,T_1^+)}$ in analogy to Eq.~\eqref{eq:BA1def3},
\begin{equation}
\begin{split}
  \widetilde{\mathbb{S}}_{\circled{0}}^{(2,0,T_1^+)} &= \left\{  D_{\mathfrak{s}\mathfrak{k}g}^{(2,0,T_1^+)},\ H_g^{(2,0,T_1^+)} \ | \ \mathfrak{s}\in \{1,\ldots,6\}  \right\}, \\
  \widetilde{\mathbb{S}}_{\circled{m}}^{(2,0,T_1^+)} &= \left\{  D_{\mathfrak{s}\mathfrak{k}g}^{(2,0,T_1^+)},\ D_{\text{m}\mathfrak{k}'g}^{(2,0,T_1^+)},\ H_g^{(2,0,T_1^+)} \ | \ \mathfrak{s}\in \{0,\ldots,6\} \setminus \text{m} \right\} ,\ \text{m} \in \{1,\ldots 6\}, \\
    \widetilde{\mathbb{S}}_{\circled{7}}^{(2,0,T_1^+)} &= \left\{  D_{\mathfrak{s}\mathfrak{k}g}^{(2,0,T_1^+)} \  | \ \mathfrak{s}\in \{0,\ldots,6\} \right\} ,
\end{split}\label{eq:BT1def3}
\end{equation}
where $g\in\{T,W\}$, $\mathfrak{k}\in\{1,\ldots,\mathcal{N}_{\mathfrak{s}}^{(0,T_1^+)}\}$, and  $\mathfrak{k}'\in\{2,\ldots,\mathcal{N}_{\text{m}}^{(0,T_1^+)}\}$.
Results for the GEVP effective FV energy shifts and fit results for GEVP energy levels with $\mathsf{n}\in\{0,\ldots,20\}$ using these interpolating-operator sets are shown in Fig.~\ref{fig:B2I0T1_badbasis_summary}.
As in the dineutron channel, it is clear that removing the interpolating-operator pair that dominantly overlaps with a particular energy level also removes the corresponding GEVP energy level from the results for the energy spectrum.
The fact that a large interpolating-operator set is not sufficient to guarantee a reliable ground-state energy extraction is demonstrated by the case of $\widetilde{\mathbb{S}}_{\circled{0}}^{(2,0,T_1^+)}$, where a set of 40 interpolating operators leads to a ground-state energy result that is $5\sigma$ larger than the corresponding result obtained using $\mathbb{S}_{\circled{0}}^{(2,0,T_1^+)}$.

\begin{figure}[!t]
	\includegraphics[width=0.47\columnwidth]{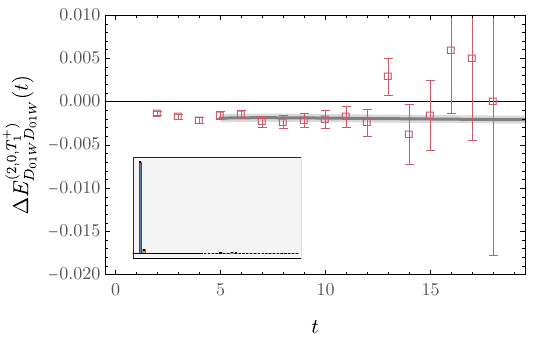}
	\includegraphics[width=0.47\columnwidth]{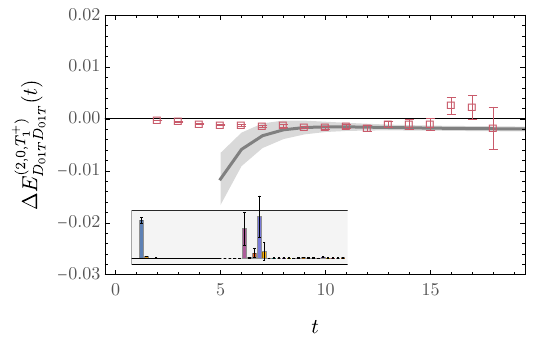}
	\includegraphics[width=0.47\columnwidth]{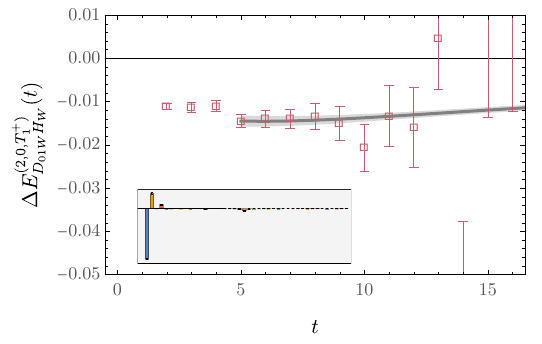}
	\includegraphics[width=0.47\columnwidth]{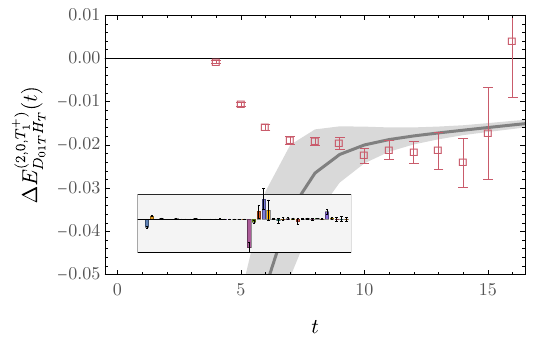}
	\includegraphics[width=0.47\columnwidth]{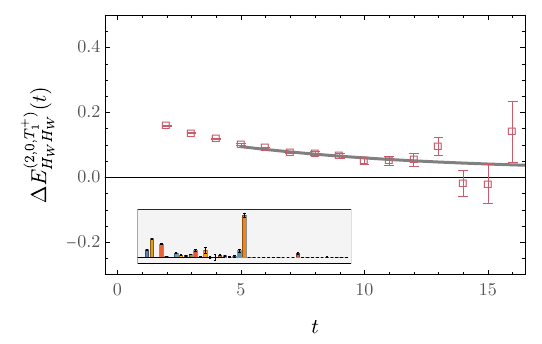}
	\includegraphics[width=0.47\columnwidth]{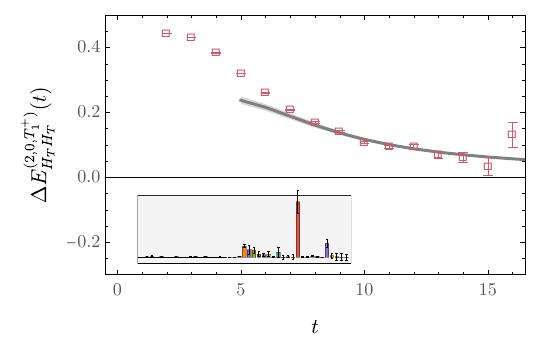}
   \caption{\label{fig:B2I0T1_reconstruction_1}
   Effective FV energy shifts and GEVP reconstructions obtained using $\mathbb{S}_{\protect \circled{0}}^{(2,0,T_1^+)}$ for $[D_{01g}, D_{01g}]$, $[D_{01g}, H_g]$, and $[H_g, H_g]$ correlation functions with wide (thin) quark-field smearing are shown in the left (right) panels. Histograms of $\widetilde{\mathcal{Z}}_{\mathsf{n}\chi\chi'}^{(2,0,T_1^+,\mathbb{S}_0)}$ are as in Fig.~\ref{fig:B2I1A1_reconstruction_1}.}
\end{figure}

Differences between FV energy shifts determined with individual $[D_{\mathfrak{s}1W}, H_W]$ correlation functions and the $\mathbb{S}_{\circled{0}}^{(2,0,T_1^+)}$ GEVP FV energy shifts from the state maximally overlapping with $D_{\mathfrak{s}1W}$, as well as the corresponding differences between $[D_{\mathfrak{s}1W}, D_{\mathfrak{s}1W}]$ FV energy shifts and $\mathbb{S}_{\circled{0}}^{(2,0,T_1^+)}$ GEVP FV energy shifts,  are shown in Fig.~\ref{fig:B2I0T1_diffs}.
As in the dineutron channel, the $[D_{\mathfrak{s}1g}, D_{\mathfrak{s}1g}]$ FV effective energy shifts approach the GEVP results from above and are indistinguishable for $t \gtrsim t_0 = 5$. Conversely, the effective energies and the FV energy shifts determined using individual $[D_{\mathfrak{s}1g}, H_g]$ correlation functions both appear (1-5)$\sigma$ lower than the corresponding GEVP results for $ t \gtrsim 5$ until statistical noise becomes larger than the differences resolved at smaller $t$.
Applying the same fitting procedure used above to combined fits of $[D_{01g}, H_g]$ correlation functions with both quark-field smearings gives a ground-state FV energy shift of $-0.0176(26)$.
Similarly, combined fits of $[D_{11g}, H_g]$ correlation functions with both quark-field smearings gives a FV energy shift of $0.0098(24)$.
Higher-statistics analyses of $[D_{01g}, H_{g'}]$ and $[D_{11g}, H_{g'}]$ correlation functions with smeared sources and both smeared and point-like sinks in Ref.~\cite{Beane:2012vq}, Ref.~\cite{Berkowitz:2015eaa}, and Ref.~\cite{Wagman:2017tmp} give more precise results of $-0.0165(26)$, $-0.0206(22)$,and  $-0.0194(23)$ for the ground-state FV energy shift, respectively, and Ref.~\cite{Beane:2013br} and Ref.~\cite{Wagman:2017tmp} give results of $0.0115(29)$ and $0.0092(23)$ for the first excited-state FV energy shift, respectively. These are consistent with the $[D_{\mathfrak{s}1g}, H_{g}]$ results of this work at the $1\sigma$ level, but are in tension with the first excited-state energy determined in Ref.~\cite{Berkowitz:2015eaa}, -0.0073(8), obtained using asymmetric correlation-functions with displaced two-nucleon sources and $D_{01g}$ sinks. 
As shown in Fig.~\ref{fig:B2I0T1_spectrum_comp}, interpolating-operator dependence among best-fit FV energy shift determinations is larger than the uncertainties of individual fit results, and it is important to emphasize that results obtained using the variational method should be interpreted as upper bounds. 

Further tests of the consistency of GEVP results using $\mathbb{S}_{\circled{0}}^{(2,0,T_1^+)}$ are obtained by comparing GEVP reconstructions of correlation functions in the original interpolating-operator set to the corresponding effective FV energy-shift results.
This comparison is shown for the GEVP energy levels with the largest overlap with the $S$-wave $D_{\mathfrak{s}1g}^{(2,0,T_1^+)}$ interpolating operators in Fig.~\ref{fig:B2I0T1_rainbow_reconstruction}.
Agreement at (1-2)$\sigma$ is seen for $[D_{\mathfrak{s}1W}, D_{\mathfrak{s}1W}]$ and $[D_{\mathfrak{s}1T}, D_{\mathfrak{s}1T}]$ for $t \gtrsim 6$. 

Analogous GEVP reconstructions for $[D_{01g}, D_{01g}]$, $[D_{01g}, H_g]$, and $[H_g, H_g]$ correlation functions are shown in Fig.~\ref{fig:B2I0T1_reconstruction_1}.
There is agreement at the level of (1-2)$\sigma$ between the GEVP reconstructions and correlation function results for $t \gtrsim  6$ with a similar pattern of cancellations between opposite-sign  ground- and excited-state contributions leading to the subthreshold behavior seen in $[D_{01g}, H_g]$ effective FV energy shifts.

\begin{figure}[!t]
	\includegraphics[width=0.47\columnwidth]{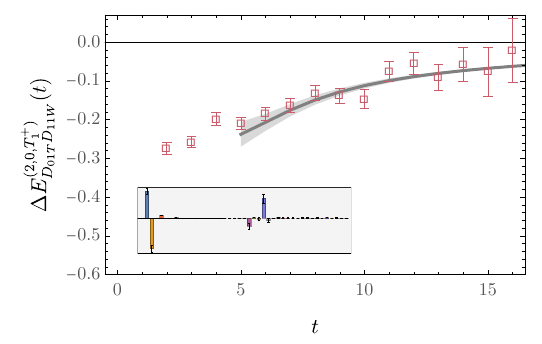}
	\includegraphics[width=0.47\columnwidth]{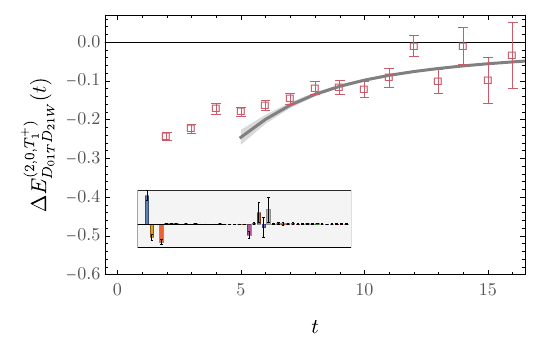}
	\includegraphics[width=0.47\columnwidth]{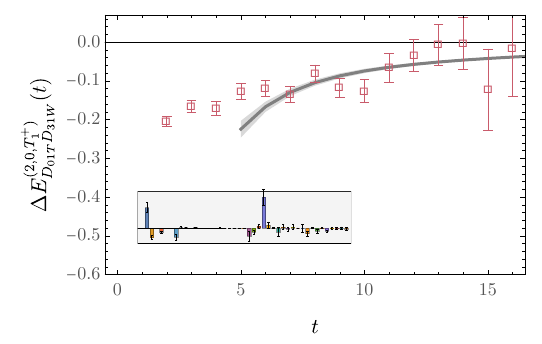}
	\includegraphics[width=0.47\columnwidth]{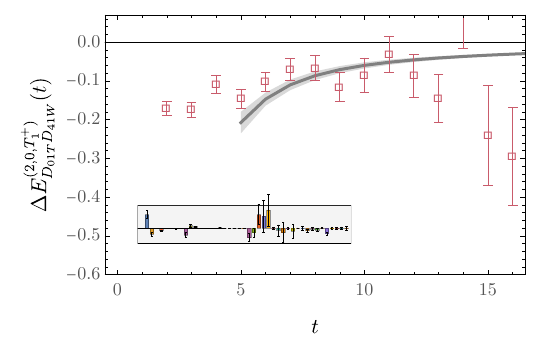}
	\includegraphics[width=0.47\columnwidth]{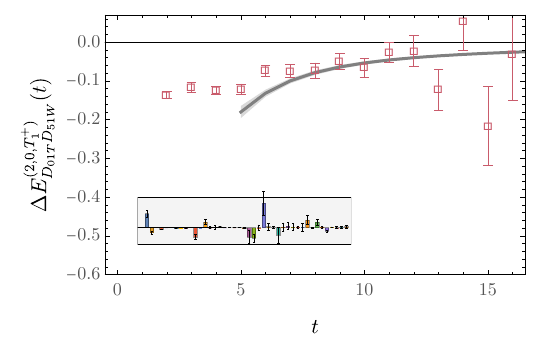}
	\includegraphics[width=0.47\columnwidth]{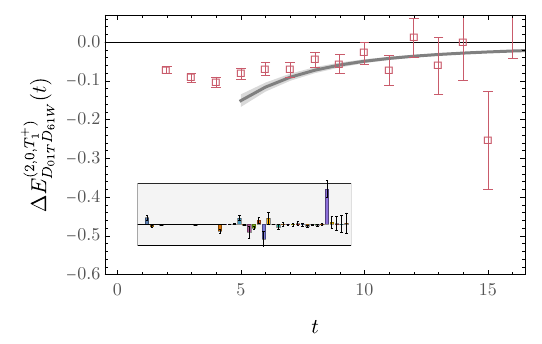}
   \caption{\label{fig:B2I0T1_reconstruction_3} Effective FV energy shifts and GEVP reconstructions obtained using $\mathbb{S}_{\protect \circled{0}}^{(2,0,T_1^+)}$ for $[D_{01T}, D_{\mathfrak{s}1W}]$ correlation functions with $\mathfrak{s} \in \{1,\ldots, 6\}$. Histograms of $\widetilde{\mathcal{Z}}_{\mathsf{n}\chi\chi'}^{(2,0,T_1^+,\mathbb{S}_0)}$ are as in Fig.~\ref{fig:B2I1A1_reconstruction_1}.}
\end{figure}

Other asymmetric correlation functions of the form $[D_{01T}, D_{\mathfrak{s}1W}]$ with $\mathfrak{s} \neq 0$ exhibit effective FV energy shifts significantly below zero over the range of $t$ studied in this calculation, as shown in Fig.~\ref{fig:B2I0T1_reconstruction_3}.
Applying the same fitting procedure described above to these correlation functions leads to $-0.074(25)$, $-0.073(15)$, $-0.073(21)$, $-0.073(12)$, $-0.045(16)$, and $-0.035(23)$ for $\mathfrak{s} = 1$ to $\mathfrak{s} = 6$, respectively.
Despite the statistically significant discrepancies between $\mathfrak{s}=2$ results and $\Delta E_{0}^{(2,0,T_1^+,\mathbb{S}_0)} = -0.00273(45)$ , the GEVP reconstruction using $\mathbb{S}_{\circled{0}}^{(2,0,T_1^+)}$ reproduces the sub-threshold behavior of effective FV energy shifts over the range of $t$ where signals can be resolved through opposite-sign linear combinations of ground- and excited-state contributions.

\begin{figure}[!t]
	\includegraphics[width=0.47\columnwidth]{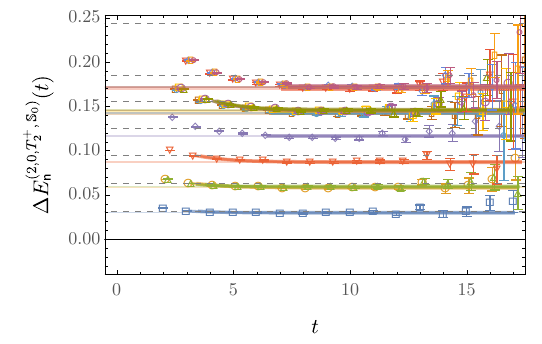}
	\includegraphics[width=0.47\columnwidth]{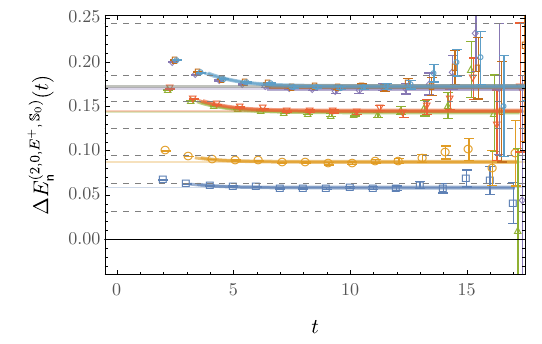}
	\includegraphics[width=0.47\columnwidth]{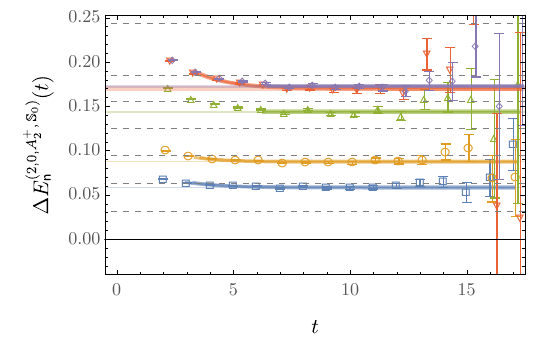}
	\includegraphics[width=0.47\columnwidth]{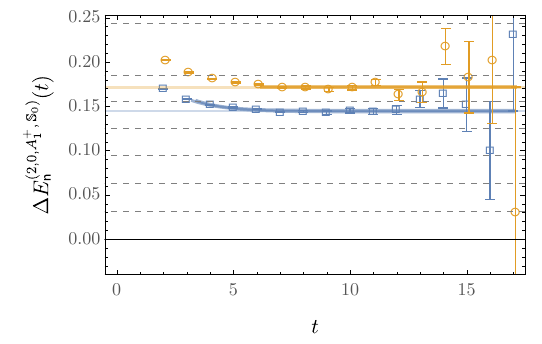}
   \caption{\label{fig:B2I0other_rainbow} GEVP effective FV energy shifts for two-nucleon systems computed using the complete sets of interpolating operators with $I=0$ and $\Gamma_J \in \{ T_2^+,E^+,A_2^+,A_1^+\}$. As in the left panel of Fig.~\ref{fig:B2I1A1_rainbow}, only GEVP energy levels below the single-nucleon first excited state are shown and the non-interacting two-nucleon energy shifts are shown as dashed gray lines. Horizontal offsets are applied to points and fit bands for clarity.}
\end{figure}

Two-nucleon correlation functions with $I=0$ and other cubic irreps associated with non-zero total angular momenta can also be constructed using linear combinations of plane-wave dibaryon operators. 
Interpolating operators with $I=0$ and $\Gamma_J \in \{T_2^+, E^+, A_2^+, A_1^+\}$ require spatial wavefunctions with $\Gamma_\ell \neq A_1^+$ and therefore only include dibaryon operators with $\mathfrak{s} > 0$.
After constructing dibaryon interpolating operators where the product of spin and orbital angular momenta transforms appropriately under the cubic group as described in Sec.~\ref{sec:FV} and Appendix~\ref{app:FV}, it is found that the full sets of dibaryon interpolating operators computed for each $\Gamma_J$ can be included in a non-degenerate interpolating-operator set where correlation-function-matrix eigenvalues and eigenvectors can be computed without issues of degeneracy.
These correlation-function matrices have rank 4, 10, 14, and 28 for the $A_1^+$, $A_2^+$, $E^+$, and $T_2^+$ irreps, respectively.
The same fitting methods applied above are used to determine energy levels from the GEVP correlation functions for each irrep; effective FV energy-shift results are shown in Fig.~\ref{fig:B2I0other_rainbow} and fitted energy results are tabulated in Appendix~\ref{app:tabs}.
In all irreps, the lower-energy half of the GEVP energy levels are in one-to-one correspondence with the non-interacting nucleon energy levels with the same $\mathfrak{s}$, which include non-trivial multiplicities $\mathcal{N}_{\mathfrak{s}}^{(0,\Gamma_J)}$ given in Table~\ref{tab:Imult}, as shown in Fig.~\ref{fig:B2I0other_rainbow}.
As in the dineutron channel, the higher-energy half of the GEVP energy levels satisfy $\Delta E_{\mathsf{n}}^{(2,0,\Gamma_J,\mathbb{S}_0)}~\gtrsim~\delta^{(1,\frac{1}{2},G_1^+,\mathbb{S}_N)}$, appearing above non-interacting energy levels not associated with interpolating operators present in this calculation, and are therefore considered unreliable and are not presented.
Results for the comparatively well-determined GEVP energy levels in all irreps are discussed further in the next section.

\subsection{Summary of spectrum and phase shift results}\label{sec:summary}

\begin{figure}[!t]
	\includegraphics[width=0.46\columnwidth]{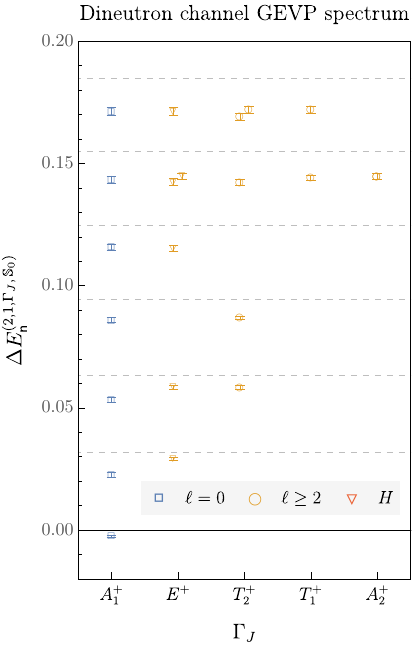}\hspace{10mm}
	\includegraphics[width=0.46\columnwidth]{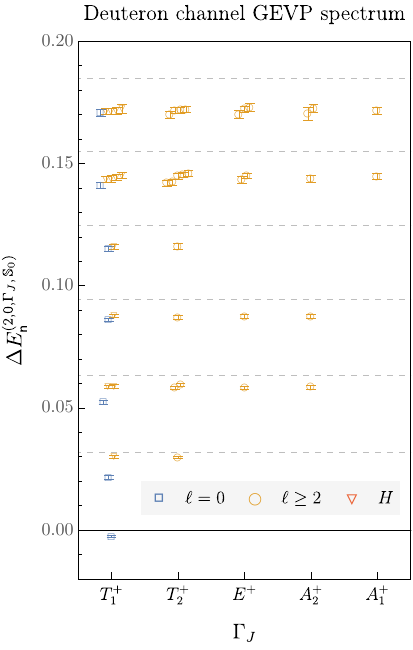}
   \caption{\label{fig:B2_summary} GEVP energy levels below the single-nucleon first excited state  with $I=1$ ($I=0$) in the left (right) panel computed using interpolating-operator set $\mathbb{S}_{\protect \circled{0}}^{(2,1,A_1^+)}$ ($\mathbb{S}_{\protect \circled{0}}^{(2,0,T_1^+)}$) for the $\Gamma_J = A_1^+$ ($\Gamma_J = T_1^+$) irrep and using all available interpolating operators for the other irreps. Points with error bars show statistical and fitting systematic uncertainties added in quadrature and are distinguished by shape and color according to whether the interpolating operator with the largest overlap with each GEVP eigenstate is a dibaryon interpolating operator with $\Gamma_\ell = A_1^+$ associated with $S$-wave states in infinite volume, a dibaryon operator with $\Gamma_\ell \neq A_1^+$ associated with $D$-wave and higher partial-wave states, or a hexaquark operator. See Figs.~\ref{fig:B2I1A1_badbasis_summary} and \ref{fig:B2I0T1_badbasis_summary} for summaries of the results, including quasi-local interpolating operators and other interpolating-operator sets. The horizontal lines indicate the positions of non-interacting two-nucleon energy levels. }
\end{figure}

The GEVP results for $\Delta E_{\mathsf{n}}^{(2,I,\Gamma_J,\mathbb{S}_0)}$ obtained using interpolating-operator sets $\mathbb{S}_{\circled{0}}^{(2,1,A_1^+)}$ and $\mathbb{S}_{\circled{0}}^{(2,0,T_1^+)}$ including all available dibaryon and hexaquark operators (but no quasi-local operators) for irreps required for studying $S$-wave nucleon-nucleon scattering, as well as GEVP results including all available dibaryon interpolating operators for other irreps needed to study $D$-wave and higher partial-wave nucleon-nucleon scattering, are shown in Fig.~\ref{fig:B2_summary}.
A total of 22 GEVP energy levels with $I=1$ and 49 GEVP energy levels with $I=0$  (including all irreps) provide a determination of the low-energy spectrum with no gaps between levels that span multiple non-interacting energy levels; as discussed above, another equal-sized set of higher-energy levels are expected to include unsuppressed contributions from excited states approximately orthogonal to the interpolating-operator sets studied here and are not included in the analysis below.
Besides the energy levels dominantly overlapping with hexaquark operators, all GEVP energy levels appear slightly below one non-interacting energy level.
Statistically significant differences are found between GEVP energy levels dominantly overlapping with $S$-wave and $D$-wave dibaryon operators with $\mathfrak{s} = 1$ in both channels, for example $\Delta E_0^{(2,1,E^+)} - \Delta E_1^{(2,1,A_1^+)} = 0.007(1)$ and $\Delta E_2^{(2,0,T_1^+)} - \Delta E_1^{(2,0,T_1^+)} = 0.008(1)$.
Smaller differences are seen between GEVP energy levels dominantly overlapping with $S$-wave and $D$-wave dibaryon operators with $\mathfrak{s} = 2$, for example $\Delta E_1^{(2,1,E^+)} - \Delta E_2^{(2,1,A_1^+)} = 0.005(1)$ and $\Delta E_2^{(2,0,T_1^+)} - \Delta E_1^{(2,0,T_1^+)} = 0.006(1)$, while analogous differences for $\mathfrak{s} \geq 3$ are consistent with zero at 1$\sigma$.

Preliminary extractions of scattering phase shifts under simplifying assumptions that partial-wave mixing can be neglected are used to facilitate comparisons with results from other works using similar but not identical LQCD actions.\footnote{We focus on comparisons of scattering phase shifts obtained by applying quantization conditions to FV energy levels in order to study interpolating-operator dependence and other systematic uncertainties of this method. Potential-method-based results from the HAL QCD collaboration predict that two-nucleon systems are unbound in both isospin channels for similar unphysical quark masses~\cite{Inoue:2011ai,Ishii:2013ira}.} 
A more detailed analysis of scattering phase shifts including partial-wave mixing that can be used to extract, for example, the $^3S_1-^3\hskip -0.03in D_1$ mixing parameters in the deuteron channel is deferred to ongoing work, where GEVP results for multiple lattice volumes can be also included.
We further note that all extracted FV energies sit far above the left-hand t-channel cut (corresponding to $k^2=-(m_\pi/2)^2$), and the use of L\"uscher's quantization condition to constrain $k \cot \delta$ functions remains valid for these energies. See Ref.~\cite{Raposo:2023nex} for further discussion of the role of the left-hand t-channel cut in constraining scattering amplitudes using L\"uscher's quantization condition.

\begin{figure}[!t]
	\includegraphics[width=\columnwidth]{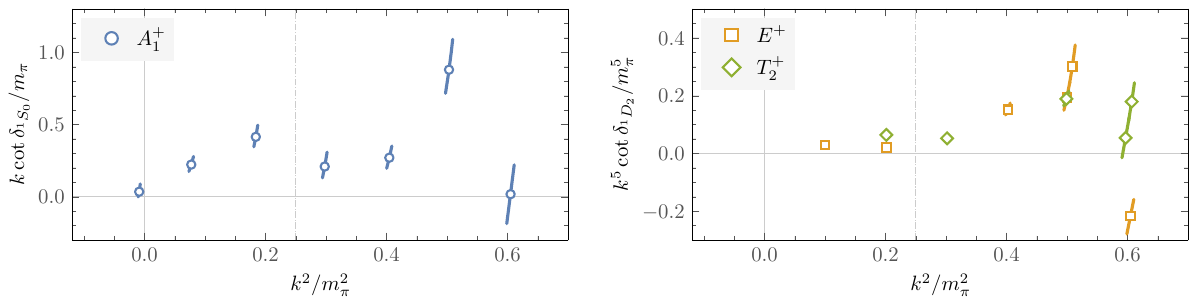}
   \caption{\label{fig:I1_phaseshift} $I=1$ phase shifts extracted from the GEVP spectrum. The dashed vertical lines (at $k^2 = m_\pi^2/4$) show the threshold above which an effective-range expansion of the $k \cot \delta$ functions is not guaranteed to converge. The phase-shift associated with $\Delta E_3^{(2,1,A_1^+)}$ is not shown because the relatively large uncertainty of this FV energy shift crosses a singularity of the quantization condition and is therefore consistent with any phase-shift value. }
\end{figure}

\begin{figure}[!t]
	\includegraphics[width=\columnwidth]{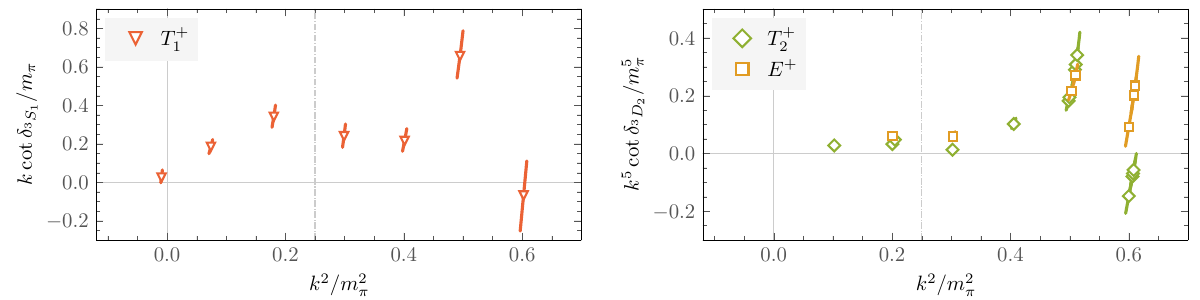}
   \caption{\label{fig:I0_phaseshift} $I=0$ phase shifts extracted from the GEVP spectrum. The dashed vertical lines (at $k^2 = m_\pi^2/4$) show the threshold above which an effective-range expansion of the $k \cot \delta$ functions is not guaranteed to converge.}
\end{figure}

For the $I=1$ channel, the quantization conditions presented in Refs.~\cite{Luu:2011ep,Briceno:2013lba}, and summarized in Appendix~\ref{app:QC}, relate the FV energy shifts in each irrep to the infinite-volume scattering phase shifts $\delta_{{}^{2S+1} \ell_J}$.
In general, the quantization condition involves the determinant of a matrix that depends on the scattering phase shifts in all partial waves.
Previous works on nucleon-nucleon scattering have truncated this matrix to a single entry corresponding to the lowest partial wave contributing to each irrep.
Future works including results with different boosts and lattice volumes can be used to test this approximation; here this same approximation is applied in order to provide direct comparisons with previous determinations.
Under this approximation, $\Gamma_J=A_1^+$ energy levels can be mapped to ${}^1S_0$ phase shifts and $\Gamma_J\in\{E^+, T_2^+\}$ energy levels can be mapped to ${}^1D_2$ phase shifts, the results of which are shown in Fig.~\ref{fig:I1_phaseshift}.
For the $I=0$ channel, wavefunctions corresponding to multiple $\Gamma_\ell$ are associated with the same $\Gamma_J$.
For consistency with the approximation of no partial-wave mixing used in this section, only the energy levels associated with the $\Gamma_\ell$ corresponding to the lowest partial wave are considered.
The quantization conditions can then be truncated to the lowest partial wave, as presented in Ref.~\cite{Briceno:2013lba} and summarized in Appendix~\ref{app:QC}, to provide maps from FV energy levels to scattering phase shifts neglecting partial-wave mixing.\footnote{For the levels from $\Gamma_J=T_1^+$, the $^3S_1$ and $^3\hskip -0.03in D_1$ partial waves mix. Here, the Blatt-Biedenharn parametrization of the scattering amplitude~\cite{Blatt:1952zza} is used, with only the $\alpha$-wave being computed, but labeled as $^3S_1$.}
Results are shown in Fig.~\ref{fig:I0_phaseshift}.
The fact that the wavefunctions for $\Gamma_\ell \in A_1^+$ and $\Gamma_\ell \in \{E^+, T_2^+\}$ transform analogously to spherical harmonics with $\ell=0$ and $\ell=2$, respectively, as discussed in Appendix~\ref{app:FV}, suggests that associating energy levels dominantly overlapping with the corresponding interpolating operators with definite $\ell$ values is reasonable.
Although the $\Gamma_\ell = T_1^+$ and $\Gamma_\ell = A_2^+$ spectra can be used to extract $G$-wave and $I$-wave phase shifts under the approximation of a single partial-wave truncation of the quantization condition, the interpolating operators for these irreps do not transform analogously to $SO(3)$ spherical harmonics with $\ell=4$ and $\ell=6$, and an identification of the corresponding energy levels with purely $G$-wave and $I$-wave states is not justified, even if partial-wave mixing is small.

\begin{figure}[!t]
\centering
\resizebox{\textwidth}{!}{\input{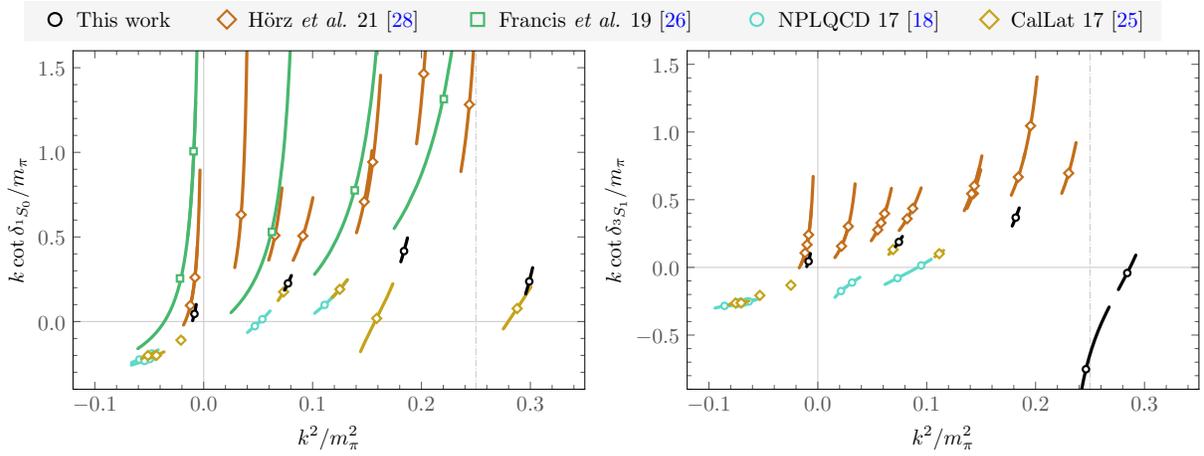}}
\caption{Comparison of the $I=1$ (left) and $I=0$ (right) two-nucleon $S$-wave phase shifts determined in this work with previous calculations using $[D,H]$ correlation functions from the NPLQCD ~\cite{Wagman:2017tmp} and CalLat collaborations~\cite{Berkowitz:2015eaa}, previous calculations using $[D,D]$ correlation functions in Ref.~\cite{Francis:2018qch}, and those using variational methods with sets of two dibaryon interpolating operators in several boosted frames in Ref.~\cite{Horz:2020zvv}. The dashed vertical lines (at $k^2 = m_\pi^2/4$) show the threshold above which an effective-range expansion of the $k \cot \delta$ functions is not guaranteed to converge.}
\label{fig:phase_sift_comparison_plots}
\end{figure}

The same approximation of neglecting partial-wave mixing has been applied in previous LQCD calculations of two-nucleon systems with similar quark masses, various lattice actions, and different interpolating-operator sets.
Results from previous LQCD calculations using $[D,H]$ correlation functions from Refs.~\cite{Wagman:2017tmp,Berkowitz:2015eaa} as well as results using $[D,D]$ correlation functions in several boosted frames from Ref.~\cite{Francis:2018qch} and GEVP results using sets of two dibaryon interpolating operators in several boosted frames from Ref.~\cite{Horz:2020zvv} are compared   in Fig.~\ref{fig:phase_sift_comparison_plots} to the GEVP results from this work using the $\mathbb{S}_{\circled{0}}^{(2,1,A_1^+)}$ and $\mathbb{S}_{\circled{0}}^{(2,0,T_1^+)}$ interpolating-operator sets.
Discrepancies of (2-3)$\sigma$ in $k\cot \delta_{{}^1S_0}$ and $k\cot \delta_{{}^3S_1}$ are seen between results associated with the ground state on this lattice volume and corresponding results from Refs.~\cite{Wagman:2017tmp,Berkowitz:2015eaa}. 
Conversely, there is agreement at the $1\sigma$ level in $k\cot \delta_{{}^1S_0}$ and $k\cot \delta_{{}^3S_1}$ between the ground state results of this work and the corresponding ground-state results of Ref.~\cite{Horz:2020zvv}, as well as $1\sigma$ agreement with the more statistically uncertain results for $k\cot \delta_{{}^1S_0}$ of Ref.~\cite{Francis:2018qch}.
This agreement is consistent with the observations that subsets of $\mathbb{S}_{\circled{0}}^{(2,1,A_1^+)}$ and $\mathbb{S}_{\circled{0}}^{(2,0,T_1^+)}$ only including $\mathfrak{s} = 0$ dibaryon operators lead to consistent ground-state energy results with results obtained using the full sets $\mathbb{S}_{\circled{0}}^{(2,1,A_1^+)}$ and $\mathbb{S}_{\circled{0}}^{(2,0,T_1^+)}$, at the statistical precision of this work.
Higher-energy phase shifts show consistency between the results of this work and that of Refs.~\cite{Francis:2018qch,Horz:2020zvv} at (1-2)$\sigma$.

Although a reliable way to determine whether a bound state is present is by determining the volume dependence of the FV energy spectrum and finding a ground-state energy below $2M_N$ in the infinite-volume limit,  statements about the likelihood of a state being bound or unbound can be made using results with a single lattice volume by invoking the analytic structure of the constrained amplitudes.
The ground-state FV energy shift is negative at $1\sigma$ in the $I=1$ and at $2\sigma$ in the $I=0$ channels, which indicates attractive interactions at very low energies in both channels, which is a necessary but not sufficient condition for a bound state to be present.
Another useful criterion is that $k\cot \delta$ should be negative (positive) for a bound (unbound) state in sufficiently large volumes~\cite{Beane:2003da,Beane:2008dv}, which given $M_N$ and the lattice volume used here corresponds to a FV energy shift below (above) the threshold $\Delta E_{BS} = \frac{4 \pi^2}{M_N L^2} d_1 = -0.003073(8)$, where $d_1 = -0.0959006$ is obtained by expanding L{\"u}scher's quantization condition~\cite{Luscher:1986pf,Luscher:1990ux} about the infinite-scattering-length limit corresponding to the transition between interactions that do and do not support a bound state~\cite{Beane:2003da,Beane:2008dv} (under the assumption that $L$ is asymptotically large and higher-order terms in $1/L$ can be neglected). 
Comparing this threshold to the results $\Delta E_{0}^{(2,1,A_1^+,\mathbb{S}_0)}=-0.00255(51)$  and  $\Delta E_{0}^{(2,0,T_1^+,\mathbb{S}_0)}=-0.00273(45)$ 
indicates a 1$\sigma$ preference for an unbound ground state in both channels.
Results using additional volumes will allow a determination of whether two-nucleon ground states are bound or unbound with higher statistical significance.
As the ground-state energies obtained with the variational method are upper bounds on the true LQCD energies,  it is also possible that a bound state exists but has small overlap with all interpolating operators used in this study.

The upper bounds for the first excited-state FV energy shifts obtained using $\mathbb{S}_0^{(2,1,A_1^+)}$ and $\mathbb{S}_0^{(2,0,T_1^+)}$ are positive, and if they provide an accurate estimate of these energy levels (in particular if there are not lower-energy states approximately orthogonal to the interpolating operators used here) than there cannot be two bound states in either the dineutron or deuteron channels.
The first excited-state energy shift is closer to the non-interacting $\mathfrak{s} = 1$ energy than zero, which is suggestive of an attractive interaction that is not strong enough to form a bound state~\cite{Beane:2003da}, but it does not rule a bound state out.

Qualitatively, the large overlap of $\mathfrak{s} = 0$ dibaryon operators onto the lowest extracted state and the relatively uniform overlap of hexaquark operators onto all of the lowest-energy states is more reminiscent of the unbound than the bound scenario in a QED model of bound-state formation~\cite{Sasaki:2006jn}.
Further high-precision variational studies of the volume dependence of $B=2$ FV energy shifts with a more extensive operator set are needed to conclusively determine whether the $B=2$ ground states at $m_\pi = 806$ MeV are bound or unbound.

\section{Discussion}\label{sec:concl}
\subsection{Strengths and weaknesses of variational~methods }\label{sec:discussion}

In the $I=1$, $\Gamma_J=A_1^+$ and $I=0$, $\Gamma_J=T_1^+$ irreps associated with $S$-wave scattering, calculations using the full sets of interpolating operators introduced in this work lead to degenerate correlation-function matrices whose determinants are consistent with zero at the present statistics.
Non-degenerate correlation-function matrices are found by removing either all 6 quasi-local operators or by removing combinations of 6 dibaryon and quasi-local interpolating operators from the set under consideration.
Where signals could be resolved, consistent results are obtained for interpolating-operator sets with this number of interpolating operators that were studied. However, removing further interpolating operators leads to the appearance of ``missing energy levels'' lower in the spectrum than other energy levels determined from GEVP results.
In particular, if neither quasi-local nor zero-relative-momentum dibaryon interpolating operators are included, then the ground-state energy determined using GEVP results is found to be consistent with an excited-state energy determined using the largest available non-degenerate interpolating-operator set.
The orthogonality of GEVP eigenvectors guarantees that GEVP results for ground-state energies are free from excited-state contamination from states that strongly overlap with any interpolating operators used to construct correlation-function matrices.
However, these examples demonstrate that the suppression of excited-state effects provided by variational methods rests on the assumption that there are no additional states whose overlap with all elements of an interpolating-operator set is sufficiently small---perturbative calculations of the toy model in Eq.~\eqref{eq:badZ} demonstrate that the ground state can be missed entirely if, for example, the overlap of all interpolating operators with one or more excited states is a factor of $e^{\Delta t/2}$ larger than the corresponding ground-state overlap where $\Delta$ is the excitation energy gap and $t$ is the time extent at which signals are resolved.

The appearance of missing energy levels using interpolating-operator sets that neglect particular operator structures has been previously observed in studies of $\pi\pi$ systems~\cite{Dudek:2012xn,Wilson:2015dqa} as well as $N\pi$ systems~\cite{Lang:2012db,Kiratidis:2015vpa}.
These studies and the results of this work indicate that a significant strength of variational methods is their ability to disentangle contributions to correlation functions from a large set of states with similar energies.
The large number of nearby energy levels arising from spin-orbit coupling in the $I=0$, $\Gamma_J = T_1^+$ irrep, for example, can be resolved using a set of 42 interpolating operators, $\mathbb{S}_{\circled{0}}^{(2,0,T_1^+)}$, with comparable precision to the more sparsely populated $I=1$, $\Gamma_J = A_1^+$ excited-state spectrum.
Variational methods can, therefore, provide precise determinations of energy spectra using source/sink separations much smaller than the inverse  energy gap between neighboring levels.
However, a significant weakness of variational methods is that GEVP solutions with finite temporal separations lead to correlation functions approximating the energy eigenstates with the largest overlap with a given interpolating-operator set rather than necessarily the lowest-energy eigenstates.
Therefore, interpolating-operator set size is not a good proxy for how reliably variational methods can identify the ground state: the set of 40 interpolating operators $\widetilde{\mathbb{S}}_{\circled{0}}^{(2,0,T_1^+)}$ misses the ground state, and the lowest-energy state identified is consistent with the first excited state identified using $\mathbb{S}_{\circled{0}}^{(2,0,T_1^+)}$,\footnote{It is clear that operator sets that miss multiple levels can also be constructed by further omissions.} while the set of two interpolating operators $\mathbb{S}_{\circled{1}}^{(2,0,T_1^+)}$ leads a to ground-state energy result that is consistent with the result obtained using the largest available non-degenerate interpolating operator set $\mathbb{S}_{\circled{0}}^{(2,0,T_1^+)}$.
Further, the toy model overlap factors discussed above provide an example for which variational methods would lead to an incorrect determination of the ground-state energy, (unless correlation functions with very large Euclidean time separations could be resolved) while examination of an asymmetric correlation function using the same interpolating-operator set would lead to a correct determination. 
Other surprising behaviors are seen in previous variational calculations using interpolating-operator sets with small ground-state overlap, for example, calculations using local $qqq\overline{q}q$ operators in Ref.~\cite{Kiratidis:2015vpa} that appear to identify the physical $N\pi$ ground-state when using partial subsets of their interpolating-operator set but not when using the full interpolating-operator set.

The most precise results for (upper bounds on) two-nucleon ground-state FV energy shifts with $L \approx 4.6$~fm calculated  in this work use interpolating-operator sets without quasi-local interpolating operators and correspond to $-3.7(6)$~MeV and $-3.5(7)$~MeV in the $I=0$ and $I=1$ channels, respectively.
These results are consistent with previous calculations using individual dibaryon correlation functions and $2\times 2$ matrices of dibaryon correlation functions: Ref.~\cite{Francis:2018qch} obtained $-4(8)$~MeV for the $I=1$ channel using a similar value of the pion mass ($m_\pi~\approx~960$~MeV), a smaller lattice spacing ($a \approx 0.066$ fm), and a smaller physical volume ($L \approx 2.1$ fm); Ref.~\cite{Horz:2020zvv} obtained  $-2.6(6)$~MeV and $-3.7(8)$~MeV for each channel using a similar physical volume ($L \approx 4.1$ fm), a similar value of the pion mass ($m_\pi~\approx~714$~MeV), and a finer lattice spacing ($a\approx 0.086$ fm) than used in this work.
However, these FV energy-shifts  are significantly closer to threshold than results obtained in Refs.~\cite{Beane:2012vq,Berkowitz:2015eaa,Wagman:2017tmp} using correlation functions with hexaquark sources and dibaryon sinks on a larger ensemble of gauge-field configurations with the same parameters as the ensemble used here, namely $-23(4)$, $-28(3)$, and $-28(4)$ MeV in the $I=0$ channel and $-15(3)$, $-17(3)$, and $-21(4)$ MeV in the $I=1$ channel, and other calculations using $[D,H]$ correlation functions at unphysically large quark masses~\cite{Beane:2010hg,Beane:2011iw,Yamazaki:2012hi,Beane:2015yha,Chang:2015qxa,Detmold:2015daa,Yamazaki:2015asa,Parreno:2016fwu,Savage:2016kon,Shanahan:2017bgi,Tiburzi:2017iux,Winter:2017bfs,Chang:2017eiq,Detmold:2020snb}.
Similar, although less precise, discrepancies are found in direct comparisons of $[D,H]$ and GEVP results computed in this work, and  discrepancies of comparable size are seen for energy levels associated with $\mathfrak{s} \in \{0,\ldots,6\}$ dibaryon interpolating operators.
These discrepancies indicate that either the asymmetric $[D,H]$ correlation functions would slowly approach the GEVP results at larger $t$ than that accessible in this calculation, or that there are additional lower-energy levels that are nearly orthogonal to the interpolating operators in the sets used in this work but that are revealed in asymmetric $[D,H]$ correlation functions.
Further work is needed to discriminate between these possibilities.

Several pieces of evidence in favor of the GEVP results of this work providing an accurate description of a low-energy sector of the LQCD spectrum without missing energy levels are described in Secs.~\ref{sec:I1}-\ref{sec:I0}.
In particular, interpolating-operator sets with quasi-local operators substituted in place of dibaryon operators dominantly overlapping with either the ground state or the first excited state reproduce the same low-lying energy spectrum, and in some cases are inconsistent with the $[D,H]$ results at high statistical significance.
It is not possible, with the currently available statistics, to obtain a non-degenerate set with quasi-local operators included in addition to all of the dibaryon operators, which indicates that any missing energy level would have to be associated with operators approximately orthogonal to those in this interpolating-operator set.
Further, GEVP correlation-function reconstructions obtained by inserting GEVP energy and overlap-factor results into the spectral representations for $[D,H]$ correlation functions are able to approximately reproduce the corresponding GEVP effective FV energy shifts for temporal separations where signals can be resolved.
This provides a proof-of-principle demonstration that linear combinations of a near-threshold ground state and above-threshold excited states can lead to $t$ dependence consistent with that observed in $[D,H]$ effective FV energy shifts.
In addition, other non-positive-definite correlation functions, for example of the form $[D_{01T}, D_{\mathfrak{s}1W}]$ with $\mathfrak{s} > 0$, lead to effective FV energy shifts that are $\sim 100\text{ MeV}$ farther below threshold than either GEVP or $[D,H]$ correlation-function results.
Some of these correlation functions show significant $t$ dependence but others are consistent with constant $t$ behavior for the range of $t$, given the statistical precision of this work.
However, the GEVP reconstructions are able to reproduce the $[D_{01T}, D_{\mathfrak{s}1W}]$ FV energy shifts and suggest that effective FV energy shifts significantly below zero for $t \lesssim 1/\delta^{(2,I,\Gamma_J)}$ may be consistent with much less negatively shifted FV energies.
The GEVP energy levels, therefore, provide a simple hypothesis for the LQCD spectrum that is consistent with all results for this volume.
Finally, these GEVP results are consistent with precise results obtained using $[D,D]$ correlation functions from Ref.~\cite{Horz:2020zvv} using similar physical parameters but a finer lattice spacing, as well as the less precise results obtained in Ref.~\cite{Francis:2018qch} using $[D,D]$ correlation functions with a smaller physical volume in the $I=1$ channel.

As stressed above, however, it is also possible for asymmetric correlation functions to reveal the presence of physical states that have small overlap with an interpolating-operator set.
The examples of missing energy levels with the $\widetilde{\mathbb{S}}_{\circled{m}}^{(2,1,A_1^+)}$ and $\widetilde{\mathbb{S}}_{\circled{m}}^{(2,0,T_1^+)}$  operator sets show that it is easy to construct operators that are effectively orthogonal to states that are visible with other operator sets using the range of $t$ studied here.
It is straightforward to construct model overlap factors realizing this scenario.
This suggests that the possibility of missing the ground state using particular interpolating-operator sets is not necessarily unlikely.
Evidence in favor of $[D,H]$ correlation functions revealing physical energy states is provided by the fact that there are high-precision calculations of $[D,H]$ correlation functions using an identical lattice action and physical volumes that find consistent results for ground-state FV energy shifts using three different volumes, including smaller and larger volumes than the one used in this work~\cite{Beane:2012vq,Beane:2013br,Berkowitz:2015eaa,Wagman:2017tmp}.
Analysis of the phase shifts associated with these FV energy levels indicates that the results are consistent with the presence of both $I=0$ and $I=1$ bound states.
It seems unlikely that linear combinations of FV energy eigenstates whose energies have inverse power-law dependence on the volume, as expected for unbound scattering states, would conspire to give FV energy-shifts with negligible volume dependence at high precision for three values of $L$ varying by a factor of 2 as further discussed in Ref.~\cite{Beane:2017edf}, especially since the excited-state contamination visible in correlation functions on the three volumes is also consistent with being volume independent.
The scattering phase-shift results determined from $[D,H]$ correlation functions in Ref.~\cite{Wagman:2017tmp} also pass all of the consistency checks described in Ref.~\cite{Iritani:2017rlk}.
Pionless EFT calculations have also used the $B =2$ and $3$  energy spectra of Refs.~\cite{Beane:2012vq,Beane:2013br} to predict the binding energies of larger nuclei at $m_\pi = 806$ MeV and find agreement with $B=4$ LQCD results using asymmetric correlation functions with multi-nucleon sinks analogous to dibaryons and local sources analogous to hexaquarks~\cite{Barnea:2013uqa}.
Furthermore, results for nuclear matrix elements of scalar currents~\cite{Chang:2017eiq} give consistent results with determinations of the same quantities from the quark-mass dependence of the FV energy shifts~\cite{Beane:2013kca}.
It is not clear why the spectrum extracted from a linear combination of ground- and excited-state correlation functions should pass these consistency checks involving nuclear matrix elements and larger-$B$ systems.
The clear instances of missing energy levels discussed above show that approximate orthogonality of interpolating-operator sets from eigenstates is easily achieved.
Given  these results and the consistency of the spectra and associated scattering amplitudes determined using $[D,H]$ correlation functions, the results of this work obtained using the variational method should most robustly be interpreted as upper bounds that do not exclude the presence of a more negatively shifted ground-state energy.

\subsection{Summary and outlook}\label{sec:outlook}

Two-nucleon systems are studied in this work using a wide variety of LQCD interpolating operators: 42 and 96 dibaryon operators in the $I=1$ and $I=0$ channels, respectively, including all products of plane-wave nucleon wavefunctions with zero center-of-mass momentum and relative momentum smaller than $\sqrt{8}\left(\frac{2\pi}{L}\right)$,  6 quasi-local operators  in each isospin channel with exponentially localized nucleon wavefunctions inspired by FV EFT bound-state wavefunctions, and 2 hexaquark operators  in each isospin channel describing six totally antisymmetrized quarks centered around a common point.
The correlation-function matrix including all of these interpolating operators as both sources and sinks is computed using sparsened timeslice-to-all quark propagators and bilocal baryon blocks. 
The resulting correlation-function matrix is then projected onto cubic irreps describing the FV analog of total angular momentum.
For each irrep, the relevant block of the correlation-function matrix is diagonalized by solving a GEVP and the FV energy spectrum is extracted from fits to the resulting approximately orthogonal correlation functions.
The ground-state energies determined by applying variational methods to correlation-function matrices involving various (although not all) combinations of  hexaquark, dibaryon, and quasi-local two-nucleon interpolating operators are consistent with results obtained using only dibaryon interpolating operators here and in previous works~\cite{Francis:2018qch,Horz:2020zvv}, while variational results from all interpolating-operator sets where signals can be resolved are inconsistent with results obtained by fitting spectral representations to asymmetric correlation functions with hexaquark sources and dibaryon sinks here and in previous works~\cite{Beane:2012vq,Beane:2013br,Berkowitz:2015eaa,Wagman:2017tmp}.
Nucleon-nucleon $S$- and $D$-wave scattering phase shifts are computed with the assumption that mixing with higher partial waves is negligible.
The results of this work are not conclusive as to whether the two-nucleon ground states are bound or unbound in both the $I=0$ and $I=1$ channels at these quark masses, and ongoing calculations using additional lattice spacings and volumes will provide further constraints. Calculations using  additional interpolating operators are needed to search for more deeply bound states with small overlap with this interpolating-operator set in order to conclusively determine the structures of the two-nucleon ground states for these quark-mass values.
Future LQCD calculations using variational methods to control  contamination from excited low-energy states, as well as using a range of quark masses, lattice spacings, and volumes to extrapolate to the continuum limit and physical quark masses,  provide the most robust known route toward controlled determinations of multi-nucleon energy spectra from QCD.

The use of LQCD with variational methods and a large set of interpolating operators is not sufficient to guarantee the reliability of the multi-nucleon spectrum extracted using LQCD calculations.
It is implausible to imagine correlation-function matrix calculations employing a genuine basis that spans the LQCD Hilbert space for any given set of quantum numbers.
Further, without novel methods for reducing the scale of exponential signal-to-noise degradation such as multilevel integration~\cite{Luscher:2001up,DellaMorte:2008jd,Ce:2016idq,Ce:2016ajy}, signal-to-noise optimization~\cite{Detmold:2014hla}, or path integral contour deformations~\cite{Alexandru:2020wrj} for noisy observables~\cite{Detmold:2020ncp,Detmold:2021ulb} successfully applied to multi-baryon systems, future LQCD calculations of two-baryon systems will be limited to source/sink separations much smaller than the inverse of the excited-state energy gap.
Additional efforts to  understand and improve the completeness of an interpolating-operator set within a low-energy sector of Hilbert space are therefore critical.
This need for physically motivated interpolating-operator sets that overlap significantly with low-energy states presents a challenge for studying multi-nucleon systems with LQCD, but systematic uncertainties related to interpolating-operator dependence can be probed to some degree by varying the interpolating-operator set that is used. Further, as calculations approach the physical values of the quark masses, nuclear EFT and models can be used to inspire new interpolating-operator structures and expand the space of states probed in future LQCD calculations.

\acknowledgments{ We are grateful to Silas Beane, Kostas Orginos, and Martin Savage for extensive discussions.
We especially thank Kostas Orginos for his role in generating the gauge-field ensemble used in this work.
We further thank Martin Savage for collaboration in the early stages of this work and for use of the University of Washington Hyak computational infrastructure to perform some parts of  the calculations that are presented.
We thank Chen~Chen, Teo~Collins, Anthony Grebe, Colin Morningstar, Robert Perry, Peng Sun, Andr{\'e} Walker-Loud, and Yi-Bo Yang for helping to identify errors in a previous version of this manuscript.
We acknowledge the Texas Advanced Computing Center (TACC) at The University of Texas at Austin for providing HPC resources on Frontera that have contributed to the research results reported within this paper (http://www.tacc.utexas.edu).
This research used resources of the Oak Ridge Leadership Computing Facility at the Oak Ridge National Laboratory, which is supported by the Office of Science of the U.S. Department of Energy under Contract number DE-AC05-00OR22725. The authors thankfully acknowledge the computer resources at MareNostrum and the technical support provided by Barcelona Supercomputing Center (RES-FI-2022-1-0040).
Computations of this work were carried out using the \verb!Chroma!~\cite{Edwards:2004sx}, \verb!Qlua!~\cite{qlua}, \verb!QUDA!~\cite{Clark:2009wm,Babich:2011np,Clark:2016rdz}, and \verb!Tiramisu!~\cite{baghdadi2020tiramisu} software libraries.

AS is supported in part by DARPA under Awards HR0011-18- 3-0007 and HR0011-20-9-0017.
ZD is supported by the Alfred P. Sloan fellowship and by Maryland Center for Fundamental Physics at the University of Maryland, College Park. She further acknowledges support from the RIKEN Center for Accelerator-based Sciences with which she was affiliated during earlier stages of this work. 
WD and PES are supported in part by the U.S.~Department of Energy, Office of Science, Office of Nuclear Physics under grant Contract WD and AVP are supported in part  by the SciDAC4 award DE-SC0018121.
WD is also supported within the framework of the TMD Topical Collaboration of the U.S.~Department of Energy, Office of Science, Office of Nuclear Physics. PES is additionally supported by the National Science Foundation under EAGER grant 2035015, by the U.S. DOE Early Career Award DE-SC0021006, by a NEC research award, and by the Carl G and Shirley Sontheimer Research Fund. WD and PES are supported by the National Science Foundation under Cooperative Agreement PHY-2019786 (The NSF AI Institute for Artificial Intelligence and Fundamental Interactions, http://iaifi.org/).
MI is supported by the Universitat de Barcelona through the scholarship APIF. MI and AP acknowledge financial support from the State Agency for Research of the Spanish Ministry of Science and Innovation through the “Unit of Excellence Mar\'{\i}a de Maeztu 2020-2023" award to the Institute of Cosmos Sciences (CEX2019-000918-M), the European FEDER funds under the contract FIS2017-87534-P, and from the EU STRONG-2020 project under the program H2020-INFRAIA-2018-1, grant agreement No. 824093. This manuscript has been authored by Fermi Research Alliance, LLC under Contract No. DE-AC02-07CH11359 with the U.S. Department of Energy, Office of Science, Office of High Energy Physics.}

\appendix

\section{Spin-color weights}\label{app:weights}

\begin{table}[h!]
  \begin{tabular}{|c||c|c|c|c|} \hline
$\alpha$ & $i(\alpha)$ & $j(\alpha)$ & $k(\alpha)$ & $w^{[N]0}_{\alpha}$   \\\hline\hline
1 & 0, 0 & 1, 1 & 0, 2 & $-2 \sqrt{2} $ \\\hline
2 & 0, 0 & 1, 2 & 0, 1 & $2 \sqrt{2} $ \\\hline
3 & 0, 1 & 1, 0 & 0, 2 & $2 \sqrt{2} $ \\\hline
4 & 0, 1 & 1, 2 & 0, 0 & $-2 \sqrt{2} $ \\\hline
5 & 0, 2 & 1, 0 & 0, 1 & $-2 \sqrt{2} $ \\\hline
6 & 0, 2 & 1, 1 & 0, 0 & $2 \sqrt{2} $ \\\hline
7 & 1, 0 & 0, 1 & 0, 2 & $2 \sqrt{2} $ \\\hline
8 & 1, 0 & 0, 2 & 0, 1 & $-2 \sqrt{2} $ \\\hline
9 & 1, 1 & 0, 0 & 0, 2 & $-2 \sqrt{2} $ \\\hline
10 & 1, 1 & 0, 2 & 0, 0 & $2 \sqrt{2} $ \\\hline
11 & 1, 2 & 0, 0 & 0, 1 & $2 \sqrt{2} $ \\\hline
12 & 1, 2 & 0, 1 & 0, 0 & $-2 \sqrt{2} $ \\\hline
\end{tabular}
  \hspace{50pt}
  \begin{tabular}{|c||c|c|c|c|} \hline
$\alpha$ & $i(\alpha)$ & $j(\alpha)$ & $k(\alpha)$ & $w^{[N]1}_{\alpha}$   \\\hline\hline
1 & 0, 0 & 1, 1 & 1, 2 & $-2 \sqrt{2} $ \\\hline
2 & 0, 0 & 1, 2 & 1, 1 & $2 \sqrt{2} $ \\\hline
3 & 0, 1 & 1, 0 & 1, 2 & $2 \sqrt{2} $ \\\hline
4 & 0, 1 & 1, 2 & 1, 0 & $-2 \sqrt{2} $ \\\hline
5 & 0, 2 & 1, 0 & 1, 1 & $-2 \sqrt{2} $ \\\hline
6 & 0, 2 & 1, 1 & 1, 0 & $2 \sqrt{2} $ \\\hline
7 & 1, 0 & 0, 1 & 1, 2 & $2 \sqrt{2} $ \\\hline
8 & 1, 0 & 0, 2 & 1, 1 & $-2 \sqrt{2} $ \\\hline
9 & 1, 1 & 0, 0 & 1, 2 & $-2 \sqrt{2} $ \\\hline
10 & 1, 1 & 0, 2 & 1, 0 & $2 \sqrt{2} $ \\\hline
11 & 1, 2 & 0, 0 & 1, 1 & $2 \sqrt{2} $ \\\hline
12 & 1, 2 & 0, 1 & 1, 0 & $-2 \sqrt{2} $ \\\hline
\end{tabular}
  \caption{ The weights $w_{\alpha}^{[N]\sigma}$ for $\sigma =0$ and $1$ are shown in the left and right tables, respectively, along with the associated index functions $i,j,k$ that map $\alpha$ to the space of spin-color indices labeled $\zeta,a$ with $\zeta\in\{0,\ldots, 3\}$ and $a\in\{0,1,2\}$. 
  \label{tab:Nweights}}
\end{table}

The nucleon weights $w_\alpha^{[N]\sigma}$ with $\alpha \in\{ 1,\ldots,\mathcal{N}_w^{[N]}=12\}$ with spin component $\sigma \in \{0,1\}$ are shown in Table~\ref{tab:Nweights}.
The hexaquark weights $w_\alpha^{[H]\rho}$ with $\alpha \in \{1,\ldots,\mathcal{N}_w^{[H]\rho}\}$ with $A_1^+ \otimes T_1^+$ spin components $\rho = 0$ and $2$ are shown in Table~\ref{tab:Hweights}; the other components $\rho = 1$ and $3$ are shown in Table~\ref{tab:Hweights13}.

\begin{table}[!t]
  \begin{tabular}{|c||c|c|c|c|c|c|c|} \hline
$\alpha$ & $i(\alpha)$ & $j(\alpha)$ & $k(\alpha)$ & $l(\alpha)$ & $m(\alpha)$ & $n(\alpha)$ & $w^{[H]0}_{\alpha}$   \\\hline\hline
1 & 0, 0 & 0, 0 & 0, 1 & 1, 1 & 1, 2 & 1, 2 & 40 \\\hline
2 & 0, 0 & 0, 0 & 0, 2 & 1, 1 & 1, 1 & 1, 2 & -40 \\\hline
3 & 0, 0 & 0, 1 & 0, 1 & 1, 0 & 1, 2 & 1, 2 & -40 \\\hline
4 & 0, 0 & 0, 1 & 0, 2 & 1, 0 & 1, 1 & 1, 2 & 16 \\\hline
5 & 0, 0 & 0, 1 & 0, 2 & 1, 0 & 1, 2 & 1, 1 & 24 \\\hline
6 & 0, 0 & 0, 1 & 0, 2 & 1, 1 & 1, 0 & 1, 2 & 24 \\\hline
7 & 0, 0 & 0, 2 & 0, 1 & 1, 0 & 1, 1 & 1, 2 & 24 \\\hline
8 & 0, 0 & 0, 2 & 0, 1 & 1, 0 & 1, 2 & 1, 1 & 16 \\\hline
9 & 0, 0 & 0, 2 & 0, 1 & 1, 1 & 1, 0 & 1, 2 & -24 \\\hline
10 & 0, 0 & 0, 2 & 0, 2 & 1, 0 & 1, 1 & 1, 1 & -40 \\\hline
11 & 0, 1 & 0, 0 & 0, 2 & 1, 0 & 1, 1 & 1, 2 & 24 \\\hline
12 & 0, 1 & 0, 0 & 0, 2 & 1, 0 & 1, 2 & 1, 1 & -24 \\\hline
13 & 0, 1 & 0, 0 & 0, 2 & 1, 1 & 1, 0 & 1, 2 & 16 \\\hline
14 & 0, 1 & 0, 1 & 0, 2 & 1, 0 & 1, 0 & 1, 2 & -40 \\\hline
15 & 0, 1 & 0, 2 & 0, 2 & 1, 0 & 1, 0 & 1, 1 & 40 \\\hline
16 & 0, 0 & 0, 0 & 1, 1 & 0, 1 & 1, 2 & 1, 2 & -40 \\\hline
17 & 0, 0 & 0, 0 & 1, 1 & 0, 2 & 1, 2 & 1, 1 & 40 \\\hline
18 & 0, 0 & 0, 1 & 1, 0 & 0, 2 & 1, 1 & 1, 2 & 24 \\\hline
19 & 0, 0 & 0, 1 & 1, 0 & 0, 2 & 1, 2 & 1, 1 & -24 \\\hline
20 & 0, 0 & 0, 1 & 1, 1 & 0, 2 & 1, 2 & 1, 0 & -16 \\\hline
21 & 0, 1 & 0, 0 & 1, 0 & 0, 1 & 1, 2 & 1, 2 & 40 \\\hline
22 & 0, 1 & 0, 0 & 1, 0 & 0, 2 & 1, 1 & 1, 2 & -24 \\\hline
23 & 0, 1 & 0, 0 & 1, 0 & 0, 2 & 1, 2 & 1, 1 & -16 \\\hline
24 & 0, 1 & 0, 0 & 1, 1 & 0, 2 & 1, 2 & 1, 0 & -24 \\\hline
25 & 0, 1 & 0, 1 & 1, 0 & 0, 2 & 1, 2 & 1, 0 & 40 \\\hline
26 & 0, 2 & 0, 0 & 1, 0 & 0, 1 & 1, 1 & 1, 2 & -16 \\\hline
27 & 0, 2 & 0, 0 & 1, 0 & 0, 1 & 1, 2 & 1, 1 & -24 \\\hline
28 & 0, 2 & 0, 0 & 1, 0 & 0, 2 & 1, 1 & 1, 1 & 40 \\\hline
29 & 0, 2 & 0, 0 & 1, 1 & 0, 1 & 1, 2 & 1, 0 & 24 \\\hline
30 & 0, 2 & 0, 1 & 1, 0 & 0, 2 & 1, 1 & 1, 0 & -40 \\\hline
31 & 0, 0 & 1, 0 & 0, 1 & 1, 1 & 0, 2 & 1, 2 & -96 \\\hline
32 & 1, 0 & 0, 0 & 1, 1 & 0, 1 & 1, 2 & 0, 2 & 96 \\\hline
\end{tabular}
  \hspace{30pt}
  \begin{tabular}{|c||c|c|c|c|c|c|c|} \hline
$\alpha$ & $i(\alpha)$ & $j(\alpha)$ & $k(\alpha)$ & $l(\alpha)$ & $m(\alpha)$ & $n(\alpha)$ & $w^{[H]2}_{\alpha}$   \\\hline\hline
1 & 0, 0 & 0, 0 & 0, 1 & 1, 1 & 1, 2 & 1, 2 & 24 \\\hline
2 & 0, 0 & 0, 0 & 0, 2 & 1, 1 & 1, 1 & 1, 2 & -24 \\\hline
3 & 0, 0 & 0, 1 & 0, 1 & 1, 0 & 1, 2 & 1, 2 & -24 \\\hline
4 & 0, 0 & 0, 1 & 0, 2 & 1, 0 & 1, 1 & 1, 2 & -16 \\\hline
5 & 0, 0 & 0, 1 & 0, 2 & 1, 0 & 1, 2 & 1, 1 & 40 \\\hline
6 & 0, 0 & 0, 1 & 0, 2 & 1, 1 & 1, 0 & 1, 2 & 40 \\\hline
7 & 0, 0 & 0, 2 & 0, 1 & 1, 0 & 1, 1 & 1, 2 & 40 \\\hline
8 & 0, 0 & 0, 2 & 0, 1 & 1, 0 & 1, 2 & 1, 1 & -16 \\\hline
9 & 0, 0 & 0, 2 & 0, 1 & 1, 1 & 1, 0 & 1, 2 & -40 \\\hline
10 & 0, 0 & 0, 2 & 0, 2 & 1, 0 & 1, 1 & 1, 1 & -24 \\\hline
11 & 0, 1 & 0, 0 & 0, 2 & 1, 0 & 1, 1 & 1, 2 & 40 \\\hline
12 & 0, 1 & 0, 0 & 0, 2 & 1, 0 & 1, 2 & 1, 1 & -40 \\\hline
13 & 0, 1 & 0, 0 & 0, 2 & 1, 1 & 1, 0 & 1, 2 & -16 \\\hline
14 & 0, 1 & 0, 1 & 0, 2 & 1, 0 & 1, 0 & 1, 2 & -24 \\\hline
15 & 0, 1 & 0, 2 & 0, 2 & 1, 0 & 1, 0 & 1, 1 & 24 \\\hline
16 & 0, 0 & 0, 0 & 1, 1 & 0, 1 & 1, 2 & 1, 2 & 24 \\\hline
17 & 0, 0 & 0, 0 & 1, 1 & 0, 2 & 1, 2 & 1, 1 & -24 \\\hline
18 & 0, 0 & 0, 1 & 1, 0 & 0, 2 & 1, 1 & 1, 2 & -40 \\\hline
19 & 0, 0 & 0, 1 & 1, 0 & 0, 2 & 1, 2 & 1, 1 & 40 \\\hline
20 & 0, 0 & 0, 1 & 1, 1 & 0, 2 & 1, 2 & 1, 0 & -16 \\\hline
21 & 0, 1 & 0, 0 & 1, 0 & 0, 1 & 1, 2 & 1, 2 & -24 \\\hline
22 & 0, 1 & 0, 0 & 1, 0 & 0, 2 & 1, 1 & 1, 2 & 40 \\\hline
23 & 0, 1 & 0, 0 & 1, 0 & 0, 2 & 1, 2 & 1, 1 & -16 \\\hline
24 & 0, 1 & 0, 0 & 1, 1 & 0, 2 & 1, 2 & 1, 0 & 40 \\\hline
25 & 0, 1 & 0, 1 & 1, 0 & 0, 2 & 1, 2 & 1, 0 & -24 \\\hline
26 & 0, 2 & 0, 0 & 1, 0 & 0, 1 & 1, 1 & 1, 2 & -16 \\\hline
27 & 0, 2 & 0, 0 & 1, 0 & 0, 1 & 1, 2 & 1, 1 & 40 \\\hline
28 & 0, 2 & 0, 0 & 1, 0 & 0, 2 & 1, 1 & 1, 1 & -24 \\\hline
29 & 0, 2 & 0, 0 & 1, 1 & 0, 1 & 1, 2 & 1, 0 & -40 \\\hline
30 & 0, 2 & 0, 1 & 1, 0 & 0, 2 & 1, 1 & 1, 0 & 24 \\\hline
31 & 0, 0 & 1, 0 & 0, 1 & 1, 1 & 0, 2 & 1, 2 & -96 \\\hline
32 & 1, 0 & 0, 0 & 1, 1 & 0, 1 & 1, 2 & 0, 2 & -96 \\\hline
\end{tabular}
  \caption{ The weights $w_{\alpha}^{[H]\rho}$ for hexaquark operators with $\rho =0$ and $2$ are shown in the left and right tables, respectively, along with the associated index functions $i,j,k,l,m,n$ that map $\alpha$ to the space of spin-color indices labeled $\zeta,a$ with $\zeta\in \{0,\ldots, 3\}$ and $a\in \{0,1,2\}$. 
  \label{tab:Hweights}}
\end{table}

\begin{table}[!t]
  \begin{tabular}{|c||c|c|c|c|c|c|c|} \hline
$\alpha$ & $i(\alpha)$ & $j(\alpha)$ & $k(\alpha)$ & $l(\alpha)$ & $m(\alpha)$ & $n(\alpha)$ & $w^{[H]1}_{\alpha}$   \\\hline\hline
1 & 0, 0 & 0, 0 & 0, 1 & 0, 1 & 1, 2 & 1, 2 & $48 \sqrt{2}$ \\\hline
2 & 0, 0 & 0, 0 & 0, 1 & 0, 2 & 1, 1 & 1, 2 & $-40 \sqrt{2}$ \\\hline
3 & 0, 0 & 0, 0 & 0, 1 & 0, 2 & 1, 2 & 1, 1 & $-8 \sqrt{2}$ \\\hline
4 & 0, 0 & 0, 0 & 0, 2 & 0, 1 & 1, 1 & 1, 2 & $-8 \sqrt{2}$ \\\hline
5 & 0, 0 & 0, 0 & 0, 2 & 0, 1 & 1, 2 & 1, 1 & $-40 \sqrt{2}$ \\\hline
6 & 0, 0 & 0, 0 & 0, 2 & 0, 2 & 1, 1 & 1, 1 & $48 \sqrt{2}$ \\\hline
7 & 0, 0 & 0, 1 & 0, 1 & 0, 2 & 1, 0 & 1, 2 & $40 \sqrt{2}$ \\\hline
8 & 0, 0 & 0, 1 & 0, 1 & 0, 2 & 1, 2 & 1, 0 & $8 \sqrt{2}$ \\\hline
9 & 0, 0 & 0, 1 & 0, 2 & 0, 2 & 1, 0 & 1, 1 & $-40 \sqrt{2}$ \\\hline
10 & 0, 0 & 0, 1 & 0, 2 & 0, 2 & 1, 1 & 1, 0 & $-8 \sqrt{2}$ \\\hline
11 & 0, 1 & 0, 0 & 0, 2 & 0, 1 & 1, 0 & 1, 2 & $8 \sqrt{2}$ \\\hline
12 & 0, 1 & 0, 0 & 0, 2 & 0, 1 & 1, 2 & 1, 0 & $40 \sqrt{2}$ \\\hline
13 & 0, 1 & 0, 0 & 0, 2 & 0, 2 & 1, 0 & 1, 1 & $-8 \sqrt{2}$ \\\hline
14 & 0, 1 & 0, 0 & 0, 2 & 0, 2 & 1, 1 & 1, 0 & $-40 \sqrt{2}$ \\\hline
15 & 0, 1 & 0, 1 & 0, 2 & 0, 2 & 1, 0 & 1, 0 & $48 \sqrt{2}$ \\\hline
16 & 0, 0 & 0, 0 & 0, 1 & 1, 1 & 0, 2 & 1, 2 & $-32 \sqrt{2}$ \\\hline
17 & 0, 0 & 0, 1 & 0, 1 & 1, 0 & 0, 2 & 1, 2 & $32 \sqrt{2}$ \\\hline
18 & 0, 0 & 0, 2 & 0, 1 & 1, 0 & 0, 2 & 1, 1 & $-32 \sqrt{2}$ \\\hline
19 & 0, 0 & 0, 0 & 1, 1 & 0, 1 & 1, 2 & 0, 2 & $-32 \sqrt{2}$ \\\hline
20 & 0, 1 & 0, 0 & 1, 0 & 0, 1 & 1, 2 & 0, 2 & $32 \sqrt{2}$ \\\hline
21 & 0, 2 & 0, 0 & 1, 0 & 0, 1 & 1, 1 & 0, 2 & $-32 \sqrt{2}$ \\\hline
\end{tabular}
  \hspace{30pt}
  \begin{tabular}{|c||c|c|c|c|c|c|c|} \hline
$\alpha$ & $i(\alpha)$ & $j(\alpha)$ & $k(\alpha)$ & $l(\alpha)$ & $m(\alpha)$ & $n(\alpha)$ & $w^{[H]3}_{\alpha}$   \\\hline\hline
1 & 0, 0 & 0, 0 & 1, 1 & 1, 1 & 1, 2 & 1, 2 & $48 \sqrt{2}$ \\\hline
2 & 0, 0 & 0, 1 & 1, 0 & 1, 1 & 1, 2 & 1, 2 & $-40 \sqrt{2}$ \\\hline
3 & 0, 0 & 0, 1 & 1, 1 & 1, 0 & 1, 2 & 1, 2 & $-8 \sqrt{2}$ \\\hline
4 & 0, 0 & 0, 2 & 1, 0 & 1, 1 & 1, 1 & 1, 2 & $40 \sqrt{2}$ \\\hline
5 & 0, 0 & 0, 2 & 1, 1 & 1, 0 & 1, 2 & 1, 1 & $8 \sqrt{2}$ \\\hline
6 & 0, 1 & 0, 0 & 1, 0 & 1, 1 & 1, 2 & 1, 2 & $-8 \sqrt{2}$ \\\hline
7 & 0, 1 & 0, 0 & 1, 1 & 1, 0 & 1, 2 & 1, 2 & $-40 \sqrt{2}$ \\\hline
8 & 0, 1 & 0, 1 & 1, 0 & 1, 0 & 1, 2 & 1, 2 & $48 \sqrt{2}$ \\\hline
9 & 0, 1 & 0, 2 & 1, 0 & 1, 0 & 1, 1 & 1, 2 & $-40 \sqrt{2}$ \\\hline
10 & 0, 1 & 0, 2 & 1, 0 & 1, 0 & 1, 2 & 1, 1 & $-8 \sqrt{2}$ \\\hline
11 & 0, 2 & 0, 0 & 1, 0 & 1, 1 & 1, 1 & 1, 2 & $8 \sqrt{2}$ \\\hline
12 & 0, 2 & 0, 0 & 1, 1 & 1, 0 & 1, 2 & 1, 1 & $40 \sqrt{2}$ \\\hline
13 & 0, 2 & 0, 1 & 1, 0 & 1, 0 & 1, 1 & 1, 2 & $-8 \sqrt{2}$ \\\hline
14 & 0, 2 & 0, 1 & 1, 0 & 1, 0 & 1, 2 & 1, 1 & $-40 \sqrt{2}$ \\\hline
15 & 0, 2 & 0, 2 & 1, 0 & 1, 0 & 1, 1 & 1, 1 & $48 \sqrt{2}$ \\\hline
16 & 0, 0 & 1, 0 & 0, 1 & 1, 1 & 1, 2 & 1, 2 & $-32 \sqrt{2}$ \\\hline
17 & 0, 0 & 1, 0 & 0, 2 & 1, 1 & 1, 1 & 1, 2 & $32 \sqrt{2}$ \\\hline
18 & 0, 1 & 1, 0 & 0, 2 & 1, 1 & 1, 0 & 1, 2 & $-32 \sqrt{2}$ \\\hline
19 & 1, 0 & 0, 0 & 1, 1 & 0, 1 & 1, 2 & 1, 2 & $-32 \sqrt{2}$ \\\hline
20 & 1, 0 & 0, 0 & 1, 1 & 0, 2 & 1, 2 & 1, 1 & $32 \sqrt{2}$ \\\hline
21 & 1, 0 & 0, 1 & 1, 1 & 0, 2 & 1, 2 & 1, 0 & $-32 \sqrt{2}$ \\\hline
\end{tabular}

  \caption{ The weights $w_{\alpha}^{[H]\rho}$ for hexaquark operators with $\rho =1$ and $3$ are shown in the left and right tables respectively along with the associated index functions $i,j,k,l,m,n$ that map $\alpha$ to the space of spin-color indices labeled $\zeta,a$ with $\zeta\in \{0,\ldots, 3\}$ and $a \in \{0,1,2\}$. 
  \label{tab:Hweights13}}
\end{table}

The dibaryon weights can be straightforwardly constructed in terms of the nucleon weights.
The first $(\mathcal{N}_w^{[N]})^2=144$ components of $w_{\alpha}^{[D]\rho}$ can be expressed using the map $\alpha(\alpha_1,\alpha_2)$ defined by
\begin{equation}
    \alpha = (\alpha_1-1)\mathcal{N}_w^{[N]}+\alpha_2,
\end{equation}
with the dibaryon index functions of $\alpha$ related to two nucleon index functions applied to $\alpha_1$ and $\alpha_2$ as
\begin{equation}
\begin{split}
    i^{[D]}(\alpha) &= i^{[N]}(\alpha_1), \hspace{20pt} j^{[D]}(\alpha) = j^{[N]}(\alpha_1), \hspace{20pt} k^{[D]}(\alpha) = k^{[N]}(\alpha_1), \\
    l^{[D]}(\alpha) &= i^{[N]}(\alpha_2), \hspace{20pt} m^{[D]}(\alpha) = j^{[N]}(\alpha_2), \hspace{20pt} n^{[D]}(\alpha) = k^{[N]}(\alpha_2),
\end{split}\label{eq:DNmaps}
\end{equation}
and dibaryon weights obtained from products of nucleon weights as
\begin{equation}
\begin{split}
    w_{\alpha}^{[D]0} &=  \frac{1}{\sqrt{2}}w_{\alpha_1}^{[N]0}  w_{\alpha_2}^{[N]1}, \\
    w_{\alpha}^{[D]1} &=  w_{\alpha_1}^{[N]0}  w_{\alpha_2}^{[N]0}, \\
    w_{\alpha}^{[D]2} &=  \frac{1}{\sqrt{2}}w_{\alpha_1}^{[N]0}  w_{\alpha_2}^{[N]1}, \\
    w_{\alpha}^{[D]3} &=  w_{\alpha_1}^{[N]1}  w_{\alpha_2}^{[N]1}.
\end{split}
\end{equation}
The dibaryon spin components $\rho = 0$ and $2$ include an additional $(\mathcal{N}_w^{[N]})^2$ weights that can be indexed as
\begin{equation}
    \alpha = \left(\mathcal{N}_w^{[N]}\right)^2 + (\alpha_1-1)\mathcal{N}_w^{[N]}+\alpha_2,
\end{equation}
with index functions defined exactly as in Eq.~\eqref{eq:DNmaps} and weights corresponding to
\begin{equation}
\begin{split}
    w_{\alpha}^{[D]0} &=  -\frac{1}{\sqrt{2}}w_{\alpha_1}^{[N]1}  w_{\alpha_2}^{[N]0}, \\
    w_{\alpha}^{[D]2} &=  \frac{1}{\sqrt{2}}w_{\alpha_1}^{[N]1}  w_{\alpha_2}^{[N]0}.
\end{split}
\end{equation}

\section{Wavefunction and spin-orbit cubic group theory factors}\label{app:FV}

A set of functions $b^{(\Gamma_\ell,\ell_z)}(\vec{x})$ provides a basis for the irrep $\Gamma_\ell$ if it satisfies the transformation law
\begin{equation}
  \begin{split}
    b^{(\Gamma_\ell,\ell_z)}(\vec{x}) \stackrel{R}{\longrightarrow}& \  b^{(\Gamma_\ell,\ell_z')}(R(\vec{x})) \equiv \sum_{\ell_z'}  b^{(\Gamma_\ell,\ell_z')}(\vec{x}) \mathcal{D}_{\ell_z' \ell_z}^{(\Gamma_\ell)}(R),
\label{eq:repdef}
  \end{split}
\end{equation}
where $R \in O_h$ is a cubic transformation, and the sum ranges over the $\ell_z$ eigenvalues corresponding to the rows of $\Gamma_\ell$.
Any matrices $\mathcal{D}_{\ell_z \ell_z'}^{(\Gamma_\ell)}(R)$ satisfying Eq.~\eqref{eq:repdef} furnish a representation of the irrep $\Gamma_\ell$, but different choices of $b^{(\Gamma_\ell,\ell_z)}(\vec{x})$ lead to $\mathcal{D}_{\ell_z \ell_z'}^{(\Gamma_\ell)}(R)$ that differ by a unitary change-of-basis transformation.
Although it is permissible to use any definition of the $\mathcal{D}_{\ell_z \ell_z'}^{(\Gamma_\ell)}(R)$ in order to construct a basis of wavefunctions, the same definition of the $\mathcal{D}_{\ell_z \ell_z'}^{(\Gamma_\ell)}(R)$ must be used to construct wavefunctions as to derive the cubic-group Clebsch-Gordan coefficients required to project $\Gamma_J = \Gamma_\ell \otimes \Gamma_S$ into irreps.
The particular form of the representations used to define a basis for each irrep can be simply specified by providing the choice of basis vectors satisfying Eq.~\eqref{eq:repdef} in terms of coordinate functions $x$, $y$, and $z$.
A basis of generators for all positive-parity cubic irreps that is compatible with the Clebsch-Gordan coefficients presented in Ref.~\cite{Basak:2005ir} is shown in Table~\ref{tab:basis}.
This choice of basis differs from the choice made in Ref.~\cite{Luu:2011ep} for the $T_1^+$ and $T_2^+$ irreps; further we have found that relative signs between wavefunctions in different rows of the $E^+$ irrep in the $\mathfrak{s}\in\{5,6\}$ shells and the coefficients defining the $\mathfrak{s}=6$ shell $T_2^+$ wavefunctions must be modified  from the values presented in Ref.~\cite{Luu:2011ep} in order to obtain consistent transformation properties for all wavefunctions transforming in a given irrep.
For these cases, we constructed basis vectors using group-theory projector techniques described for example in Ref.~\cite{Dresselaus:2008} and applied to LQCD interpolating-operator construction in Refs.~\cite{Basak:2005aq,Dudek:2010wm,Morningstar:2013bda}.

The basis vectors shown in Table~\ref{tab:basis} are used to construct a representation of the cubic group $\mathcal{D}_{\ell_z \ell_z'}^{(\Gamma_\ell)}(R)$ using  Eq.~\eqref{eq:repdef} by constructing linear combinations of plane-wave wavefunctions  $\psi_{\mathfrak{s} \mathfrak{k}}^{(\Gamma_\ell,\ell_z)}(\vec{x}_1,\vec{x}_2) = \sum_{\mathfrak{m}} G^{(\Gamma_\ell,\ell_z)}_{\mathfrak{s} \mathfrak{k} \mathfrak{m}}  \psi_{\mathfrak{m}}^{[D]}(\vec{x}_1, \vec{x}_2)$ as in Eq.~\eqref{eq:cubicwvfdef} that satisfy
\begin{equation}
  \mathcal{D}_{\ell_z \ell_z'}^{(\Gamma_\ell)}(R) = \int d^3x_1\  d^3x_2\ \left[ \psi_{\mathfrak{s} \mathfrak{k}}^{(\Gamma_\ell,\ell_z)}(\vec{x}_1,\vec{x}_2) \right]^* \psi_{\mathfrak{s} \mathfrak{k}}^{(\Gamma_\ell,\ell_z')}(R(\vec{x}_1), R(\vec{x}_2) ). \label{eq:Dpsi}
\end{equation}
The basis functions are chosen to have definite $\ell_z$ and follow the usual Condon-Shortley phase convention for spherical harmonics so that the Clebsch-Gordan coefficients of Ref.~\cite{Basak:2005ir} can be used to form spin-orbit products $\Gamma_J$ with definite cubic transformation properties.
  The $A_1^+$, $E^+$, and $T_2^+$ basis vectors are proportional to complex conjugated spherical harmonics $(Y_l^m)^*$ (so that states created by operators with wavefunctions $ \psi_{\mathfrak{s} \mathfrak{k}}^{(\Gamma_\ell,\ell_z)}(\vec{x}_1,\vec{x}_2)^*$ transform as spherical harmonics $Y_l^m$): $(Y_0^0)^*$ for $A_1^+$; $(Y_2^0)^*$ and $(Y_2^2 + Y_2^{-2})^*/\sqrt{2}$ for $E^+$; and $(Y_2^1)^*$, $(Y_2^2 - Y_2^{-2})^*/\sqrt{2}$, and $(Y_2^{-1})^*$ for $T_2^+$. All basis vectors including those for the $T_1^+$ and $A_2^+$ irreps can be obtained by transforming the Cartesian basis vectors in Ref.~\cite{Dresselaus:2008} into a basis of $\ell_z$ eigenstates and adding appropriate normalization and phase factors.
  Although cubic transformation properties are unaffected by multiplying the wavefunctions for all rows of an irrep by a common phase, the relative phases of wavefunctions in different irreps do affect dibaryon operators that include these wavefunctions in spin-orbit products. Phase factors have therefore been chosen to ensure that all dibaryon correlation functions transform identically under charge conjugation.

In the $\mathfrak{s} \in \{5,6\}$ shells, there are multiple linearly independent wavefunctions that transform identically.
Group-theory projector techniques must be augmented by a method for choosing a particular wavefunction basis in cases where a given irrep appears with non-unit multiplicity for a given $\mathfrak{s}$: for the $E^+$ irrep and $\mathfrak{s}=5$ shell, the basis wavefunctions of Ref.~\cite{Luu:2011ep} are used, while for the $T_2^+$ irrep and $\mathfrak{s}=6$ shell, it is straightforward to construct an orthogonal pair of wavefunctions by choosing $\psi_{6 1}^{(T_2^+,2)}$ and $\psi_{6 2}^{(T_2^+,2)}$ to be linear combinations of disjoint sets of the $\psi_{\mathfrak{m}}^{(D)}$.
The results for the change-of-basis coefficients $G^{(\Gamma_\ell,\ell_z)}_{\mathfrak{s} \mathfrak{k} \mathfrak{m}}$ leading to a set of $\mathfrak{s} \in \{0,\ldots,6\}$ wavefunctions that satisfy Eq.~\eqref{eq:Dpsi} with $\mathcal{D}_{\ell_z \ell_z'}^{(\Gamma_\ell)}(R)$ defined by  Eq.~\eqref{eq:repdef}, with the basis vectors presented in Table~\ref{tab:basis}, are shown in Tables~\ref{tab:CG1}-\ref{tab:CG6}.

\begin{table}[!t]
\begin{ruledtabular}
\begin{tabular}{c|ccccc} 
            $\ell_z$     & $A_1^+$  & $E^+$               & $T_2^+$       & $T_1^+$                                                                    & $A_2^+$ \\\hline
            $0$ &     $1$        & $2z^2 - x^2 - y^2$  & $-$           & $i\sqrt{2}xy(x^2-y^2)$                                                      & $-$ \\ 
            & & & & & \\
            $1$ &     $-$        & $-$                 & $-zx + i yz$  & \begin{tabular}{@{}c@{}}$- zx(z^2-x^2)$ \\ $-i yz(y^2-z^2)$ \end{tabular}  & $-$ \\ 
              & & & & &  \\
            $2$ &     $-$        & $\sqrt{3}(x^2 - y^2)$         & $-i\sqrt{2}xy$ & $-$                                                                        &  \begin{tabular}{c} $x^4(y^2 - z^2)$ \\ $+ y^4(z^2-x^2)$ \\ $+ z^4(x^2-y^2)$ \end{tabular}  \hspace{10pt}  \\ 
              & & & & &  \\
            $3$ &     $-$        &  $-$                & $zx + i yz$   & \begin{tabular}{@{}c@{}}$- zx(z^2-x^2)$ \\ $+i yz(y^2-z^2)$ \end{tabular}   & $-$ \\
  \end{tabular}
  \caption{ Basis functions used to construct wavefunctions $\psi_{\mathfrak{s}\mathfrak{k}}^{(\Gamma_\ell,\ell_z)}(\vec{x}_1,\vec{x}_2)$ transforming according to each irrep of the cubic group, $\Gamma_\ell$, and the corresponding row labeled by $\ell_z$ (note that for the cubic group, $\ell_z$ is only defined modulo 4). Products of $x,\ y,\ z$ denote symmetrized tensor products. The relative signs and normalizations of generators within each column are chosen so that products of spin and orbital angular momentum cubic irreps can be projected in total-angular-momentum cubic irreps using the spinor conventions presented in Appendix~\ref{app:weights} and the Clebsch-Gordan coefficients presented in Ref.~\cite{Basak:2005ir}. 
       \label{tab:basis}}
   \end{ruledtabular}
\end{table}

\begin{table}[!h]
\begin{tabular}{|c||c|} \hline
   $\mathfrak{s} = 0$ & \scalebox{0.71}{$(0,0,0)$} \\\hline\hline
 $(A_1^+,0,1)$ & $1$ \\\hline
  \end{tabular}
  \hspace{50pt}
\begin{tabular}{|c||c|c|c|} \hline
   $\mathfrak{s} = 1$ & \scalebox{0.71}{$(1,0,0)$} & \scalebox{0.71}{$(0,1,0)$} & \scalebox{0.71}{$(0,0,1)$}  \\\hline\hline
 $(A_1^+,0,1)$ & $\frac{1}{\sqrt{3}}$ & $\frac{1}{\sqrt{3}}$ & $\frac{1}{\sqrt{3}}$ \\\hline
 $(E^+,0,1)$ & $\frac{1}{\sqrt{6}}$  & $\frac{1}{\sqrt{6}}$ & $-\sqrt{\frac{2}{3}}$ \\\hline
 $(E^+,2,1)$ & $-\frac{1}{\sqrt{2}}$  & $\frac{1}{\sqrt{2}}$ & 0 \\\hline
  \end{tabular}
  \caption{ The coefficients $G^{(\Gamma_\ell,\ell_z)}_{\mathfrak{s} \mathfrak{k} \mathfrak{m}}$ with $(\Gamma_\ell,\ell_z,\mathfrak{k})$ shown in the left column and $\vec{n}_{\mathfrak{m}}$ shown in the top row corresponding to the $\mathfrak{s}=0$ and $\mathfrak{s}=1$ shells. \label{tab:CG1}}
\end{table}

\begin{table}[h!]
  \begin{tabular}{|c||c|c|c|c|c|c|} \hline
   $\mathfrak{s} = 2$ & \scalebox{0.71}{$(1,1,0)$} & \scalebox{0.71}{$(1,0,1)$} & \scalebox{0.71}{$(1,0,-1)$} & \scalebox{0.71}{$(1,-1,0)$} & \scalebox{0.71}{$(0,1,1)$} & \scalebox{0.71}{$(0,1,-1)$}  \\\hline\hline
 $(A_1^+,0,1)$ & $\frac{1}{\sqrt{6}}$ & $\frac{1}{\sqrt{6}}$ & $\frac{1}{\sqrt{6}}$ & $\frac{1}{\sqrt{6}}$ & $\frac{1}{\sqrt{6}}$ & $\frac{1}{\sqrt{6}}$ \\\hline
 $(E^+,0,1)$ & $\frac{1}{\sqrt{3}}$ & $-\frac{1}{2\sqrt{3}}$ &  $-\frac{1}{2\sqrt{3}}$ & $\frac{1}{\sqrt{3}}$ & $-\frac{1}{2\sqrt{3}}$ &  $-\frac{1}{2\sqrt{3}}$ \\\hline
 $(E^+,2,1)$ & $0$ & $-\frac{1}{2}$ &  $-\frac{1}{2}$ & $0$ & $\frac{1}{2}$ &  $\frac{1}{2}$ \\\hline
 $(T_2^+,1,1)$ & $0$ & $\frac{1}{2}$ &  $-\frac{1}{2}$ & $0$ & $-\frac{i}{2}$ &  $\frac{i}{2}$ \\\hline
 $(T_2^+,2,1)$ & $\frac{i}{\sqrt{2}}$ & $0$ &  $0$ & $-\frac{i}{\sqrt{2}}$ & $0$ &  $0$ \\\hline
 $(T_2^+,3,1)$  & $0$ & $-\frac{1}{2}$ &  $\frac{1}{2}$ & $0$ & $-\frac{i}{2}$ &  $\frac{i}{2}$ \\\hline
  \end{tabular}
  \caption{ The coefficients $G^{(\Gamma_\ell,\ell_z)}_{\mathfrak{s} \mathfrak{k} \mathfrak{m}}$ with $(\Gamma_\ell,\ell_z,\mathfrak{k})$ shown in the left column and $\vec{n}_{\mathfrak{m}}$ shown in the top row corresponding to the $\mathfrak{s} = 2$ shell. 
  \label{tab:CG2}}
\end{table}

\begin{table}[!h]
\begin{tabular}{|c||c|c|c|c|} \hline
   $\mathfrak{s} = 3$ & \scalebox{0.71}{$(1,1,1)$} & \scalebox{0.71}{$(1,1,-1)$} & \scalebox{0.71}{$(1,-1,1)$} & \scalebox{0.71}{$(1,-1,-1)$} \\\hline\hline
 $(A_1^+,0,1)$ & $\frac{1}{2}$ & $\frac{1}{2}$ & $\frac{1}{2}$ & $\frac{1}{2}$ \\\hline
  $(T_2^+,1,1)$  &  $-\frac{1-i}{2\sqrt{2}}$ &  $\frac{1-i}{2\sqrt{2}}$  &   $-\frac{1+i}{2\sqrt{2}}$  &  $\frac{1+i}{2\sqrt{2}}$  \\\hline
  $(T_2^+,2,1)$  &  $-\frac{i}{2}$ & $-\frac{i}{2}$ & $\frac{i}{2}$ & $\frac{i}{2}$ \\\hline
  $(T_2^+,3,1)$  &  $\frac{1+i}{2\sqrt{2}}$ &  $-\frac{1+i}{2\sqrt{2}}$  &   $\frac{1-i}{2\sqrt{2}}$  &  $-\frac{1-i}{2\sqrt{2}}$  \\\hline
  \end{tabular}
  \hspace{50pt}
\begin{tabular}{|c||c|c|c|} \hline
   $\mathfrak{s} = 4$ & \scalebox{0.71}{$(2,0,0)$} & \scalebox{0.71}{$(0,2,0)$} & \scalebox{0.71}{$(0,0,2)$}  \\\hline\hline
 $(A_1^+,0,1)$ & $\frac{1}{\sqrt{3}}$ & $\frac{1}{\sqrt{3}}$ & $\frac{1}{\sqrt{3}}$ \\\hline
 $(E^+,0,1)$ & $\frac{1}{\sqrt{6}}$  & $\frac{1}{\sqrt{6}}$ & $-\sqrt{\frac{2}{3}}$ \\\hline
 $(E^+,2,1)$ & $-\frac{1}{\sqrt{2}}$  & $\frac{1}{\sqrt{2}}$ & 0 \\\hline
  \end{tabular}
  \caption{ The coefficients $G^{(\Gamma_\ell,\ell_z)}_{\mathfrak{s} \mathfrak{k} \mathfrak{m}}$ with $(\Gamma_\ell,\ell_z,\mathfrak{k})$ shown in the left column and $\vec{n}_{\mathfrak{m}}$ shown in the top row corresponding to the $\mathfrak{s}=3$ and $\mathfrak{s}=4$ shells. \label{tab:CG3}}
\end{table}

\begin{table}[h!]
  \begin{tabular}{|c||c|c|c|c|c|c|c|c|c|c|c|c|} \hline
   $\mathfrak{s} = 5$ & \scalebox{0.71}{$(2,1,0)$} & \scalebox{0.71}{$(2,-1,0)$}  & \scalebox{0.71}{$(2,0,1)$} & \scalebox{0.71}{$(2,0,-1)$} & \scalebox{0.71}{$(1,2,0)$} & \scalebox{0.71}{$(1,-2,0)$}  & \scalebox{0.71}{$(1,0,2)$} & \scalebox{0.71}{$(1,0,-2)$} & \scalebox{0.71}{$(0,1,2)$} & \scalebox{0.71}{$(0,1,-2)$}  & \scalebox{0.71}{$(0,2,1)$} & \scalebox{0.71}{$(0,2,-1)$}  \\\hline\hline
  $(A_1^+,0,1)$ & $\frac{1}{2\sqrt{3}}$ & $\frac{1}{2\sqrt{3}}$ & $\frac{1}{2\sqrt{3}}$ & $\frac{1}{2\sqrt{3}}$ & $\frac{1}{2\sqrt{3}}$ & $\frac{1}{2\sqrt{3}}$ & $\frac{1}{2\sqrt{3}}$ & $\frac{1}{2\sqrt{3}}$ & $\frac{1}{2\sqrt{3}}$ & $\frac{1}{2\sqrt{3}}$ & $\frac{1}{2\sqrt{3}}$ & $\frac{1}{2\sqrt{3}}$ \\\hline
  $(A_2^+,2,1)$ & $-\frac{1}{2\sqrt{3}}$ & $-\frac{1}{2\sqrt{3}}$ & $\frac{1}{2\sqrt{3}}$ & $\frac{1}{2\sqrt{3}}$ & $\frac{1}{2\sqrt{3}}$ & $\frac{1}{2\sqrt{3}}$ & $-\frac{1}{2\sqrt{3}}$ & $-\frac{1}{2\sqrt{3}}$ & $\frac{1}{2\sqrt{3}}$ &
  $\frac{1}{2\sqrt{3}}$ &  $-\frac{1}{2\sqrt{3}}$ & $-\frac{1}{2\sqrt{3}}$ \\\hline
  $(E^+,0,1)$ & $\frac{5}{2\sqrt{78}}$ & $\frac{5}{2\sqrt{78}}$ & $\frac{1}{\sqrt{78}}$ & $\frac{1}{\sqrt{78}}$ & $\frac{5}{2\sqrt{78}}$ & $\frac{5}{2\sqrt{78}}$ & $-\frac{7}{2\sqrt{78}}$ & $-\frac{7}{2\sqrt{78}}$ & $-\frac{7}{2\sqrt{78}}$ & $-\frac{7}{2\sqrt{78}}$ & $\frac{1}{\sqrt{78}}$ & $\frac{1}{\sqrt{78}}$ \\\hline 
  $(E^+,0,2)$ & $\frac{3}{2\sqrt{26}}$ & $\frac{3}{2\sqrt{26}}$ & $-\sqrt{\frac{2}{13}}$ & $-\sqrt{\frac{2}{13}}$ & $\frac{3}{2\sqrt{26}}$ & $\frac{3}{2\sqrt{26}}$ & $\frac{1}{2\sqrt{26}}$ & $\frac{1}{2\sqrt{26}}$ & $\frac{1}{2\sqrt{26}}$ & $\frac{1}{2\sqrt{26}}$ & $-\sqrt{\frac{2}{13}}$ & $-\sqrt{\frac{2}{13}}$ \\\hline
  $(E^+,2,1)$ & $-\frac{3}{2\sqrt{26}}$ & $-\frac{3}{2\sqrt{26}}$ & $-\sqrt{\frac{2}{13}}$ & $-\sqrt{\frac{2}{13}}$ & $\frac{3}{2\sqrt{26}}$ & $\frac{3}{2\sqrt{26}}$ & $-\frac{1}{2\sqrt{26}}$ & $-\frac{1}{2\sqrt{26}}$ & $\frac{1}{2\sqrt{26}}$ & $\frac{1}{2\sqrt{26}}$ & $\sqrt{\frac{2}{13}}$ & $\sqrt{\frac{2}{13}}$ \\\hline 
  $(E^+,2,2) $ & $\frac{5}{2\sqrt{78}}$ & $\frac{5}{2\sqrt{78}}$ & $-\frac{1}{\sqrt{78}}$ & $-\frac{1}{\sqrt{78}}$ & $-\frac{5}{2\sqrt{78}}$ & $-\frac{5}{2\sqrt{78}}$ & $-\frac{7}{2\sqrt{78}}$ & $-\frac{7}{2\sqrt{78}}$ & $\frac{7}{2\sqrt{78}}$ & $\frac{7}{2\sqrt{78}}$ & $\frac{1}{\sqrt{78}}$ & $\frac{1}{\sqrt{78}}$ \\\hline
  $(T_1^+,0,1)$ & $\frac{i}{2}$ & $-\frac{i}{2}$ & 0 & 0 & $-\frac{i}{2}$ & $\frac{i}{2}$ & 0 & 0 & 0 & 0 & 0 & 0 \\\hline
  $(T_1^+,1,1)$ & 0 & 0 & $\frac{1}{2\sqrt{2}}$ & $-\frac{1}{2\sqrt{2}}$ & 0 & 0 & $-\frac{1}{2\sqrt{2}}$ & $\frac{1}{2\sqrt{2}}$ & $\frac{i}{2\sqrt{2}}$ & $-\frac{i}{2\sqrt{2}}$ & $-\frac{i}{2\sqrt{2}}$ & $\frac{i}{2\sqrt{2}}$ \\\hline
  $(T_1^+,3,1)$ & 0 & 0 & $\frac{1}{2\sqrt{2}}$ & $-\frac{1}{2\sqrt{2}}$ & 0 & 0 & $-\frac{1}{2\sqrt{2}}$ & $\frac{1}{2\sqrt{2}}$ & $-\frac{i}{2\sqrt{2}}$ & $\frac{i}{2\sqrt{2}}$ & $\frac{i}{2\sqrt{2}}$ & $-\frac{i}{2\sqrt{2}}$ \\\hline
  $(T_2^+,1,1)$ & 0 & 0 & $-\frac{1}{2\sqrt{2}}$ & $\frac{1}{2\sqrt{2}}$ & 0 & 0 & $-\frac{1}{2\sqrt{2}}$ & $\frac{1}{2\sqrt{2}}$ & $\frac{i}{2\sqrt{2}}$ & $-\frac{i}{2\sqrt{2}}$ & $\frac{i}{2\sqrt{2}}$ & $-\frac{i}{2\sqrt{2}}$ \\\hline
  $(T_2^+,2,1)$ & $-\frac{i}{2}$ & $\frac{i}{2}$ & 0 & 0 & $-\frac{i}{2}$ & $\frac{i}{2}$ & 0 & 0 & 0 & 0 & 0 & 0 \\\hline
  $(T_2^+,3,1)$ & 0 & 0 & $\frac{1}{2\sqrt{2}}$ & $-\frac{1}{2\sqrt{2}}$ & 0 & 0 & $\frac{1}{2\sqrt{2}}$ & $-\frac{1}{2\sqrt{2}}$ & $\frac{i}{2\sqrt{2}}$ & $-\frac{i}{2\sqrt{2}}$ & $\frac{i}{2\sqrt{2}}$ & $-\frac{i}{2\sqrt{2}}$ \\\hline
  \end{tabular}
  \caption{ The coefficients $G^{(\Gamma_\ell,\ell_z)}_{\mathfrak{s} \mathfrak{k} \mathfrak{m}}$ with $(\Gamma_\ell,\ell_z,\mathfrak{k})$ shown in the left column and $\vec{n}_{\mathfrak{m}}$ shown in the top row corresponding to the $\mathfrak{s} = 5$ shell. 
  \label{tab:CG5}}
\end{table}

\begin{table}[h!]
\resizebox{\columnwidth}{!}{
  \begin{tabular}{|c||c|c|c|c|c|c|c|c|c|c|c|c|} \hline
   $\mathfrak{s} = 6$ & \scalebox{0.71}{$(2,1,1)$} & \scalebox{0.71}{$(2,1,-1)$}  & \scalebox{0.71}{$(2,-1,1)$} & \scalebox{0.71}{$(2,-1,-1)$} & \scalebox{0.71}{$(1,2,1)$} & \scalebox{0.71}{$(1,2,-1)$}  & \scalebox{0.71}{$(1,-2,1)$} & \scalebox{0.71}{$(1,-2,-1)$} & \scalebox{0.71}{$(1,1,2)$} & \scalebox{0.71}{$(1,1,-2)$}  & \scalebox{0.71}{$(1,-1,2)$} & \scalebox{0.71}{$(1,-1,-2)$}  \\\hline\hline
  $(A_1^+,0,1)$ & $\frac{1}{2\sqrt{3}}$ & $\frac{1}{2\sqrt{3}}$ & $\frac{1}{2\sqrt{3}}$ & $\frac{1}{2\sqrt{3}}$ & $\frac{1}{2\sqrt{3}}$ & $\frac{1}{2\sqrt{3}}$ & $\frac{1}{2\sqrt{3}}$ & $\frac{1}{2\sqrt{3}}$ & $\frac{1}{2\sqrt{3}}$ & $\frac{1}{2\sqrt{3}}$ & $\frac{1}{2\sqrt{3}}$ & $\frac{1}{2\sqrt{3}}$ \\\hline
  $(E^+,0,1) $ & $\frac{1}{2\sqrt{6}}$ & $\frac{1}{2\sqrt{6}}$ & $\frac{1}{2\sqrt{6}}$ & $\frac{1}{2\sqrt{6}}$ & $\frac{1}{2\sqrt{6}}$ & $\frac{1}{2\sqrt{6}}$ & $\frac{1}{2\sqrt{6}}$ & $\frac{1}{2\sqrt{6}}$ & $-\frac{1}{\sqrt{6}}$ & $-\frac{1}{\sqrt{6}}$ & $-\frac{1}{\sqrt{6}}$ & $-\frac{1}{\sqrt{6}}$ \\\hline
  $(E^+,2,1)$ & $-\frac{1}{2\sqrt{2}}$ & $-\frac{1}{2\sqrt{2}}$ & $-\frac{1}{2\sqrt{2}}$ & $-\frac{1}{2\sqrt{2}}$ & $\frac{1}{2\sqrt{2}}$ & $\frac{1}{2\sqrt{2}}$ & $\frac{1}{2\sqrt{2}}$ & $\frac{1}{2\sqrt{2}}$ & 0 & 0 & 0 & 0 \\\hline
  $(T_1^+,0,1)$ & $\frac{i}{2\sqrt{2}}$ & $\frac{i}{2\sqrt{2}}$ & $-\frac{i}{2\sqrt{2}}$ & $-\frac{i}{2\sqrt{2}}$ & $-\frac{i}{2\sqrt{2}}$ & $-\frac{i}{2\sqrt{2}}$ & $\frac{i}{2\sqrt{2}}$ & $\frac{i}{2\sqrt{2}}$ & 0 & 0 & 0 & 0 \\\hline
  $(T_1^+,1,1)$ & $\frac{1}{4}$ & $-\frac{1}{4}$ & $\frac{1}{4}$ & $-\frac{1}{4}$ & $-\frac{i}{4}$ & $\frac{i}{4}$ & $\frac{i}{4}$ & $-\frac{i}{4}$ & $-\frac{1-i}{4}$ & $\frac{1-i}{4}$ & $-\frac{1+i}{4}$ & $\frac{1+i}{4}$ \\\hline
  $(T_1^+,3,1)$ & $\frac{1}{4}$ & $-\frac{1}{4}$ & $\frac{1}{4}$ & $-\frac{1}{4}$ & $\frac{i}{4}$ & $-\frac{i}{4}$ & $-\frac{i}{4}$ & $\frac{i}{4}$ & $-\frac{1+i}{4}$ & $\frac{1+i}{4}$ & $-\frac{1-i}{4}$ & $\frac{1-i}{4}$ \\\hline
  $(T_2^+,1,1)$ &  $-\frac{1}{4}$ &  $\frac{1}{4}$ &  $-\frac{1}{4}$ &   $\frac{1}{4}$ &  $\frac{i}{4}$ &  $-\frac{i}{4}$ &  $-\frac{i}{4}$ &  $\frac{i}{4}$  &  $-\frac{1-i}{4}$   &  $\frac{1-i}{4}$  &  $-\frac{1+i}{4}$  &  $\frac{1+i}{4}$   \\\hline 
  $(T_2^+,1,2)$ &  $\frac{i}{2\sqrt{2}}$ &  $-\frac{i}{2\sqrt{2}}$ &  $-\frac{i}{2\sqrt{2}}$ &   $\frac{i}{2\sqrt{2}}$ &   $-\frac{1}{2\sqrt{2}}$  &  $\frac{1}{2\sqrt{2}}$ &  $-\frac{1}{2\sqrt{2}}$  &  $\frac{1}{2\sqrt{2}}$  &  $0$  &  $0$ &  $0$ &  $0$  \\\hline 
  $(T_2^+,2,1)$ &  $-\frac{i}{2\sqrt{2}}$ &  $-\frac{i}{2\sqrt{2}}$  &  $\frac{i}{2\sqrt{2}}$ &    $\frac{i}{2\sqrt{2}}$ &  $-\frac{i}{2\sqrt{2}}$ &  $-\frac{i}{2\sqrt{2}}$ &  $\frac{i}{2\sqrt{2}}$ &  $\frac{i}{2\sqrt{2}}$  &  $0$ &  $0$ &  $0$ &  $0$ \\\hline 
  $(T_2^+,2,2)$ &  $0$ &  $0$ &  $0$ &   $0$ &  $0$ &  $0$ &  $0$ &  $0$ &  $-\frac{i}{2}$   &  $-\frac{i}{2}$ &  $\frac{i}{2}$ &  $\frac{i}{2}$ \\\hline 
  $(T_2^+,3,1)$ & $\frac{1}{4}$ &  $-\frac{1}{4}$ &  $\frac{1}{4}$ &   $-\frac{1}{4}$ &  $\frac{i}{4}$ &  $-\frac{i}{4}$ &  $-\frac{i}{4}$ &  $\frac{i}{4}$  &  $\frac{1+i}{4}$   &  $-\frac{1+i}{4}$  &  $\frac{1-i}{4}$  &  $-\frac{1-i}{4}$   \\\hline 
  $(T_2^+,3,2)$ &  $\frac{i}{2\sqrt{2}}$ &  $-\frac{i}{2\sqrt{2}}$ &  $-\frac{i}{2\sqrt{2}}$ &   $\frac{i}{2\sqrt{2}}$ &   $\frac{1}{2\sqrt{2}}$  &  $-\frac{1}{2\sqrt{2}}$ &  $\frac{1}{2\sqrt{2}}$  &  $-\frac{1}{2\sqrt{2}}$  &  $0$  &  $0$ &  $0$ &  $0$  \\\hline 
  \end{tabular}}
  \caption{ The coefficients $G^{(\Gamma_\ell,\ell_z)}_{\mathfrak{s} \mathfrak{k} \mathfrak{m}}$ with $(\Gamma_\ell,\ell_z,\mathfrak{k})$ shown in the left column and $\vec{n}_{\mathfrak{m}}$ shown in the top row corresponding to the $\mathfrak{s} = 6$ shell.
  \label{tab:CG6}}
\end{table}

For positive-parity two-nucleon systems with $I=1$, the spin-orbit tensor product is trivial and the wavefunctions above provide a complete basis for $\mathfrak{s} \in \{0,\ldots,6\}$ wavefunctions that transform irreducibly under cubic transformations.
The corresponding $I=0$ systems have $\Gamma_S = T_1^+$, and the spin-orbit product representation must be decomposed into direct sums of irreps.
The cubic group Clebsch-Gordan coefficients required for this decomposition can be obtained from the $SO(3)$ Clebsch-Gordan coefficients required to decompose products of the generators shown in Table~\ref{tab:basis} (which are related to spherical harmonics) into direct sums of  $SO(3)$ irreps and are presented in Ref.~\cite{Basak:2005ir}.
The non-zero Clebsch-Gordan coefficients $\mathcal{C}^{(\Gamma_J,J_z,\Gamma_\ell,\ell_z)}_\rho$ appearing in Eq.~\eqref{eq:D0def} that are required for constructing $I=0$ correlation functions are reproduced here for completeness.

For the $A_1^+$ irrep, the only non-zero Clebsch-Gordan coefficients are given by
\begin{equation}
    \mathcal{C}_{1}^{(A_1^+,0,T_1^+,3)} = -\mathcal{C}_{2}^{(A_1^+,0,T_1^+,0)} = \mathcal{C}_{3}^{(A_1^+,0,T_1^+,1)} = \frac{1}{\sqrt{3}},
\label{eq:CA1}
\end{equation}
where it should be noted that for the cubic group, $J_z$ is only defined modulo 4, and a basis is used in which two-nucleon spin components $\rho = 1,2,3$ correspond to $J_z=1$, $J_z=0$, and $J_z=-1 = 3 \text{ mod } 4$, respectively.
To complete the construction of dibaryon operators with $I=0$ and $\Gamma_J = A_1^+$, it  remains only to construct the multiplicity-label tensor $M^{(\Gamma_J,\Gamma_\ell)}_{\mathfrak{s}\mathfrak{k}\mathfrak{k}'}$  introduced in Eq.~\eqref{eq:D0def}.
The multiplicity-label tensor is defined to provide a definite ordering $\mathfrak{k} \in \{1,\ldots,\mathcal{N}_{\mathfrak{s}}^{(0,\Gamma_J)} \}$ for all wavefunctions in the same irrep and momentum shell, and for a given $\Gamma_J$ and $\mathfrak{s}$, it is equal to one for a single $\Gamma_\ell$ and $\mathfrak{k}'$ and equal to zero otherwise.
The only momentum shells considered here which include $\Gamma_\ell = T_1^+$ wavefunctions are $\mathfrak{s} \in \{5,6\}$; since only a single $\Gamma_J = A_1^+$ wavefunction can be constructed for each case, the  multiplicity-label tensor $M^{(A_1^+,T_1^+)}_{\mathfrak{s}\mathfrak{k}\mathfrak{k}'}$  is simply  $M^{(A_1^+,T_1^+)}_{\mathfrak{s}11} = 1$ for $\mathfrak{s} \in \{5,6\}$.

For the $A_2^+$ irrep, the non-zero Clebsch-Gordan coefficients are given by
\begin{equation}
    \mathcal{C}_{1}^{(A_2^+,2,T_2^+,1)} = \mathcal{C}_{2}^{(A_2^+,2,T_2^+,2)} = -\mathcal{C}_{3}^{(A_2^+,2,T_2^+,3)} = \frac{1}{\sqrt{3}}.
\label{eq:CA2}
\end{equation}
The $\mathfrak{s} \in \{2,3,5\}$ shells each include one $T_2^+$ wavefunction, while the $\mathfrak{s}=6$ shell includes two.
No other $\Gamma_\ell$ contributes to $\Gamma_J = A_2^+$ dibaryon operators, and so the non-zero elements of the multiplicity-label tensor for this irrep is  $M^{(A_2^+,T_2^+)}_{\mathfrak{s} 11} = 1$ for $\mathfrak{s} \in \{2,3,5,6\}$ and $M^{(A_2^+,T_2^+)}_{622} = 1$.

For the $E^+$ irrep, the non-zero Clebsch-Gordan coefficients are given by
\begin{equation}
\begin{split}
    \mathcal{C}_{1}^{(E^+,0,T_1^+,3)} &=  \mathcal{C}_{3}^{(E^+,0,T_1^+,1)} = \frac{1}{\sqrt{6}}, \hspace{25pt} \mathcal{C}_{2}^{(E^+,0,T_1^+,0)} = \sqrt{\frac{2}{3}}, \\
    \mathcal{C}_{1}^{(E^+,2,T_1^+,1)} &= \mathcal{C}_{3}^{(E^+,2,T_1^+,3)} = \frac{1}{\sqrt{2}}, \\
    \mathcal{C}_{1}^{(E^+,0,T_2^+,3)} &= -\mathcal{C}_{3}^{(E^+,0,T_2^+,1)} = \frac{1}{\sqrt{2}},  \\
    \mathcal{C}_{1}^{(E^+,2,T_2^+,1)} &= -\mathcal{C}_{3}^{(E^+,2,T_2^+,3)} = \frac{1}{\sqrt{6}}, \hspace{25pt} \mathcal{C}_{2}^{(E^+,2,T_2^+,2)} = -\sqrt{\frac{2}{3}}.
\end{split}\label{eq:CE}
\end{equation}
The $\mathfrak{s} \in \{5,6\}$ shells include wavefunctions with both $\Gamma_\ell \in \{T_1^+, T_2^+\}$, and the multiplicity-label tensor $M^{(E^+,\Gamma_\ell)}_{\mathfrak{s}\mathfrak{k}\mathfrak{k}'}$ is chosen so that lower (higher) values of $\mathfrak{k}$ are associated with $T_1^+$ ($T_2^+$) wavefunctions.
The non-zero multiplicity-label tensor elements have value unity, as always, and are $M^{(E^+,T_2^+)}_{\mathfrak{s}11}$ for the $\mathfrak{s} \in \{2,3\}$ shells; 
$M^{(E^+,T_1^+)}_{511}$ and $M^{(E^+,T_2^+)}_{521}$ for the $\mathfrak{s}=5$ shell; and $M^{(E^+,T_1^+)}_{611}$, $M^{(E^+,T_2^+)}_{621}$, and $M^{(E^+,T_2^+)}_{632}$ for the $\mathfrak{s}=6$ shell.

For the $T_1^+$ irrep, the non-zero Clebsch-Gordan coefficients are given by
\begin{equation}
\begin{split}
    \mathcal{C}_{1}^{(T_1^+,1,A_1^+,0)} &= \mathcal{C}_{2}^{(T_1^+,0,A_1^+,0)} = \mathcal{C}_{3}^{(T_1^+,3,A_1^+,0)} = 1, \\
    \mathcal{C}_{1}^{(T_1^+,1,E^+,0)} &= \mathcal{C}_{3}^{(T_1^+,3,E^+,0)} = \frac{1}{2}, \hspace{25pt} \mathcal{C}_{2}^{(T_1^+,0,E^+,0)} = -1,  \\
    \mathcal{C}_{3}^{(T_1^+,1,E^+,2)} &= \mathcal{C}_{1}^{(T_1^+,3,E^+,2)} = \frac{\sqrt{3}}{2}, \\
    \mathcal{C}_{2}^{(T_1^+,1,T_1^+,1)} & = -\mathcal{C}_{1}^{(T_1^+,1,T_1^+,0)} = \mathcal{C}_{3}^{(T_1^+,0,T_1^+,1)} = -\mathcal{C}_{1}^{(T_1^+,0,T_1^+,3)} = 
    \frac{1}{\sqrt{2}}, \\
    \mathcal{C}_{3}^{(T_1^+,3,T_1^+,0)} &= -\mathcal{C}_{2}^{(T_1^+,3,T_1^+,3)} =  \frac{1}{\sqrt{2}}, \\
    \mathcal{C}_{2}^{(T_1^+,1,T_2^+,1)} &=  -\mathcal{C}_{3}^{(T_1^+,1,T_2^+,2)}  = -\mathcal{C}_{1}^{(T_1^+,0,T_2^+,3)} = -\mathcal{C}_{3}^{(T_1^+,0,T_2^+,1)} = \frac{1}{\sqrt{2}}, \\
    \mathcal{C}_{1}^{(T_1^+,3,T_2^+,2)} &=   \mathcal{C}_{2}^{(T_1^+,3,T_2^+,3)} = \frac{1}{\sqrt{2}}.
\end{split}\label{eq:CT1}
\end{equation}
All $\mathfrak{s}$ shells include $\Gamma_J = T_1^+$ operators with multiple $\Gamma_\ell$, and $M^{(T_1^+,\Gamma_\ell)}_{\mathfrak{s}\mathfrak{k}\mathfrak{k}'}$ is chosen so that increasing $\mathfrak{k}$ is assigned to the ordering $\Gamma_\ell \in \{ A_1^+, E^+, T_1^+, T_2^+\}$.
The non-zero multiplicity-label tensor elements have value unity and are as follows: $M^{(T_1^+,A_1^+)}_{011}$ for the $\mathfrak{s}=0$ shell; $M^{(T_1^+,A_1^+)}_{111}$ and $M^{(T_1^+,E^+)}_{121}$ for the $\mathfrak{s}=1$ shell; $M^{(T_1^+,A_1^+)}_{211}$, $M^{(T_1^+,E^+)}_{221}$, and $M^{(T_1^+,T_2^+)}_{231}$ for the $\mathfrak{s}=2$ shell; $M^{(T_1^+,A_1^+)}_{311}$ and $M^{(T_1^+,E^+)}_{321}$ for the $\mathfrak{s}=3$ shell; $M^{(T_1^+,A_1^+)}_{411}$ and $M^{(T_1^+,E^+)}_{421}$ for the $\mathfrak{s}=4$ shell; $M^{(T_1^+,A_1^+)}_{511}$, $M^{(T_1^+,E^+)}_{521}$, $M^{(T_1^+,E^+)}_{532}$, $M^{(T_1^+,T_1^+)}_{541}$, and $M^{(T_1^+,T_2^+)}_{551}$ for the $\mathfrak{s}=5$ shell; and  $M^{(T_1^+,A_1^+)}_{611}$, $M^{(T_1^+,E^+)}_{621}$, $M^{(T_1^+,T_1^+)}_{631}$, $M^{(T_1^+,T_2^+)}_{641}$,  and $M^{(T_1^+,T_2^+)}_{652}$ for the $\mathfrak{s}=6$ shell.

For the $T_2^+$ irrep, the non-zero Clebsch-Gordan coefficients are given by
\begin{equation}
\begin{split}
    \mathcal{C}_{3}^{(T_2^+,1,A_2^+,2)} &= -\mathcal{C}_{2}^{(T_2^+,2,A_2^+2,)} = -\mathcal{C}_{1}^{(T_2^+,3,A_2^+,2)} = 1, \\
    \mathcal{C}_{1}^{(T_2^+,1,E^+,0)} &= - \mathcal{C}_{3}^{(T_2^+,3,E^+,0)} = \frac{\sqrt{3}}{2}, \hspace{25pt}  \mathcal{C}_{2}^{(T_2^+,2,E^+,2)} = -1, \\
    \mathcal{C}_{3}^{(T_2^+,1,E^+,2)} &= -\mathcal{C}_{1}^{(T_2^+,3,E^+,2)} = -\frac{1}{2},  \\
    \mathcal{C}_{1}^{(T_2^+,1,T_1^+,0)} &= \mathcal{C}_{2}^{(T_2^+,1,T_1^+,1)} = \mathcal{C}_{1}^{(T_2^+,2,T_1^+,1)}   = -\mathcal{C}_{3}^{(T_2^+,2,T_1^+,3)} = \frac{1}{\sqrt{2}}, \\
    \mathcal{C}_{3}^{(T_2^+,3,T_1^+,0)} &= \mathcal{C}_{2}^{(T_2^+,3,T_1^+,3)} = \frac{1}{\sqrt{2}}, \\
    \mathcal{C}_{2}^{(T_2^+,1,T_2^+,1)} &= \mathcal{C}_{3}^{(T_2^+,1,T_2^+,2)} = -\mathcal{C}_{1}^{(T_2^+,2,T_2^+,1)} = -\mathcal{C}_{3}^{(T_2^+,2,T_2^+,3)}  =  \frac{1}{\sqrt{2}}, \\ 
     \mathcal{C}_{1}^{(T_2^+,3,T_2^+,2)} &= -\mathcal{C}_{2}^{(T_2^+,3,T_2^+,3)} = \frac{1}{\sqrt{2}}.
\end{split}\label{eq:CT2}
\end{equation}
Multiple $\Gamma_\ell$ lead to $\Gamma_J = T_2^+$ operators in several $\mathfrak{s}$ shells and $M^{(T_2^+,\Gamma_\ell)}_{\mathfrak{s}\mathfrak{k}\mathfrak{k}'}$ is chosen so that increasing $\mathfrak{k}$ is assigned to the ordering $\Gamma_\ell \in \{ A_2^+, E^+, T_1^+, T_2^+\}$.
The non-zero multiplicity-label tensor elements have value unity and are as follows: $M^{(T_2^+,E^+)}_{111}$ for the $\mathfrak{s}=1$ shell; $M^{(T_2^+,E^+)}_{211}$ and $M^{(T_2^+,T_2^+)}_{221}$ for the $\mathfrak{s}=2$ shell; $M^{(T_2^+,E^+)}_{311}$ for the $\mathfrak{s}=3$ shell; $M^{(T_2^+,E^+)}_{411}$ for the $\mathfrak{s}=4$ shell; $M^{(T_2^+,A_2^+)}_{511}$, $M^{(T_2^+,E^+)}_{521}$, $M^{(T_2^+,E^+)}_{532}$, $M^{(T_2^+,T_1^+)}_{541}$, and $M^{(T_2^+,T_2^+)}_{551}$ for the $\mathfrak{s}=5$ shell; and  $M^{(T_2^+,E^+)}_{611}$, $M^{(T_2^+,T_1^+)}_{621}$, $M^{(T_2^+,T_2^+)}_{631}$,  and $M^{(T_2^+,T_2^+)}_{642}$ for the $\mathfrak{s}=6$ shell.

\section{Contraction algorithm implementation}\label{app:contractions}

We have implemented the contraction algorithm described in Sec.~\ref{sec:contractions} using  \verb!Tiramisu! \cite{baghdadi2020tiramisu}, a domain-specific language (DSL) embedded in C++. 
\verb!Tiramisu! provides a C++ API that allows users to write high level, architecture-independent algorithms, and a set of API calls to specify how the code should be optimized.
\verb!Tiramisu! is designed for efficiently expressing algorithms that operate over dense arrays using loop nests, which in computer-science and applied-math contexts are often found in the areas of dense linear algebra, tensor operations, stencil computations, image processing, and deep learning.
Both the bilocal baryon-block construction and correlation-function calculation described in Sec.~\ref{sec:contractions} involve tensor operations with many nested sums over space, spin, and color indices and are therefore amenable to similar optimization techniques.

\verb!Tiramisu! uses a mathematical model known as the polyhedral model internally \cite{feautrier_array_1988,pencil_paper,bondhugula_practical_2008,pencil,baghdadi2020tiramisu} to represent code, code transformations, and to reason about the correctness of code transformations.
Optimizations applied using \verb!Tiramisu! include loop parallelization (with OpenMP style parallelism), vectorization (taking advantage of instruction sets such as AVX2 and AVX512 as well as fused-multiply-add hardware instructions), loop reordering, and loop fusion.
A simple example of using loop reordering to improve memory access patterns and reduce computational redundancy via  common-subexpression elimination is to order the  loop over spatial wavefunction types innermost during block construction, which allows the product of three quark propagators to be computed once and then reused during construction of blocks with all spatial wavefunctions. 
Loop fusion is useful for eliminating the need for large temporary arrays as well as improving data locality and  is critical to the feasibility of the calculations presented here.
In particular, by fusing the $\vec{x}_1$ and $\vec{x}_2$ loops required to evaluate $\mathcal{E}^{(q\pm)ijk}_{g\sigma \mathfrak{m}'g'}(\vec{x}_1,\vec{x}_2,t)$ using Eq.~\eqref{eq:exblock1} with the loops required to evaluate the $\vec{x}_1$ and $\vec{x}_2$ sums in Eq.~\eqref{eq:NN-NNblocks}, it is possible to reduce the intermediate array storage required for each bilocal baryon block to a single $(2\times 3)^3 = 216$ component tensor for each baryon-level spin and spatial wavefunction.
Without loop fusion, storing all entries of $\mathcal{E}^{(\pm q)ijk}_{s \rho^\prime \mathfrak{m}^\prime g'}(\vec{x}_1,\vec{x}_2)$ for the calculations described in Sec.~\ref{sec:numerics} would require constructing 8 TB arrays for each quark smearing type, but with loop fusion only 60 kB of memory is required. 
Although loop fusion can introduce redundant computation, the resultant increase in computational cost (a sub-percent increase in the number of computations to be performed in this case) can be justified in cases of sufficiently large memory-use optimization such as this.
This approach further permits distributed parallelization over the $\vec{x}_1$ and $\vec{x}_2$  sums with no communication required and facilitates practical GPU execution that will be used in future work.

The advantages of a contraction algorithm based on bilocal baryon block construction and of the \verb!Tiramisu! compiler can be seen by example.
With $\vsp = 8^3$ and $N_{\rm src} = 50$, evaluating Eq.~\eqref{eq:NN-NNcontract} without introducing baryon blocks requires $16\times 36 N_W^{[D]}N_{\rm src}^2  \vsp^4~\sim~ 10^{19}$ products of six complex numbers for each choice of source and sink smearing.
Instead computing bilocal baryon blocks defined by Eq.~\eqref{eq:exblock1} using the same lattice volume and interpolating-operator set requires $6 \times 216 N_W^{(B)} N_{\rm src} \vsp^3~\sim~10^{14}$ products of six complex numbers.
Subsequently evaluating Eq.~\eqref{eq:NN-NNblocks} using local and bilocal baryon blocks requires $16 \times 4 \times 36 N_W^{[D]} N_{\rm src}^2 \vsp^2~\sim~10^{14}$ products of two complex numbers.
Introduction of bilocal baryon blocks and use of Eqs.~\eqref{eq:exblock1}-\eqref{eq:NN-NNblocks} therefore leads to a reduction in computational cost by $10^5$ for this example that would be even larger for larger $\vsp$.
Practically calculating bilocal baryon blocks for this lattice volume and interpolating-operator set then requires fusion of  loops over $\vec{x}_1$ and $\vec{x}_2$ in order to avoid excessive memory requirements as discussed above.
In comparison to a C++ code implementing this loop fusion and the loop ordering optimization discussed above, additional optimizations performed internally by \verb!Tiramisu! lead to a further order of magnitude speedup for $\vsp = 512$. 

In order to test the validity of our codes, we have explicitly tested that applying all LQCD symmetry transformations to the smeared-smeared sparsened timeslice-to-all quark propagators computed for one $m_\pi = 806$ MeV gauge-field configuration and used as input to the \verb!Tiramisu! hadron correlation-function calculation code leads to the expected transformations of all hadron correlation functions computed.\footnote{ We have also verified that applying symmetry transformations to a small random gauge field before calculating quark propagators leads to the expected transformation properties within inverter tolerances.  }
The transformations of gauge fields $U_\mu(x)$ and quark propagators $S(\vec{x},t;\vec{y},0) = \left< q(\vec{x},t) \overline{q}(\vec{y},0) \right>$ under all the symmetries of the LQCD action including $SU(3)_C$ gauge transformations, the discrete symmetries $C$, $P$, and $T$, and spacetime isometries valid for a cubic finite volume with periodic boundary conditions, as well as $t$-dependent transformations associated with $U(1)_B$ and $U(1)_{u-d}$ symmetries, are listed in Table~\ref{tab:symmetries}.
The 48 elements of the cubic group $R \in O_h$ can be described as products of reflections $r_k$ about the $\hat{e}_k$ axis and permutations $p_{ij}$ of the $\hat{e}_k$ and $\hat{e}_j$ axes.
Equivalently, cubic-symmetry transformations $R \in O_h$ can be described as products of rotations $R(\theta \hat{e}_k)$ of $\theta$ radians about the $\hat{e}_k$ axis and the parity operation taking $\vec{x}\rightarrow P\vec{x} = -\vec{x}$, with the two descriptions related by\footnote{There are multiple descriptions of cubic group permutations as rotations, another is obtained for example using $p_{ij} = P R( \frac{\pi}{2} \hat{e}_i \times \hat{e}_j) P( \pi \hat{e}_i)$, that give distinct spin-$1/2$ representations of these cubic group transformations. The distinct spin-$1/2$ representations provided by these multiple descriptions are related by unitary changes of basis and therefore equivalent.}
\begin{equation}
\begin{split}
    r_k &= P R(\pi \hat{e}_k), \\
    p_{ij} &= P R(\pi \hat{e}_j ) R\left( \frac{\pi}{2} \hat{e}_{i} \times \hat{e}_j \right).
\end{split}\label{eq:cubicdict}
\end{equation}
Quark fields transform under the spin-$1/2$ representation of (the double cover of) the $SO(3)$ rotation group defined by~\cite{Weinberg:1995mt}
\begin{equation}
\begin{split}
    q(\vec{x},t) &\stackrel{R}{\longrightarrow} \mathcal{D}^{(1/2)}\left(  R \right)^\dagger  q(R(\vec{x}),t) = e^{ \frac{1}{8} \omega_{ij} [\gamma^i,\gamma^j]} q(R(\vec{x}),t), \\
    \overline{q}(\vec{x},t) &\stackrel{R}{\longrightarrow}  \overline{q}(R(\vec{x}),t)\mathcal{D}^{(1/2)}\left(  R \right) = \overline{q}(R(\vec{x}),t) e^{ -\frac{1}{8} \omega_{ij} [\gamma^i,\gamma^j]}, \label{eq:spin1/2}
\end{split}
\end{equation}
where $\omega_{ij} = \theta \epsilon_{ijk}$ for $R = R(\theta \hat{e}_k)$.
The representations of quark fields under cubic group transformations can be derived by combining Eq.~\eqref{eq:cubicdict} with Eq.~\eqref{eq:spin1/2} to obtain
\begin{equation}
\begin{split}
    \mathcal{D}^{(1/2)}(r_k)^\dagger &= \gamma_k \gamma_5, \\
    \mathcal{D}^{(1/2)}(p_{ij})^\dagger &= \frac{1}{\sqrt{2}} \gamma_5 \gamma_j\left(1 + \gamma_4 \sum_k \varepsilon_{ijk}  \gamma_{k} \gamma_5\right),
\end{split}\label{eq:quarkcubic}
\end{equation}
with the  spin-$1/2$ representation of the 24 transformations in $O_h \setminus \{ P \}$ obtained by taking products of representations of reflections $\{1, r_2, r_3, r_2 r_3\}$ and representations of permutations $\{1, p_{12}, p_{23}, p_{31}, p_{123}, p_{132} \}$ where $p_{123} \equiv p_{31} p_{12}$ and $p_{132} \equiv p_{12} p_{31} = p_{123}^{-1}$.
This representation is unitary, $\mathcal{D}^{(1/2)}(R)^{-1} = D^{(1/2)}(R)^\dagger$. The adjoint representation can be obtained using products of $\mathcal{D}^{(1/2)}(r_k)^\dagger = -\mathcal{D}^{(1/2)}(r_k)$ for reflections and $\mathcal{D}^{(1/2)}(p_{ij})^\dagger = -\mathcal{D}^{(1/2)}(p_{ij})$ for permutations.
Comparing the traces of these representations to character tables for the cubic group demonstrates the well-known fact that quark fields transform in the $G_1^+ \oplus~G_1^-$ representation, that is $\mathcal{D}^{(G_1^+\oplus G_1^-)}(R) = \mathcal{D}^{(1/2)}(R)$ where $\mathcal{D}^{(\Gamma_J)}(R)$ denotes the representation $\Gamma_J$.\footnote{The two parts of the direct sum $G_1^+ \oplus G_1^-$ correspond to the upper and lower spinor components of the quark field in the Dirac basis.}
Quark propagators therefore transform under cubic-group elements as shown in Table~\ref{tab:symmetries}.

\begin{table}[!t]
\begin{ruledtabular}
\begin{tabular}{c|ccc} 
            Transformation &  $U_\mu(z)$ & $S(\vec{x},t;\vec{y},0)$      & $C^{(B,I,\Gamma_J)}_{\chi\chi'}(t)$ \\\hline
            $SU(3)_C$ &  $\Omega(z)^\dagger  U_\mu(z) \Omega(z+\hat{e}_\mu)$ & $\Omega(\vec{x},t) S(\vec{x},t;\vec{y},0) \Omega(\vec{y},0)^\dagger$ & $C^{(B,I,\Gamma_J)}_{\chi\chi'}(t)$ \\
            $U(1)_B$ & $\{ U_i(z),e^{i \theta/3} U_4(z) \}$ & $e^{i \theta t / 3} S(\vec{x},t;\vec{y},0)$ & $e^{i B \theta t} C^{(B,I,\Gamma_J)}_{\chi\chi'}(t)$ \\
            $U(1)_{u-d}$ & $\{ U_i(z), e^{i \theta_{u-d}} U_4(z) \}$ & $e^{i \theta_{u-d} t } S(\vec{x},t;\vec{y},0)$ & $ e^{i I_z \theta_{u-d} t} C^{(B,I,\Gamma_J)}_{\chi\chi'}(t)$ \\
             $C$ & $U_\mu(z)^*$ & $\gamma_2\gamma_4 \gamma_5 S(\vec{x},t;\vec{y},0)^* \gamma_5  \gamma_4 \gamma_2 $ & $C^{(B,I,\Gamma_J)}_{\chi\chi'}(t)^*$ \\
           $P$ & $\{ U_{-i}(-\vec{z},z_4), U_4(-\vec{z},z_4)\}$ & $\gamma_4  S(\vec{x},t;\vec{y},0) \gamma_4$ & $C^{(B,I,\Gamma_J)}_{\chi\chi'}(t)$ \\
           $T$ &  $\{ U_{i}(\vec{z},-z_4), U_{-4}(\vec{z},-z_4)\}$ & $-\gamma_4 \gamma_5  S(\vec{x},-t;\vec{y},0) \gamma_5 \gamma_4$ & $C^{(B,I,\Gamma_J)}_{\chi\chi'}(-t)$ \\
            Translation $\hat{e}_k$ & $U_{\mu}(\vec{x}+\hat{e}_k,t)$ & $ S(\vec{x}+\hat{e}_k,t;\vec{y}+\hat{e}_k,0) $  & $C^{(B,I,\Gamma_J)}_{\chi\chi'}(t)$ \\
           $p_{ij}$  & $U_{p_{ij}(\mu)}(p_{ij}(\vec{x}),t)$ & $\mathcal{D}(p_{ij})  S(p_{ij}(\vec{x}),t;p_{ij}(\vec{y}),0) \mathcal{D}(p_{ij})^\dagger$  & $C^{(B,I,\Gamma_J)}_{\chi\chi'}(t)$ \\
           $r_{k}$  & $U_{r_k(\mu)}(r_k (\vec{x}),t)$ & $\mathcal{D}(r_k)  S(r_k(\vec{x}),t;r_k(\vec{y}),0) \mathcal{D}(r_k)^\dagger$  & $C^{(B,I,\Gamma_J)}_{\chi\chi'}(t)$ \\
  \end{tabular}
  \caption{ Transformations associated with symmetries of the (Wilson-like) LQCD action and results of transforming gauge fields, quark propagators, and hadron two-point correlation functions, respectively.
  In order to obtain expressions in terms of the original timeslice-to-all propagator, $\gamma_5$-Hermiticity of the propagator, $S(\vec{y},0;\vec{x},t) = \gamma_5 S(\vec{x},t;\vec{y},0)^\dagger \gamma_5$, has been used in conjunction with charge conjugation.
  Opposite-oriented gauge links are defined by $U_{-\mu}(z) = U_\mu(z - \hat{e}_\mu)^\dagger$.
  A minus sign appropriate for antiperiodic boundary conditions has been included with $T$.
  Translation $\hat{e}_\mu$ represents a shift by one lattice site in the $\hat{e}_\mu$ direction, or for sparsened quark propagators a shift by $\mathcal{S}$ lattice units in the $\hat{e}_\mu$ direction. 
  $p_{ij}$ and $r_k$ denote cubic symmetry transformations $R \in \{p_{ij},\ r_k\} = O_h$ corresponding to permuting the $\hat{e}_i$ and $\hat{e}_j$ axes and reflecting with respect to  the $\hat{e}_k$ axis respectively; the corresponding representations of quark fields are denoted $\mathcal{D}(R) \equiv \mathcal{D}^{(G_1^+ \oplus G_1^-)}(R)$ and explicitly presented in Eq.~\eqref{eq:quarkcubic}. Transformation results are independent of quark-field smearing, and propagator smearing indices are therefore suppressed. For the case of $U(1)_{u-d}$ isospin, $\theta_{u-d}$ denotes that a $U(1)$ transformation of $e^{i\theta_{u-d} t}$ ($e^{-i\theta_{u-d}t}$) should be applied to $u$-quark propagators ($d$-quark propagators) with a corresponding factor of $e^{\pm i \theta_{u-d}}$ applied to gauge fields involved in their calculation.
  \label{tab:symmetries}}
   \end{ruledtabular}
\end{table}

Correlation functions involving interpolating operators $\chi^{(B,I,\Gamma_J,J_z)}$ with definite $J_z$ transform as
\begin{equation}
\begin{split}
    \left< \chi^{(B,I,\Gamma_J,J_z)}(t) \overline{\chi}^{(B,I,\Gamma_J,J_z')}(0) \right> \stackrel{R}{\longrightarrow} & \sum_{J_z'',J_z'''} \mathcal{D}^{(\Gamma_J)}_{J_z'' J_z }(R)^* \left< \chi^{(B,I,\Gamma_J,J_z'')}(t) \overline{\chi}^{(B,I,\Gamma_J,J_z''')}(0) \right> \\
    &\hspace{20pt} \times \mathcal{D}^{(\Gamma_J)}_{ J_z''' J_z'}(R),  
\end{split}\label{eq:wigek}
\end{equation}
where $\overline{\chi} \equiv \chi^\dagger$ for bosonic states with even $B$.
It follows from Eq.~\eqref{eq:wigek} and the orthogonality relation~\cite{Dresselaus:2008}
\begin{equation}
    \sum_{R \in O_h} \mathcal{D}_{J_z J_z'}^{(\Gamma_J)}(R) \mathcal{D}_{J_z'' J_z'''}^{(\Gamma_J')}(R)^* = \frac{48}{d_{\Gamma_J}} \delta_{\Gamma_J \Gamma_J'} \delta_{J_z J_z''} \delta_{J_z' J_z'''},
\label{eq:orthogonality}
\end{equation}
that correlation functions involving different $\Gamma_J$ or $J_z$ at the source and sink vanish after averaging over cubic-symmetry transformations, and further that $J_z$-averaged correlation functions are invariant under cubic-symmetry transformations,
\begin{equation}
    C^{(B,I,\Gamma_J)}_{\chi\chi'}(t) \stackrel{R}{\longrightarrow} C^{(B,I,\Gamma_J)}_{\chi\chi'}(t).
\end{equation}
Analogous arguments demonstrate the invariance of hadron correlation functions under translations, $C$, $P = r_1 r_2 r_3$, $T = r_4$ (a reflection about the Euclidean time axis analogous to $r_k$), and $SU(3)_C$ gauge transformations, as well as the $U(1)_B$ and $U(1)_{u-d}$ transformations shown in Table~\ref{tab:symmetries}.
Applying each of these transformations to sparsened timeslice-to-all propagators with $t = T/2$ leads to exact symmetries of correlation-function matrices containing arbitrary sets of dibaryon and hexaquark operators (up to complex conjugation for $C$ or multiplication by an overall phase for $t$-dependent $U(1)_B$ and $U(1)_{u-d}$ transformations). We have verified that the expected symmetries hold up to machine precision for all of the interpolating operators described in this work.\footnote{Correlation functions involving quasi-local interpolating operators $\kappaOp_{\rho\mathfrak{q}g}$ are computed using factorized operators $F_{\rho\mathfrak{q}g}$ as sources. Although $[\kappaOp, F]$ and $[\kappaOp, \kappaOp]$ correlation functions have identical expectation values, they are not equivalent before ensemble averaging. Since $F$ does not share all of the symmetries of $\kappaOp$ (in particular translation invariance), correlation functions with factorized quasi-local sources are not invariant under all of the quark propagator symmetry transformations described here.}

The $C$ and $T$ symmetries provide constraints on hadron correlation functions: $T$ symmetry gives $C_{\chi\chi'}^{(B,I,\Gamma_J)}(t) = C_{\chi\chi'}^{(B,I,\Gamma_J)}(-t)$, which justifies averaging of correlation functions with sinks at $\pm t$, and $C$ symmetry gives that correlation functions matrix elements are real $C_{\chi\chi'}^{(B,I,\Gamma_J)}(t) = C_{\chi\chi'}^{(B,I,\Gamma_J)}(t)^*$.
Provided a symmetric set of interpolating operators is used at the source and sink, it follows from $C$ symmetry that correlation-function matrices are real and symmetric.
Besides verifying that applying the transformations shown in Table~\ref{tab:symmetries} to a set of quark propagators from one gauge-field configuration leads to the correct transformations of correlation functions, we have verified that the ensemble average imaginary and antisymmetric parts of all hadron correlation-function matrices are consistent with zero.

\section{Correlation function fitting procedure}\label{app:fits}

This work uses the same procedure for sampling over possible fitting analysis choices and estimating the associated systematic uncertainties introduced in Ref.~\cite{Beane:2020ycc} and compared to previous analysis strategies for multi-baryon correlation functions in Ref.~\cite{Illa:2020nsi}.
Analysis results are completely fixed by the specification of several tolerances and other hyper-parameters of the fitting algorithm.
Details of this procedure and relevant features of the fitting algorithm are described for completeness below; for more details see Refs.~\cite{Beane:2020ycc,Illa:2020nsi}.

The GEVP correlation functions $\widehat{C}_{\mathsf{n}}^{(B,I,\Gamma_J)}(t)$ are obtained using the eigenvectors of $C_{\chi\chi'}^{(B,I,\Gamma_J)}(t)$ as described in Sec.~\ref{sec:GEVP}.
In order to provide a uniform normalization for all interpolating operators and avoid spuriously increasing the condition number of $C_{\chi\chi'}^{(B,I,\Gamma_J)}(t)$, a normalization factor of $1/ \sqrt{ C_{\chi\chi}^{(B,I,\Gamma_J)}(0) C_{\chi'\chi'}^{(B,I,\Gamma_J)}(0) }$ is applied to the correlation-function matrix before solving the GEVP, Eq.~\eqref{eq:GEVP}.
Single- and multi-exponential fits to truncations of the spectral representation in Eq.~\eqref{eq:GEVPspectral} are performed for all  $\widehat{C}_{\mathsf{n}}^{(B,I,\Gamma_J)}(t)$ for $t$ in the range $[t_{\rm min},\, t_{\rm max}]$, where $t_{\rm min}$ is varied over all values $t_{\rm min} \geq 2$ for which at least $tol_{\rm plateau} = 5$ source/sink separations are available for fitting. Here, $t_{\rm max}$ is chosen to be the largest source/sink separation $\leq tol_{\rm therm}$ for which  $\sqrt{ \text{Var}\left[ E_{\mathsf{n}}^{(B,I,\Gamma_J,\mathbb{S}_0)}(t) \right] } / E_{\mathsf{n}}^{(B,I,\Gamma_J,\mathbb{S}_0)}(t) < tol_{\rm noise}$, where $tol_{\rm therm} = \frac{3}{8}T$ is used as in Ref.~\cite{Beane:2020ycc} and $tol_{\rm noise} = 0.1$.
For the single- and two-nucleon ground-state GEVP correlation functions, the condition $t_{\rm max} \leq \frac{3}{8}T$ used to avoid contamination from thermal effects is more restrictive than the signal-to-noise cutoff and identical results are obtained for any choice of $tol_{\rm noise} \gtrsim 0.05$.
The $\chi^2 / N_{\rm dof}$ for one-state fits with a variety of $t_{\rm min}$ increase significantly if $t_{\rm max} > \frac{3}{8}T$ is chosen, which suggests that $tol_{\rm therm}$ cannot be increased without introducing non-negligible thermal effects.

For each choice of $t_{\rm min}$, a one-state fit is first performed in which Eq.~\eqref{eq:GEVPspectral} is truncated to include a single exponential.
After performing one-state fits, two-state fits are performed and taken to be preferred if they improve the Akaike information criterion (AIC)~\cite{AkaikeAIC} by at least $tol_{\rm AIC} = -0.5$ times the number of degrees of freedom in the one-state fit. 
If the two-state fit is rejected, the one-state fit is taken to be the optimal truncation of Eq.~\eqref{eq:GEVPspectral} for this choice of $t_{\rm min}$.
This procedure is repeated with three- and more-state fits if necessary in order to determine the optimal truncation of Eq.~\eqref{eq:GEVPspectral} for each $t_{\rm min}$.
All fits are performed using correlated $\chi^2$-minimization with optimal shrinkage~\cite{stein1956,Ledoit:2004} implemented using the diagonal part of the covariance matrix as the shrinkage target as in Refs.~\cite{Rinaldi:2019thf,Beane:2020ycc}.
Fits are then repeated $N_{\rm boot} = 200$ times using bootstrap ensembles sampled from the GEVP correlation functions computed on each gauge-field configuration.
The $68\%$ empirical bootstrap confidence interval is used to determine the statistical uncertainties in the energy levels and overlap factors for this fit range~\cite{davison_hinkley_1997}.
The GEVP eigenvectors used in Eq.~\eqref{eq:GEVPcorrelators} are not varied during bootstrap resampling in order to avoid numerical instabilities from poorly conditioned resampled correlation-function matrices; analogous strategies are used for example in Ref.~\cite{Bulava:2016mks}.
Several checks are then applied to assess the consistency of the multi-exponential fit solution: two nonlinear solvers, implemented in the \verb!Julia!~language~\cite{Julia-2017} and using the \verb!Optim!~optimization package~\cite{mogensen2018optim}, are verified to agree on $E_{\mathsf{n}}^{(B,I,\Gamma_J,\mathbb{S}_0)}$ within a tolerance of $tol_{\rm sol} = 10^{-5}$, correlated and uncorrelated fits are verified to agree within $tol_{\rm corr} = 5\sigma$, and bootstrap median and sample mean fits are verified to agree within $tol_{\rm med} = 2\sigma$.
Fits passing these checks are then averaged using a weighted average with weights proportional to the estimated $p$-value/variance for each energy-level result as used in Refs.~\cite{Rinaldi:2019thf,Beane:2020ycc}.
The lowest energy level obtained by the fit is identified with $E_{\mathsf{n}}^{(B,I,\Gamma_J,\mathbb{S}_0)}$ since the orthogonality of GEVP solutions provides correlation functions orthogonal to states $\mathsf{m}$ with $\mathsf{m} < \mathsf{n}$ assuming that they strongly overlap with an interpolating operator in the set under consideration.\footnote{Relatively high-energy GEVP correlation functions above the single-nucleon first excited state are too noisy to obtain 3 or more acceptable fits with $tol_{\rm noise} = 0.1$. In these cases, the noise tolerance is doubled and the fitting procedure is repeated with successive noise tolerance doublings repeated until there are at least 3 acceptable fits. Systematic fitting uncertainties may be underestimated for the relatively high-energy levels where only $\sim 3$-$5$ acceptable fits are used for weighted averaging. However, these results are expected to be contaminated by small but non-zero mixing with lower-energy states and are only used as inputs for GEVP reconstructions of correlation-function matrix elements  such as Fig.~\ref{fig:B2I1A1_rainbow_reconstruction}. In particular, their values are not interpreted as reliable estimates of energy levels. \vspace{50pt}}
Positive-definiteness of  $E_{\mathsf{n}}^{(B,I,\Gamma_J,\mathbb{S}_0)}$ and $Z_{\mathsf{n}\chi}^{(B,I,\Gamma_J)}$ are enforced by using their logarithms as optimization parameters.

Determinations of the FV energy shifts $\Delta E_{\mathsf{n}}^{(2,I,\Gamma_J,\mathbb{S}_0)}$ apply the same fitting procedure above in order to determine estimates of both $E_{\mathsf{n}}^{(2,I,\Gamma_J,\mathbb{S}_0)}$ and $E_0^{(1,\frac{1}{2},G_1^+)}$ using each choice of $t_{\rm min}$ sampled. Correlated differences of these fit results are used during bootstrap resampling in order to determine statistical uncertainties of $\Delta E_{\mathsf{n}}^{(2,I,\Gamma_J,\mathbb{S}_0)}$. The central value and statistical plus fitting systematic uncertainty for $\Delta E_{\mathsf{n}}^{(2,I,\Gamma_J,\mathbb{S}_0)}$ is then obtained from a weighted average of all acceptable fit results (where the acceptable fit cuts are applied to the two-nucleon and single-nucleon fits independently) with the same weights described for individual energies above. 

After determining $\Delta E_{\mathsf{n}}^{(2,I,\Gamma_J,\mathbb{S}_0)}$ for each GEVP correlation function $\widehat{C}_{\mathsf{n}}^{(B,I,\Gamma_J)}$, the resulting energies and associated GEVP correlation functions are reordered so that $\Delta E_{\mathsf{n}}^{(2,I,\Gamma_J,\mathbb{S}_0)} < \Delta E_{\mathsf{m}}^{(2,I,\Gamma_J)}$ for $\mathsf{n} < \mathsf{m}$. In particular, energy levels that are degenerate within statistical uncertainties are ordered by the central values of $\Delta E_{\mathsf{n}}^{(2,I,\Gamma_J,\mathbb{S}_0)}$.

\section{Tests of variational analysis stability}\label{app:plots}

\begin{figure}[!t]
	\includegraphics[width=0.47\columnwidth]{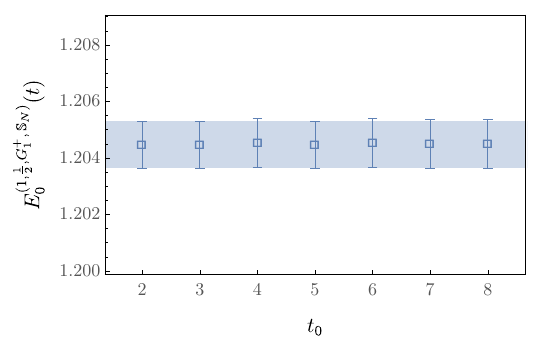}
	\includegraphics[width=0.47\columnwidth]{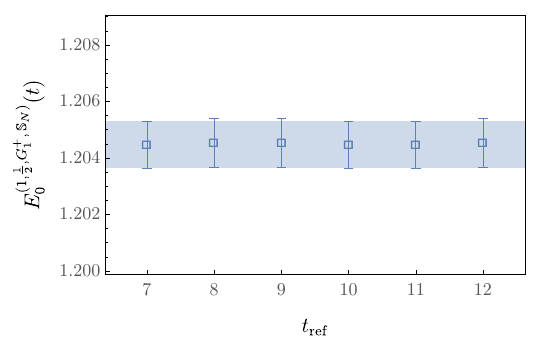}
   \caption{\label{fig:B1G1_stability} Stability of the single-nucleon GEVP ground-state energy with respect to variation of $t_0$ and $t_{\rm ref}$. The blue bands show the ground-state energy determined using $t_0 = 5$ and $t_{\rm ref} = 10$. In the left plot $t_{\rm ref}=10$ is held fixed while $t_0$ is varied, while in the right plot $t_0 = 5$ is held fixed with $t_{\rm ref}$ is varied.}
\end{figure}

\begin{figure}[!t]
	\includegraphics[width=0.47\columnwidth]{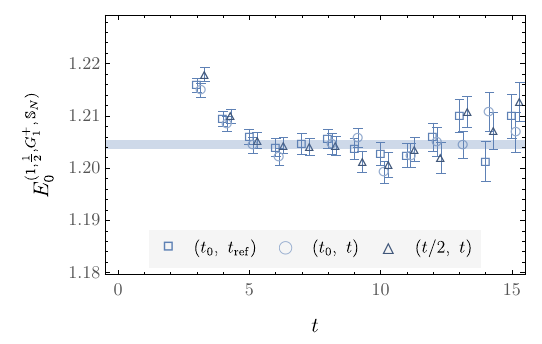}
	\includegraphics[width=0.47\columnwidth]{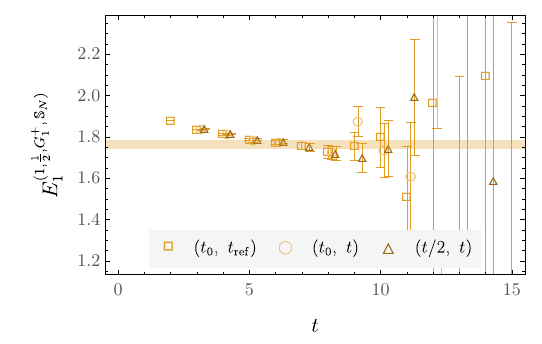}
   \caption{\label{fig:B1G1_GEVP_comparison} GEVP effective energies determined using different schemes for choosing $t_0$ and $t_{\rm ref}$ to obtain GEVP eigenvectors. Results with fixed $t_0 = 5$ and $t_{\rm ref} = 10$ are identical to those in Fig.~\ref{fig:B1G1_fits} and detailed in the main text, while the other results shown use fixed $t_0 = 5$ and variable $t_{\rm ref} = t$ (omitting the singular point $t = t_0 = 5$) and variable $t_0 = t/2$ (rounded down to the nearest integer in lattice units) and $t_{\rm ref} = t$, respectively. In the latter two cases, GEVP eigenvectors are obtained for each bootstrap sample and the corresponding bootstrap eigenvalues are sorted by similarity with the eigenvectors of the sample-mean correlation-function matrix as described in the main text. Colored bands display the fit results for the fixed $t_0$ and $t_{\rm ref}$ GEVP correlation functions (whose effective energies are shown as squares) with the fit range sampling and averaging procedure described in the main text and Appendix~\ref{app:fits} and also  shown in Fig.~\ref{fig:B1G1_fits}. }
\end{figure}

Fit results for the nucleon mass $M_N = E_0^{(1,\frac{1}{2},G_1^+)}$ are stable with respect to variation of $t_0$ and $t_{\rm ref}$ and both the central values and uncertainties are compatible over a 1 fm range of variation as shown in Fig.~\ref{fig:B1G1_stability}.
Results for the first excited-state energy are similarly stable.
Results using the GEVP eigenvalues in order to define effective energies are also consistent within $1\sigma$ uncertainties for both the ground and first excited state as shown in Fig.~\ref{fig:B1G1_GEVP_comparison}.
Effective energies using GEVP eigenvalues with $t_0 = t/2$ have similar uncertainties for small $t$ and larger uncertainties for large $t$ in comparison to those obtained using fixed $t_0 = 5$. The effective energy defined in Eq.~\eqref{eq:GEVPEM} has similar uncertainties for small $t$ and smaller uncertainties for large $t$ than either eigenvalue-based definition.
The GEVP correlation functions defined using eigenvectors with fixed $t_0$ and $t_{\rm ref}$ and associated effective energies in Eq.~\eqref{eq:GEVPEM} are therefore used in the main text.

An issue that arises with bootstrap resampling of eigenvalue-based effective energies is that closely-spaced eigenvalues associated with different eigenvectors may swap ordering between bootstrap samples.
This issue does not arise in practice for the single-nucleon correlation-function matrix studied here; however, it does arise for two-nucleon correlation functions.
To avoid this issue in general, the eigenvalues associated with the sample-mean correlation-function matrix are ordered    for a fixed source/sink separations, chosen here as $t_{\rm ref} = 10$, simply by enforcing $\lambda_{\mathsf{n}}^{(B,I,\Gamma_J)}(t_{\rm ref}, t_0) > \lambda_{\mathsf{m}}^{(B,I,\Gamma_J)}(t_{\rm ref}, t_0)$ for $\mathsf{n} < \mathsf{m}$, and the corresponding eigenvectors $v_{\mathsf{n} \chi}^{(B,I,\Gamma_J)}(t_{\rm ref}, t_0)$ are computed.
For the sample-mean eigenvalues with $t \neq t_{\rm ref}$ and the bootstrap resampled eigenvalues for all $t$, eigenvalues are then ordered by enforcing that $  \left| \sum_{\chi}  v_{\mathsf{m} \chi}^{(B,I,\Gamma_J)}(t, t_0)^* v_{\mathsf{n} \chi}^{(B,I,\Gamma_J)}(t_{\rm ref}, t_0) \right|$ is maximized for $\mathsf{m} = \mathsf{n}$.
Analogous issues are discussed for example in Ref.~\cite{Dudek:2007wv}.

\begin{figure}[!t]
	\includegraphics[width=0.47\columnwidth]{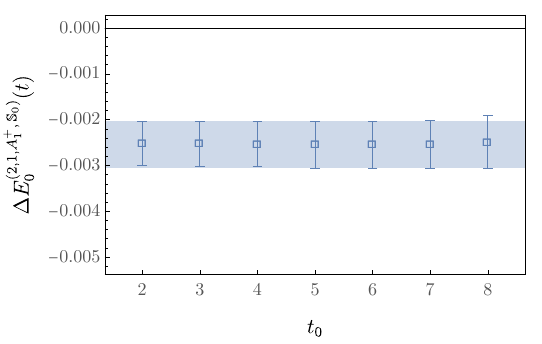}
	\includegraphics[width=0.47\columnwidth]{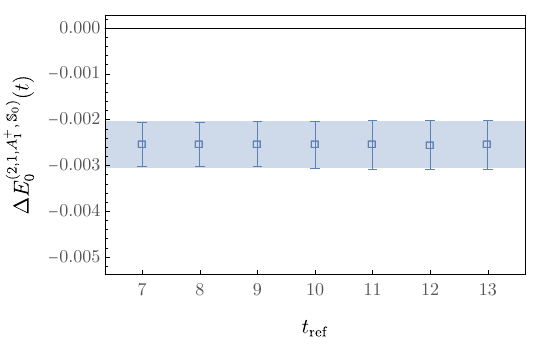}
   \caption{\label{fig:B2I1A1_stability} Stability of $I=1, \Gamma_J = A_1^+$ GEVP two-nucleon ground-state FV energy shifts using interpolating-operator set $\mathbb{S}_{\protect \circled{0}}^{(2,1,A_1^+)}$ with respect to variation of $t_0$ and $t_{\rm ref}$. Details are as in Fig.~\ref{fig:B1G1_stability}. }
\end{figure}

\begin{figure}[!t]
	\includegraphics[width=0.47\columnwidth]{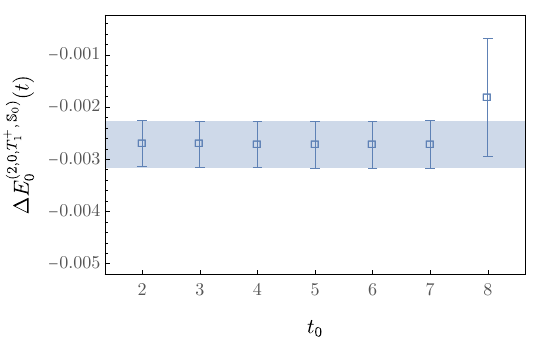}
	\includegraphics[width=0.47\columnwidth]{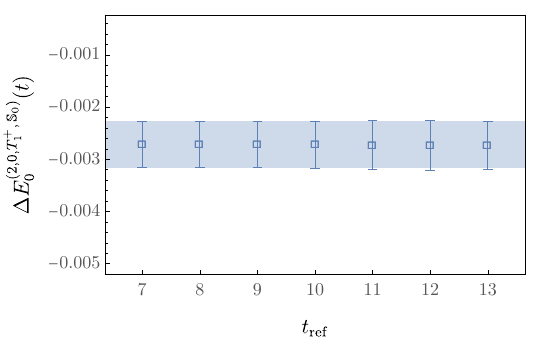}
   \caption{\label{fig:B2I0T1_stability} Stability of $I=0, \Gamma_J = T_1^+$ GEVP two-nucleon ground-state FV energy shifts using interpolating-operator set $\mathbb{S}_{\protect \circled{0}}^{(2,0,T_1^+)}$ with respect to variation of $t_0$ and $t_{\rm ref}$. Details are as in Fig.~\ref{fig:B1G1_stability}. } 
\end{figure}

A similar analysis can be performed for two-nucleon systems.
In the dineutron channel, results using $\mathbb{S}_{\circled{0}}^{(2,1,A_1^+)}$ are insensitive to the choice of $t_0$ and $t_{\rm ref}$ used to solve the GEVP, as shown for the ground state in Fig.~\ref{fig:B2I1A1_stability}.
Higher energy states show similar levels of insensitivity to $t_0$ and $t_{\rm ref}$ as long as the size of the fit results for $E_\mathsf{n}^{(2,1,A_1^+,\mathbb{S}_0)}$ are used to order the energy levels.
Analogous results for the deuteron channel are shown in Fig.~\ref{fig:B2I0T1_stability}.

\begin{figure}[!t]
	\includegraphics[width=0.46\columnwidth]{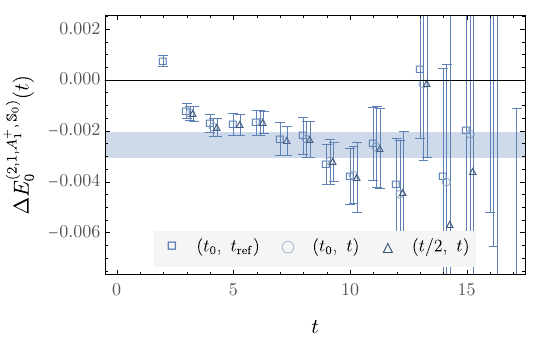}
	\includegraphics[width=0.46\columnwidth]{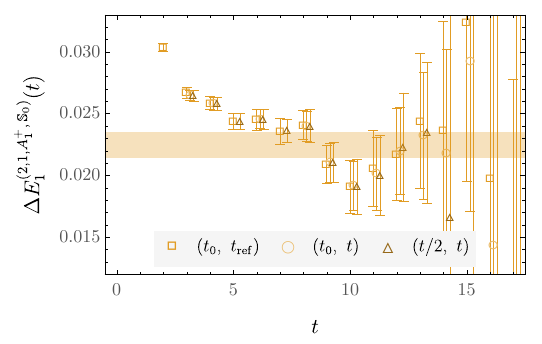}
	\includegraphics[width=0.46\columnwidth]{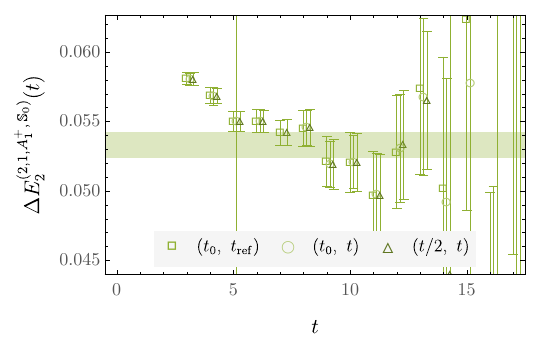}
	\includegraphics[width=0.46\columnwidth]{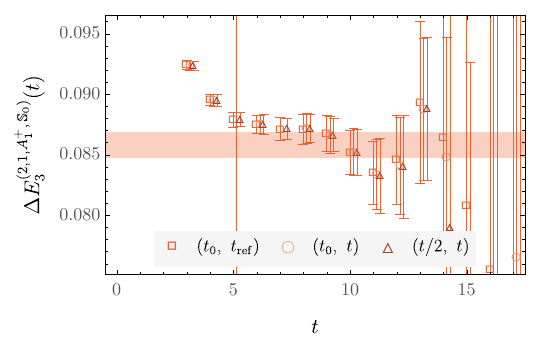}
	\includegraphics[width=0.46\columnwidth]{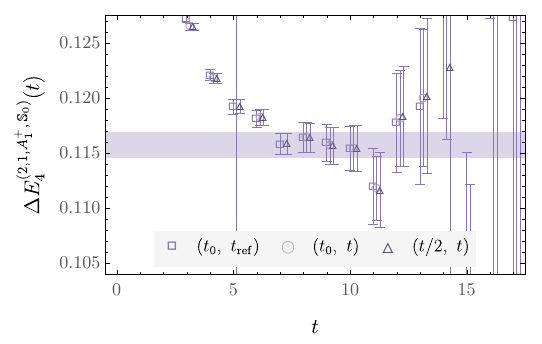}
	\includegraphics[width=0.46\columnwidth]{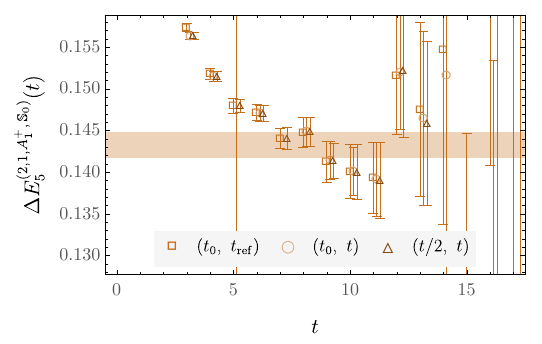}
	\includegraphics[width=0.46\columnwidth]{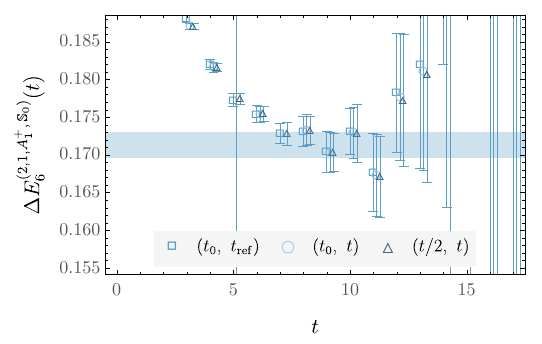}
	\includegraphics[width=0.46\columnwidth]{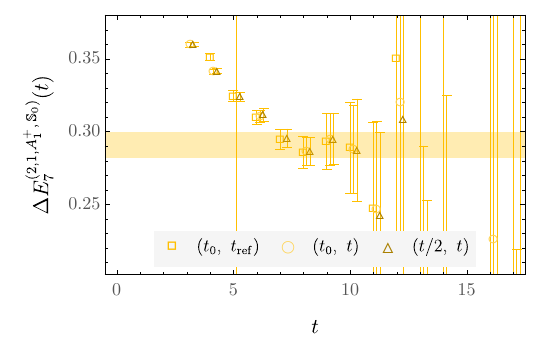}
   \caption{\label{fig:B2I1A1_GEVP_comparison} The GEVP effective FV energy shifts determined using different schemes for choosing $t_0$ and $t_{\rm ref}$ to obtain GEVP eigenvectors using the interpolating-operator set $\mathbb{S}_{\protect \circled{0}}^{(2,1,A_1^+)}$. Results with fixed $t_0 = 5$ and $t_{\rm ref} = 10$ are identical to those in Fig.~\ref{fig:B2I1A1_FV_fits} and detailed in the main text, while the other results shown use fixed $t_0 = 5$ and variable $t_{\rm ref} = t$ and variable $t_0 = t/2$ and $t_{\rm ref} = t$ in analogy to Fig.~\ref{fig:B1G1_GEVP_comparison}. }
\end{figure}

\begin{figure}[!t]
	\includegraphics[width=0.47\columnwidth]{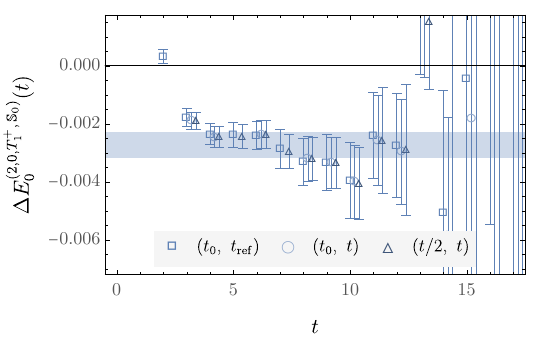}
	\includegraphics[width=0.47\columnwidth]{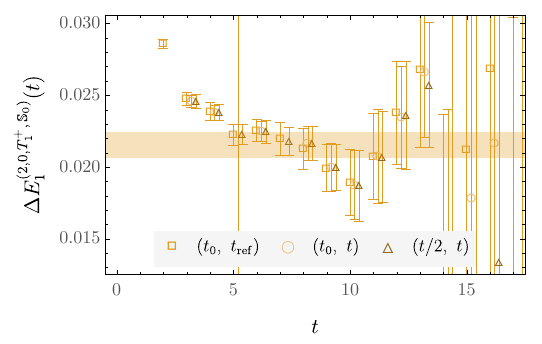}
	\includegraphics[width=0.47\columnwidth]{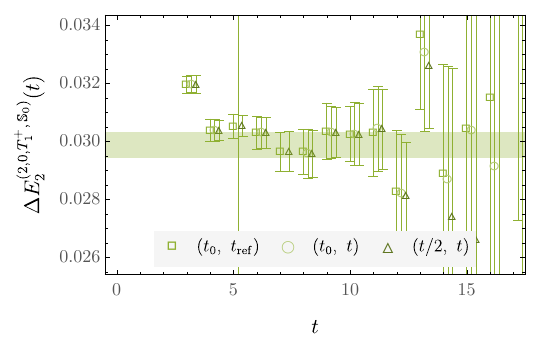}
	\includegraphics[width=0.47\columnwidth]{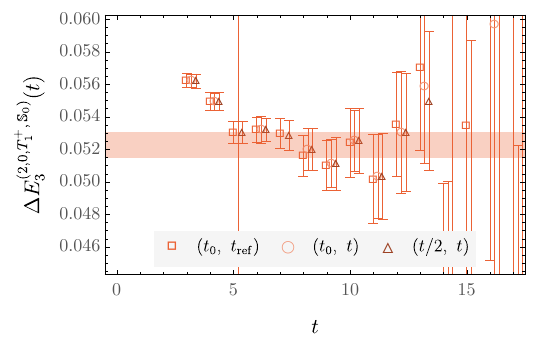}
	\includegraphics[width=0.47\columnwidth]{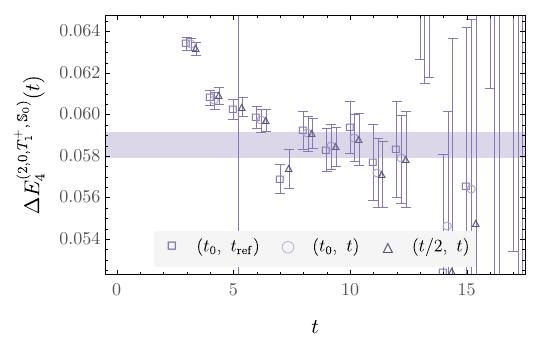}
	\includegraphics[width=0.47\columnwidth]{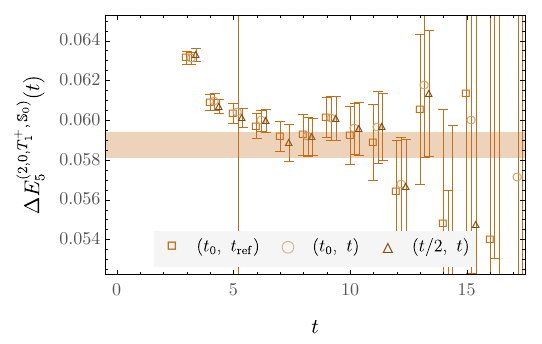}
	\includegraphics[width=0.47\columnwidth]{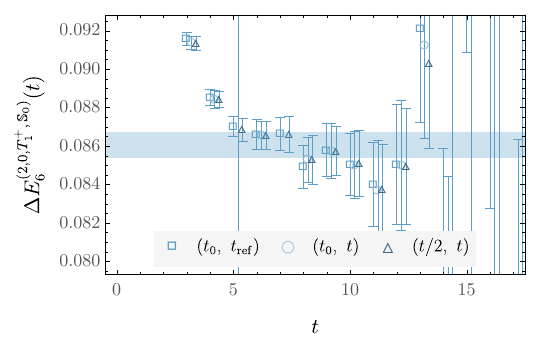}
	\includegraphics[width=0.47\columnwidth]{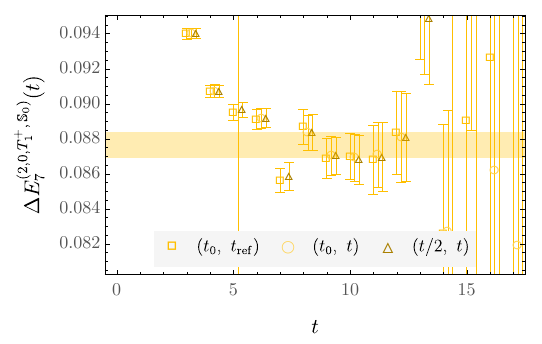}
   \caption{\label{fig:B2I0T1_GEVP_comparison} Analogous results to Fig.~\ref{fig:B2I1A1_GEVP_comparison} for the deuteron channel using interpolating-operator set $\mathbb{S}_{\protect \circled{0}}^{(2,0,T_1^+)}$. }
\end{figure}

Results for GEVP effective FV energy shifts using the eigenvalue-based definition, $ \ln \left( \frac{ \lambda_{\mathsf{n}}^{(B,I,\Gamma_J)}(t,t_0) }{ \lambda_{\mathsf{n}}^{(B,I,\Gamma_J)}(t+1, t_0) } \right)$,
are compared to results using the eigenvector-based definition in Eq.~\eqref{eq:GEVPcorrelators} for the dineutron channel in Fig.~\ref{fig:B2I1A1_GEVP_comparison}.
The effective FV energy shifts are seen to be insensitive to which definition is used.
Analogous results for the deuteron channel are shown in Fig.~\ref{fig:B2I0T1_GEVP_comparison}.
In this case most but not all of the effective FV energy shifts are seen to be insensitive to which definition is used.
The GEVP correlation functions computed using the eigenvalue-based definition are significantly noisier than those computed using Eq.~\eqref{eq:GEVPcorrelators} for the $\mathsf{n}\in\{4,5\}$ states, which as seen in Fig.~\ref{fig:B2I0T1_FV_fits} involve mixtures of interpolating operators and have relatively noisy overlap factor determinations.
The eigenvector-based definition with fixed $t_0$ and $t_{\rm ref}$ in Eq.~\eqref{eq:GEVPcorrelators} therefore appears useful for resolving states involving mixtures of interpolating operators that are not well determined at small $t$.

\begin{figure}[!t]
	\includegraphics[width=0.47\columnwidth]{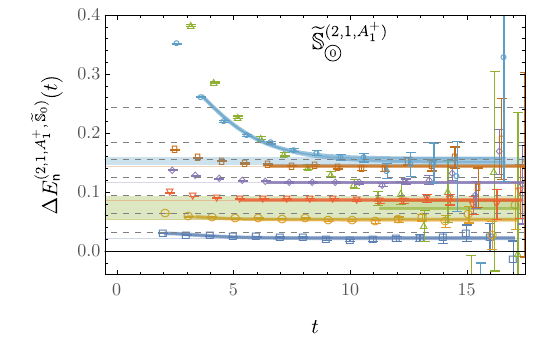}
	\includegraphics[width=0.47\columnwidth]{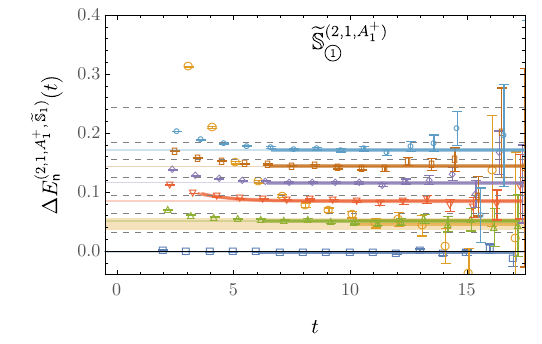}
	\includegraphics[width=0.47\columnwidth]{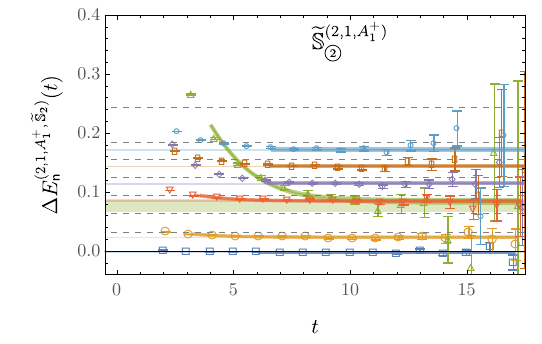}
	\includegraphics[width=0.47\columnwidth]{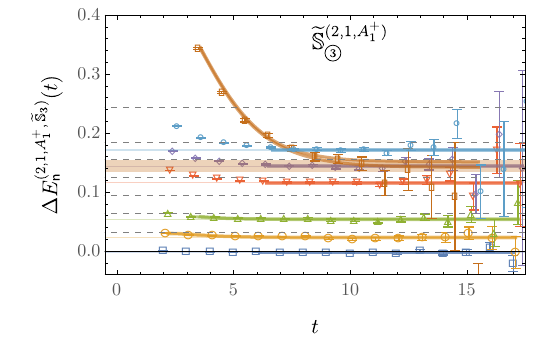}
	\includegraphics[width=0.47\columnwidth]{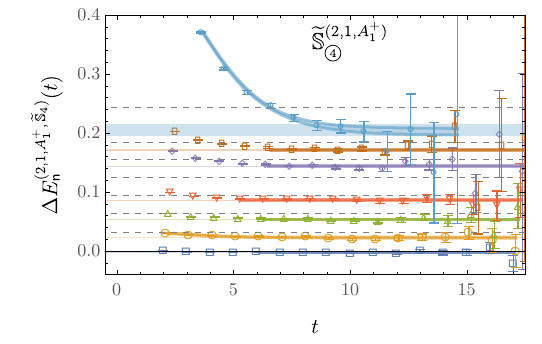}
	\includegraphics[width=0.47\columnwidth]{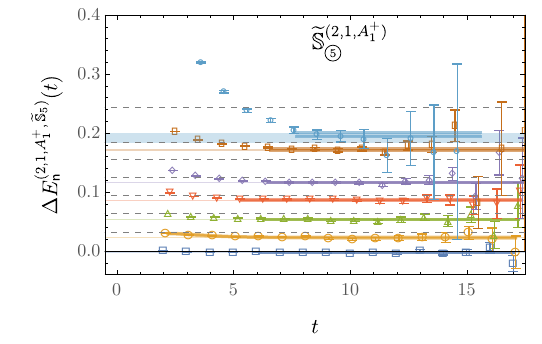}
	\includegraphics[width=0.47\columnwidth]{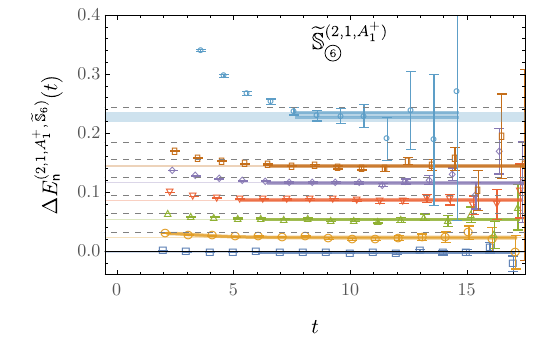}
	\includegraphics[width=0.47\columnwidth]{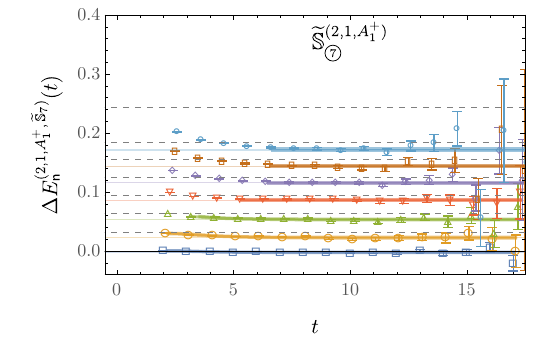}
   \caption{\label{fig:B2I1A1_badbasis_fits}  The GEVP effective FV energy shifts analogous to Fig.~\ref{fig:B2I1A1_rainbow} for interpolating-operator sets $\widetilde{\mathbb{S}}_{\protect\circled{0}}^{(2,1,A_1^+)},\ldots,  \widetilde{\mathbb{S}}_{\protect\circled{7}}^{(2,1,A_1^+)}$ defined in Eq.~\eqref{eq:BA1def3}.}
\end{figure}

\begin{figure}[!t]
	\includegraphics[width=0.47\columnwidth]{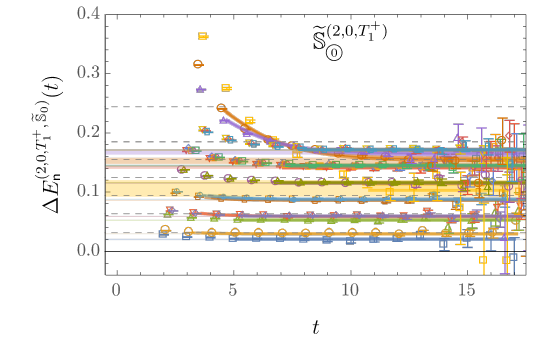}
	\includegraphics[width=0.47\columnwidth]{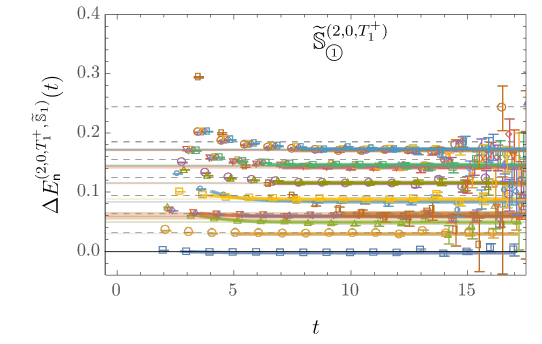}
	\includegraphics[width=0.47\columnwidth]{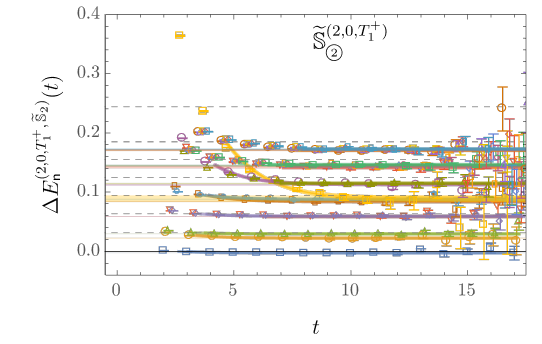}
	\includegraphics[width=0.47\columnwidth]{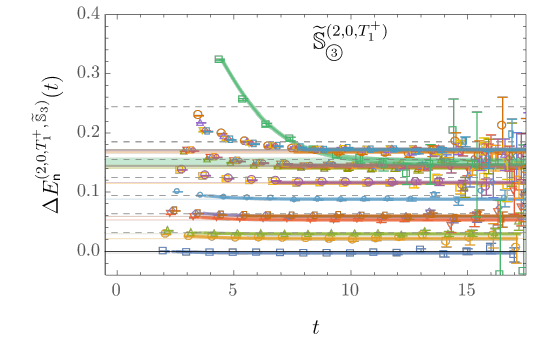}
	\includegraphics[width=0.47\columnwidth]{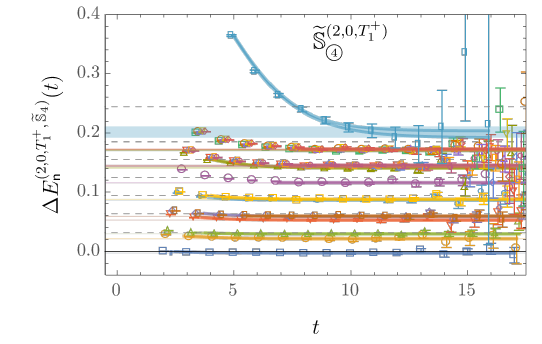}
	\includegraphics[width=0.47\columnwidth]{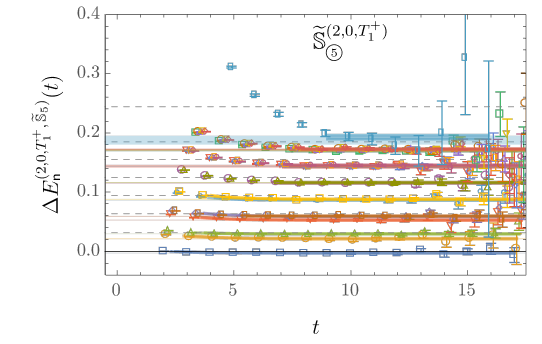}
	\includegraphics[width=0.47\columnwidth]{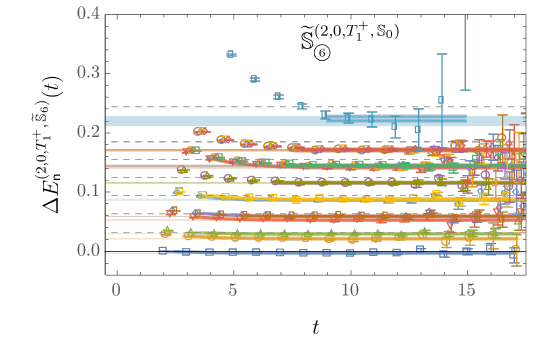}
	\includegraphics[width=0.47\columnwidth]{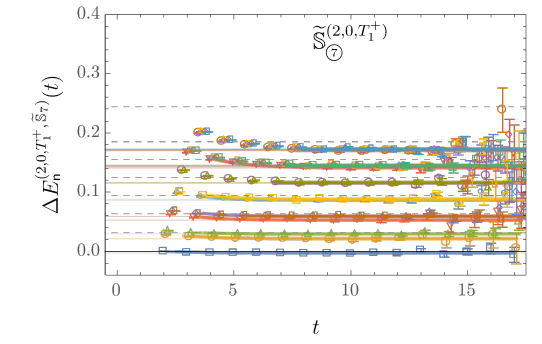}
   \caption{\label{fig:B2I0T1_badbasis_fits} Analogous results to Fig.~\ref{fig:B2I1A1_badbasis_fits} for the deuteron channel.}
\end{figure}

The effective FV energy shifts for low-energy states obtained using the interpolating-operator sets $\widetilde{\mathbb{S}}_{\protect\circled{0}}^{(2,1,A_1^+)},\ldots,  \widetilde{\mathbb{S}}_{\protect\circled{7}}^{(2,1,A_1^+)}$ and $\widetilde{\mathbb{S}}_{\protect\circled{0}}^{(2,0,T_1^+)},\ldots,  \widetilde{\mathbb{S}}_{\protect\circled{7}}^{(2,0,T_1^+)}$ are shown in Figs.~\ref{fig:B2I1A1_badbasis_fits}-\ref{fig:B2I0T1_badbasis_fits}.
The missing energy levels seen in Fig.~\ref{fig:B2I1A1_badbasis_summary} for the dineutron channel and Fig.~\ref{fig:B2I0T1_badbasis_summary} are the deuteron channel are clearly visible in the effective FV energy shifts.
Besides the absence of particular energy levels, the effective FV energy shifts present in  Figs.~\ref{fig:B2I1A1_badbasis_fits}-\ref{fig:B2I0T1_badbasis_fits} do not show any unusual behavior that could be interpreted as a signature of excited-state contamination from missing energy levels comparable to or below the energy levels present.

\clearpage

\section{Variational energy-level results}\label{app:tabs}

Results for the single-nucleon ground- and first-excited-state energies obtained using the interpolating-operator set $\{N_W,N_T\}$ and the fitting methods adapted from Ref.~\cite{Beane:2020ycc} and discussed in Sec.~\ref{sec:nucresults} are shown in Table~\ref{tab:B1}.
The uncertainties shown include systematic uncertainties associated with the variation in fit results obtained with different choices of $t_{\rm min}$ added in quadrature to statistical uncertainties calculated using bootstrap methods.
Results for two-nucleon energy levels $E_{\mathsf{n}}^{(2,I,\Gamma_J,\mathbb{S}_0)}$ and FV energy shifts $\Delta E_{\mathsf{n}}^{(2,I,\Gamma_J,\mathbb{S}_0)}$ are shown in Tables~\ref{tab:B2I1}-\ref{tab:B2I0}.
The interpolating-operator sets $\mathbb{S}_{\circled{0}}^{(2,1,A_1^+)}$ and $\mathbb{S}_{\circled{0}}^{(2,0,T_1^+)}$ used to obtain results in the $I=1$, $\Gamma_J=A_1^+$ and $I=0$, $\Gamma_J=T_1^+$ irreps include all $D_{\mathfrak{s}\mathfrak{k}g}$ operators with $\mathfrak{s} \leq 6$ and $H_g$ operators (but not $\kappaOp_{\mathfrak{q}g}$ operators).
The interpolating-operator sets used for the other irreps only include $D_{\mathfrak{s}\mathfrak{k}g}$ operators with $\mathfrak{s} > 0$ as described in Sec.~\ref{sec:FV}.
Results are presented in both lattice and physical units, and in addition the values of $\Delta E_{\mathsf{n}}^{(2,I,\Gamma_J,\mathbb{S}_0)}$ are used to obtain results for the corresponding center-of-mass momenta, $k_{\mathsf{n}}^{(2,I,\Gamma_J)}$ , for scattering states and the dimensionless ratios $k_{\mathsf{n}}^{(2,I,\Gamma_J)^2} / m_\pi^2$ are presented in Tables~\ref{tab:B2I1}-\ref{tab:B2I0}.

\begin{table}[h!]
\begin{tabular}{|c|c|c|} \hline
$n$ & $a E_{\mathsf{n}}^{(2,\frac{1}{2},G_1^+,\mathbb{S}_N)}$ & $E_{\mathsf{n}}^{(2,\frac{1}{2},G_1^+,\mathbb{S}_N)}\text{ [GeV]}$ \\\hline
0 & $1.2045(8)$ & $1.636(18)$ \\\hline
1 & $1.7704(143)$ & $2.404(33)$ \\\hline
\end{tabular}
\caption{Single-nucleon energy levels in lattice and physical units obtained from a weighted average of single- and multi-exponential fits to GEVP correlation functions as described in the main text. \label{tab:B1}}
\end{table}

\begin{table}[h!]
\begin{tabular}{|c|c|c|c|c|} \hline
$n$ & $a E_{\mathsf{n}}^{(2,1,A_1^+,\mathbb{S}_0)}$ & $a \Delta E_{\mathsf{n}}^{(2,1,A_1^+,\mathbb{S}_0)}$ & $\Delta E_{\mathsf{n}}^{(2,1,A_1^+,\mathbb{S}_0)}\text{ [MeV]}$ & $k_{\mathsf{n}}^{(2,1,A_1^+,\mathbb{S}_0)^2} / m_\pi^2$ \\\hline
0 & $2.4068(19)$ & $-0.0026(5)$ & $-3.5(0.7)$ & $-0.0087(17)$ \\\hline
1 & $2.4315(23)$ & $0.0225(11)$ & $30.5(1.5)$ & $0.0770(36)$ \\\hline
2 & $2.4624(21)$ & $0.0533(9)$ & $72.4(1.5)$ & $0.1839(32)$ \\\hline
3 & $2.4948(24)$ & $0.0858(11)$ & $116.6(1.9)$ & $0.2980(38)$ \\\hline
4 & $2.5251(28)$ & $0.1158(12)$ & $157.2(2.4)$ & $0.4044(42)$ \\\hline
5 & $2.5523(30)$ & $0.1433(16)$ & $194.6(3.0)$ & $0.5034(57)$ \\\hline
6 & $2.5803(29)$ & $0.1713(17)$ & $232.7(3.5)$ & $0.6053(62)$ \\\hline
7 & $2.6989(97)$ & $0.2909(89)$ & $395.1(12.9)$ & $1.0523(342)$ \\\hline
\end{tabular} \\\vspace{20pt}
\begin{tabular}{|c|c|c|c|c|} \hline
$n$ & $a E_{\mathsf{n}}^{(2,1,E^+,\mathbb{S}_0)}$ & $a \Delta E_{\mathsf{n}}^{(2,1,E^+,\mathbb{S}_0)}$ & $\Delta E_{\mathsf{n}}^{(2,1,E^+,\mathbb{S}_0)}\text{ [MeV]}$ & $k_{\mathsf{n}}^{(2,1,E^+,\mathbb{S}_0)^2} / m_\pi^2$ \\\hline
0 & $2.4383(18)$ & $0.0291(5)$ & $39.6(0.9)$ & $0.1000(19)$ \\\hline
1 & $2.4675(20)$ & $0.0584(7)$ & $79.4(1.3)$ & $0.2018(25)$ \\\hline
2 & $2.5244(23)$ & $0.1152(11)$ & $156.5(2.3)$ & $0.4025(39)$ \\\hline
3 & $2.5514(27)$ & $0.1423(14)$ & $193.2(2.9)$ & $0.4997(52)$ \\\hline
4 & $2.5540(26)$ & $0.1448(13)$ & $196.7(2.8)$ & $0.5090(46)$ \\\hline
5 & $2.5801(28)$ & $0.1712(16)$ & $232.5(3.4)$ & $0.6048(60)$ \\\hline
\end{tabular} \\\vspace{20pt}
\begin{tabular}{|c|c|c|c|c|} \hline
$n$ & $a E_{\mathsf{n}}^{(2,1,T_2^+,\mathbb{S}_0)}$ & $a \Delta E_{\mathsf{n}}^{(2,1,T_2^+,\mathbb{S}_0)}$ & $\Delta E_{\mathsf{n}}^{(2,1,T_2^+,\mathbb{S}_0)}\text{ [MeV]}$ & $k_{\mathsf{n}}^{(2,1,T_2^+,\mathbb{S}_0)^2} / m_\pi^2$ \\\hline
0 & $2.4676(19)$ & $0.0583(7)$ & $79.2(1.2)$ & $0.2014(23)$ \\\hline
1 & $2.4960(19)$ & $0.0869(7)$ & $118.0(1.6)$ & $0.3017(26)$ \\\hline
2 & $2.5509(22)$ & $0.1421(12)$ & $193.0(2.7)$ & $0.4990(45)$ \\\hline
3 & $2.5777(25)$ & $0.1690(16)$ & $229.5(3.3)$ & $0.5967(58)$ \\\hline
4 & $2.5809(25)$ & $0.1718(13)$ & $233.3(3.2)$ & $0.6070(50)$ \\\hline
\end{tabular} \\\vspace{20pt}
\begin{tabular}{|c|c|c|c|c|} \hline
$n$ & $a E_{\mathsf{n}}^{(2,1,T_1^+,\mathbb{S}_0)}$ & $a \Delta E_{\mathsf{n}}^{(2,1,T_1^+,\mathbb{S}_0)}$ & $\Delta E_{\mathsf{n}}^{(2,1,T_1^+,\mathbb{S}_0)}\text{ [MeV]}$ & $k_{\mathsf{n}}^{(2,1,T_1^+,\mathbb{S}_0)^2} / m_\pi^2$ \\\hline
0 & $2.5532(23)$ & $0.1441(11)$ & $195.7(2.6)$ & $0.5062(40)$ \\\hline
1 & $2.5811(25)$ & $0.1719(13)$ & $233.4(3.1)$ & $0.6073(48)$ \\\hline
\end{tabular} \\\vspace{20pt}
\begin{tabular}{|c|c|c|c|c|} \hline
$n$ & $a E_{\mathsf{n}}^{(2,1,A_2^+,\mathbb{S}_0)}$ & $a \Delta E_{\mathsf{n}}^{(2,1,A_2^+,\mathbb{S}_0)}$ & $\Delta E_{\mathsf{n}}^{(2,1,A_2^+,\mathbb{S}_0)}\text{ [MeV]}$ & $k_{\mathsf{n}}^{(2,1,A_2^+,\mathbb{S}_0)^2} / m_\pi^2$ \\\hline
0 & $2.5538(26)$ & $0.1447(12)$ & $196.5(2.7)$ & $0.5084(45)$ \\\hline
\end{tabular}
\caption{Two-nucleon $I=1$ energy levels in all positive-parity total-angular-momentum cubic irreps $\Gamma_J \in \{A_1^+,E^+,T_2^+,T_1^+,A_2^+\}$ in lattice and physical units obtained from a weighted average of single- and multi-exponential fits to GEVP correlation functions as described in the main text. Finite-volume energy differences obtained using correlated differences of GEVP fit results and the associated center-of-mass momenta are also provided. \label{tab:B2I1}}
\end{table}

\begin{table}[h!]
\begin{tabular}{|c|c|c|c|c|} \hline
$n$ & $a E_{\mathsf{n}}^{(2,0,T_1^+,\mathbb{S}_0)}$ & $a \Delta E_{\mathsf{n}}^{(2,0,T_1^+,\mathbb{S}_0)}$ & $\Delta E_{\mathsf{n}}^{(2,0,T_1^+,\mathbb{S}_0)}\text{ [MeV]}$ & $k_{\mathsf{n}}^{(2,0,T_1^+,\mathbb{S}_0)^2} / m_\pi^2$ \\\hline
0 & $2.4067(19)$ & $-0.0027(4)$ & $-3.7(0.6)$ & $-0.0093(15)$ \\\hline
1 & $2.4308(22)$ & $0.0216(9)$ & $29.3(1.3)$ & $0.0739(31)$ \\\hline
2 & $2.4392(18)$ & $0.0299(4)$ & $40.6(0.8)$ & $0.1026(15)$ \\\hline
3 & $2.4614(20)$ & $0.0523(8)$ & $71.0(1.3)$ & $0.1803(28)$ \\\hline
4 & $2.4678(18)$ & $0.0585(6)$ & $79.5(1.2)$ & $0.2021(22)$ \\\hline
5 & $2.4678(19)$ & $0.0588(7)$ & $79.8(1.2)$ & $0.2029(23)$ \\\hline
6 & $2.4952(20)$ & $0.0861(7)$ & $116.9(1.6)$ & $0.2988(23)$ \\\hline
7 & $2.4969(20)$ & $0.0877(7)$ & $119.0(1.7)$ & $0.3044(26)$ \\\hline
8 & $2.5240(22)$ & $0.1149(10)$ & $156.1(2.2)$ & $0.4015(36)$ \\\hline
9 & $2.5249(23)$ & $0.1158(10)$ & $157.3(2.2)$ & $0.4046(36)$ \\\hline
10 & $2.5499(25)$ & $0.1410(14)$ & $191.5(2.8)$ & $0.4952(49)$ \\\hline
11 & $2.5525(22)$ & $0.1435(11)$ & $194.8(2.6)$ & $0.5040(39)$ \\\hline
12 & $2.5526(26)$ & $0.1438(15)$ & $195.3(2.9)$ & $0.5053(52)$ \\\hline
13 & $2.5533(22)$ & $0.1443(11)$ & $196.0(2.6)$ & $0.5070(39)$ \\\hline
14 & $2.5542(25)$ & $0.1451(12)$ & $197.0(2.7)$ & $0.5098(44)$ \\\hline
15 & $2.5796(24)$ & $0.1706(15)$ & $231.7(3.2)$ & $0.6026(53)$ \\\hline
16 & $2.5803(26)$ & $0.1713(13)$ & $232.7(3.1)$ & $0.6052(46)$ \\\hline
17 & $2.5803(25)$ & $0.1713(13)$ & $232.7(3.1)$ & $0.6053(47)$ \\\hline
18 & $2.5803(25)$ & $0.1715(13)$ & $232.8(3.1)$ & $0.6057(46)$ \\\hline
19 & $2.5810(28)$ & $0.1723(17)$ & $234.0(3.4)$ & $0.6088(61)$ \\\hline
20 & $2.6896(100)$ & $0.2816(92)$ & $382.4(13.2)$ & $1.0167(350)$ \\\hline
\end{tabular}
\caption{Two-nucleon $I=0$ energy levels in the deuteron total-angular-momentum cubic irrep $\Gamma_J = T_1^+$ in lattice and physical units obtained from a weighted average of single- and multi-exponential fits to GEVP correlation functions as described in the main text. Finite-volume energy differences obtained using correlated differences of GEVP fit results and the associated center-of-mass momenta are also provided. \label{tab:B2I0T1}}
\end{table}

\begin{table}[h!]
\begin{tabular}{|c|c|c|c|c|} \hline
$n$ & $a E_{\mathsf{n}}^{(2,0,T_2^+,\mathbb{S}_0)}$ & $a \Delta E_{\mathsf{n}}^{(2,0,T_2^+,\mathbb{S}_0)}$ & $\Delta E_{\mathsf{n}}^{(2,0,T_2^+,\mathbb{S}_0)}\text{ [MeV]}$ & $k_{\mathsf{n}}^{(2,0,T_2^+,\mathbb{S}_0)^2} / m_\pi^2$ \\\hline
0 & $2.4391(18)$ & $0.0298(5)$ & $40.4(0.8)$ & $0.1022(17)$ \\\hline
1 & $2.4674(20)$ & $0.0581(7)$ & $78.9(1.2)$ & $0.2006(23)$ \\\hline
2 & $2.4687(21)$ & $0.0593(6)$ & $80.5(1.2)$ & $0.2048(21)$ \\\hline
3 & $2.4961(19)$ & $0.0870(7)$ & $118.1(1.6)$ & $0.3021(26)$ \\\hline
4 & $2.5254(25)$ & $0.1161(11)$ & $157.7(2.3)$ & $0.4056(39)$ \\\hline
5 & $2.5510(25)$ & $0.1419(13)$ & $192.8(2.7)$ & $0.4985(46)$ \\\hline
6 & $2.5515(27)$ & $0.1423(13)$ & $193.3(2.8)$ & $0.4998(47)$ \\\hline
7 & $2.5540(26)$ & $0.1448(12)$ & $196.6(2.7)$ & $0.5087(44)$ \\\hline
8 & $2.5547(27)$ & $0.1453(11)$ & $197.3(2.6)$ & $0.5104(39)$ \\\hline
9 & $2.5552(26)$ & $0.1459(12)$ & $198.2(2.7)$ & $0.5128(43)$ \\\hline
10 & $2.5789(25)$ & $0.1699(14)$ & $230.7(3.2)$ & $0.5999(52)$ \\\hline
11 & $2.5806(26)$ & $0.1716(13)$ & $233.1(3.1)$ & $0.6063(47)$ \\\hline
12 & $2.5809(28)$ & $0.1718(12)$ & $233.3(3.1)$ & $0.6069(45)$ \\\hline
13 & $2.5812(24)$ & $0.1721(12)$ & $233.7(3.0)$ & $0.6080(43)$ \\\hline
\end{tabular} \\\vspace{20pt}
\begin{tabular}{|c|c|c|c|c|} \hline
$n$ & $a E_{\mathsf{n}}^{(2,0,E^+,\mathbb{S}_0)}$ & $a \Delta E_{\mathsf{n}}^{(2,0,E^+,\mathbb{S}_0)}$ & $\Delta E_{\mathsf{n}}^{(2,0,E^+,\mathbb{S}_0)}\text{ [MeV]}$ & $k_{\mathsf{n}}^{(2,0,E^+,\mathbb{S}_0)^2} / m_\pi^2$ \\\hline
0 & $2.4674(19)$ & $0.0581(7)$ & $79.0(1.3)$ & $0.2007(24)$ \\\hline
1 & $2.4967(23)$ & $0.0873(9)$ & $118.6(1.8)$ & $0.3032(31)$ \\\hline
2 & $2.5527(28)$ & $0.1433(14)$ & $194.7(2.9)$ & $0.5035(51)$ \\\hline
3 & $2.5543(24)$ & $0.1451(10)$ & $197.0(2.6)$ & $0.5097(38)$ \\\hline
4 & $2.5794(32)$ & $0.1700(16)$ & $230.9(3.3)$ & $0.6004(58)$ \\\hline
5 & $2.5815(27)$ & $0.1722(12)$ & $233.8(3.1)$ & $0.6084(45)$ \\\hline
6 & $2.5823(39)$ & $0.1728(17)$ & $234.6(3.5)$ & $0.6105(61)$ \\\hline
\end{tabular} \\\vspace{20pt}
\begin{tabular}{|c|c|c|c|c|} \hline
$n$ & $a E_{\mathsf{n}}^{(2,0,A_2^+,\mathbb{S}_0)}$ & $a \Delta E_{\mathsf{n}}^{(2,0,A_2^+,\mathbb{S}_0)}$ & $\Delta E_{\mathsf{n}}^{(2,0,A_2^+,\mathbb{S}_0)}\text{ [MeV]}$ & $k_{\mathsf{n}}^{(2,0,A_2^+,\mathbb{S}_0)^2} / m_\pi^2$ \\\hline
0 & $2.4678(22)$ & $0.0584(9)$ & $79.3(1.5)$ & $0.2017(31)$ \\\hline
1 & $2.4968(23)$ & $0.0875(9)$ & $118.8(1.8)$ & $0.3037(31)$ \\\hline
2 & $2.5528(31)$ & $0.1436(14)$ & $195.0(2.9)$ & $0.5044(53)$ \\\hline
3 & $2.5801(74)$ & $0.1702(27)$ & $231.2(4.5)$ & $0.6013(101)$ \\\hline
4 & $2.5817(27)$ & $0.1725(15)$ & $234.3(3.3)$ & $0.6095(54)$ \\\hline
\end{tabular} \\\vspace{20pt}
\begin{tabular}{|c|c|c|c|c|} \hline
$n$ & $a E_{\mathsf{n}}^{(2,0,A_1^+,\mathbb{S}_0)}$ & $a \Delta E_{\mathsf{n}}^{(2,0,A_1^+,\mathbb{S}_0)}$ & $\Delta E_{\mathsf{n}}^{(2,0,A_1^+,\mathbb{S}_0)}\text{ [MeV]}$ & $k_{\mathsf{n}}^{(2,0,A_1^+,\mathbb{S}_0)^2} / m_\pi^2$ \\\hline
0 & $2.5540(29)$ & $0.1446(13)$ & $196.4(2.8)$ & $0.5081(46)$ \\\hline
1 & $2.5804(27)$ & $0.1714(14)$ & $232.7(3.2)$ & $0.6054(50)$ \\\hline
\end{tabular}
\caption{Two-nucleon $I=0$ energy levels in total-angular-momentum cubic irreps $\Gamma_J \in \{T_2^+,E^+,A_2^+,A_1^+\}$ in lattice and physical units obtained from a weighted average of single- and multi-exponential fits to GEVP correlation functions as described in the main text. Finite-volume energy differences obtained using correlated differences of GEVP fit results and the associated center-of-mass momenta are also provided. \label{tab:B2I0}}
\end{table}

\clearpage

\section{Quantization conditions}\label{app:QC}

The truncated quantization conditions used to obtain the $k\cot\delta$ values plotted in Figs.~\ref{fig:I1_phaseshift}-\ref{fig:I0_phaseshift} can be found in Refs.~\cite{Luu:2011ep,Briceno:2013lba}, and are listed below for completeness. For the $I=1$ system,
\begin{equation}
\begin{split}
A_1^+:& \quad k\cot\delta_{^1S_0} = \frac{2}{\sqrt{\pi} L} \ \mathcal{Z}_{0,0}(1;\tilde k^2)\, , \\
E^+:& \quad k^5\cot\delta_{^1D_2} = \left(\frac{2\pi}{L}\right)^5 \frac{1}{\pi^{3/2}}
\left( \tilde k^4 \mathcal{Z}_{0,0}(1;\tilde k^2) + \frac{6}{7} \mathcal{Z}_{4,0}(1;\tilde k^2) \right) , \\
  T_2^+:& \quad k^5\cot\delta_{^1D_2} = \left(\frac{2\pi}{L}\right)^5 \frac{1}{\pi^{3/2}} \left( \tilde{k} ^4  \mathcal{Z}_{0,0}\left(1;\tilde k^2 \right) - \frac{4}{7} \mathcal{Z}_{4,0}\left(1;\tilde{k}^2\right) \right) ,
\end{split}
\end{equation}
while for the $I=0$ system,
\begin{equation}
\begin{split}
T_1^+:& \quad k\cot\delta_{^3S_1}=\frac{2}{\sqrt{\pi} L}\ \mathcal{Z}_{0,0}(1;\tilde k^2)\, , \\
T_2^+:& \quad k^5\cot\delta_{^3D_2} = \left(\frac{2\pi}{L}\right)^5\frac{1}{\pi^{3/2}} \left(\tilde k^4 \mathcal{Z}_{0,0}(1;\tilde k^2) +\frac{8}{21} \mathcal{Z}_{4,0}(1;\tilde k^2)\right) ,\\
E^+:& \quad k^5\cot\delta_{^3D_2} = \left(\frac{2\pi}{L}\right)^5\frac{1}{\pi^{3/2}} \left(\tilde k^4 \mathcal{Z}_{0,0}(1;\tilde k^2) -\frac{4}{7} \mathcal{Z}_{4,0}(1;\tilde k^2)\right) ,
\end{split}
\end{equation}
where $\tilde k \equiv kL/(2\pi)$ and $\mathcal{Z}_{l,m}(1,\tilde k^2)=\sum_{\vec{n}\in\mathbb{Z}^3}|\vec{n}|^l Y^m_l(\hat{n})/(|\vec{n}|^2-\tilde k^2)$ is the $\mathcal{Z}$-function~\cite{Luscher:1986pf,Luscher:1990ux}. Methods for efficient numerical evaluation of the $\mathcal{Z}$-function are discussed in Refs.~\cite{Luscher:1990ux,NPLQCD:2011htk,Leskovec:2012gb}.

\clearpage 

\section{Glossary of notation}\label{app:glossary}

\begin{table}[!h]
\begin{ruledtabular}
\begin{tabular}{ccc} 
            $x=(\vec{x},t)$ & Euclidean spacetime coordinates & Eq.~\eqref{eq:Ninterp} \\
            $\sigma,\sigma' $ & Nucleon spin indices (rows of $G_1^+$) & Eq.~\eqref{eq:Ninterp} \\
            $N_\sigma(x)$ & Nucleon interpolating operator & Eq.~\eqref{eq:Ninterp} \\
            $i,j,k,\ldots $ & Spin-color indices & Eq.~\eqref{eq:Ndef} \\
            $q^i(x)$ & Quark field & Eq.~\eqref{eq:Ndef} \\
            $\alpha,\alpha'$ & Spin-color weight indices &  Eq.~\eqref{eq:Ndef} \\
            $w_\alpha^{[N]\sigma}$ & Nucleon spin-color weights & Eq.~\eqref{eq:Ndef} \\
            $\mathcal{N}_w^{[N]\sigma}$ & Nucleon weight multiplicity & Eq.~\eqref{eq:Ndef} \\
            $g $ & Smearing width index & Eq.~\eqref{eq:qsmear_def} \\
            $q^i_g(x)$ & Smeared quark fields & Eq.~\eqref{eq:qsmear_def} \\
            $N_{\sigma g}(x)$ & Smeared nucleon field & Eq.~\eqref{eq:Nsmear_def} \\
            $L$ & Lattice extent in spatial directions & Eq.~\eqref{eq:sparse_lattice} \\
            $T$ & Lattice extent in temporal direction & Eq.~\eqref{eq:propdef} \\
            $\vec{x}, \vec{x}_1, \vec{x}_2$ & Spatial positions of sink operators & Eq.~\eqref{eq:sparse_lattice} \\
            $\vec{y}, \vec{y}_1, \vec{y}_2$ & Spatial positions of source operators & Eq.~\eqref{eq:propdef} \\
            $t$ & Euclidean time separation between source (at $t=0$) and sink & Eq.~\eqref{eq:sparse_lattice} \\
            $\mathcal{S}$ & Sparsening factor & Eq.~\eqref{eq:sparse_lattice} \\
            $\Lambda_\mathcal{S}$ & Set of sparse lattice sites & Eq.~\eqref{eq:sparse_lattice} \\
            $V_\mathcal{S}$ & Sparse lattice volume $(L/\mathcal{S})^3$ & Eq.~\eqref{eq:sparse_lattice} \\
            $\mathfrak{c}$ & Center-of-mass momentum index & Eq.~\eqref{eq:psiNdef} \\
            $\psi_{\mathfrak{c}}^{[h]}(\vec{x})$ & Spatial wavefunction for hadron $h \in \{N, H\}$ & Eq.~\eqref{eq:psiNdef} \\
            $N_{\sigma\mathfrak{c}g}(t)$ & Nucleon interpolating operator (momentum-projected) & Eq.~\eqref{eq:Nprojdef} \\
            $C_{\sigma\mathfrak{c}g\sigma^\prime\mathfrak{c}^\prime g^\prime}^{[N,N]}(t)$ & Nucleon correlation function  & Eq.~\eqref{eq:CNsparse} \\
            $M_N$ & Nucleon mass  & Eq.~\eqref{eq:nucSpec} \\
            $H_{\rho\mathfrak{c}g}(t)$ & Hexaquark interpolating operator & Eq.~\eqref{eq:H0def} \\
            $w_\alpha^{[h]\rho}$ & Two-nucleon spin-color weights for $h \in \{H,D,Q\}$ & Eq.~\eqref{eq:Hweightdef} \\
            $\mathcal{N}_w^{[h]\rho}$ & Two-nucleon weight multiplicity for $h \in \{H,D,Q\}$ & Eq.~\eqref{eq:Hweightdef} \\
            $D_{\rho\mathfrak{m}g}(t)$ & Dibaryon interpolating operator & Eq.~\eqref{eq:Dinterp} \\
            $v_{\sigma\sigma'}^\rho$ & Weights for spin product $G_1^+ \otimes G_1^+ = A_1^+ \oplus T_1^+$ & Eq.~\eqref{eq:veights} \\
            $\psi_{\mathfrak{c}}^{[h]}(\vec{x}_1, \vec{x}_2)$ & Dibaryon spatial wavefunction & Eq.~\eqref{eq:psiDdef} \\
            $\mathfrak{m}$ & Two-nucleon relative momentum index & Eq.~\eqref{eq:psiDdef} \\
            $\vec{P}_\mathfrak{m}$ & Two-nucleon relative momentum vector & Eq.~\eqref{eq:psiDdef} \\
            $\vec{n}_\mathfrak{m}$ & Relative-momentum integer $\vec{n}_\mathfrak{m} = \left( \frac{L}{2\pi} \right) \vec{P}_\mathfrak{m}$ & Eq.~\eqref{eq:psiDdef} \\
            $\mathfrak{s}$ & Relative-momentum shell $\mathfrak{s} \equiv \mathfrak{s}({\mathfrak{m}}) = |\vec{n}_\mathfrak{m}|^2$ & Eq.~\eqref{eq:psiDdef} \\
  \end{tabular}
  \caption{  Glossary of notation including symbols used throughout this work (left), a description of their meaning (center), and the first equation where the notation is first introduced or the closest equation to where notation is introduced inline (right). 
       \label{tab:glossary}}
   \end{ruledtabular}
\end{table}

\begin{table}[!h]
\begin{ruledtabular}
\begin{tabular}{ccc} 
            $\mathfrak{q}$  & Quasi-local spatial wavefunction index & Eq.~\eqref{eq:Einterp} \\
            $\kappa_\mathfrak{q}$  & Quasi-local wavefunction exponential localization scale  & Eq.~\eqref{eq:Einterp} \\
            $\kappaOp_{\rho\mathfrak{q}g}(t)$ & Quasi-local interpolating operator & Eq.~\eqref{eq:Einterp} \\
            $\psi_{\mathfrak{q}}^{[\kappaOp]}(\vec{x}_1,\vec{x}_2,\vec{R})$  & Quasi-local spatial wavefunction & Eq.~\eqref{eq:Edef} \\
            $C_{\sigma\mathfrak{t}g\sigma^\prime\mathfrak{t}^\prime g^\prime}^{[\mathcal{T},\mathcal{T}']}(t)$ & Correlation function with $\mathcal{T},\mathcal{T}' \in \{H, D, Q\}$ and $\mathfrak{t},\mathfrak{t}' \in \{ \mathfrak{c}, \mathfrak{m}, \mathfrak{q} \}$  & Eq.~\eqref{eq:C2Ndef} \\
            $S_{gg'}^{ij}(\vec{x},t;\vec{y},0)$ & Quark propagator &  Eq.~\eqref{eq:propdef} \\
            $\mathcal{P}$ & Element of permutation group & Eq.~\eqref{eq:N-Ncontract} \\
            $\mathcal{B}^{ (\pm)ijk}_{g\sigma \mathfrak{m}'g'}(\vec{x},t)$ & Local baryon block & Eq.~\eqref{eq:blockdef} \\
            $\mathcal{E}^{(q\pm)ijk}_{g\sigma \mathfrak{m}'g'}(\vec{x}_1,\vec{x}_2,t)$ & Bilocal baryon block & Eq.~\eqref{eq:exblock1}   \\
            $J_z$ &  Related to eigenvalue $e^{i J_z \pi/2}$ of rotation by $\pi/2$ about $\hat{z}$-axis & Eq.~\eqref{eq:Szrho} \\
            $\ell_z$ & Analogous phase for rotations applied only to spatial wavefunctions & Eq.~\eqref{eq:Szrho} \\
            $S_z(\rho)$  & Analogous phase for rotations applied to two-nucleon spin with row $\rho$ & Eq.~\eqref{eq:Szrho} \\
            $\Gamma_J$ & Cubic irrep associated with total angular momentum $J$ &  Eq.~\eqref{eq:JLS} \\
            $\Gamma_\ell$ & Cubic irrep associated with orbital angular momentum $\ell$ &  Eq.~\eqref{eq:JLS} \\
            $\Gamma_S$ & Cubic irrep associated with spin $S$ &  Eq.~\eqref{eq:JLS} \\
            $H^{(2,I,\Gamma_J,J_z)}_{g}(t)$ & Hexaquark operator with definite quantum numbers & Eq.~\eqref{eq:cubicwvfdef} \\
            $\kappaOp^{(2,I,\Gamma_J,J_z)}_{\mathfrak{q}g}(t)$ & Quasi-local operator with definite quantum numbers & Eq.~\eqref{eq:cubicwvfdef} \\
            $\psi^{[D](\Gamma_\ell,\ell_z)}_{\mathfrak{s} \mathfrak{k} }(\vec{x}_1,\vec{x}_2)$  & Cubic-irrep projected dibaryon wavefunctions & Eq.~\eqref{eq:cubicwvfdef} \\
            $G^{(\Gamma_\ell,\ell_z)}_{\mathfrak{s} \mathfrak{k}\mathfrak{m}}$ & Coefficient projecting dibaryon wavefunctions to cubic irreps & Eq.~\eqref{eq:cubicwvfdef} \\
            $D^{(2,I,\Gamma_J,J_z)}_{\mathfrak{s}\mathfrak{k}g}(t)$ & Dibaryon operator with definite quantum numbers & Eq.~\eqref{eq:D1def} \\
            $M^{(\Gamma_J,\Gamma_\ell)}_{\mathfrak{s}\mathfrak{k}\mathfrak{k}^\prime}$ & Spin-orbit multiplicity-label tensor & Eq.~\eqref{eq:D0def} \\
            $\mathcal{C}^{(\Gamma_J,J_z,\Gamma_\ell,\ell_z)}_\rho$ & Spin-orbit cubic group Clebsch-Gordon coefficients  & Eq.~\eqref{eq:D0def} \\
            $d_\Gamma$ & Dimension of representation $\Gamma$ & Eq.~\eqref{eq:CDDJz} \\
            $(B,I,\Gamma_J)$ & Quantum numbers of zero-momentum states & Eq.~\eqref{eq:spectral} \\
            $\mathsf{n}$ & Label for energy eigenstates $\mathsf{n}=0,1,\ldots$ with fixed quantum numbers & Eq.~\eqref{eq:spectral} \\
            $E_{\mathsf{n}}^{(B,I,\Gamma_J,\mathbb{S})}$ & Energy of state $\mathsf{n}$ determined using interpolating-operator set $\mathbb{S}$ & Eq.~\eqref{eq:spectral} \\
            $Z_{\mathsf{n}\chi}^{(B,I,\Gamma_J,\mathbb{S})}$ & Overlap of interpolating operator $\chi$ with state $\mathsf{n}$ & Eq.~\eqref{eq:spectral} \\
            $C_{\chi \chi^\prime}^{(B,I,\Gamma_J)}(t)$ & Correlation function with sink/source operators $\chi/\chi'$ & Eq.~\eqref{eq:spectral} \\
            $\widehat{C}_{\mathsf{n}}^{(B,I,\Gamma_J,\mathbb{S})}(t)$ & GEVP correlation function ordered so that energy increases with $\mathsf{n}$ & Eq.~\eqref{eq:GEVPcorrelators} \\
            $\delta^{(B,I,\Gamma_J,\mathbb{S})}$ & Energy gap between ground and first excited states & Eq.~\eqref{eq:GEVPspectral} \\

            $E_{\chi\chi'}^{(B,I,\Gamma_J)}(t)$ & Effective energy of $C_{\chi \chi^\prime}^{(B,I,\Gamma_J)}(t)$ & Eq.~\eqref{eq:Echichip} \\
            $E_{\mathsf{n}}^{(B,I,\Gamma_J,\mathbb{S})}(t)$ & Effective energy of $\widehat{C}_{\mathsf{n}}^{(B,I,\Gamma_J)}(t)$ & Eq.~\eqref{eq:GEVPEM} \\
            $\Delta E_{\mathsf{n}}^{(2,I,\Gamma_J,\mathbb{S})}(t)$ & Effective FV energy shift from GEVP correlation functions  & Eq.~\eqref{eq:GEVPDeltaEM} \\
            $\Delta E_{\mathsf{n}}^{(2,I,\Gamma_J,\mathbb{S})}$ & FV energy shift from fits to GEVP correlation functions & Eq.~\eqref{eq:GEVPDeltaM} \\
            $\mathcal{Z}_{\mathsf{n}\chi}^{(B,I,\Gamma_J,\mathbb{S})}$ & Relative overlap of interpolating operator $\chi$ with state $\mathsf{n}$ & Eq.~\eqref{eq:GEVPZrelative} \\
            $\widetilde{\mathcal{Z}}_{\mathsf{n}\chi}^{(B,I,\Gamma_J,\mathbb{S})}$ & Relative contribution from state $\mathsf{n}$ to $C_{\chi \chi^\prime}^{(B,I,\Gamma_J)}(t)$ & Eq.~\eqref{eq:GEVPZtilde} \\
  \end{tabular}
  \caption{  The continuation of Table~\ref{tab:glossary}. 
       \label{tab:glossary2}}
   \end{ruledtabular}
\end{table}

\clearpage 

\bibliography{variational}

\end{document}